\documentclass{aastex63}
\usepackage{xspace}
\usepackage{amsmath}

\newcommand{\OIII}{[\ion{O}{3}]\xspace}

\newcommand{\SII}{[\ion{S}{2}]\xspace}
\newcommand{\NII}{[\ion{N}{2}]\xspace}

\received{2021 March 22}
\revised{2021 April 18}
\accepted{2021 May 4}
\submitjournal{ApJ}

\shorttitle{PNLF-MUSE}
\shortauthors{Roth et al.}

\graphicspath{{./}{figures/}}

\begin{document}

\title{Towards Precision Cosmology With Improved PNLF Distances Using VLT-MUSE\\
I. Methodology and Tests
}

\correspondingauthor{Martin M. Roth}
\email{mmroth@aip.de}

\author[0000-0003-2451-739X] {Martin M. Roth}
\affiliation{Leibniz Institute for Astrophysics Potsdam (AIP), An der Sternwarte 16, 14482 Potsdam, Germany}

\author[0000-0001-7970-0277]{George H. Jacoby}   
\affiliation{NSF’s NOIRLab, 950 N. Cherry Ave., Tucson, AZ 85719, USA}

\author[0000-0002-1328-0211]{Robin Ciardullo}   
\affiliation{Department of Astronomy \& Astrophysics, The Pennsylvania State University, University Park, PA 16802, USA}
\affiliation{Institute for Gravitation and the Cosmos, The Pennsylvania State University, University Park, PA 16802}

\author[0000-0002-8994-6489]{Brian D. Davis}   
\affiliation{Department of Astronomy \& Astrophysics, The Pennsylvania State University, University Park, PA 16802, USA}

\author[0000-0002-0304-5701]{Owen Chase}   
\affiliation{Department of Astronomy \& Astrophysics, The Pennsylvania State University, University Park, PA 16802, USA}

\author[0000-0003-4766-902X] {Peter M. Weilbacher}
\affiliation{Leibniz Institute for Astrophysics Potsdam (AIP), An der Sternwarte 16, 14482 Potsdam, Germany}

\begin{abstract}
The \OIII $\lambda 5007$ Planetary Nebula Luminosity Function (PNLF) is an established distance indicator that has been used for more than 30 years to measure the distances of galaxies out to $\sim15$ Mpc. With the advent of the Multi-Unit Spectroscopic Explorer on the Very Large Telescope (MUSE) as an efficient wide-field integral field spectrograph, the PNLF method is due for a renaissance, as the spatial and spectral information contained in the instrument's datacubes provides many advantages over classical narrow-band imaging.  Here we use archival MUSE data to explore the potential of a novel differential emission-line filter (DELF) technique to produce  spectrophotometry that is more accurate and more sensitive than other methods.  We show that DELF analyses are superior to classical techniques in high surface brightness regions of galaxies and we validate the method both through simulations and via the analysis of data from two early-type galaxies (NGC\,1380 and NGC\,474) and one late-type spiral (NGC\,628).   We demonstrate that with adaptive optics support or under excellent seeing conditions, the technique is capable of producing precision ($\lesssim 0.05$~mag) \OIII photometry out to distances of 40~Mpc while providing discrimination between planetary nebulae and other emission-line objects such as \ion{H}{2} regions, supernova remnants, and background galaxies.  These capabilities enable us to use MUSE to measure precise PNLF distances beyond the reach of Cepheids and the tip of the red giant branch method, and become an additional tool for constraining the local value of the Hubble constant.
 \end{abstract}


\section{Introduction} \label{sec:intro}

\citet{Ciardullo+89} demonstrated that the \OIII $\lambda 5007$ luminosity function (LF) of planetary nebulae (PNe) in nearby galaxies has a bright upper limit. That limit, which is $\sim 640 L_{\odot}$, is nearly universal across all galaxies and can therefore be exploited as a distance indicator \citep[e.g.,][]{jacoby+90, Ciardullo+02}. In fact, a careful comparison of the distances to $\sim 20$ galaxies within $\sim 10$~Mpc shows that the accuracy and precision of the planetary nebula luminosity function (PNLF) method is comparable to that obtainable from the tip of the red giant branch (TRGB) and Cepheids \citep{Ciardullo12, Ciardullo13}.  However, as initially implemented with narrow-band filter imaging, the PNLF technique begins to have difficulties beyond $\sim 10$~Mpc and reaches its effective limit by $\sim 20$~Mpc.  Consequently, for the past couple of decades, the application of the PNLF to cosmological questions has been limited.

With the increasing tension between the Hubble Constant ($H_0$) derived from the distance ladder \citep[e.g.,][]{Riess+19, Breuval+20, Freedman+20} and $H_0$ derived from early Universe measurements \citep{Hinshaw+13,Planck+18}, a revitalization of the PNLF is worth considering as the method could be used as an alternative to Cepheid and the TRGB measurements.  However, to be competitive in this era of precision cosmology, the method's accuracy beyond 10~Mpc must be improved, and its effective limit pushed beyond 20~Mpc.

Ideally, one would like to obtain PNLF distances to galaxies in a clean Hubble flow.  This presents a problem for the method, as the technique is most easily applied to early-type galaxies which are preferentially found in clusters.  Unfortunately, the local potential of a typical cluster introduces a non-cosmological component to the radial velocity that is roughly $\sim$1000 km~s$^{-1}$ \citep{Ruel+14}.  Even if a dozen cluster galaxies are observed, this peculiar motion would still introduce a major uncertainty into any $H_0$ calculation.  The alternative is to target isolated field galaxies, where the bulk velocity uncertainty is much smaller, of the order of $\sim 300$~km~s$^{-1}$ \citep{Scrimgeour+16}. Most large field galaxies are spirals, and, though the PNLF method can be applied to these systems \citep{Ciardullo+02}, care must be taken to remove \ion{H}{2} regions and supernova remnants from the PN sample.  Moreover, a 1\% determination of the Hubble Constant requires measuring field galaxies at a distance of $\sim 400$~Mpc; this is far beyond the reach of the PNLF\null.    

Nevertheless, there are a number of relatively nearby early-type galaxies that are not within large galaxy clusters, and, with care, PNLF distances can be obtained to spirals and other star-forming galaxies.  It is therefore reasonable to try to extend the technique with modern telescopes and instrumentation.  For example, a galaxy at $\sim 50$\,Mpc will have a $\sim$2.5\% error in $H_0$ due solely to peculiar motion. If a typical PNLF distance carries a statistical uncertainty of 5\%, then the total error associated with 10 $D \sim50$\,Mpc galaxies would be roughly 2\%. Such a precision would be interesting with regard to the problem of the Hubble Constant tension.  Moreover, if the PNLF can be shown to be reliable at these larger distances, then it can be used to calibrate the luminosities of Type Ia supernovae (SN\,Ia) in early-type galaxies and in systems beyond the reach of Cepheids and the TRGB.

The Multi-Unit Spectroscopic Explorer (MUSE) optical integral field spectrograph (IFS) \citep{Bacon+10} on the 8.2~m Very Large Telescope (VLT) enables this type of observation.  There are several ways in which MUSE improves
upon previous PNLF studies:

\begin{enumerate}
    \item {The VLT offers a larger aperture, as it has four times the collecting area of the 4-m class telescopes used in most earlier PNLF works.}
    \item {The Paranal Observatory site frequently delivers much better seeing than $1\farcs 2$, which was the typical image quality of the previous work. The ground-layer adaptive optics system (GLAO) available for MUSE further enhances the capability to deliver data with high image quality.}
    \item {MUSE delivers an effective bandpass that is more than five times narrower than what is typically provided by interference filters. Since PNLF measurements are background dominated, the reduced noise substantially improves the detectability and photometric accuracy of planetary nebulae.}
    \item {Since MUSE covers the spectral range between 4800\,\AA\ and 9300\,\AA\ and has a resolution of $R \sim 2000$ at 5000\,\AA, it produces a spectrum for every emission-line object in its field.  Contaminating objects such as \ion{H}{2} regions, supernova remnants (SNRs) and background galaxies (such as Ly$\alpha$ emitters) can immediately be identified, thus preventing them from skewing the PNLF statistics.}
    \item  {MUSE spectra can allow spatial blends to be identified, enabling the  emission from two merged sources, such as PN pairs, to be disentangled.}
    \item {Because MUSE does not require a narrow-band filter, all PNe have the same photometric throughput, independent of their velocity. In contrast, narrow-band filters, when placed in the fast beams of large telescopes, generate a system throughput that depends on the velocity of the emission-line object being observed. This introduces a photometric error that depends on a galaxy's rotation curve and velocity dispersion \citep{Jacoby+89}.}   
\end{enumerate}
 
PNLF distances rely on accurate \OIII $\lambda 5007$ photometry of planetary nebulae superimposed on the bright continuum surface brightness of their host galaxy. \citet{Roth+04} have demonstrated that an IFS is capable of delivering accurate spectrophotometry of point sources by observing PNe in the bulge of M31 with the PMAS at the Calar Alto 3.5\,m telescope \citep{Roth+05}, the MPFS at the 6\,m BTA in Selentchuk \citep{Sil'chenko+00}, and INTEGRAL at the WHT \citep{Arribas+98}. In the M31 pilot study, it was also serendipitously discovered that spectral information, specifically the H$\alpha$ and the \SII $\lambda\lambda 6717,6731$ emission lines, facilitates the identification and exclusion of interloping supernova remnants.  However, the sizes of the first generation integral field units (IFUs) were far too small to cover the field-of-view needed to obtain PNLF measurements (e.g., the PMAS field-of-view was only $8\times8$~arcsec$^2$). 

With its much larger field of view  of 1~arcmin$^2$, MUSE overcomes this limitation.  For example, \citet{Kreckel+17} (henceforth Kr2017) used
$\sim 45$~min MUSE exposures to identify 63~PNe in a small section (three $1\arcmin \times 1\arcmin$ pointings) of the large face-on spiral NGC\,628. These authors  reported a PNLF distance modulus of $(m-M)_0 = 29.91^{+0.08}_{-0.13}$ ($9.6^{+0.4}_{-0.6}$~Mpc), which is 0.26~mag larger than that found by \citet{Herrmann+08} using PNe identified with narrow-band filters.  The authors ascribed the offset to MUSE's ability to discriminate PNe from supernova remnants.  

More recently, \citet{Spriggs+20} (hereafter Sp2020) extended PNLF measurements with MUSE out to the Fornax cluster, and obtained distances to the early-type galaxies of NGC\,1380 and NGC\,1404 (FCC 167 and FCC 219 in their nomenclature). Using Moffat profile PSF-fitting photometry as introduced by \citet{Kamann+13}, these authors obtained PN magnitudes that are on average 0.4~mag fainter than corresponding measurements by \citet{Feldmeier+07} and \citet{McMillan+93}, and hence inferred larger PNLF distances than the previous studies. Given these developments, it is worthwhile to explore the potential of PN observations with MUSE across a larger sample of galaxies.
 
In this work, we demonstrate the effectiveness of MUSE for improving distances to previously studied galaxies.  We will show consistency with earlier work, we will derive distances to galaxies that were previously beyond the reach of the PNLF, and we demonstrate that it is possible to reliably measure distances to late type galaxies, thus extending the calibration of Type Ia supernovae beyond that performed by \citet{Feldmeier+07}. This study focuses on the methodology. In a forthcoming paper, we will address the large set of galaxies currently in the MUSE archives. For now, we concentrate on two galaxies for which recent PNLF results exist in the literature, NGC\,628, and NGC\,1380.  The analyses of these objects will allow us to benchmark the capabilities of MUSE PN observations against the results obtained by other distance scale techniques. In addition, we also examine the archival data for NGC\,474, to confirm that MUSE can obtain PNLF distances to galaxies that are beyond the reach of Cepheid and TRGB measurements.  

\begin{deluxetable*}{lccccllc}[h]
\tablecaption{MUSE Exposures of the Galaxies\label{tab:archive}}
\tablewidth{0pt}
\tablehead{
\colhead{Galaxy} &
\colhead{Date} &
\colhead{Time} &
\colhead{seeing} &
\colhead{T$_{exp}$} &
\colhead{Program ID} & 
\colhead{Object ID} &
\colhead{Notes}  \\
\colhead{Name} &
\colhead{} &
\colhead{} &
\colhead{(arcsec)} &
\colhead{(s)} &
\colhead{} &
\colhead{} &
\colhead{}
}
\startdata
NGC~628 &   2014-10-31   &    03:39:58	   &   0.77   &    2535	  &       094.C-0623   &   	NGC628-1     & (1) \\
        &   2014-10-31   &    04:43:34	   &   0.83   &    2535   &       094.C-0623   &   	NGC628-2     & (1)      \\
        &   2015-09-15   &    05:00:36	   &   0.76   &    2970   &   	   095.C-0473   &   NGC0628-P1    & (1)   \\
        &   2017-07-22   &    07:36:36	   &   1.14   &    2970   &   	   098.C-0484   &   NGC0628-P2    & (1)   \\
        &   2017-11-13   &    03:43:55	   &   0.95   &    2970   &   	   098.C-0484   &   NGC0628-P3    & (1)   \\
        &   2017-09-16   &    04:17:21	   &   1.08   &    2970   &   	   098.C-0484   &   NGC0628-P4    & (1)   \\
        &   2016-12-30   &    01:01:36	   &   1.05   &    2970   &   	   098.C-0484   &   NGC0628-P5    & (1)   \\
        &   2016-10-01   &    04:56:15	   &   0.69   &    2970   &   	   098.C-0484   &   NGC0628-P6    & (1)   \\
        &   2016-10-01   &    06:08:16	   &   0.70   &    2970   &       098.C-0484   &   	NGC0628-P7    & (1)   \\
        &   2017-07-21   &    08:25:54	   &   0.82   &    2970   &   	   098.C-0484   &   NGC0628-P8    & (1)   \\
        &   2017-11-13   &    01:22:45	   &   0.96   &    2970   &   	   098.C-0484   &   NGC0628-P9    & (1)   \\
        &   2017-11-13   &    02:33:10    &    0.75   &    2970   &   	   098.C-0484   &   NGC0628-P12   & (1)   \\
\hline
NGC~1380 &  2016-12-31   &    03:19:12     &   0.74   &    	720   &    	296.B-5054   &   FCC167\_CENTER & (2)      \\
        &   2016-12-31   &    03:37:29	   &   0.88   &   	720   &   	296.B-5054   &   FCC167\_CENTER & (2)      \\
        &   2016-12-31   &    03:51:26	   &   0.76   &   	720   &   	296.B-5054   &   FCC167\_CENTER & (2)      \\
        &   2016-12-31   &    04:09:45     &   0.55   &    	720   &   	296.B-5054   &   FCC167\_CENTER & (2)      \\
        &   2016-12-31   &    04:23:44     &   0.45   &   	720   &   	296.B-5054   &   FCC167\_CENTER & (2)      \\
        &   2017-01-20   &    02:18:59     &   0.93   &   	600   &   	296.B-5054   &   FCC167\_MIDDLE & (2)      \\
        &   2017-01-20   &    02:52:43	   &   0.91   &   	600   &   	296.B-5054   &   FCC167\_MIDDLE & (2)      \\
        &   2017-01-20   &    03:09:31	   &   0.83   &   	600   &   	296.B-5054   &   FCC167\_MIDDLE & (2)     \\
        &   2017-11-10   &    04:03:05	   &   1.46   &   	600   &   	296.B-5054   &   FCC167\_MIDDLE & (2)      \\
        &   2017-11-10   &    04:15:03     &   1.43   &   	600   &   	296.B-5054   &   FCC167\_MIDDLE & (2)      \\
        &   2016-12-30   &    02:29:24	   &   0.85   &   	600   &   	296.B-5054   &   FCC167\_HALO & (2)   \\
        &   2016-12-30   &    02:45:41	   &   1.05   &   	600   &   	296.B-5054   &   FCC167\_HALO & (2)    \\
        &   2016-12-30   &    02:57:37     &   0.96   &   	600   &   	296.B-5054   &   FCC167\_HALO & (2)   \\
        &   2017-01-20   &    01:11:54	   &   1.02   &   	600   &   	296.B-5054   &   FCC167\_HALO & (2)    \\
        &   2017-01-20   &    01:28:14	   &   0.81   &   	600   &   	296.B-5054   &   FCC167\_HALO & (2)    \\
        &   2017-01-20   &    01:40:14     &   0.73   &   	600   &   	296.B-5054   &   FCC167\_HALO & (2)    \\
\hline
NGC~474	&   2019-01-02   &     10:34:21   &   0.65   &  37312  &     099.B-0328        &    WFM-NGC474-S    & (1)  \\
\enddata
\tablecomments{(1) fully reduced data cube retrieved, (2) raw data retrieved and re-reduced.}
\end{deluxetable*}

\section{Observations} 
\label{sec:observations}
For this initial demonstration, we used the publicly available, fully reduced MUSE data cubes \citep[see][]{Romaniello+18} in the ESO Archive to derive PNLFs across a range of galaxy types and distances. We did not sift through the archive completely, but rather, we selected three representative systems that are amenable for analysis and validation of our methodology. A follow-up paper will address more galaxies. Our archive search was facilitated by the graphical user interface with ancillary data that is accessible through the ESO Portal for registered users. Because the MUSE data cubes were obtained from a variety of programs that were executed at the VLT between 2016 and 2019, the data are inhomogeneous and sample a wide range of observing conditions, with seeing measurements between $0\farcs 65$ and $1\farcs 1$ and exposure times between 0.25 and 10 hours. This heterogeneity is particularly useful for exploring the capabilities and limitations of MUSE for PNLF measurements. In one case (NGC\,1380), we discovered that the semi-automatic pipeline used for creating the reduced archival data products had not worked as expected, due to the lack of bright field stars available for positional reference.  This caused the individual exposures to be combined with incorrect offsets, and produced a final data cube whose image quality was a factor of two worse than those of the individual frames.  For this particular data set, we retrieved the raw FITS files that are also available from the ESO Archive and re-reduced a subset of the data to restore the expected quality. Table~\ref{tab:archive} summarizes the archival data sets used for this paper.

\section{Data Reduction and Analysis} \label{sec:datareduction}
Most classical PNLF measurements were performed with direct imaging cameras employing narrow-band filters, and most were mounted at the prime focus of 4\,m class telescopes. As we wish to investigate the capabilities and limitations of IFS for PNLF distance determinations, it is useful to remember that the images that will be shown in this paper have been extracted from MUSE data cubes.  These cubes were created via a complex process which involved the data reduction and analysis of roughly 90$\,$000 raw spectra that were projected onto 24 CCD cameras and mounted to 24 spectrograph modules. In what follows, we describe this process in some detail.

\pagebreak

\subsection{Data Reduction Pipeline}
\label{subsec:pipeline}
A description of the most recent version of the MUSE data reduction software (DRS) for science users is given in \citet{Weilbacher+20}. Previous versions and software development aspects are discussed in \citet{Weilbacher+14}. The DRS pipeline chiefly consists of {\it basic processing\/} and {\it post-processing.} {\it Basic processing} includes the tasks of marking bad pixels, bias subtraction, master/sky flatfield correction, and wavelength calibration. Unlike other pipelines for fiber-based IFUs, there is no step for the tracing and extraction of spectra. Instead of performing multiple interpolations, the MUSE DRS creates a {\it pixel table\/} that maintains the integrity of CCD pixels by assigning each one a unique wavelength and sky coordinate. The pixel table is the output of {\it basic processing}. The {\it post-processing\/} step merges the data sets from the 24 spectrograph modules into one file, performs the sky subtraction, applies velocity corrections, performs the astrometric and flux calibration, and mosaics the different exposures (which may have different ditherings and rotations) into one dataset. A resampling algorithm then creates the final data product as an NAXIS=3 FITS format data cube. The output FITS file comes with two extensions: the first contains the actual data, and the second provides the variance. A summary of performance parameters is given in Table~\ref{tab:DRSperformance}. For a full discussion, see \citet{Weilbacher+20}.

\begin{deluxetable*}{ll}[h]
\tablecaption{MUSE Data Reduction Software Performance\label{tab:DRSperformance}}
\tablewidth{100pt}
\tablehead{\colhead{Parameter} & \colhead{Value} }
\startdata
Bias subtraction residuals                            &  $\leq$ 0.1~e$^-$h$^{-1}$,  Note (1)   \\
Pixel table wavelength calibration accuracy &  0.01 --- 0.024~\AA\  (0.4 -- 1.0 km\,s$^{-1}$)       \\
Data cube wavelength calibration accuracy &  0.06 --- 0.08~\AA\  (2.5 --- 4.0 km\,s$^{-1}$)  \\
Sky subtraction accuracy                              &  1\% at 500~nm      \\
Flux calibration accuracy                               &  2\% at 500~nm      \\
Astrometric accuracy  (relative)                    &  $0\farcs 05$ in RA, $0\farcs 03$ in DEC    \\
\enddata 
\tablecomments{(1) measured in units of dark current per pixel.}
\end{deluxetable*}

\subsection{Differential Emission Line Filtering on Data Cube Layers}
\label{subsec:onoff}

Although PNe are intrinsically bright, and a large fraction of their luminosity is radiated in the \OIII $\lambda 5007$ emission line, their signal is totally swamped in broadband images by the continuum surface brightness of their host galaxy.  To detect this line, the PNLF distance technique has relied on direct imaging through narrow band filters that suppress most of the continuum while transmitting the light within the passband of the filter.  \citet{Jacoby+89} have explained that by creating a difference ({\it diff}) image by subtracting a scaled continuum off-band ({\it off}) image from a corresponding \OIII on-band ({\it on}) image, the continuum surface brightness is conveniently removed, and the PNe become detectable as faint point sources on a flat noise floor. \citet{Roth+18} have demonstrated that MUSE data cubes allow a synthetic implementation of the on-band/off-band technique:  by co-adding selected data cube layers for a few wavelength bins around a given emission line (Doppler shifted to the systemic velocity of the galaxy) and comparing these data to an appropriately chosen off-band image, the effective filter bandwidth of the classical direct imaging technique can be reduced from 30 to 50\,\AA\ down to 4 or 5\,\AA\null. As a result, the photon shot noise contribution from the underlying galaxy is reduced by the square root of the ratio of those numbers, thereby increasing the signal-to-noise of a PN detection by a factor of $\sim$2.5.  In other words, exposures with conventional narrow-band filters would have to be 6 to 10 times longer to achieve the same signal-to-noise.  Obviously, this would not be practical for observing distant galaxies.  Moreover, another significant improvement can be obtained via a spaxel-to-spaxel approach to flux calibration. This novel procedure can only be achieved with an IFU, and will be described below.

For the purpose of precision PN photometry, we have refined the on-band/off-band technique by creating stacks of 15 single data cube layers, each having a width of 1.25\,\AA\null.  These stacks are grouped around the wavelength of the Doppler-shifted \OIII line, in order to account for a range of PN radial velocities of $\pm$500~km\,s$^{-1}$ centered on the systemic velocity of the galaxy. We also create a 125\,\AA\  intermediate-bandwidth continuum image by co-adding 100 data cube layers redward of the redshifted 5007\,\AA\ emission line.  From each of the 15 {\it on}~layers, we subtract the normalized continuum {\it off\/} image to form a total of 15 {\it diff\/} images. 

\begin{figure}[h!]
\centerline{ 
\includegraphics[width=0.5\hsize,bb=10 80 700 460,clip]{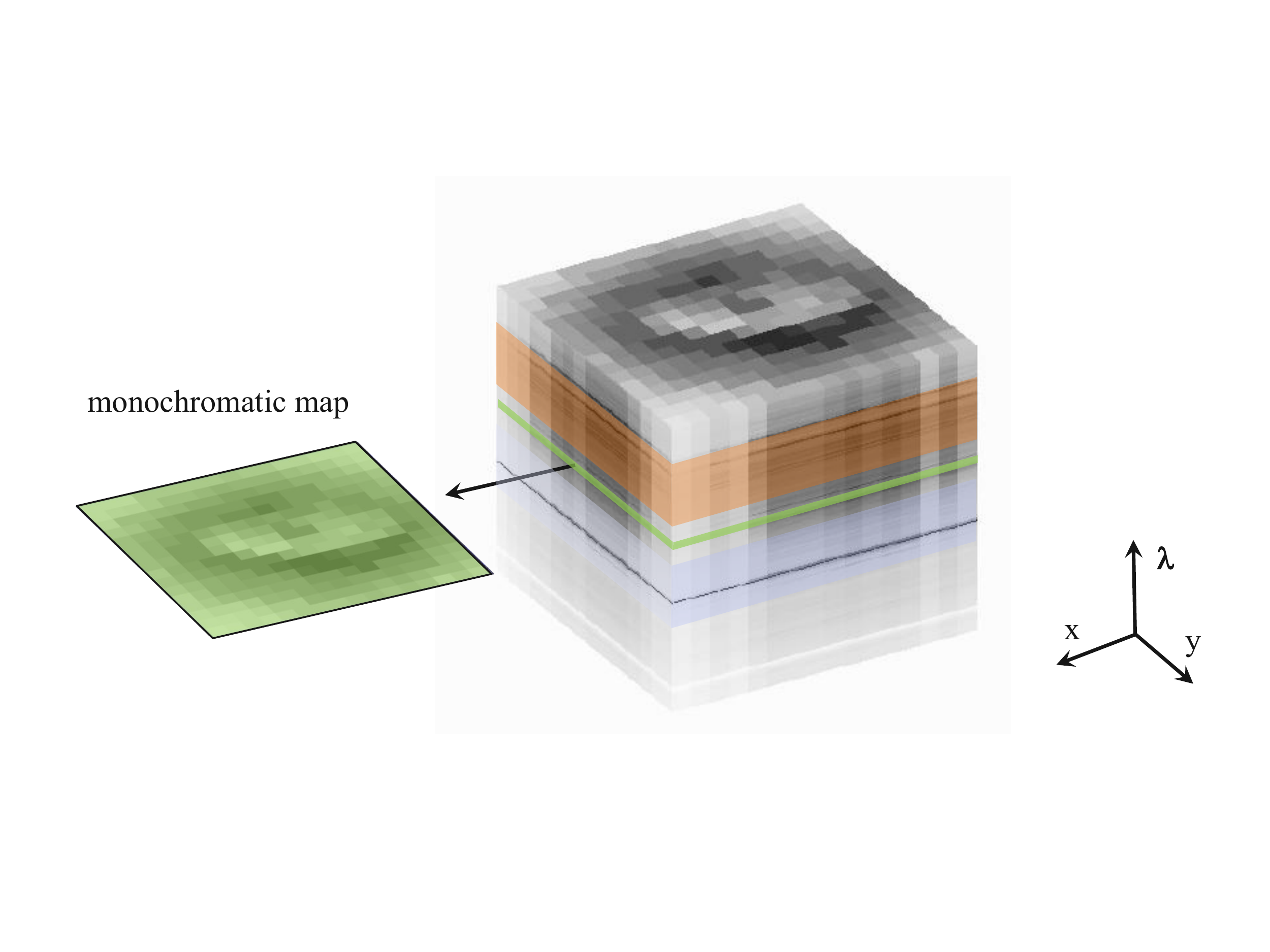}
\includegraphics[width=0.5\hsize,bb=0 80 750 500,clip]{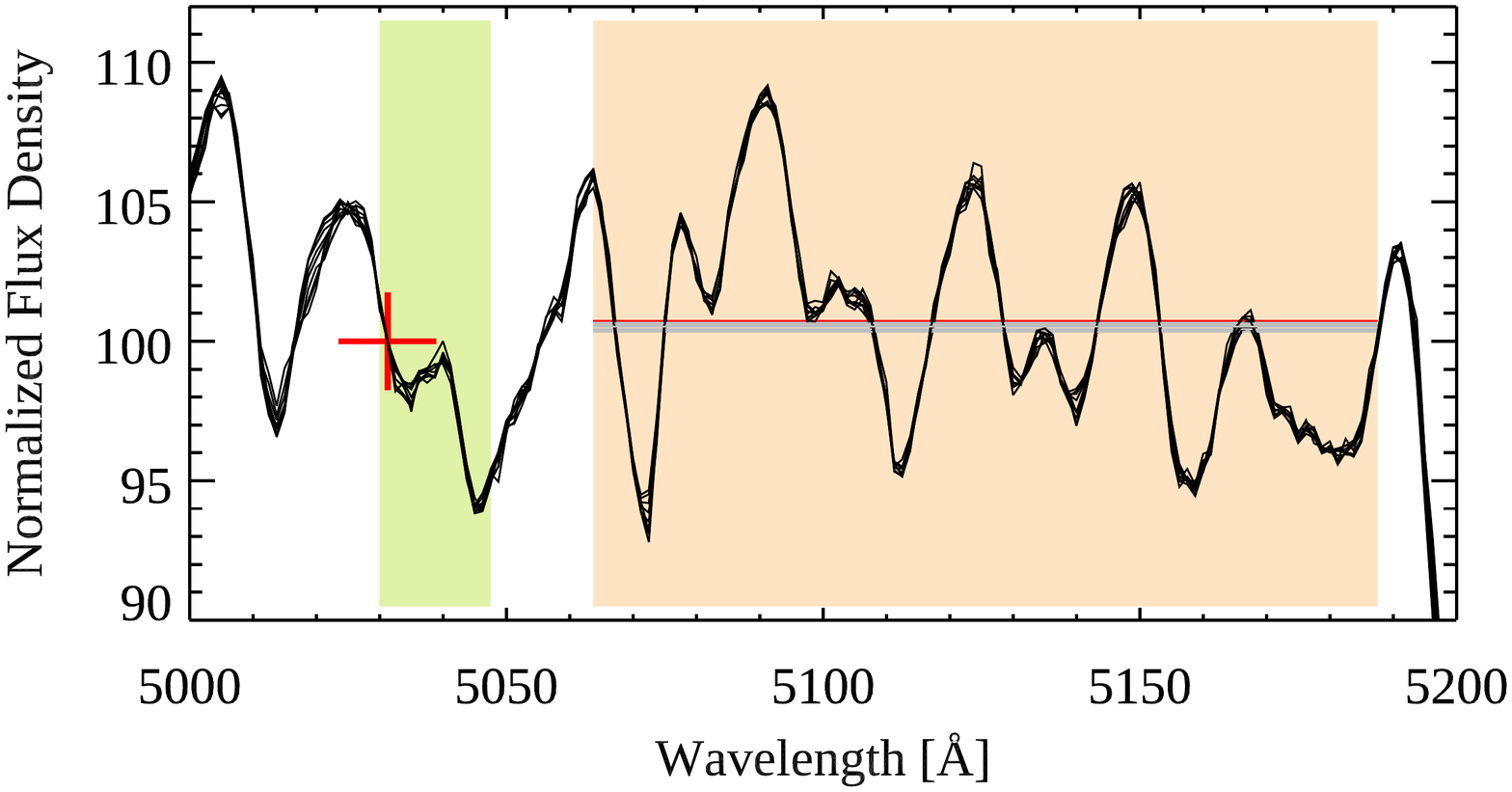}}
\caption{Left: a data cube with the on-band image (green) and the adjacent stack of off-band images (beige). Another filter variant with an additional offband stack is indicated in blue hues (see Section~\ref{subsec:spectro}). Right: 
The relative flux of the pseudo continuum for NGC\,1380, sampled at 10 different regions in the galaxy, with the wavelength intervals for on-band images (between 5029.7 and 5047.2\,\AA) shown in green, and the off-band image (5063.5 to 5187.2\,\AA) displayed in beige. The spectra show very little variation over these different regions.}
 \label{fig:NGC1380_on_off_bands}
\end{figure}

As an example, the right-hand panel of Figure~\ref{fig:NGC1380_on_off_bands} shows the normalized spectrum of NGC\,1380 measured over ten $51\times 51$ spaxel apertures that extend radially from the galaxy's nucleus, with offsets of 10 spaxels between each sample. The on-band layers, shaded in green, are tightly related to the mean of the continuum in the off-band region, which is highlighted in beige. The latter was chosen to be close to the \OIII doublet, but away from  the strong Mg\,{\it b} absorption feature between rest frame 5160 and 5192~\AA\null. For reference, the wavelength of the first on-band layer is shown with a red cross. In this plot, the spectra for the 10 different apertures, normalized to the flux density at the red cross, lie almost on top of each other. The mean continuum flux, averaged over the off-band and plotted as horizontal lines, has an aperture-to-aperture standard deviation of just 0.13\,\%. For the calibration of each wavelength bin in the {\it on\/} bandpass, the continuum background at the wavelengths of Doppler-shifted \OIII emission lines can therefore be tied to the off-band with extremely high accuracy. Over the seeing disk of a point-source PN, the relation is essentially constant and even robust against surface brightness fluctuations \citep{Tonry+88,Mitzkus+18}. Galaxy rotation and stellar population differences can lead to systematic shifts of the calibration constant, but as these effects generally occur on spatial scales much larger than relevant for point source photometry, they only introduce a small, locally constant residual in the background and cancel out. As will be shown below, the principle of self-referencing in each data cube spaxel is uniquely efficient for removing residual fixed pattern noise and therefore preferable over the technique of subtracting a model spectrum of the galaxy.

\begin{figure}[h!]
\centerline{ \includegraphics[width=15cm,bb=0 850 1000 1300,clip]{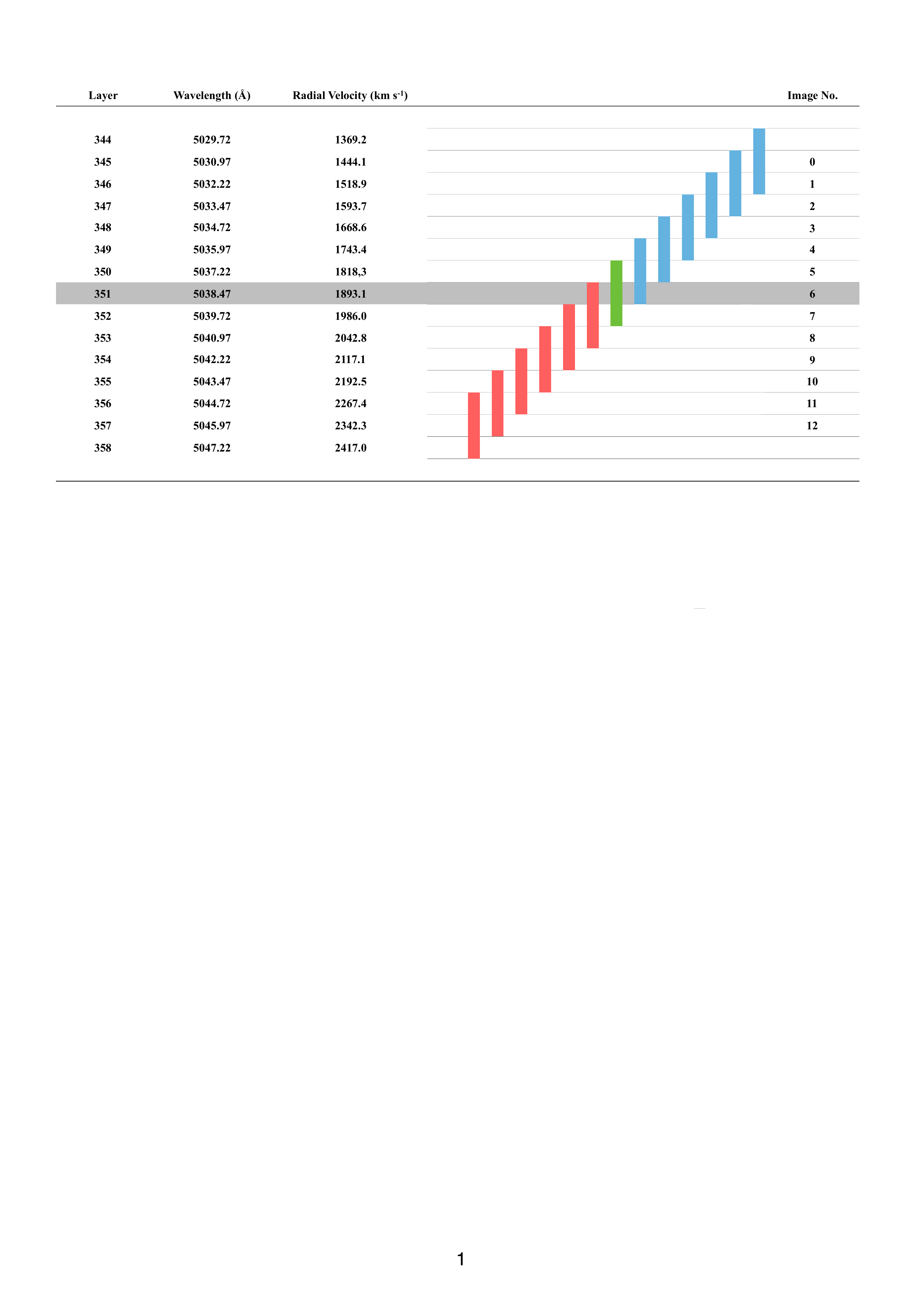}}
\caption{Data cube layers and narrow band images for NGC1380.  Each layer is 1.25\,\AA\ wide, and each narrow-band image for PN detection consists of 3 adjacent layers.
} \label{fig:NGC1380_layers}
\end{figure}

Since the wavelength resolution of MUSE is roughly twice the dispersion of the data cube, the \OIII $\lambda 5007$ emission from a planetary nebulae will typically be distributed over 2 or 3 wavelength bins (layers).  Thus, our method for PN detection involves summing the \OIII flux from three adjacent layers of the cube.  Figure~\ref{fig:NGC1380_layers} illustrates this process for the Fornax lenticular galaxy NGC\,1380.  Here the galaxy's systemic velocity of 1877~km\,s$^{-1}$ happens to fall within the 351st wavelength bin (data cube layer) and is shown as the grey shaded row.  The galaxy's internal motions then shift the 5007\,\AA\  emission of individual PNe to wavelengths between $\sim 5030$\,\AA\ (wavelength bin 345) and $\sim 5045$\,\AA\ (wavelength bin 357), depending on the exact velocity of the object. The red and blue bars illustrate that by co-adding three adjacent layers of the data cube, 13 images are formed with effective bandpasses of 3.75\,\AA\null.  This collects all the \OIII emission from all the PNe while greatly increasing the contrast of the PNe over the continuum, allowing the detection of the faintest emission-line objects relative to a narrow-band (e.g., 40\,\AA) image. The final photometry, however, is executed on the single layer images (Section~\ref{subsec:photometry}).

\begin{figure}[h!]
    \begin{minipage}{1.0\linewidth}
    \centerline{
    \includegraphics[width=0.9\hsize,bb=0 225 600 850,clip]{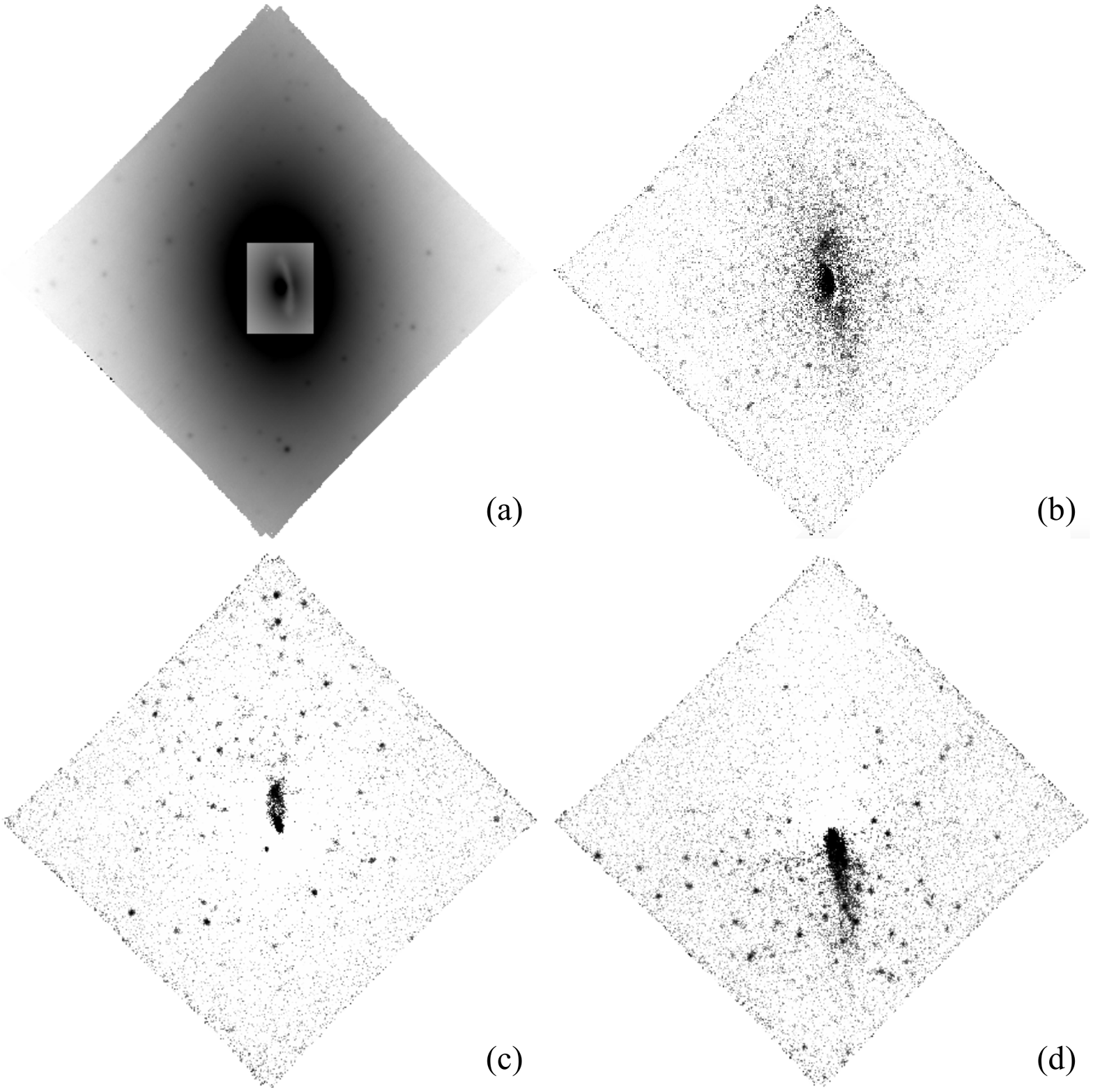}
    }
    \end{minipage}  
   \caption{Continuum and narrow-band {\it diff\/} images of the high surface brightness nuclear region of NGC\,1380, orientation: north up, east left. (a) Off-band image, note the insert that highlights a ring of dust around the nucleus, (b) {\it diff\/} image with a filter bandwidth of 40\,\AA\ as typically used for classical PNLF observations, (c) DELF image, blue-shifted with respect to the systemic velocity by approximately $-200$\,km\,s$^{-1}$, (d) DELF image, red-shifted by $+200$\,km\,s$^{-1}$. While in (c) and (d) the point-like sources are identified as PN candidates, their signal is almost completely washed out in the conventional diff image (b). The presence of a large ionized gas disk is hinted in the {\it diff\/} images and further discussed in Section~\ref{subsubsec:NGC1380spectro}. }
 \label{fig:NGC1380c_blue_red}
\end{figure}

The images shown in Figure~\ref{fig:NGC1380c_blue_red} illustrate how this scheme compensates for the variation of radial velocities. The $\sim 200$~km\,s$^{-1}$ rotation of the galaxy \citep{DOnofrio+95} is clearly seen, as the PNe north of the nucleus are systematically blue-shifted, while those in the southern part of the galaxy are primarily redshifted.  The effects of the galaxy's $\sim 200$~km\,s$^{-1}$ velocity dispersion \citep{DOnofrio+95, Vanderbeke+11} are also immediately visible, as there exist a few counter-rotating objects in both regions.

To better understand the efficacy of the result, we must consider the sources of noise in the data cube.  Accurate background subtraction has long been known to be a challenge for faint object spectroscopy, and the systematic errors associated with flatfield corrections are an important reason why the limit imposed by photon statistics is seldom achieved. For example, as pointed out by \citet{Cuillandre+94} for the case of long-slit spectroscopy, the limit for long exposures is not photon shot noise, but systematic multiplicative errors, caused by the CCD flatfield error $\epsilon$ (for the continuum) and slit alignment errors, $\omega$ (for strong sky line residuals).  Numerically, the signal-to-noise near a sky line can be expressed as
\begin{equation} 
\left( \frac{S}{N} \right)_\mathrm{LS} = \frac{I_O T}{\sqrt{I_S T + \sigma^2_\mathrm{CCD}} +(\epsilon+\omega)I_S T}
\end{equation}
where $T$ is the exposure time,  $I_O$ is the object flux,  $I_S$ is the sky background flux, and $\sigma^2_\mathrm{CCD}$ is the detector noise. If $\alpha$ is the ratio between the object and sky fluxes, this term converges for long exposure times to
\begin{equation} 
\left( \frac{S}{N} \right)_\mathrm{LS, limit} = \frac{\alpha}{\epsilon+\omega}
\label{eqn:SNRlimit}
\end{equation}

\begin{figure}[h!]
    \begin{minipage}{1.0\linewidth}
    \centerline{
    \includegraphics[width=55mm]{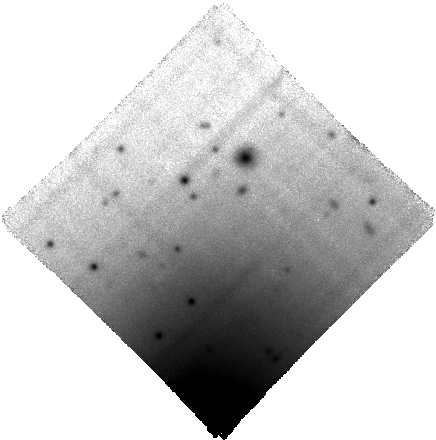}
    \includegraphics[width=55mm]{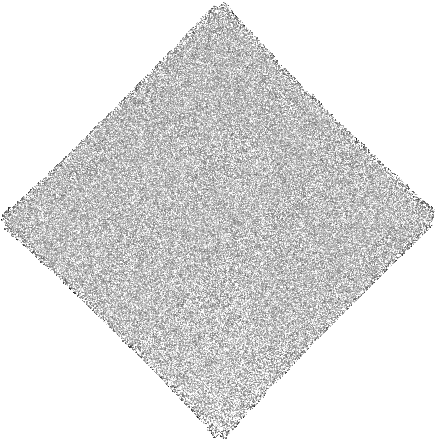}  
    \includegraphics[width=55mm]{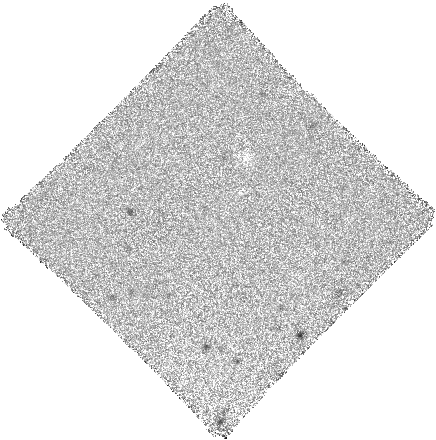}
    }
    \end{minipage}  
   \caption{Example of efficient suppression of background flatfield residuals with differential emission line filtering in the HALO field of NGC\,1380 (see also Fig.~\ref{fig:NGC1380_pointings}, north is up and east to the left). Left: broad-band continuum image obtained from co-adding data cube layers over the wavelength interval $5063.75 \leq \lambda \leq 5188.75$\,\AA\null.  The image exhibits flatfield residuals of up to 10\% of the background intensity. Middle: a single data cube layer at 5045\,\AA, continuum-subtracted using the scaled mean galaxy spectrum to remove the pixel-to-pixel flatfield variations. Right: a continuum-subtracted single data cube layer at 5035\,\AA\null. At this wavelength, the redshifted \OIII line at 5007\,\AA\ is transmitted for PNe with a radial velocity of 1687\,km s$^{-1}$, i.e., a projected velocity of $-190$\,km s$^{-1}$ relative to systemic. These PNe appear as faint point sources in the image. Note that there is a hint of a local zero-point offset at the location of an unrelated $z=0.3355$ background galaxy (north of the frame center) that appears as slightly oversubtracted. This small continuum correction factor mismatch does not affect the overall flatfield of the frame nor our ability to detect PNe.}
 \label{fig:DELF}
\end{figure}

Adopting Equation~\ref{eqn:SNRlimit} for our MUSE data, it follows that the errors from \OIII emission line spectrophotometry are strongly affected by residual flatfielding errors. As illustrated in Figure~\ref{fig:DELF}, these systematic uncertainties are visible as a criss-cross pattern of brightness-enhanced streaks throughout the image, with $\epsilon$ assuming values as large as 10\%. This limitation was already discovered in the course of the MUSE surveys for faint Ly$\alpha$-emitting galaxies \citep[e.g.,][]{Bacon+17,Herenz+17,Wisotzki+18,Bacon+21}. Moreover, MUSE integral field spectroscopy is affected by residual errors that are more complex than the ones for long-slit spectroscopy. As pointed out by \citet{Soto+16}, since the light path varies from slice to slice and from IFU to IFU, small discontinuous variations are introduced into the line-spread-function (LSF) and the wavelength solution of the final reconstructed data cube.  These issues are then further exacerbated by systematic flatfielding and bias subtraction errors. 

In order to remove the resulting residual patterns from deep MUSE exposures, \citet{Soto+16} invoked the ZAP filter, which is based on principal component analysis (PCA)\null. ZAP constructs a sky residual spectrum for each individual spaxel, which can then be subtracted from the original data cube.  While the method does reduce the sky residuals, it has the potential drawback that its eigenspectra, which characterize the residuals, are unable to distinguish between astronomical signals and the background.  Thus the method requires very careful treatment of the filter parameters and interpretation of the filtered data. 

In our application, the host galaxy background is orders of magnitude brighter than the sky, and the spectral region of interest is not plagued by bright night-sky emission lines.  Under these conditions, a simple generalization of the on-band/off-band direct imaging technique is very efficient at accurately subtracting the background and suppressing the $\epsilon$-term in Equation~\ref{eqn:SNRlimit}.

Ignoring for a moment the statistical errors that add in quadrature as well as the slit alignment error $\omega$, we can write the influence of the multiplicative systematic error $\epsilon$ on the flux measurement $F(\lambda)$ as
\begin{equation}
    F(\lambda) = \sum_{i,j}^{N_{aper}}I_O(x_i,y_j,\lambda) - c \sum_{m,n}^{N_{skyrad}}I_S(x_m,y_n,\lambda)
\end{equation}
The first term in this equation sums the apparent fluxes, $I_0$, in the spaxels of an object within an aperture {\it aper}; the second term does the same for the apparent spaxel fluxes, $I_S$, in the source's sky annulus, {\it skyrad}.  The normalizing constant $c$, which is typically $\sim 0.1$, accounts for the greater number of spaxels in the sky region. If $o$ and $s$ are the true spaxel fluxes in the object aperture and sky annulus, and $F(x_i,y_j,\lambda)$ is the flux per spaxel ($i$,$j$) contributing to the total, then, assuming the residual flatfield errors $\epsilon_{ij}$ and $\epsilon_{mn}$ are not varying with wavelength,
\begin{equation}
    \begin{split}
    F(x_i.y_j,\lambda) = (1+\epsilon_{ij}) \left( o(x_i,y_j,\lambda) + s(x_i,y_j,\lambda) \right) - c/N_{aper} \sum_{m,n}^{N_{skyrad}}(1+\epsilon_{mn})s(x_m,y_n,\lambda)\\
     = o(x_i,y_j,\lambda) + s(x_i,y_j,\lambda) + \epsilon_{ij} o(x_i,y_j,\lambda)  + \epsilon_{ij} s(x_i,y_j,\lambda) - c/N_{aper} \sum_{m,n}^{N_{skyrad}}s(x_m,y_n,\lambda) \\ - c/ N_{aper} \sum_{m,n}^{N_{skyrad}}\epsilon_{mn} s(x_m,y_n,\lambda)
\end{split}
\end{equation}
Under the assumption of a flat, or to first order, constant gradient in the background surface brightness, two terms cancel to zero, so that
\begin{equation}
    s(x_i,y_j,\lambda) - c/N_{aper} \sum_{m,n}^{N_{skyrad}}s(x_m,y_n,\lambda) = 0
\end{equation}
This leaves us with three remaining terms that add a systematic error to the object flux $o(x_i,y_j,\lambda)$:
\begin{equation}
F(x_i.y_j,\lambda) = o(x_i,y_j,\lambda) + \epsilon_{ij} o(x_i,y_j,\lambda)  + \epsilon_{ij} s(x_i,y_j,\lambda) - c/ N_{aper} \sum_{m,n}^{N_{skyrad}}\epsilon_{mn} s(x_m,y_n,\lambda)
\end{equation}

The term $\epsilon_{ij} o(x_i,y_j,\lambda)$, which is no more than $\sim 10\%$ of the object flux, does not scale with the background surface brightness and can therefore be neglected. The two remaining terms are directly proportional to the background, meaning that deviations of $\epsilon_{ij}$ from zero can contribute a significant residual to the extracted point source spectrum whenever the background surface brightness is high.

By contrast, the difference frame method applies a scaled continuum flux subtraction within identical spaxels that are subject to the same error, $\epsilon_{ij}$.  The term therefore cancels out as shown in Equation~\ref{eqn:DELF}:
\begin{equation}
    \begin{split}
F(x_i.y_j,\lambda) = (1+\epsilon_{ij})  o(x_i,y_j,\lambda) + (1+\epsilon_{ij}) s(x_i,y_j,\lambda)  - k (1+\epsilon_{ij})  \sum_{\lambda=\lambda_1}^{\lambda_n} s(x_i,y_j,\lambda) \\
= (1+\epsilon_{ij}) \left[ o(x_i,y_j,\lambda) + s(x_i,y_j,\lambda)  - k \sum_{\lambda=\lambda_1}^{\lambda_n} s(x_i,y_j,\lambda) \right]
    \label{eqn:DELF}
\end{split}
\end{equation}
The scaling factor $k$ can be accurately measured from the data cube itself, such that the background terms vanish:
\begin{equation}
    s(x_i,y_j,\lambda)  - k \sum_{\lambda=\lambda_1}^{\lambda_n} s(x_i,y_j,\lambda) = 0
\end{equation}
and we are merely left with a small error on the flux:
\begin{equation}
F(x_i.y_j,\lambda) = (1+\epsilon_{ij})  o(x_i,y_j,\lambda)
\end{equation}
As we have verified in various tests (Section~\ref{subsec:tests}), the continuum band can be scaled to the adjacent \OIII $\lambda 5007$ window with very high accuracy, as long as there are no dramatic changes in the underlying stellar population or kinematics of the host galaxy. Such changes do not often occur over the MUSE field-of-view in early-type systems (such as NGC\,1380), so the error in the scaling factor $k$ is held well below 1\%. (For aperture photometry, the term is even less important:  as long as there is no strong population gradient, the factor will cancel.) The middle and right panels in Fig.~\ref{fig:DELF} illustrate how well the background subtraction is accomplished in practice. 

Based on the above analysis, we have expanded the technique of extracting a small number of continuum-subtracted {\it diff\/} images and have processed the entire data cube with a tool that replaces each layer with a continuum-subtracted {\it diff\/} frame; this step isolates any emission features and produces a high signal-to-noise measurement with practically no background residuals. This tool has also been instrumental for measuring the emission lines of H$\alpha$ and [\ion{S}{2}] that are important for the reliable classification of PNe (see Section~\ref{subsec:classification} below).

To summarize, the generalization of the classical on-band/off-band technique to MUSE data cubes not only provides an advantage of much smaller filter bandwidths, hence less background flux from the host galaxy, but also reduces spaxel-to-spaxel flatfield residuals that would otherwise produce overwhelmingly large errors in regions of high host galaxy surface brightness.  These two advantages allow PNLF studies to be made with greater precision and to much greater distances than previous studies.  In fact, this {\it differential emission line filter (DELF)\/} technique is conceptually similar to using beam-switching in the near infrared, or to the nod-shuffle option in optical spectroscopy \citep{Cuillandre+94,Glazebrook+01,Roth+02}, in the sense that identical pixels are used for comparing object + sky with sky.  The main difference is that no extra exposure time is spent on sky frames.

\subsection{Source detection and photometry}
\label{subsec:photometry}
As described in the previous section, the detection of PN candidates and the measurement of \OIII\ magnitudes were accomplished by adapting the methods of classical on-band/off-band photometry to the set of {\it diff\/} images at different wavelengths produced from the MUSE data cubes.  Our procedure consisted of four major steps. 

First, the identification of PN candidates was performed visually by scanning through the set of images that were co-added over three wavelength bins.  We found that loading the images into 13 consecutive frames of the DS9 tool \citep{Joye+03} and ``tabbing'' through the images provided a generalization of the classical blinking technique. We required all valid PN candidates to appear in at least three successive frames and have a point-source appearance. In a future tool, this requirement can be implemented in software via an source detection algorithm.  However for the present work, this level of automation was not necessary. Once found, the position of each PN candidate was recorded and used as a first estimate for the next step in our analysis.

With the initial coordinate estimates in hand, the centroids of the PN candidates were measured using the {\tt GCNTRD} routine, available in the NASA IDL Astronomy User's Library\footnote{\url{https://idlastro.gsfc.nasa.gov/}}.  This routine fits a Gaussian to the image in all 15 frames of the unbinned {\it diff\/} images. Because the \OIII $\lambda 5007$ emission only appears in those few images of the stack that correspond to the PN's radial velocity, we only measured the centroid when the flux from rough aperture photometry exceeded some threshold value above the noise. The centroid obtained from the brightest image of the series was adopted as the best ($x$,$y$) position of the PN candidate.

\begin{figure}[h!]
\begin{minipage}{1.0\linewidth}
    \begin{minipage}{1.0\linewidth}
    \centerline{
       \includegraphics[width=10mm,bb=100 460 180 800,clip]{ 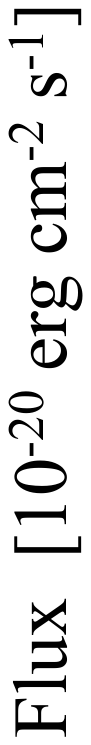}
    \includegraphics[width=90mm,bb=40 500  620 703,clip]{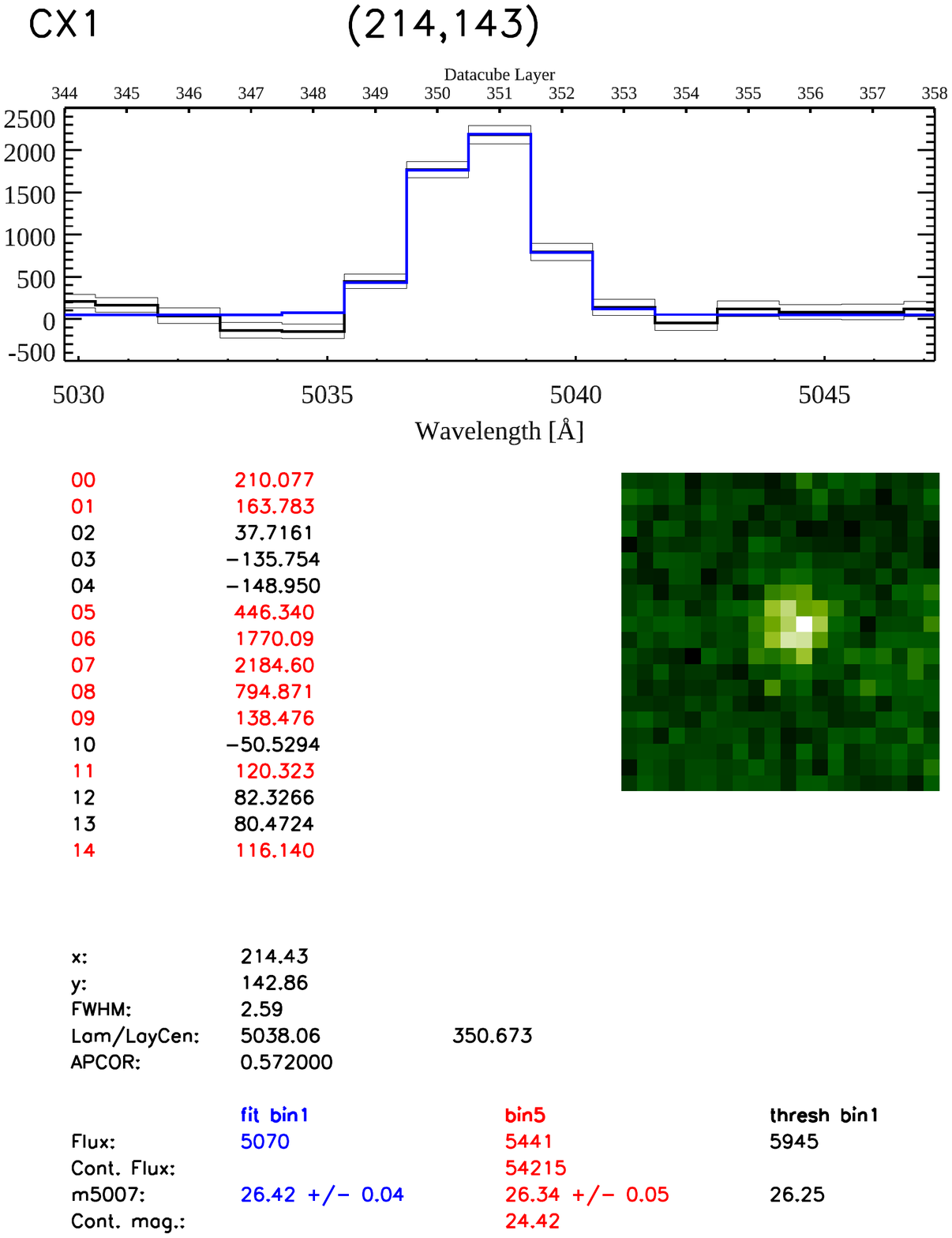}
    \includegraphics[width=30mm,bb=0 -50  400 400,clip]{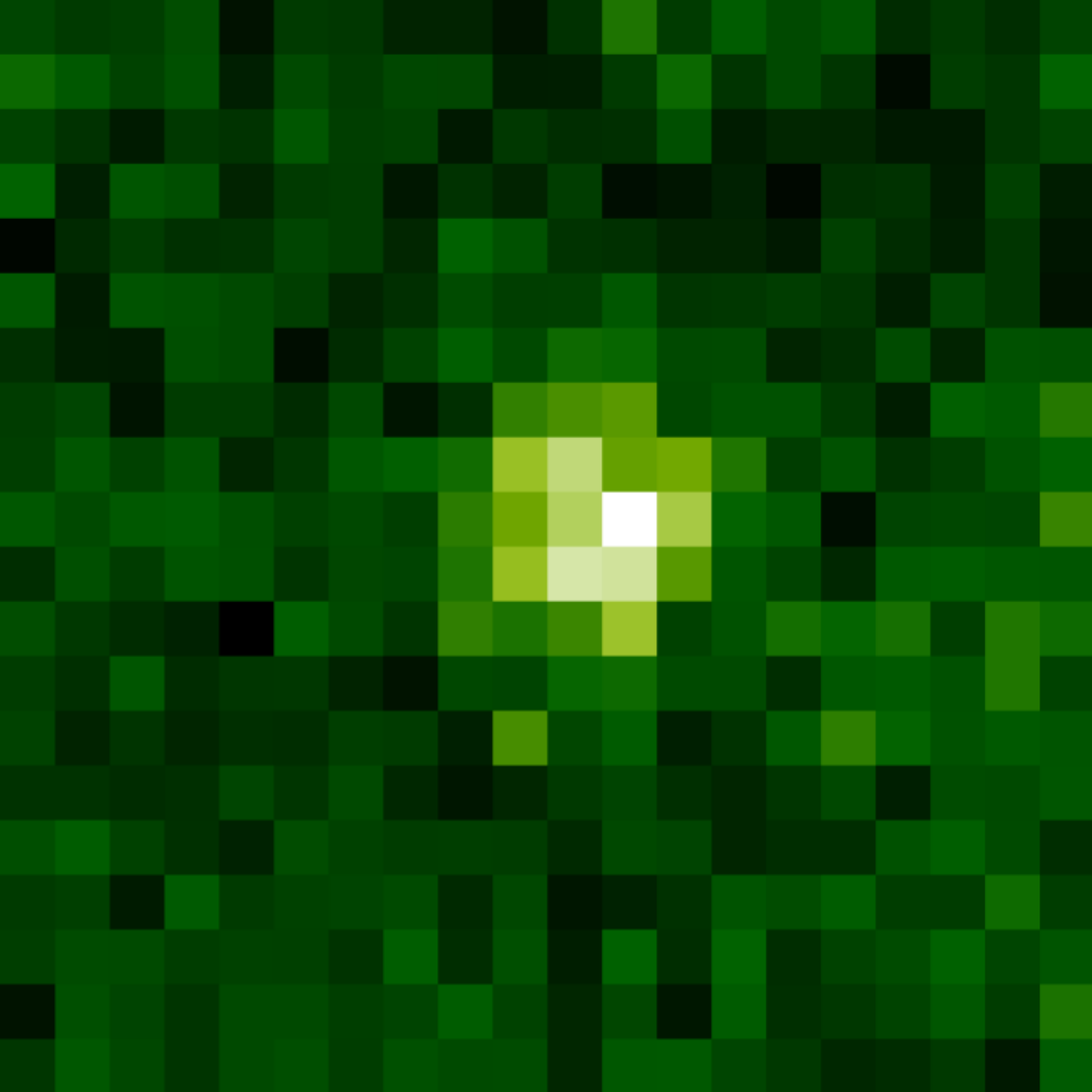}
    }
    \end{minipage} 
    \begin{minipage}{1.0\linewidth}
    \centerline{
     \includegraphics[width=10mm,bb=100 460 180 800,clip]{ Ordinate_Label_Minispectrum.pdf}
     \includegraphics[width=90mm,bb=40 500  620 703,clip]{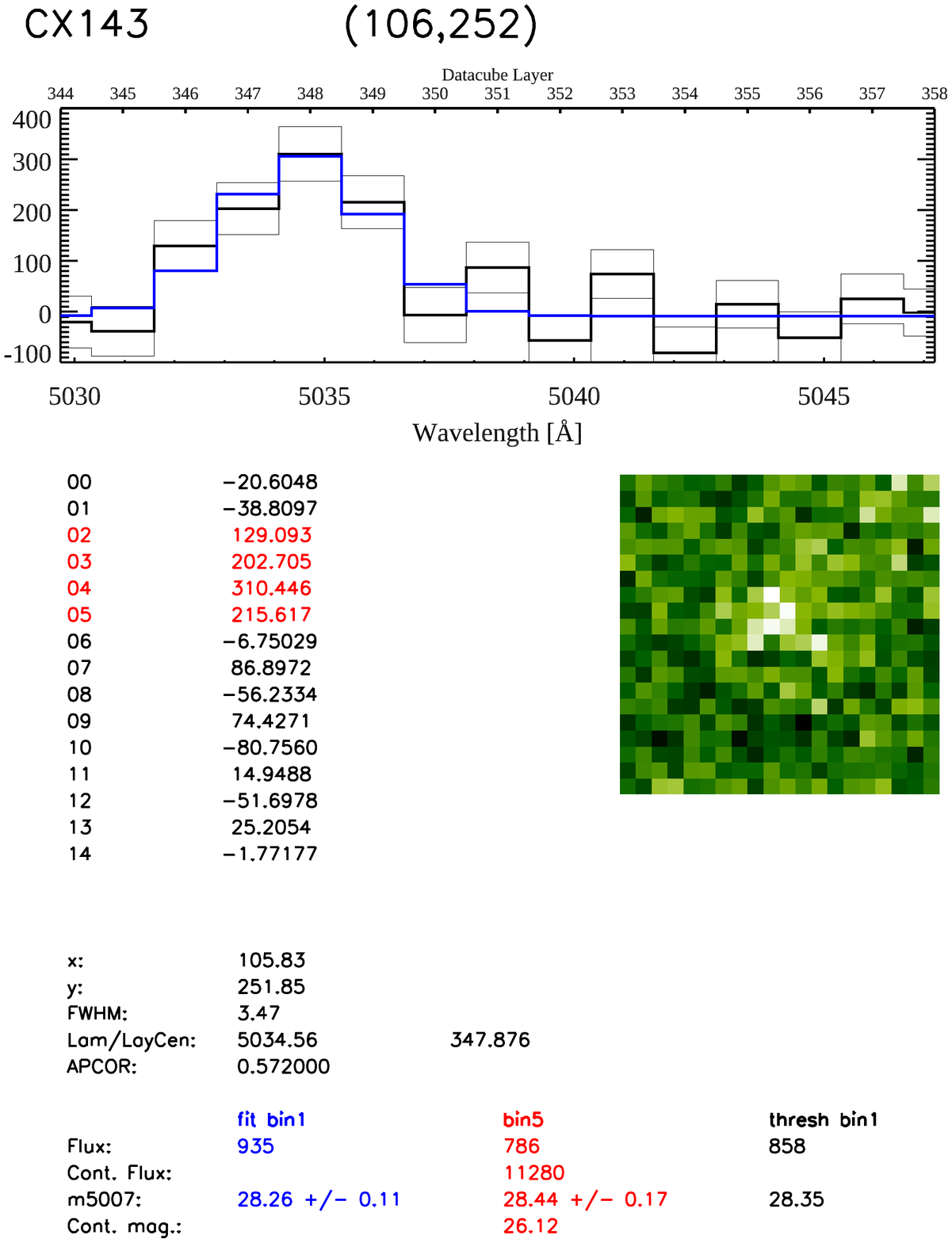}
    \includegraphics[width=30mm,bb=0 -50  400 400,clip]{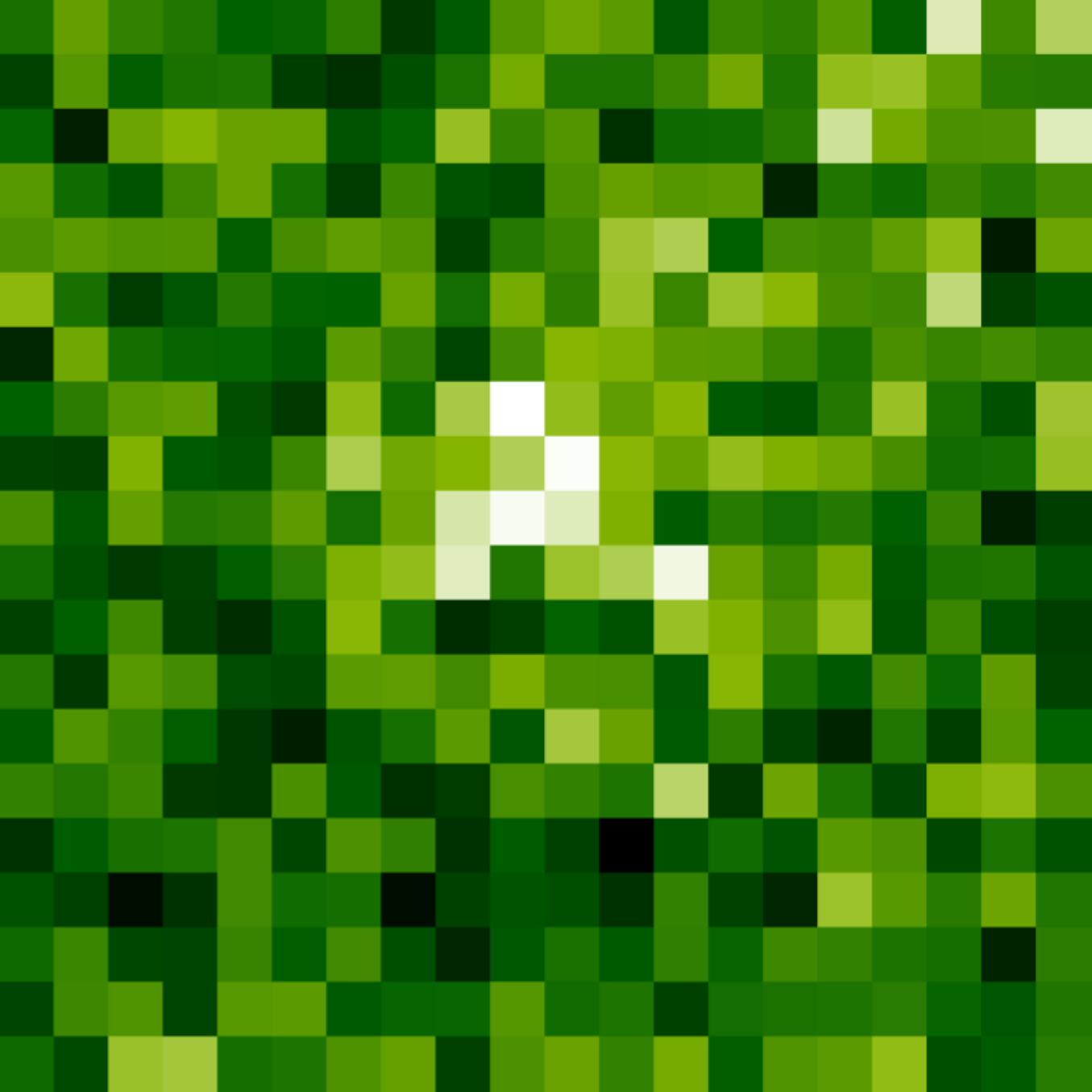}  
    }
    \end{minipage}  
     \end{minipage}  
   \caption{The short spectra for one bright and one faint PN in NGC\,1380.  Images of the PNe formed by co-adding 5 data cube layers centered on the brightest bin are shown on the right; the spectra formed from 15 layers of the data cube are displayed on the left. The thick black histograms show the measured DAOPHOT fluxes in each bin, while the thin black lines illustrate the photometric uncertainties. The blue histograms display the Gaussian fits.  The abscissa is labeled in wavelength on the bottom and layer numbers on the top.  Top row: CX1, a PN with $m_{5007}=26.43\pm0.05$. Bottom row: PN CX143, with $m_{5007}=28.25\pm0.17$. Gaussian fits to the two emission lines yield central wavelengths and corresponding radial velocities of 5038.06\,\AA /$1865.9$\,km\,s$^{-1}$ and 5034.56\,\AA /$1657.6\,$km\,s$^{-1}$, respectively.}
   \label{fig:minispec}
\end{figure}

In step 3 of our procedure, we used the Gaussian-based centroids to perform aperture photometry in all 15 frames of the unbinned {\it diff\/} images, thereby creating a short 15-pixel spectrum in the relevant wavelength region of interest. We validated the photometry with two different tools (see Section~\ref{subsec:tests}).   Depending on the image quality of a given data cube, we chose a DAOPHOT aperture radius of 3 pixels, or slightly larger, and a sky annulus with inner and outer radii of typically 12 and 15 pixels.  Examples  of two of these short spectra are shown in Figure~\ref{fig:minispec}.  

In step four, we measured the \OIII\ flux by first fitting a Gaussian to the resulting emission line in the short spectrum (with the central wavelength, line full width at half maximum (FWHM), and normalization as free parameters), and then either recording the integrated flux over the fitted profile, or co-adding the individual flux measurements from the closest 5 bins around the peak of the Gaussian fit. For well-behaved profiles, the two methods yield values that agree within a few hundredths of a magnitude. However, in cases of double-lined or broadened profiles, which occurred when the images of two PNe with different radial velocities happen to overlap with each other, the flux from summing over 5 bins is significantly larger. We resolved this problem with an interactive deblending tool that fits two separate Gaussians to the data.  This produced a more accurate magnitude for each component of the blend (see Section~\ref{subsubsec:NGC1380delf}).


These steps were supplemented by visual inspection of both the original {\it diff\/} images and the short spectra produced in step four.  This allowed us to immediately identify overlapping objects and assess sources corrupted by unrelated emission features or spurious signals.  

To account for the flux beyond our 3 pixel radius measuring aperture, we need to apply an aperture correction using stars in the field. Given the small field of view of MUSE, most galaxy exposures in the ESO archive contain too few stars bright enough for a reliable PSF measurement using just the data layers of interest.  Moreover, within the body of a nearby galaxy, it is often difficult to discriminate foreground stars from one of the galaxy's semi-resolved globular clusters.  Consequently, we need another method to measure the aperture correction for our PN measurements.

To address this challenge, we created broadband (200\,\AA) images from co-added data cube layers centered on the redshifted \OIII line. Point source candidates of sufficient brightness were identified with DAOPHOT {\tt FIND} and ordered by apparent magnitude from their DAOPHOT {\tt APER} measurements. After identifying up to 10 of the brightest stars in the field, we applied the Levenberg-Marquardt algorithm {\tt mpfit} to fit their PSFs with Gaussian and Moffat functions \citep{Markwardt09}. For each star, two different approaches were used for the fit:  one in the {\it off\/} broadband wavelength region to a create high signal-to-noise measurement of the PSF. As an independent check, and also to address the wavelength dependence of aperture corrections, another fit was applied in each layer of the data cube, albeit at the expense of noisier results.

\begin{figure}[h!]
\begin{minipage}{1.0\linewidth}
    \begin{minipage}{1.0\linewidth}
    \centerline{
    \includegraphics[width=55mm,bb=0 0  660 680,clip]{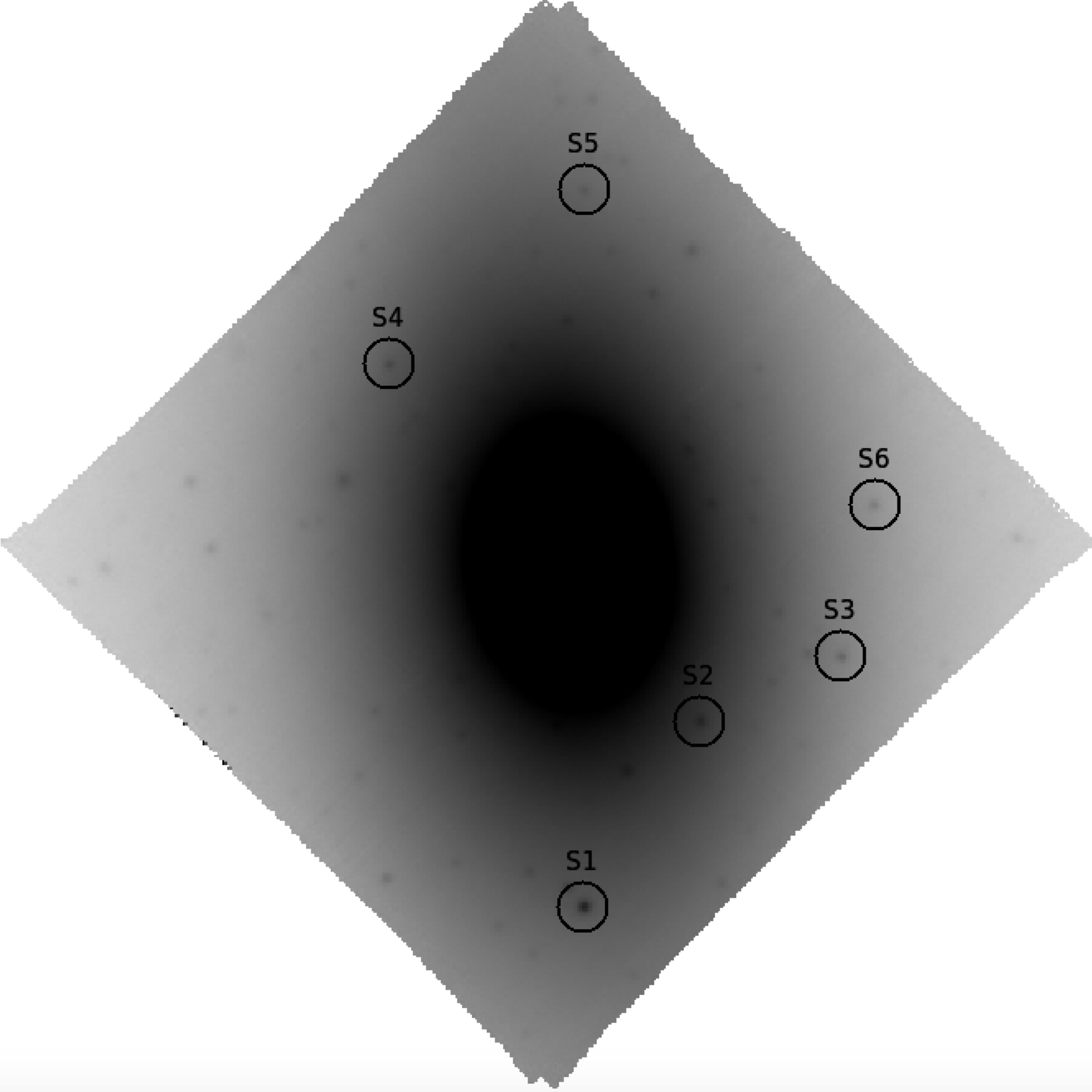}
    \includegraphics[width=55mm,bb=0 0  660 680,clip]{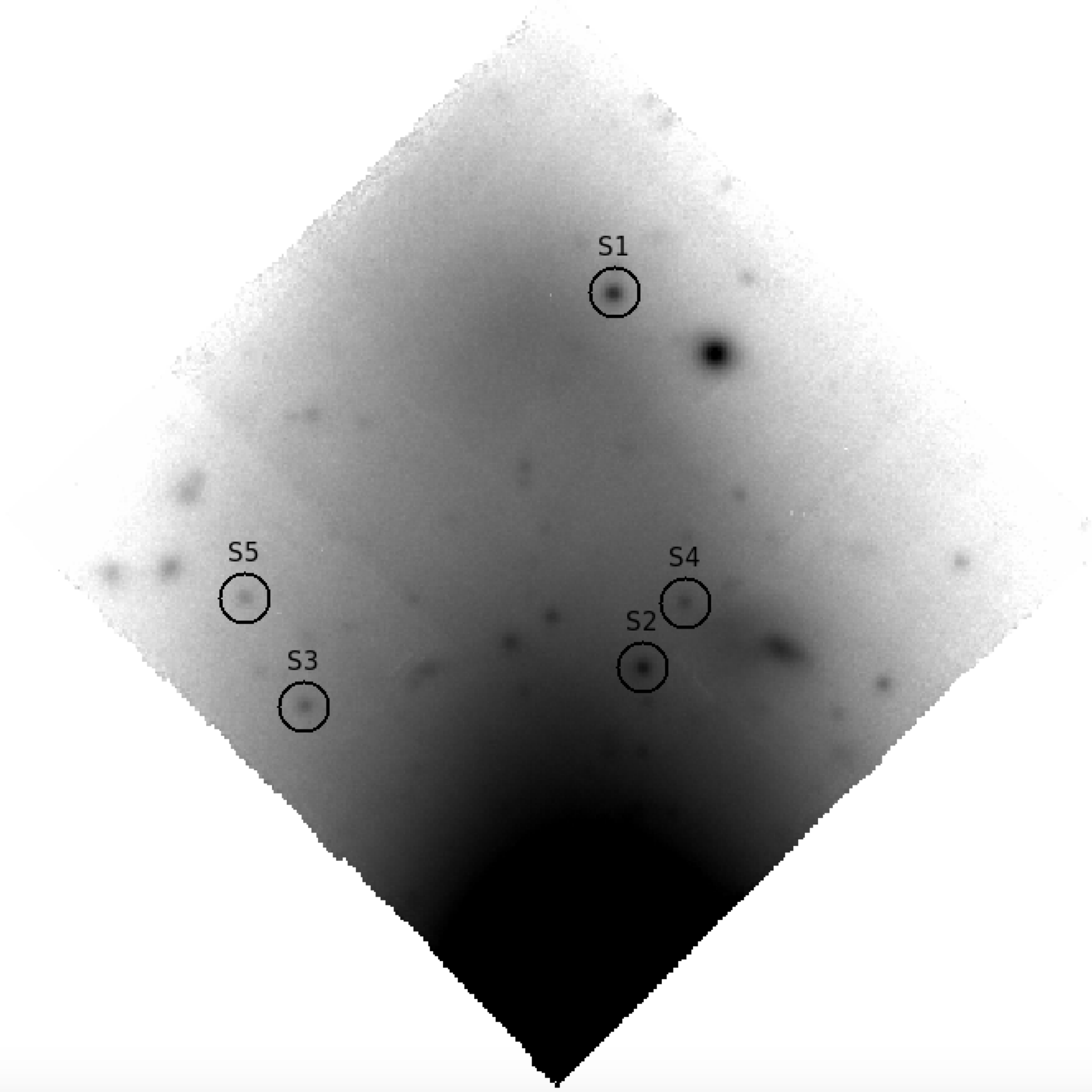}  
    \includegraphics[width=55mm,bb=0 0  660 680,clip]{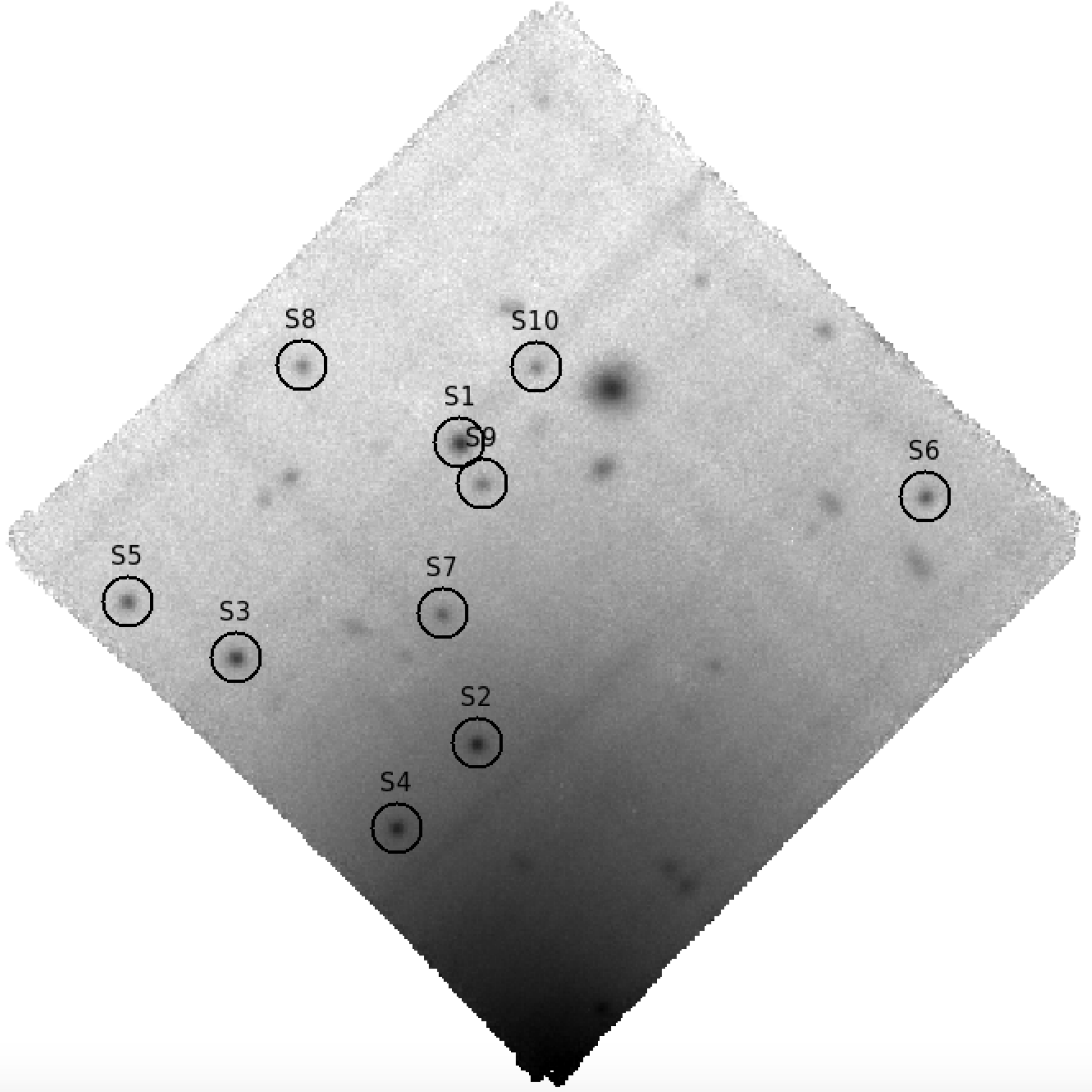}
    }
    \end{minipage} 
    \vspace{5mm}\\
    \begin{minipage}{1.0\linewidth}
    \centerline{
    \includegraphics[width=56mm,bb=0 0  760 470,clip]{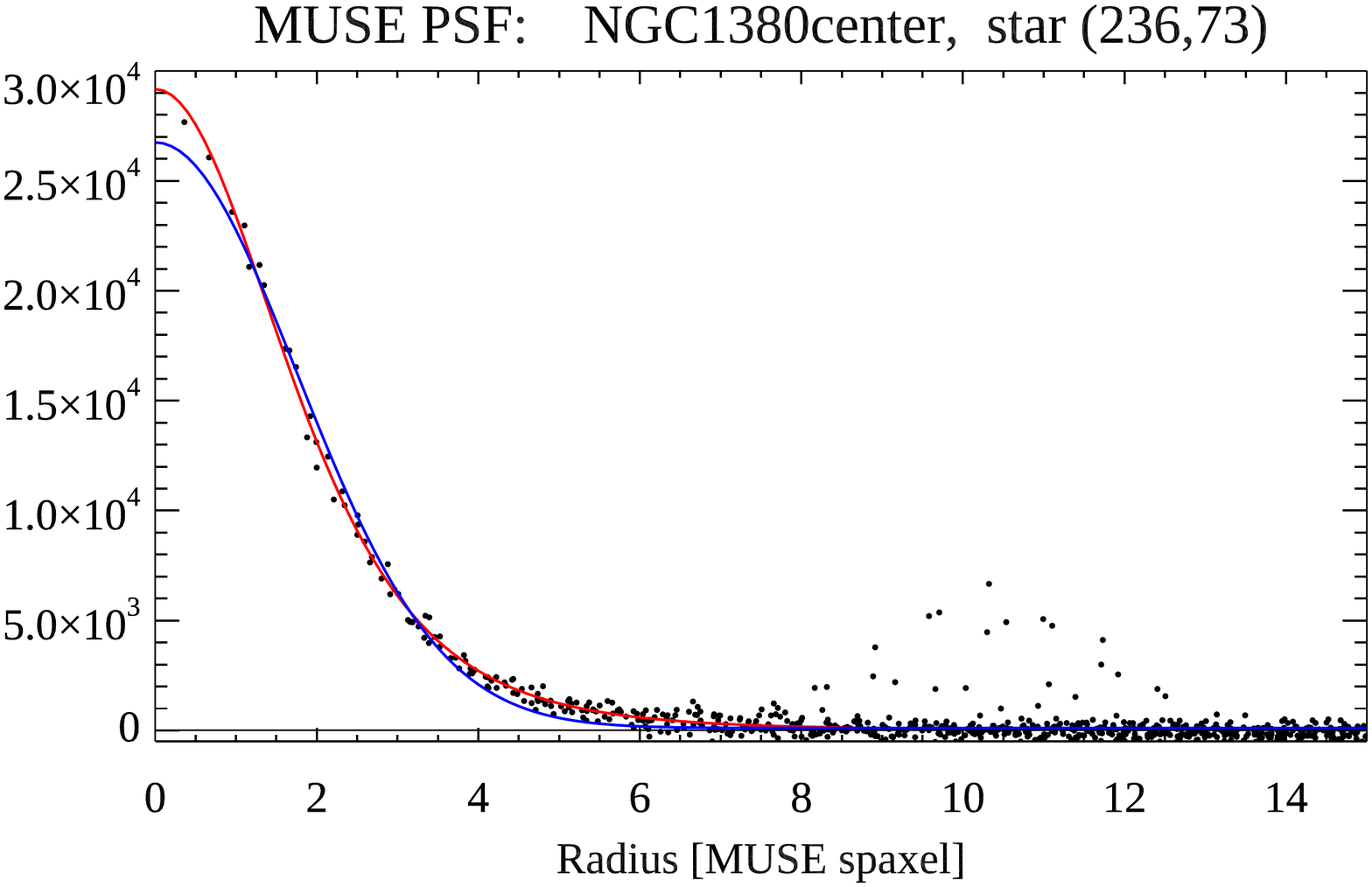}
    \includegraphics[width=55mm,bb=20 0  770 470,clip]{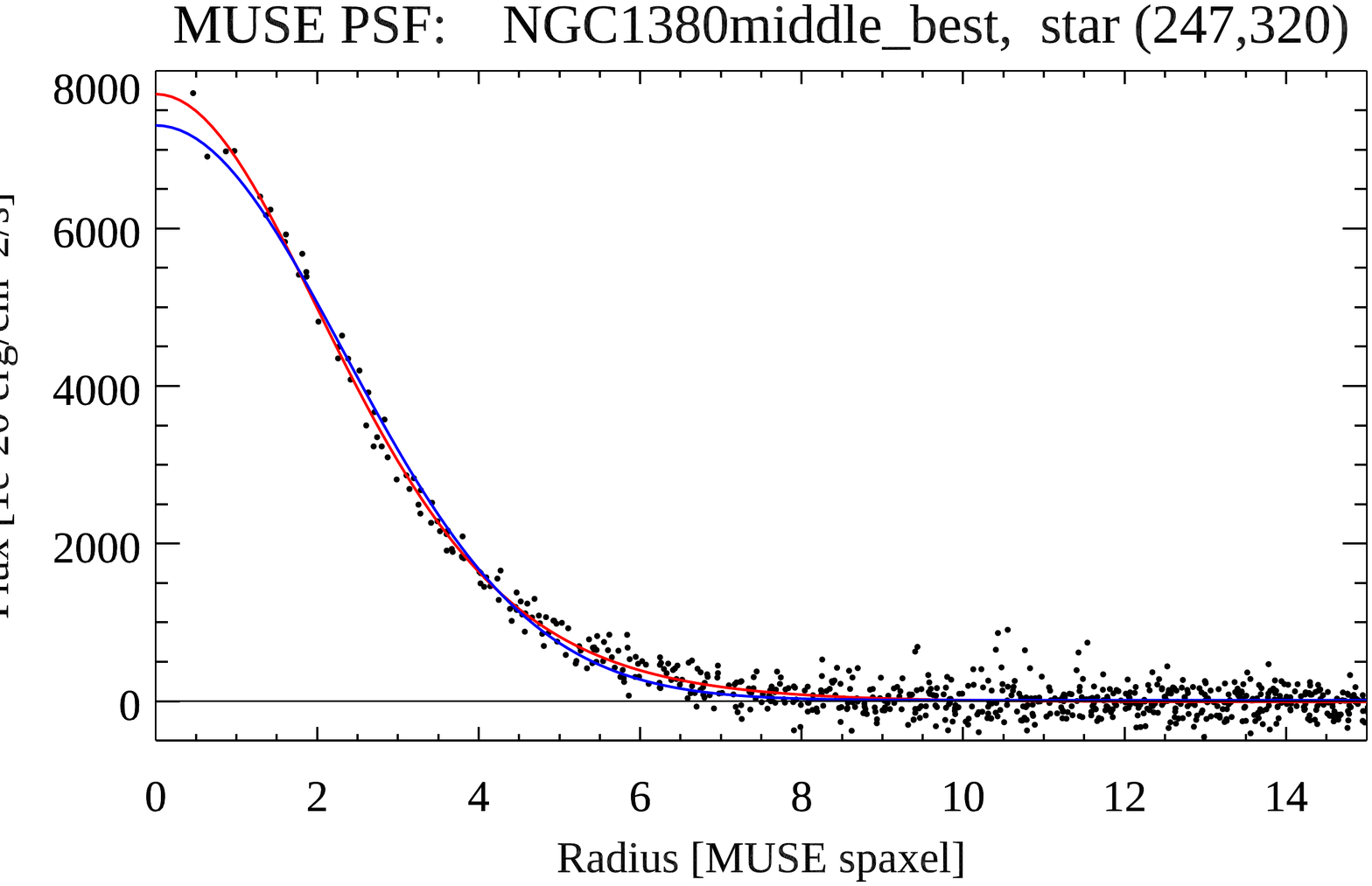}  
    \includegraphics[width=55mm,bb=20 0  770 470,clip]{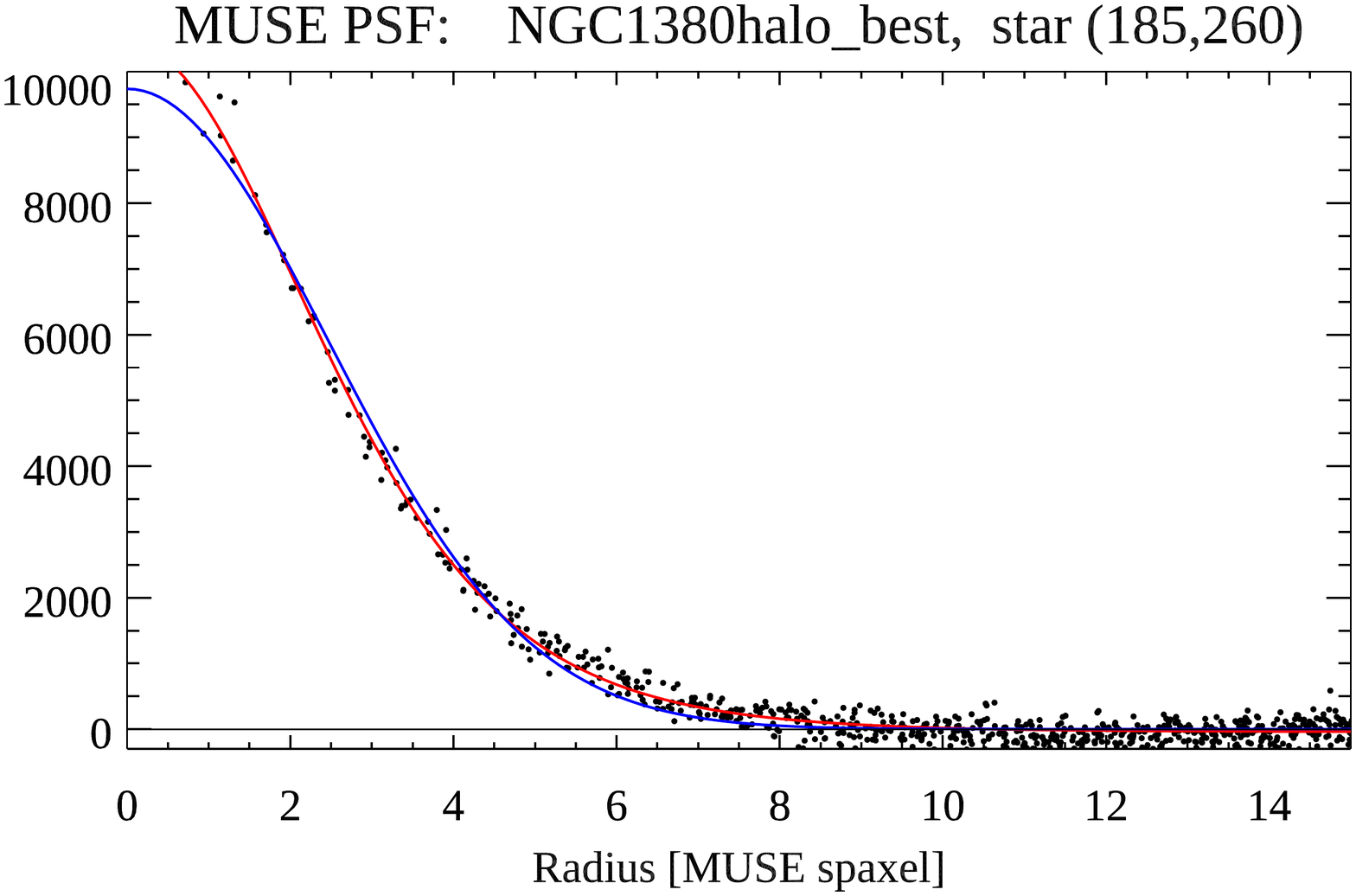} 
    }
    \end{minipage}  
    \begin{minipage}{1.0\linewidth}
    \centerline{
    \includegraphics[width=55mm,bb=80 350  560 700,clip]{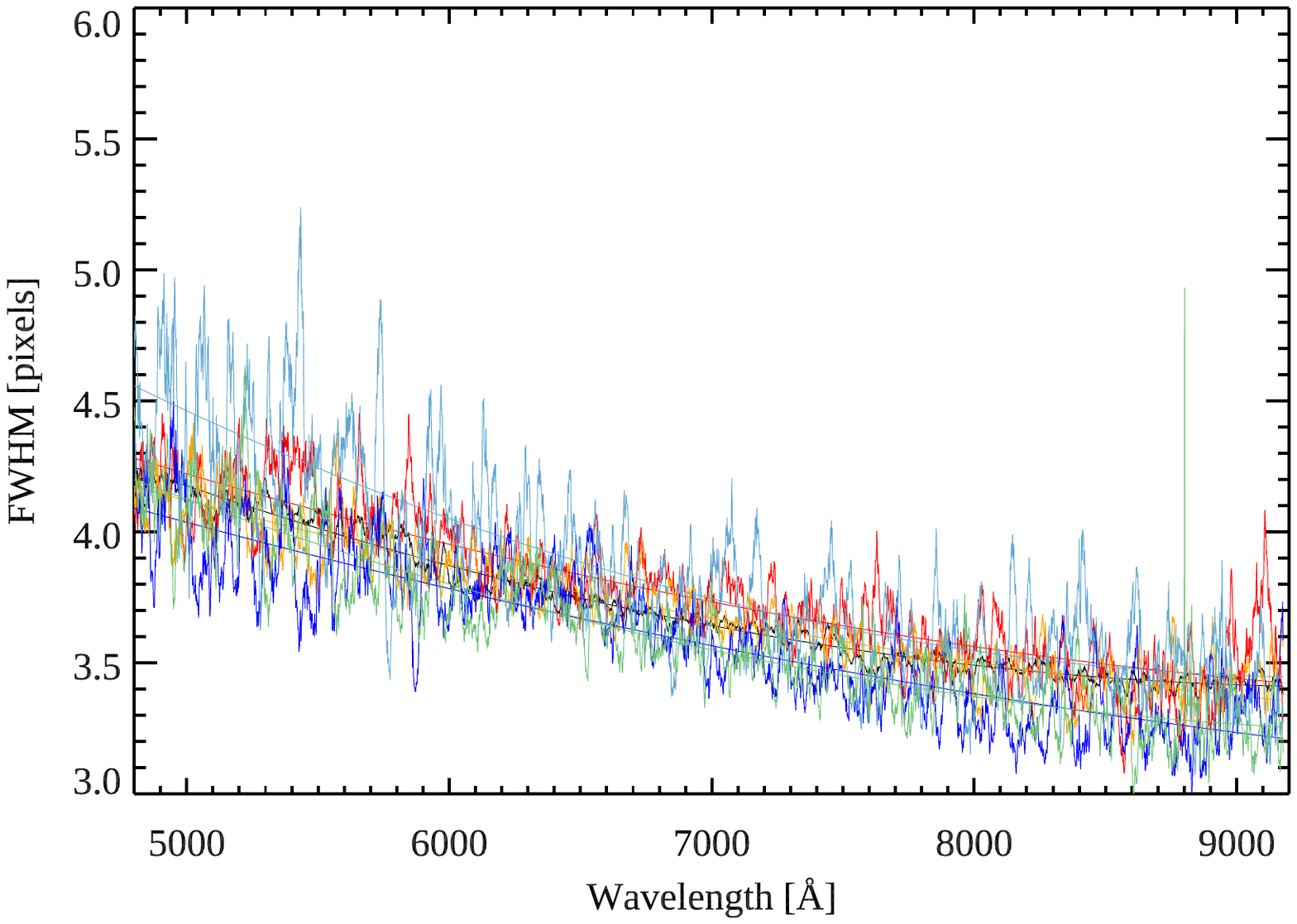}
    \includegraphics[width=55mm,bb=80 350  560 700,clip]{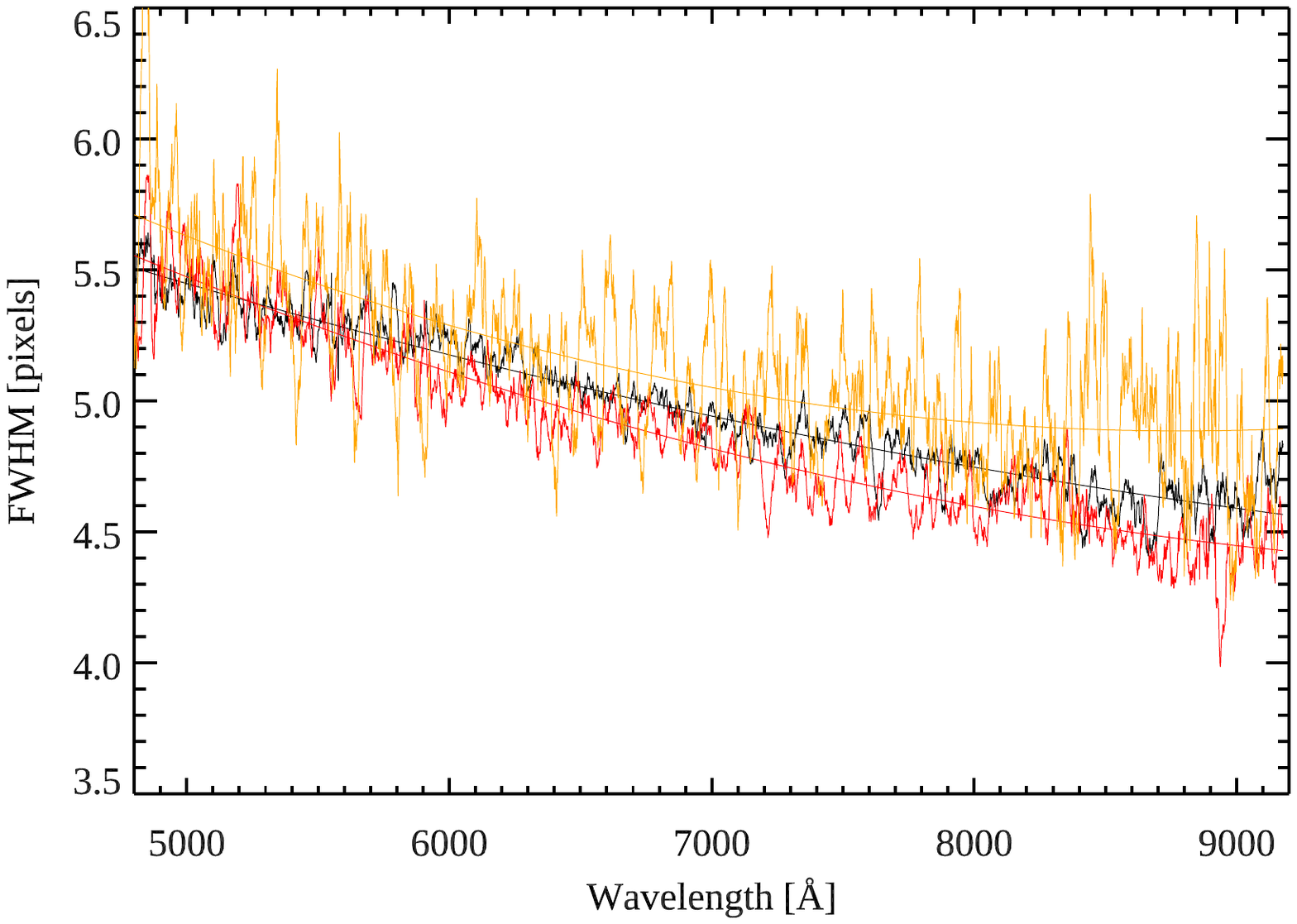}  
    \includegraphics[width=55mm,bb=80 350  560 700,clip]{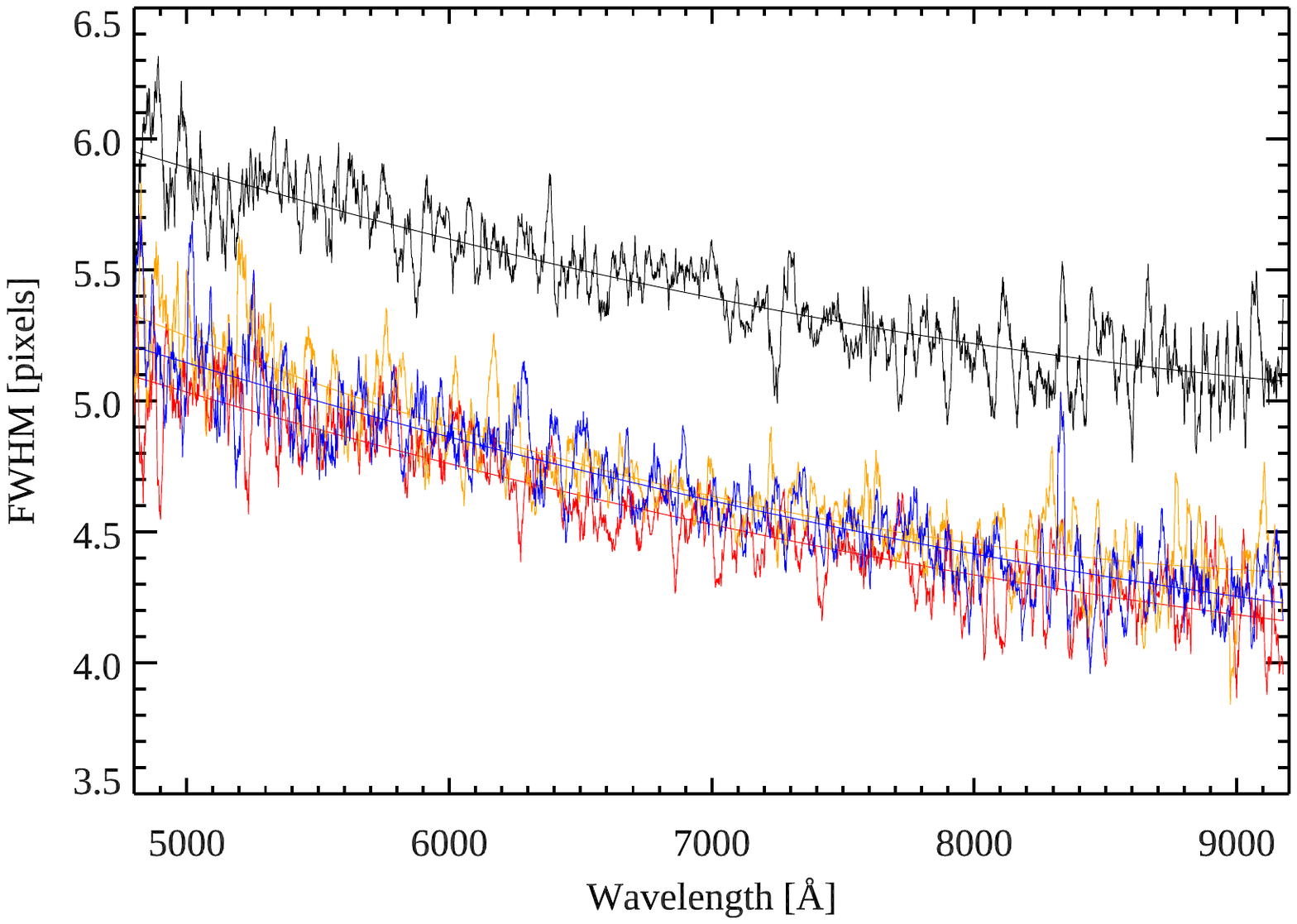} 
    }
    \end{minipage} 
    \vspace{2mm}\\         
    \begin{minipage}{1.0\linewidth}
    \hspace{0.1mm}
    \centerline{
    \includegraphics[width=55mm,bb=90 350  560 700,clip]{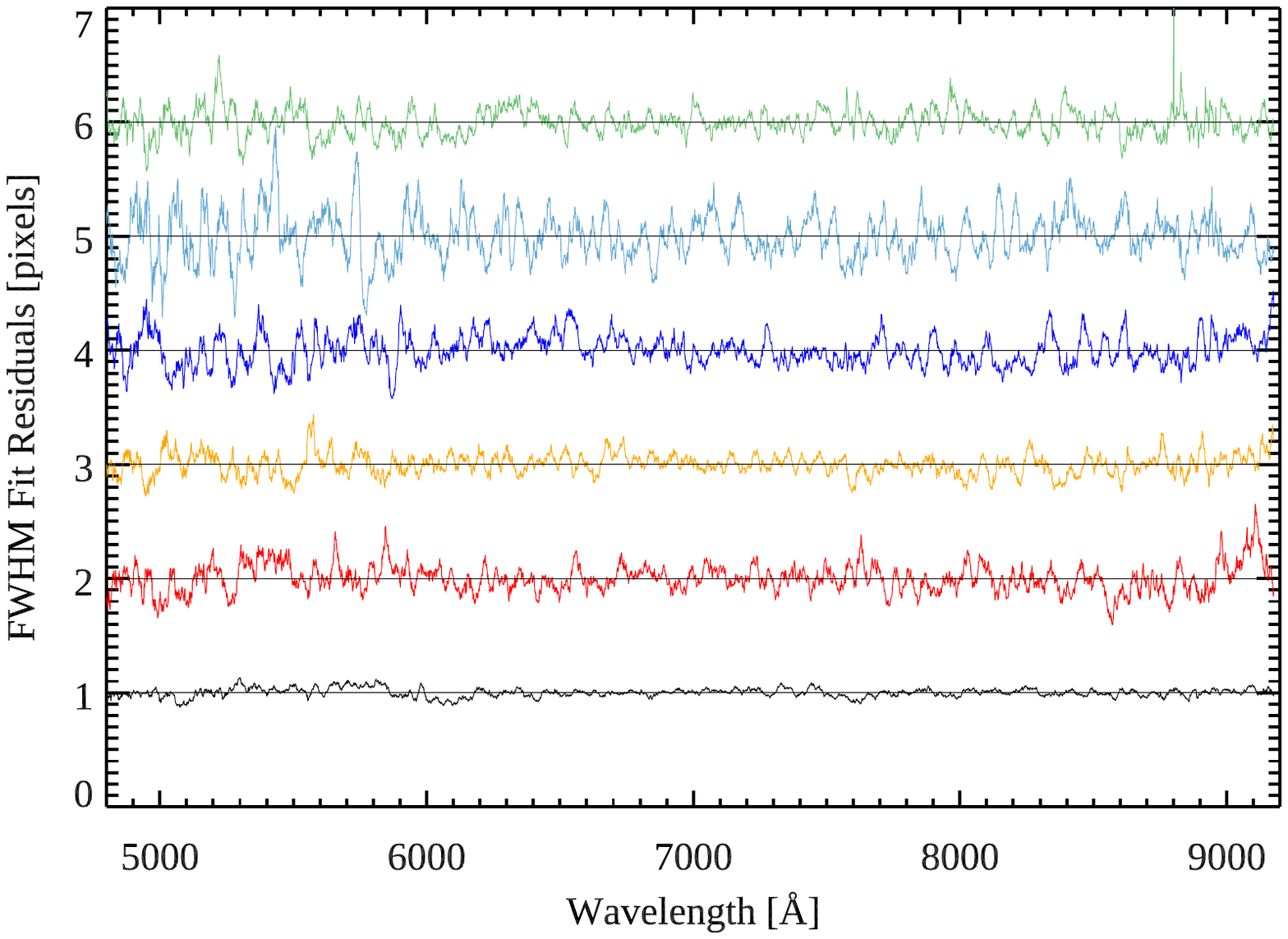}
    \includegraphics[width=55mm,bb=90 350  560 700,clip]{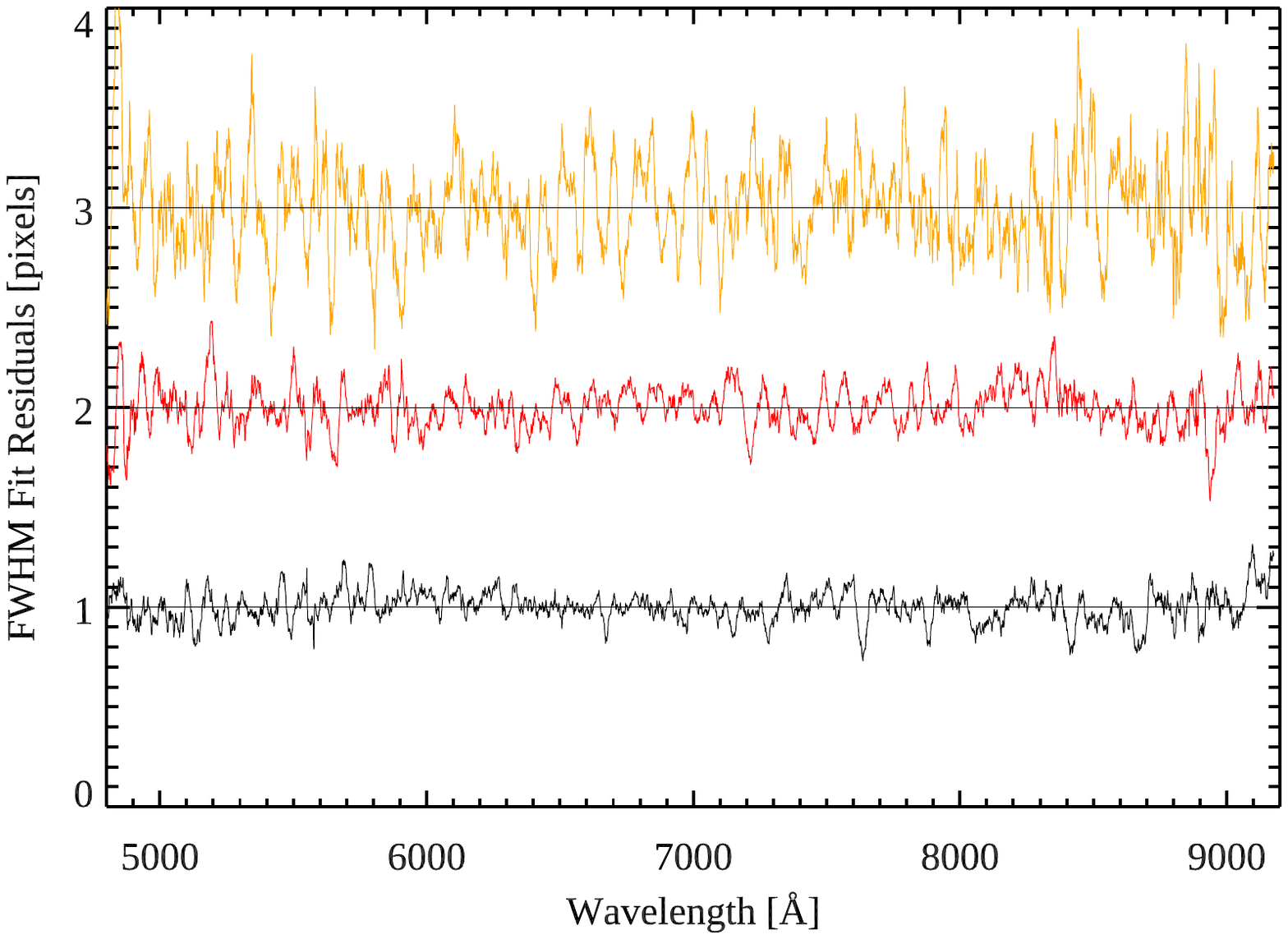}  
    \includegraphics[width=55mm,bb=90 350  560 700,clip]{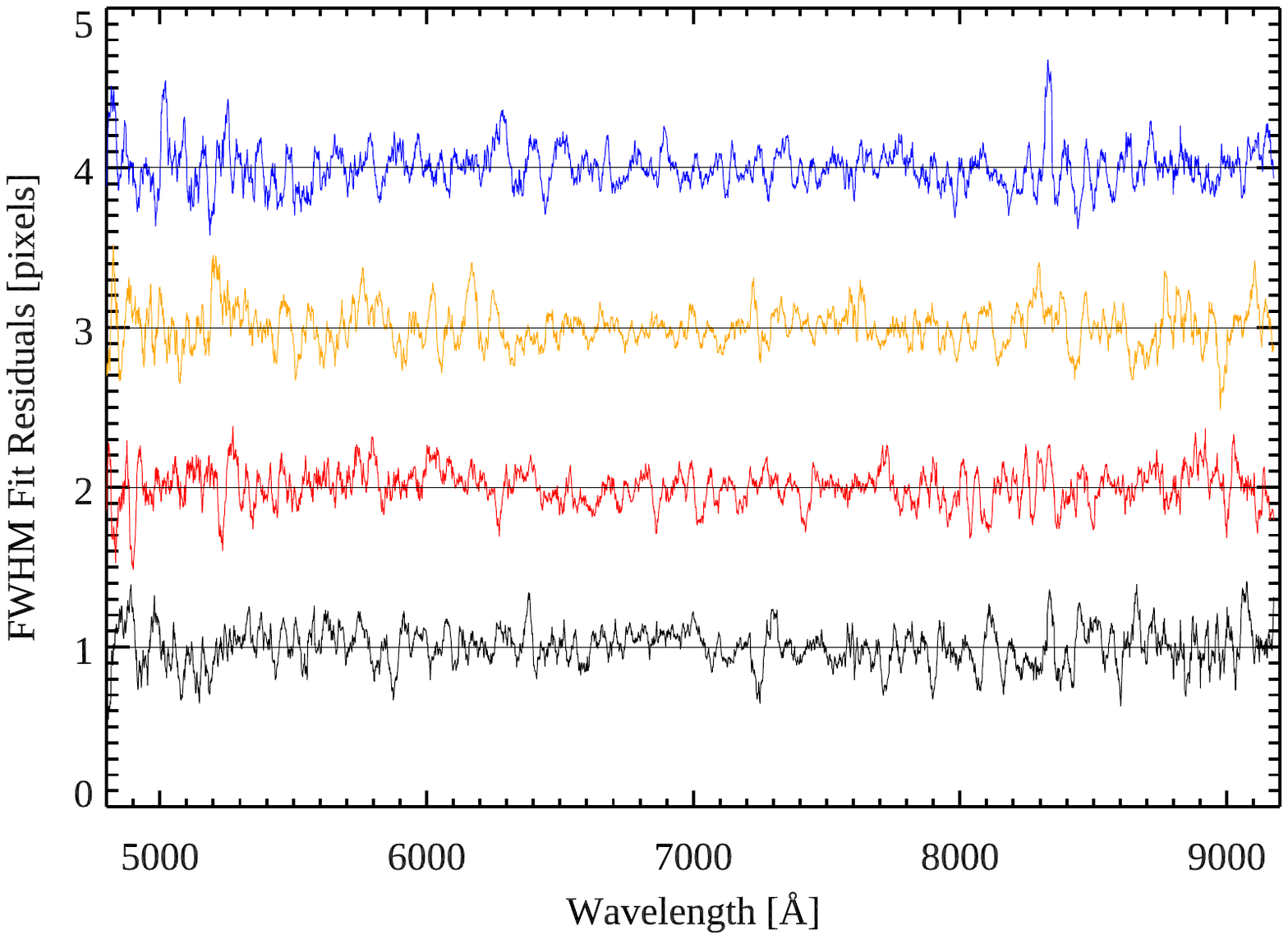} 
    }
    \end{minipage} 
     \end{minipage}  
   \caption{Examples of PSF measurements in the MUSE data cubes of NGC\,1380's CENTER (left column), MIDDLE (middle column), and HALO (right column) - see also Fig.~\ref{fig:NGC1380_pointings}.
    The first row shows continuum images  centered on the redshifted \OIII line, with a bandwidth of 200~\AA\null. Point sources 
    are indicated with circles and labeled S1, S2, etc. Radial plots for the brightest point source in each field, overplotted with a 
    Gaussian fit (blue) and a Moffat fit (red), are shown in the second row (ordinate flux units: $10^{-20}$\,erg cm$^{-2}$ s$^{-1}$). Row~3 presents the FWHM for all field stars that were bright enough to 
    yield a successful PSF fit, plotted as a function of wavelength across the data cube (ordinate units: pixels). The scatter of these values, shown as 
    residuals from second order polynomial fits (the full drawn lines in Row~3) are plotted in Row~4 (with incremental offsets of 1 for clarity). The difference in image quality from the
    CENTER (4.1 pixels FWHM, $0\farcs 82$), DISK (5.4 pixels FWHM, $1\farcs 08$), and HALO (5.1 pixels FWHM, $1\farcs 02$) is   
    obvious. Note that HALO star S1 is a resolved globular cluster and has a FWHM of 5.8 pixels, significantly larger than the other 
    objects in this field (black curve in Row~3, right column). This source cannot be used to determine an aperture correction.}
   \label{fig:PSFapcor}
\end{figure}

As pointed out by \citet{Kamann+13}, PSF fitting parameters are expected to vary smoothly with wavelength, with the FWHM monotonically decreasing towards the red.  We used this {\it a priori} knowledge to fit a second order polynomial to the measured FWHM as a function of wavelength; this model proved to be satisfactory even for very faint objects. The scatter of the residuals at the nominal wavelength of \OIII, Doppler-shifted to the systemic velocity of the galaxy in question, was taken as a measure for the uncertainty of the PSF determination. Finally, for the well-behaved point sources from this analysis, aperture photometry was performed within incrementing radii, typically ranging from 3 pixels up to 12 pixels, where, as before, the sky annulus was defined using inner and outer radii of 12 and 15 pixels, respectively. The difference between the flux at 3 pixels and the asymptotic value at large radii was adopted as the aperture correction for the given data cube. The formal error for this value was estimated from the standard deviation of the residuals, again from a polynomial fit, over an interval of $\pm 100$\,\AA\ around the redshifted \OIII line.

The corresponding curves for a number of stars superposed on NGC\,1380 are shown in Figure~\ref{fig:PSFapcor}. This approach is, unfortunately, sensitive to contaminants:  in NGC\,1380, some of the  ``stars'' are resolved on {\sl HST\/} frames and are listed by \citet{Jordan+15} as candidate globular clusters.  Thus, one must be cautious about using a blind analysis of field objects.  Without further information, such as high-resolution {\sl HST\/} imaging, a MUSE frame's PSF may be overestimated.  Object S1 in the HALO field of NGC\,1380 is actually a globular cluster, as it has a significantly larger FWHM than other sources in the field and is therefore unsuitable for measuring the frame's PSF\null.   We note that for future targeted PNLF observations with MUSE, the pointings should be planned to ensure the presence of PSF template stars in the field.

\begin{figure}[h!]
\begin{minipage}{1.0\linewidth}
    \centerline{
    \includegraphics[width=40mm,bb=0 -200 800 800,clip]{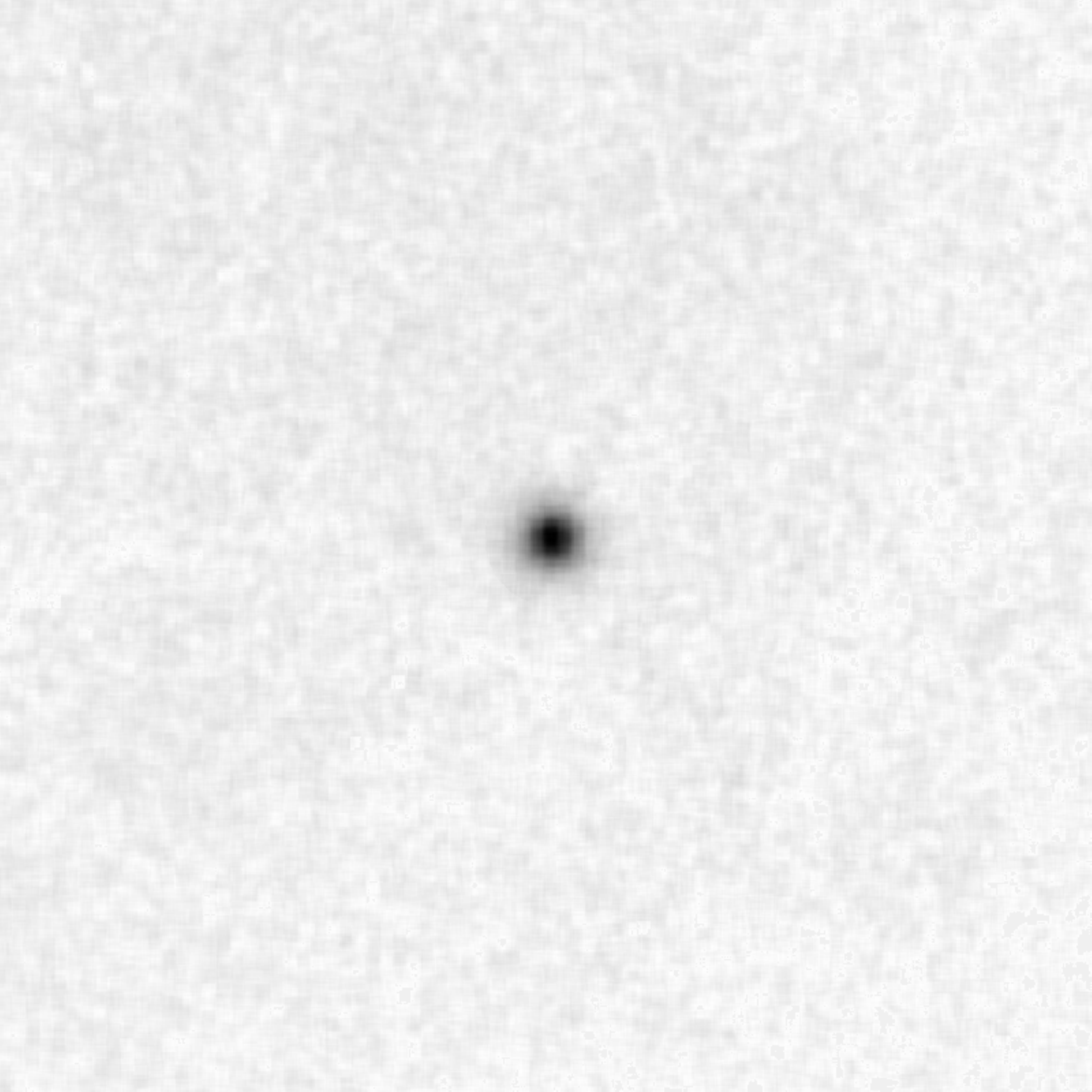}
    \hspace{10mm}
    \includegraphics[width=90mm,bb=0 350 600 695,clip]{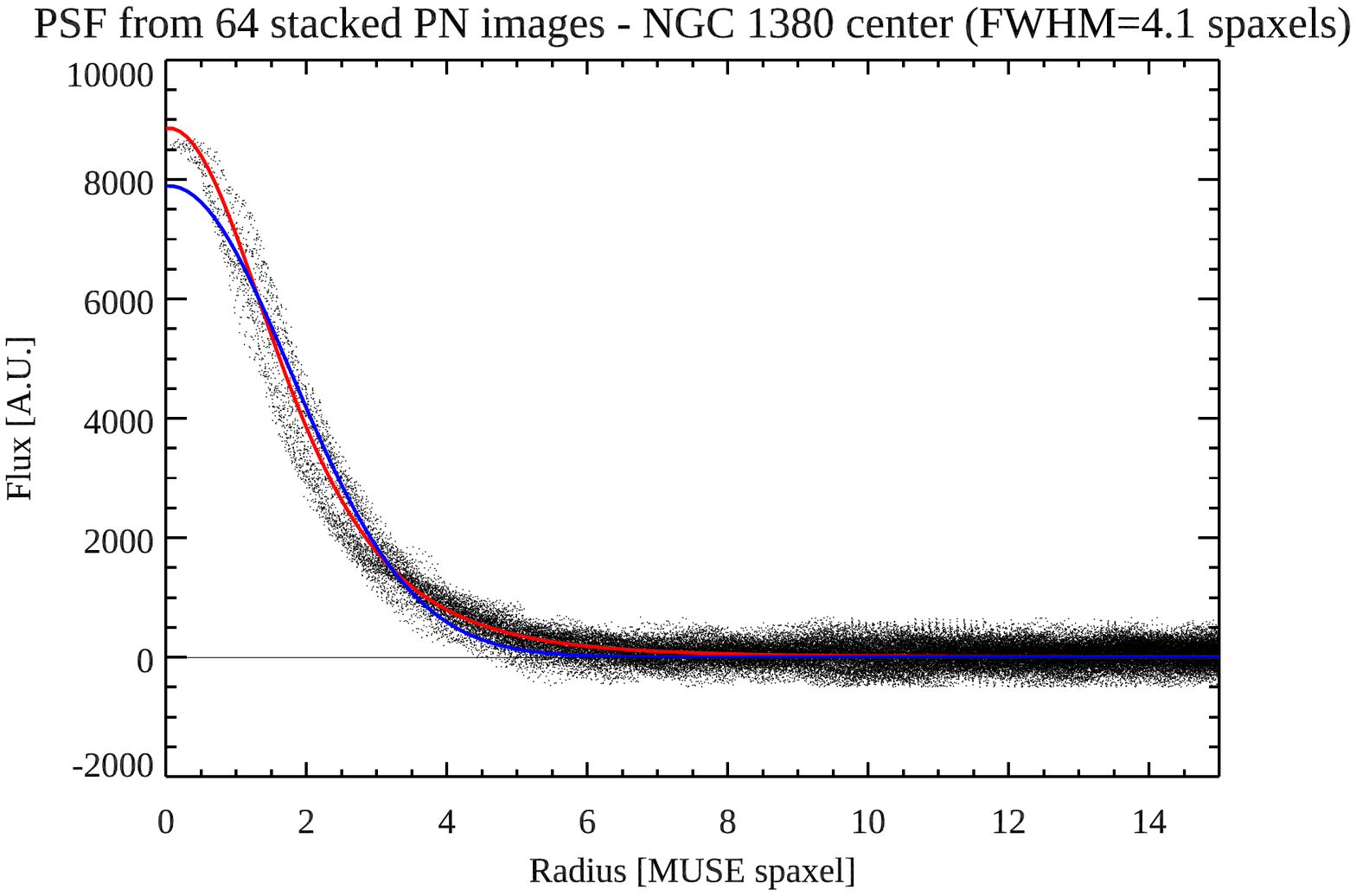} 
    }
    \end{minipage} 
   \caption{Left: image of a stack of the 64 brightest PNe in NGC\,1380 CENTER field. Right: radial plot of the PSF\null. A Gaussian fit to the PSF (shown in blue) yields a FWHM of 4.1~pixel, in agreement with the analysis of the field stars. For comparison, a Moffat fit is shown in red.}
 \label{fig:PSF_PN}
\end{figure}

In order to validate the stellar PSF determination as applicable for the PNe,  we stacked several tens of the brightest PN images from the 3-wavelength-bin series of frames. To account for the sub-pixel offset between the point source positions, the sub-images around each object were rescaled by a factor of 10 and shifted to a common centroid. This allowed for registration to a common center that is accurate to within one tenth of a pixel ($0\farcs 02$), i.e., small enough to have a negligible affect on the PSF and aperture correction.  As an example, Figure~\ref{fig:PSF_PN} shows the stacked PN images for the central field in NGC\,1380 along with a radial plot of the resulting PSF.  

\begin{figure}[t!]
\begin{minipage}{1.0\linewidth}
    \centerline{
    \includegraphics[width=9cm,bb=70 230  700 400,clip]{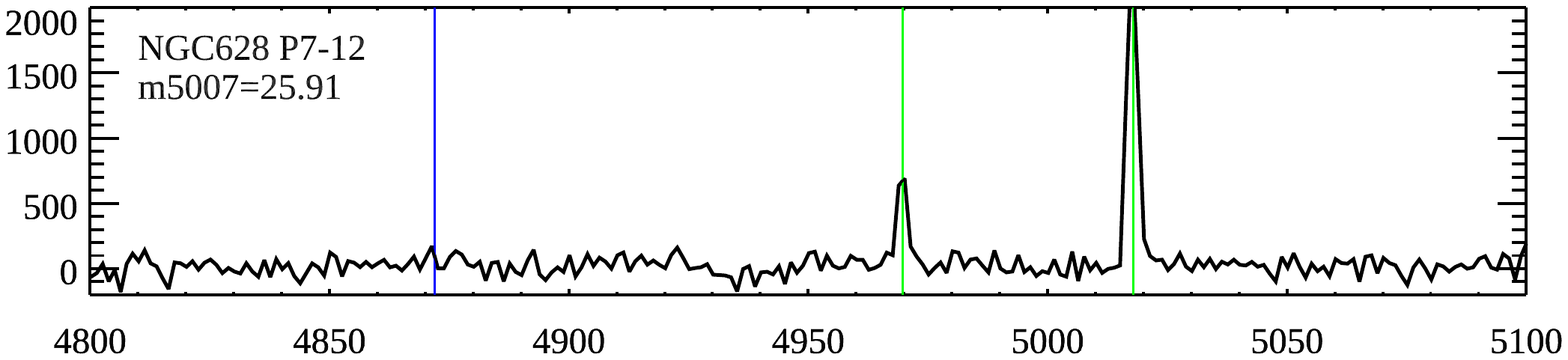}
    \includegraphics[width=9cm,bb=70 230  700 400,clip]{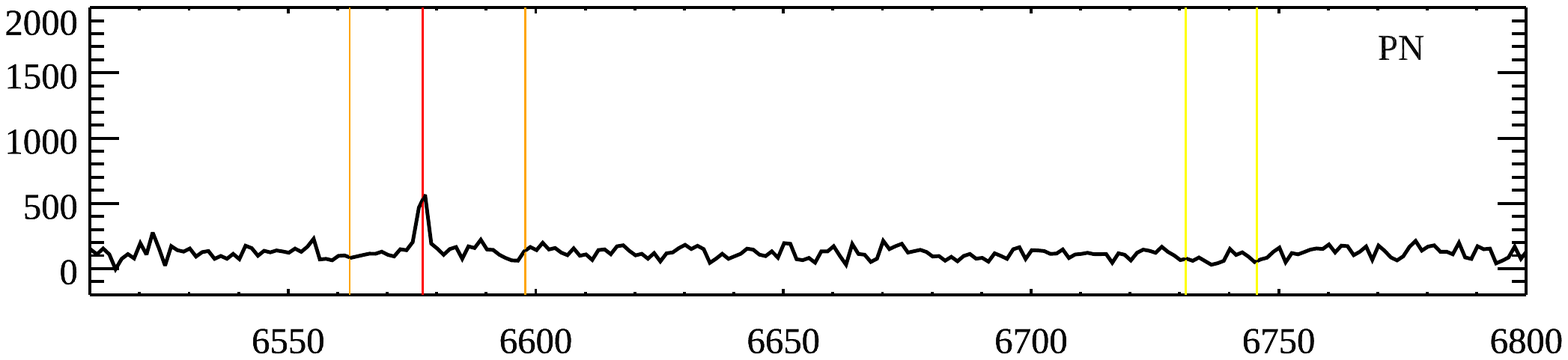}
    }
    \end{minipage}
\begin{minipage}{1.0\linewidth}
    \centerline{
    \includegraphics[width=9cm,bb=70 230  700 380,clip]{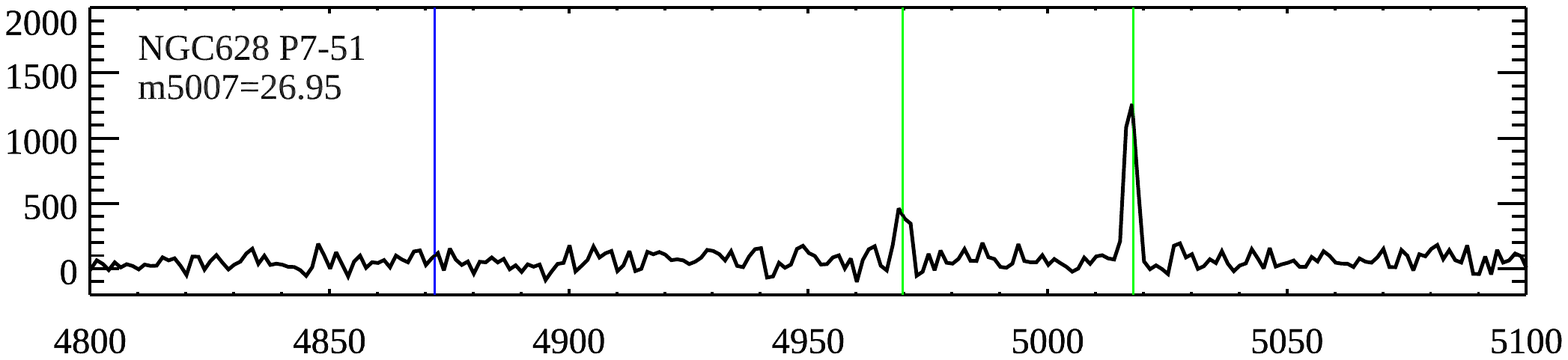}
    \includegraphics[width=9cm,bb=70 230  700 380,clip]{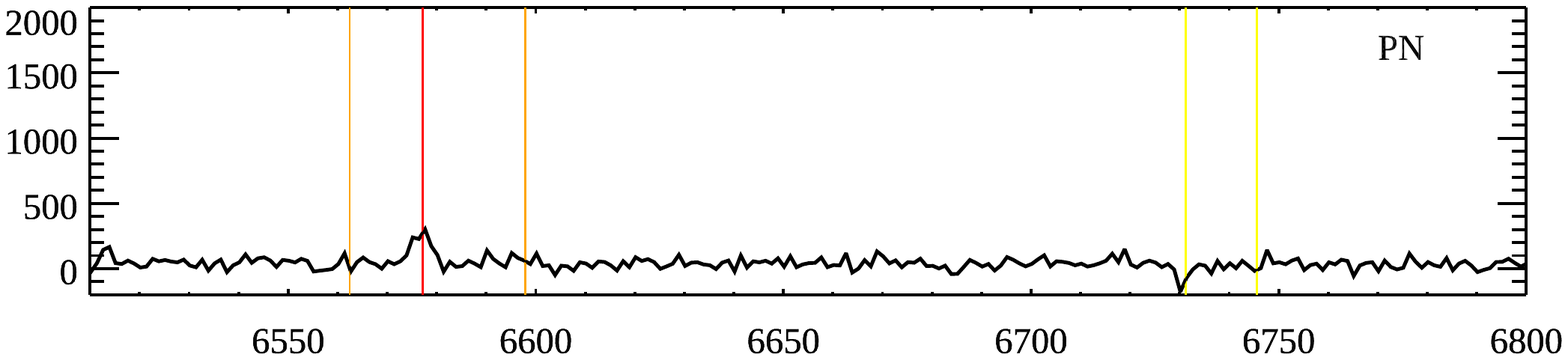}
    }
    \end{minipage}
\begin{minipage}{1.0\linewidth}
    \centerline{
    \includegraphics[width=9cm,bb=70 230  700 380,clip]{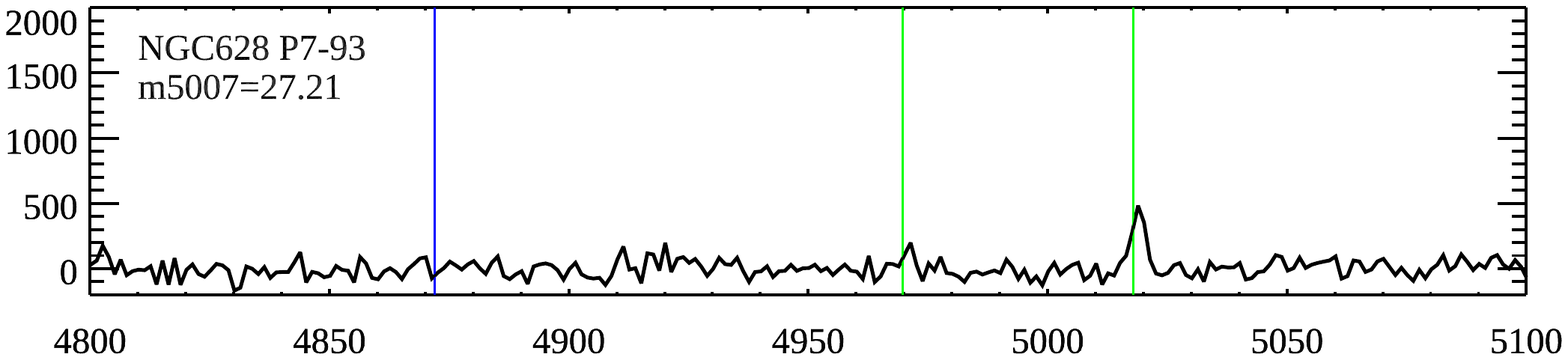}
    \includegraphics[width=9cm,bb=70 230  700 380,clip]{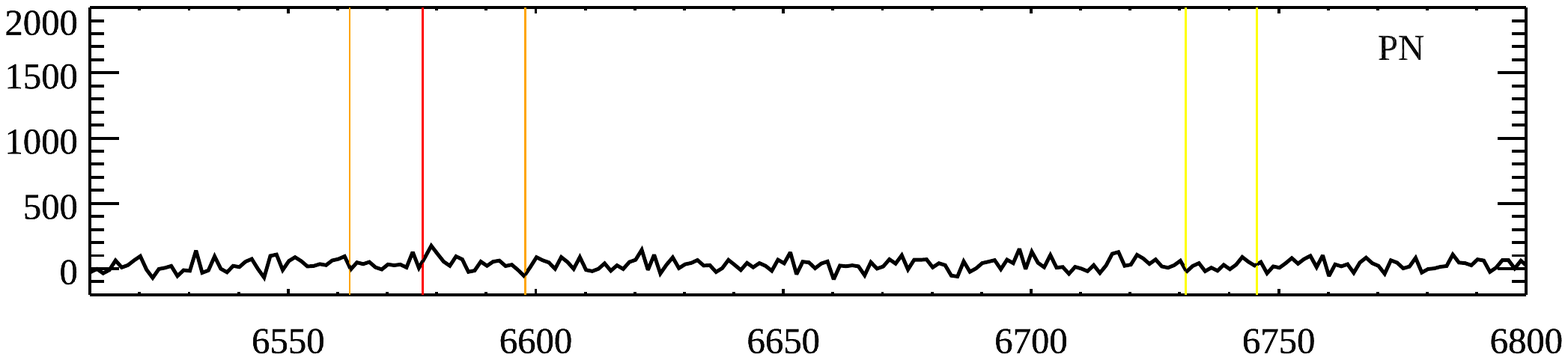}
    }
    \end{minipage}
\begin{minipage}{1.0\linewidth}
    \centerline{
    \includegraphics[width=9cm,bb=70 230  700 380,clip]{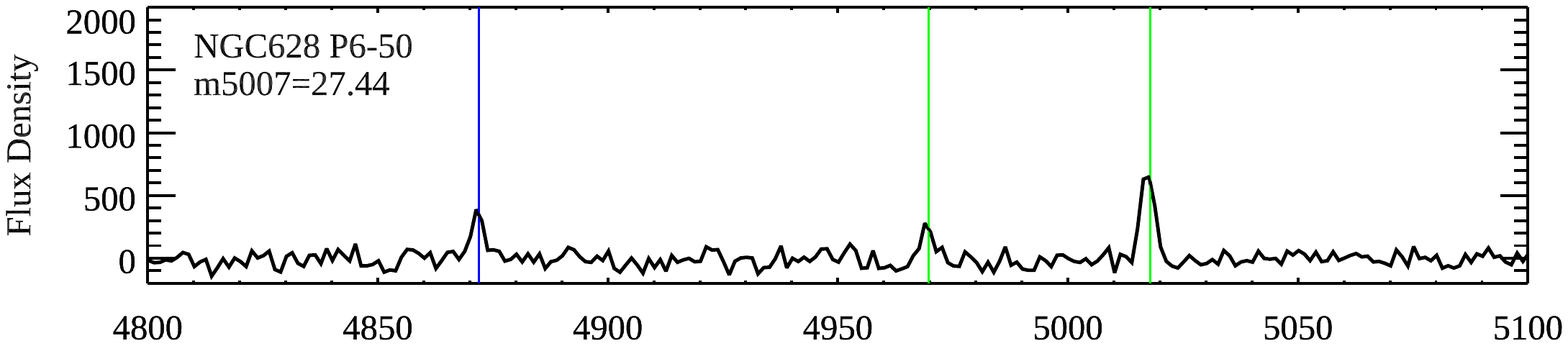}
    \includegraphics[width=9cm,bb=70 230  700 380,clip]{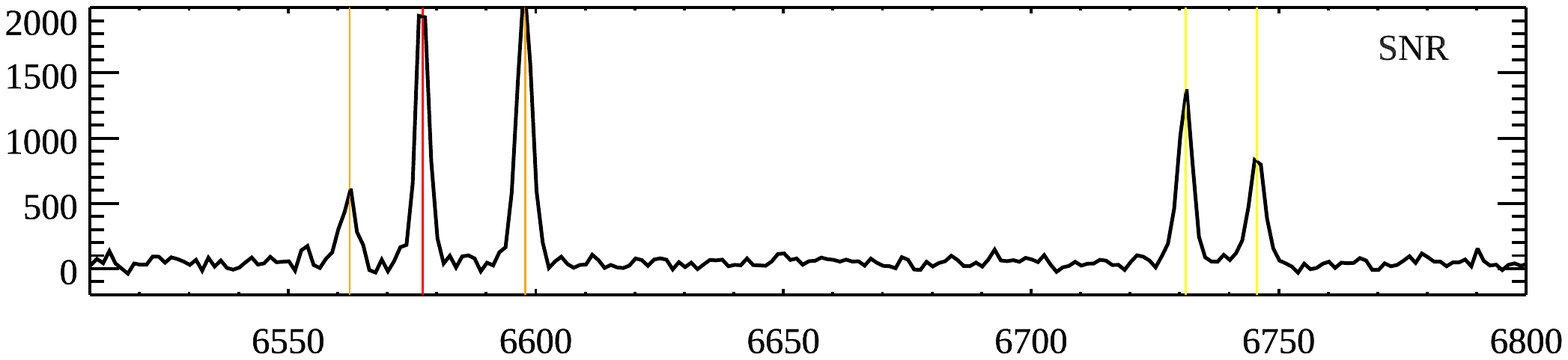}
    }
    \end{minipage}
\begin{minipage}{1.0\linewidth}
    \centerline{
    \includegraphics[width=9cm,bb=70 230  700 380,clip]{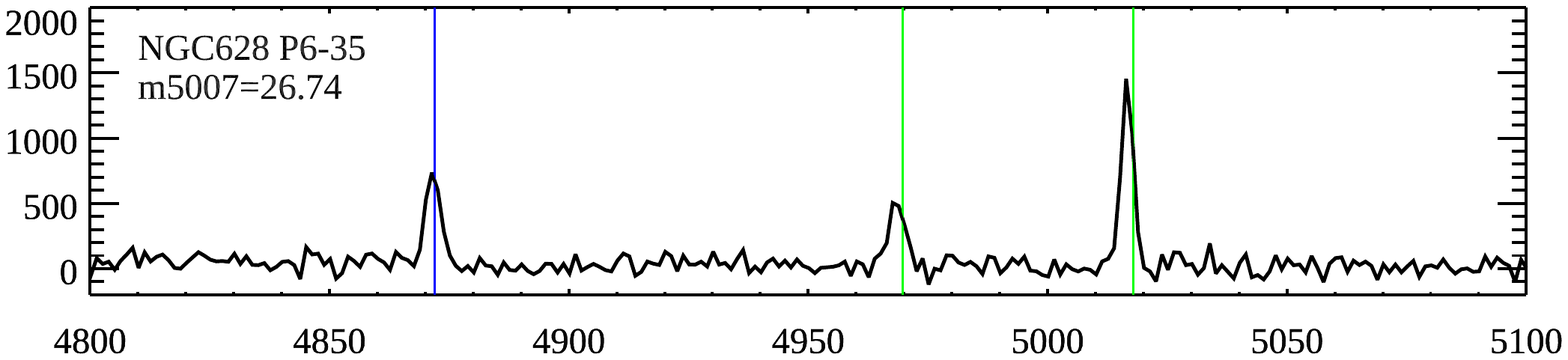}
    \includegraphics[width=9cm,bb=70 230  700 380,clip]{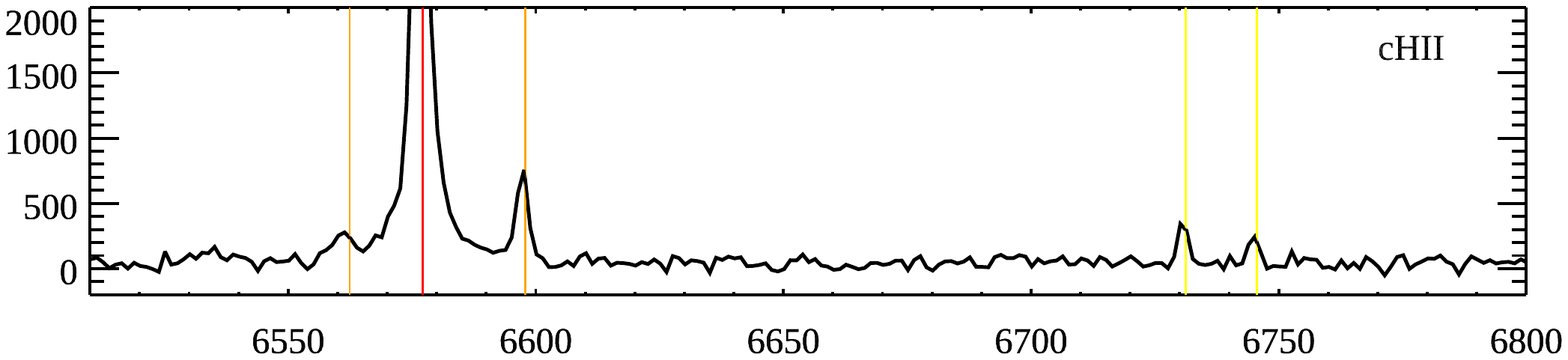}
    }
    \end{minipage}
\begin{minipage}{1.0\linewidth}
    \centerline{
    \includegraphics[width=9cm,bb=70 230  700 380,clip]{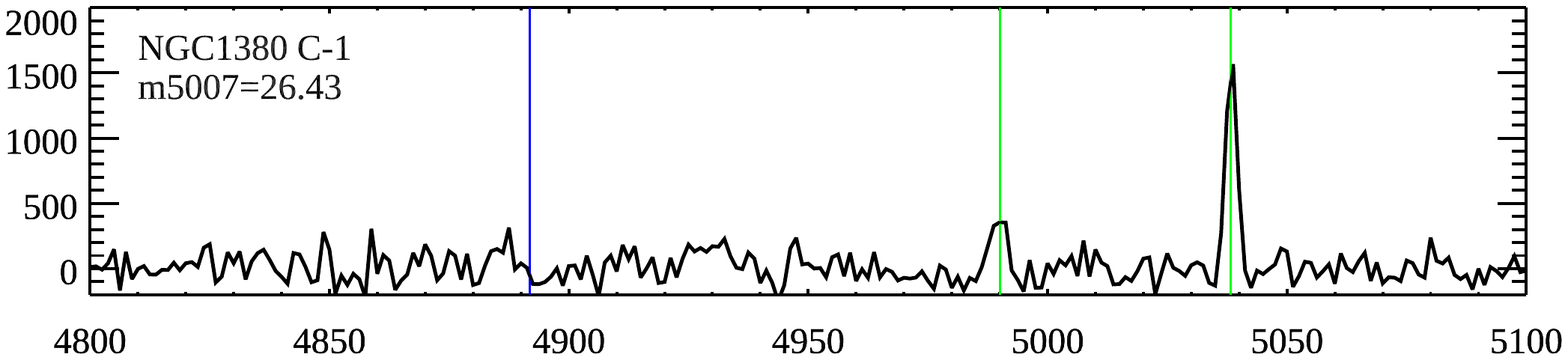}
    \includegraphics[width=9cm,bb=70 230  700 380,clip]{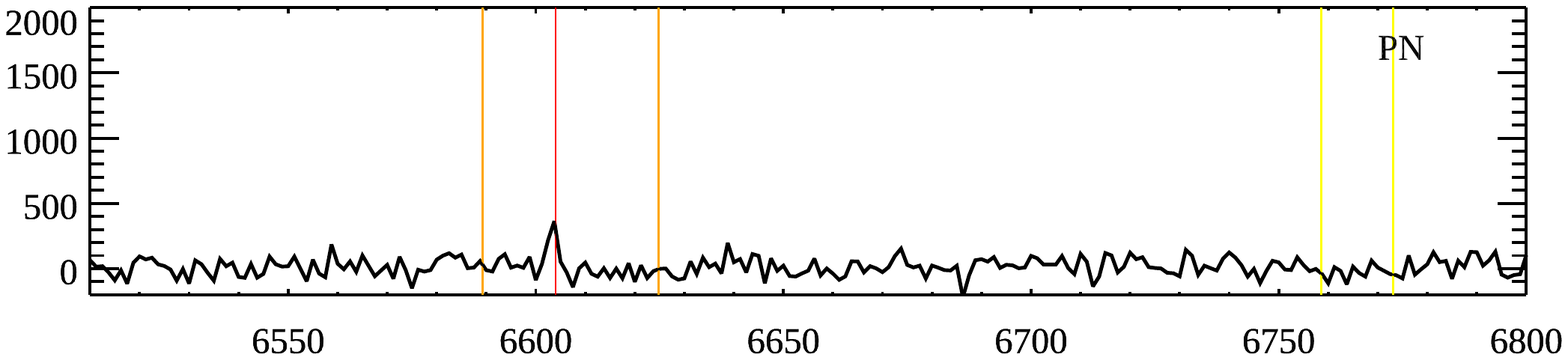}
    }
    \end{minipage}
\begin{minipage}{1.0\linewidth}
    \centerline{
    \includegraphics[width=9cm,bb=70 200  700
    380,clip]{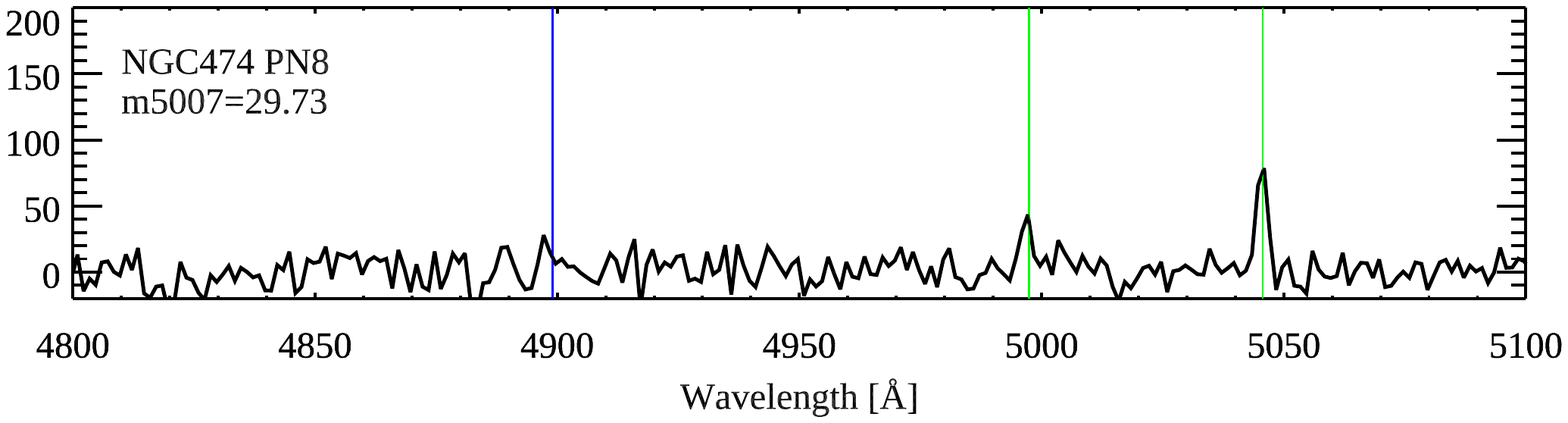}
    \includegraphics[width=9cm,bb=70 200  700 380,clip]{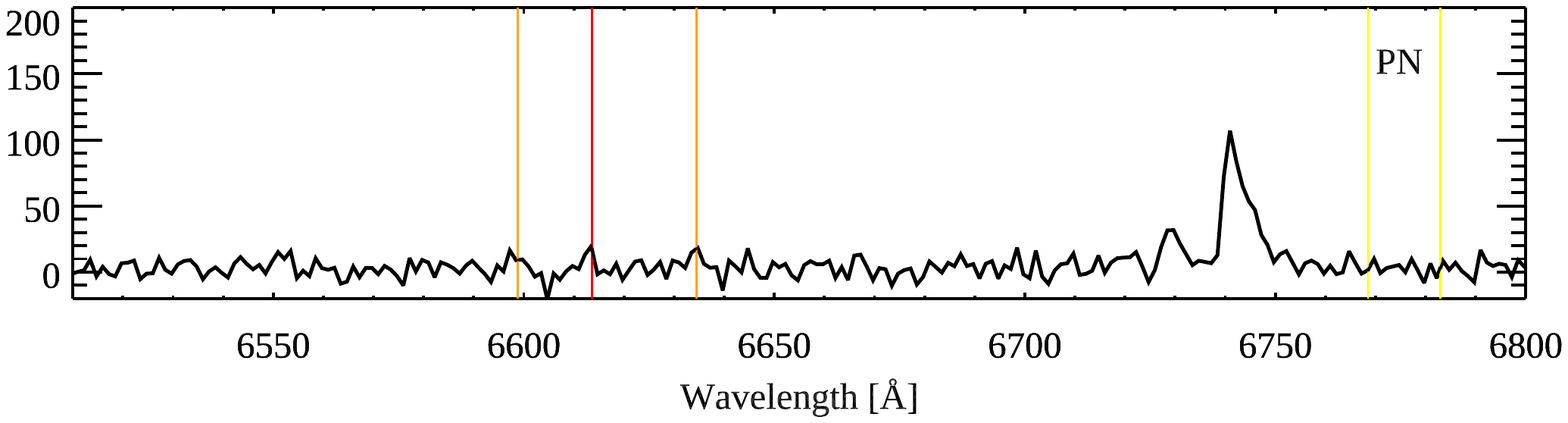}
    }
    \end{minipage}
\caption{Example spectra, illustrating the range of brightness, and different classes of emission line objects. Flux density is plotted in units of $10^{-20}$\,erg\,cm$^{-2}$\,s$^{-1}$\,\AA$^{-1}$. Note that for the faintest object (Row 7) the scale is stretched by a factor of 10. The wavelengths of H$\beta$, \OIII, H$\alpha$, \NII, and \SII, shifted to the systemic velocity of the host galaxy, are indicated with colored lines. The spectrum of the overluminous PN C-1 in NGC\,1380, shown in Row 6, presents no peculiarities, such as split or broadened emission lines that otherwise might indicate a blend of two objects. As a curiosity, the spectrum of PN8 in NGC\,474 (Row 7) shows a broad, asymmetric emission line at a wavelength of $\sim 6740$\,\AA. The corresponding datacube layers reveal a chance alignment of PN8 with one of two Ly$\alpha$ emitting galaxies at a redshift of $z=4.551$ (see Fig.~\ref{fig:NGC474_off_diff}).
 \label{fig:PN_HII_SNR_spectra}}
\end{figure}

\subsection{Spectroscopy}
\label{subsec:spectro}

To confirm a PN candidate, it is necessary (though not sufficient) to detect a point source at the wavelength of the \OIII $\lambda 5007$ emission line, Doppler-shifted to the systemic velocity of the galaxy, and allowing for orbital motion within the system's gravitational potential. To rule out interlopers such as high redshift background galaxies, one generally needs to detect another emission line at the correct wavelength, typically  \OIII $\lambda 4959$, or H${\alpha}$. Also, depending on the Hubble type of the host galaxy, it may be necessary to distinguish PNe from supernova remnants or \ion{H}{2} regions --- a task that is particularly critical in late type galaxies. 

For this purpose, we extracted the spectrum of each PN candidate by performing DAOPHOT aperture photometry in each layer of the data cube, using the same procedure as for the determination of $m_{5007}$ magnitudes, including  centroiding the line and applying (wavelength-dependent) aperture corrections. To this end, we expanded the scope of the DELF to the entire wavelength range of the data cube. This version of our code samples the continuum underlying a targeted emission line using two wavelength intervals which bracket the line of interest.  This option is illustrated by highlighting the data cube layers of Fig.~\ref{fig:NGC1380_on_off_bands} in blue and red hues. We note that bright night sky emission lines in the regions which define the continuum can create a bias and must therefore be masked from the analysis.  This feature has not yet be implemented in our software.   However, since the current study does not extend to wavelengths beyond $\sim$\,7500\,\AA, masking was not necessary for our analysis and satisfactory results were found with the former version of the code. In problematic cases, e.g., for objects located near the edge of the field, we were able to measure spectra interactively using the P3D visualization tool\footnote{\url{https://p3d.sourceforge.io/}}.  This program allows the user to select individual spaxels to represent objects and background by defining arbitrary geometries in the data cube \citep{Sandin+10}. 

Figure~\ref{fig:PN_HII_SNR_spectra} presents a few representative examples of the emission line sources detected by our analysis, including PNe of different brightness,  a supernova remnant, a compact \ion{H}{2} region, and a high redshift interloper.  


\subsection{Classification}
\label{subsec:classification}
The confirmation of PN candidates as true planetaries is an important task, as background galaxies, supernova remnants, compact \ion{H}{2} regions, and even some AGN may contaminate a sample of PN candidates.  If left unrecognized, these objects can distort the bright-end cutoff of the PNLF and cause an underestimate of a galaxy's distance.   While the (rare) cases of single emission line interlopers are easily flagged owing to the absence of \OIII $\lambda 4959$ and/or H$\alpha$, the spectrum of a SNR or a compact \ion{H}{2} region is not necessarily that different from that of a PN\null. Furthermore, spatially overlapping PNe, which may sometimes appear as a single overly bright object in narrow-band images can be deblended spectroscopically, only if the radial velocities of the two objects differ by more than $\sim 100$~km\,s$^{-1}$.

\begin{figure}[h!]
    \begin{minipage}{1.0\linewidth}
    \includegraphics[width=90mm,bb=120 30  600 530,clip]{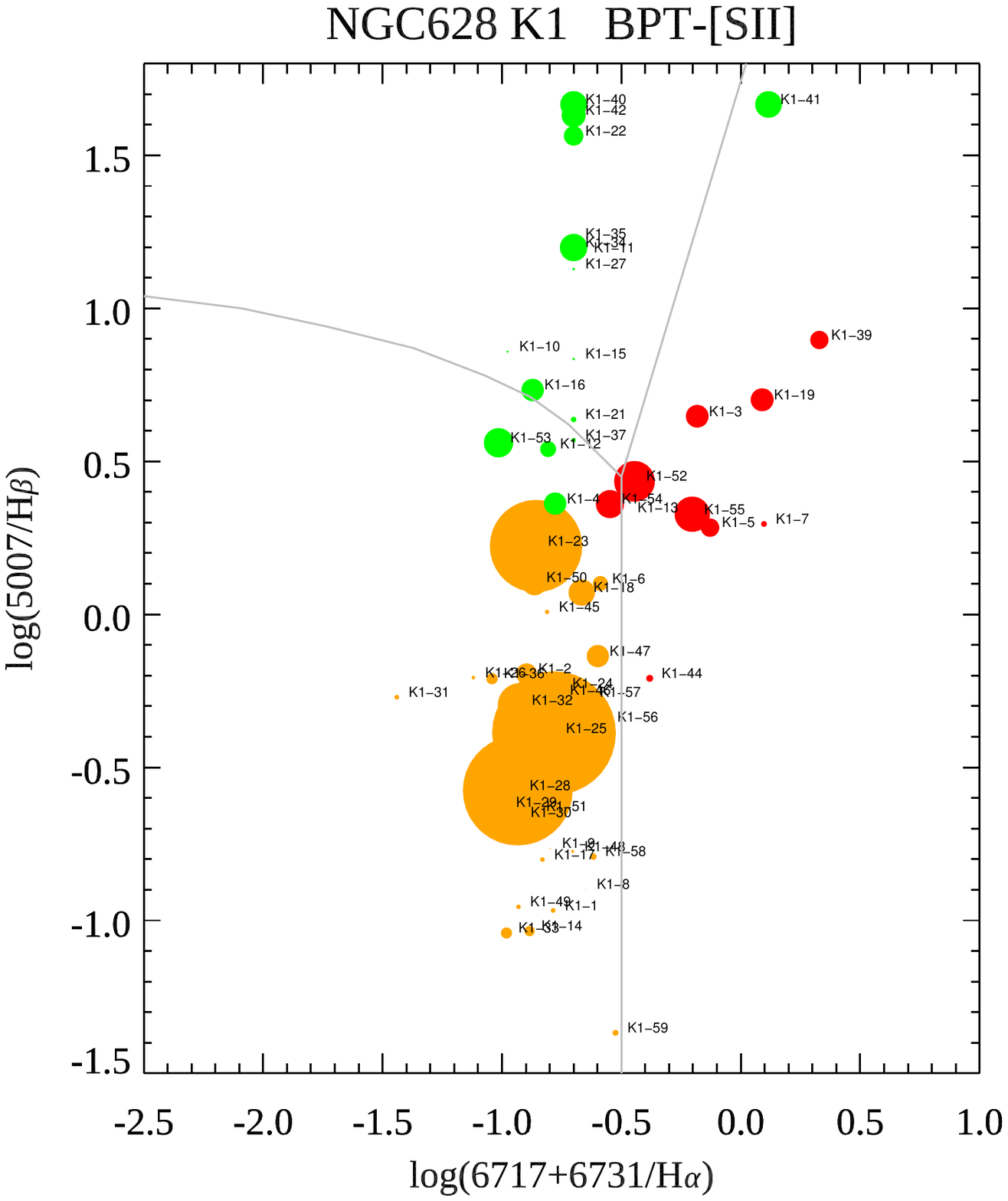}
    \includegraphics[width=90mm,bb=120 30  600 530,clip]{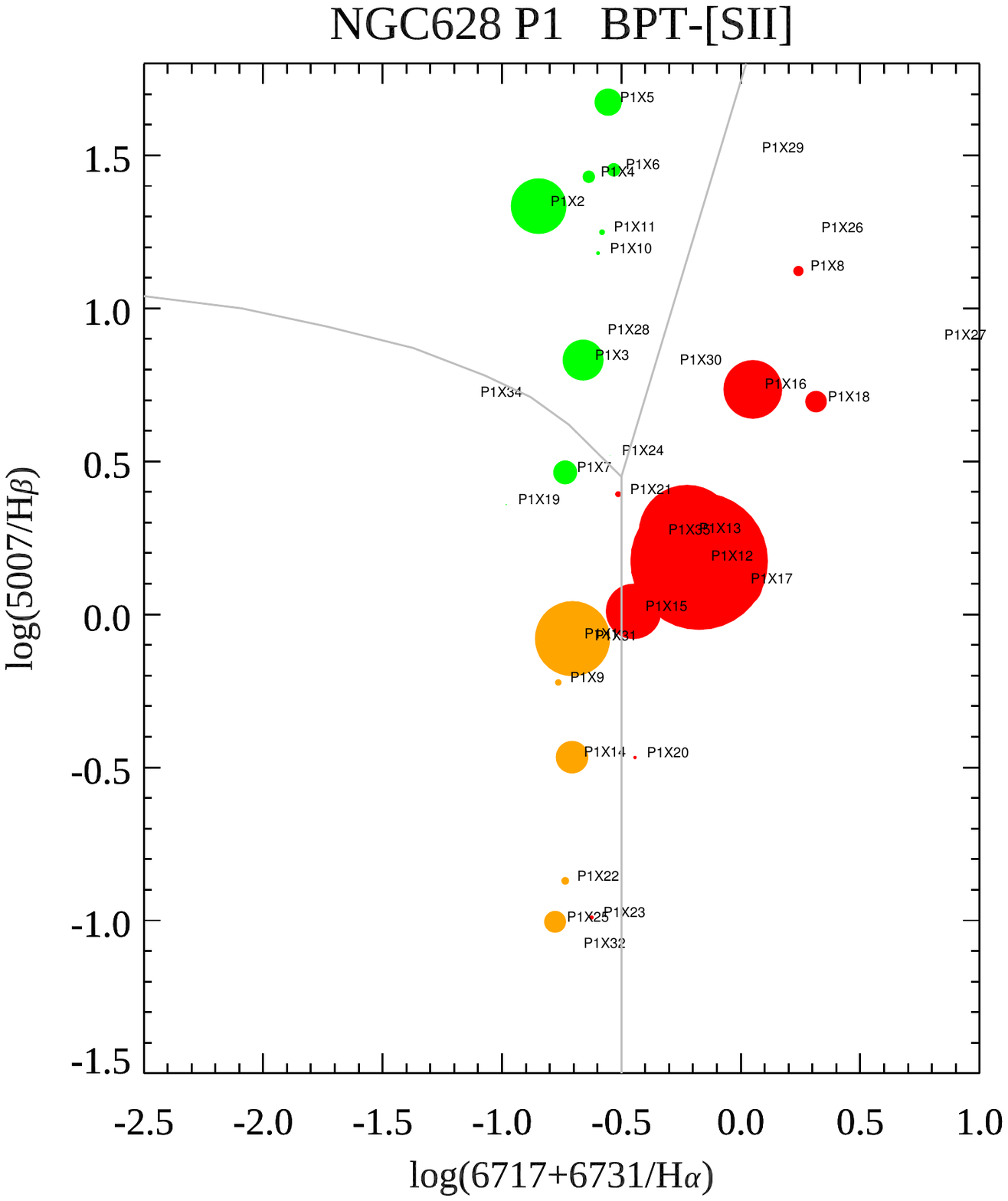} 
     \end{minipage}  
   \caption{BPT-diagrams \citep{Baldwin+81} for pointings K1 (left), and P1 (right)in NGC\,628. The green circles are PNe, orange indicates \ion{H}{2} regions, and red represent supernova remnants. The symbol sizes indicate the $m_{5007}$ magnitude, with larger circles representing brighter objects.}
   \label{fig:BPT}
\end{figure}

To distinguish PNe from \ion{H}{2} regions and SNRs, we have adopted the classification scheme of \citet{Frew+10}, who plotted the \OIII/H$\beta$ line ratio against \SII/H$\alpha$, where \SII refers to the sum of the \SII doublet $\lambda\lambda$\,6717, 6731\,\AA\null. This adaptation of the  ``BPT-diagram'' \citep{Baldwin+81} has the advantage of being insensitive to dust extinction, while still being very effective at discriminating various classes of emission-lines objects. 

Figure~\ref{fig:BPT} shows our BPT diagram for two fields in the face-on spiral galaxy NGC\,628.  The curves show the regions occupied by PNe, \ion{H}{2} regions, and supernova remnants; three very bright \ion{H}{2} regions in the K1 field have been deliberately included for illustration.  In the figure, the symbol sizes are directly proportional to the luminosity of the \OIII\ $\lambda 5007$ line.  We note that while the dividing line between PNe and SNR is rather well-defined and based on empirical data \citep{Sabin+13}, the separation between PNe and \ion{H}{2} regions as defined by \citet{Kewley+01} is less clear.  Although there is a sharp upper limit to the \OIII/H$\beta$ line ratio of \ion{H}{2} regions, a considerable number of Galactic PNe fall below this line \citep{Frew+10}.  However, the \citet{Frew+10} study included PNe of all luminosities, while the line ratios of PNe in the brightest $\sim 1$~mag of the PNLF are considerably more homogenous \citep{Ciardullo+02, Richer+08}. This fact is illustrated in the figure:  objects with \OIII\ luminosities significantly brighter than the PNLF cutoff fall securely in the \ion{H}{2} region area of the diagram.  In contrast, the cluster of objects identified as PNe all have \OIII/H$\beta$ ratios above 5.  Nevertheless, because of the PN/\ion{H}{2} region ambiguity, we can create two versions of the PNLF, one with, and another without the borderline cases.  A comparison of the resultant two distances would yield a systematic component in the uncertainty in the corresponding distance determination.  

\pagebreak

\subsection{Tests}
\label{subsec:tests}
To convince ourselves that the photometry on DELF processed data cubes delivers the expected accuracy, we conducted several internal and external tests, including embedding artificial PNe of known fluxes into the observed data cubes. These tests allowed us to search for systematic photometric errors or inaccurate error estimates that could enter the PNLF distance determination algorithm, and potentially produce biased distances.  

\pagebreak

\subsubsection{Differential emission line filtering}
\label{subsec:DELFtest}
\begin{figure}[h!]
\begin{minipage}{1.0\linewidth}
    \centerline{
    \includegraphics[width=90mm,bb=50 50  740 480,clip]{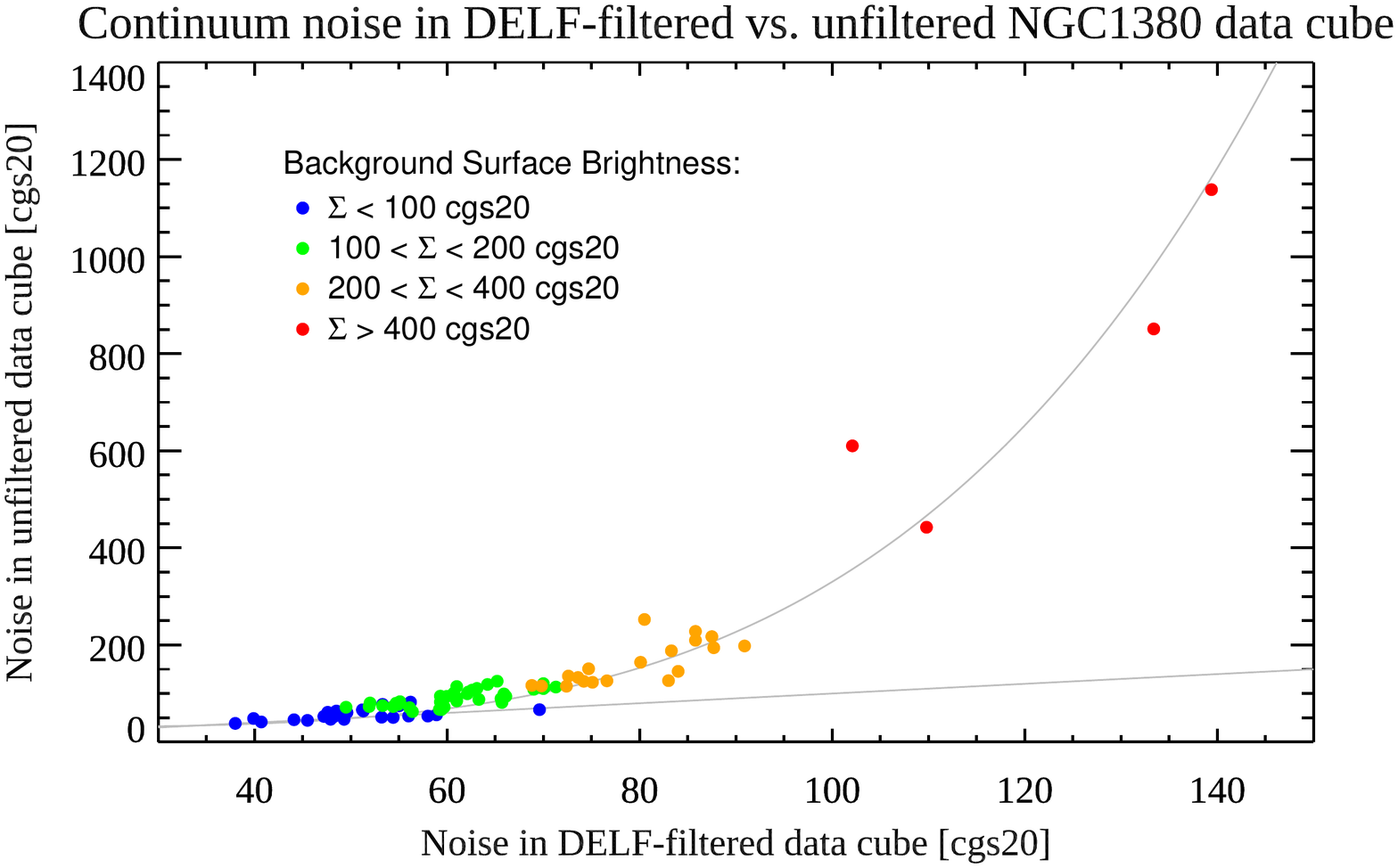}
    \includegraphics[width=90mm,bb=50 50  740 480,clip]{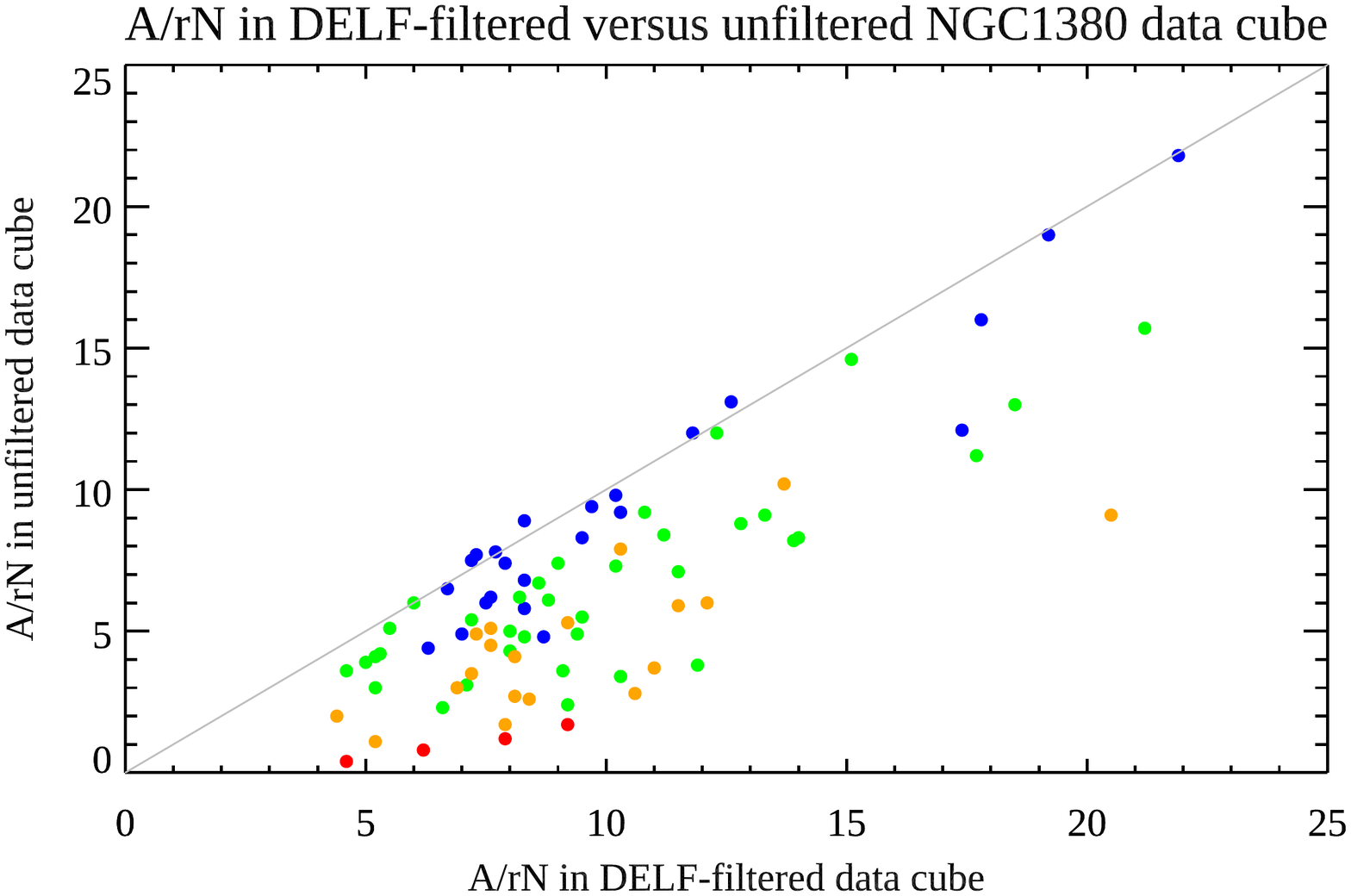}
    }
    \end{minipage} 
   \caption{Test of DELF photometry on PNe projected near the center of NGC\,1380.
   The objects are located in regions with a variety of surface brightnesses, with blue representing $\Sigma < 100$, green $100 \leq \Sigma < 200$, orange $200 \leq \Sigma < 400$, and red $\Sigma \geq 400$, in units of $10^{-20}$ erg\,cm$^{-2}$\,s$^{-1}$\,\AA$^{-1}$\,[cgs20], as integrated over the aperture.  Left: comparison of continuum noise of the unfiltered data cube versus DELF-filtered cube.   Right: amplitude/residual noise ratio (ArN) for PNe measured in the unfiltered data cube versus the DELF-filtered cube.}
 \label{fig:ArN}
\end{figure}

We validated the performance of the DELF technique using data cubes centered on the nuclei of galaxies.  These regions have a wide range of continuum surface brightnesses, and are therefore excellent locations for testing the efficiency of our photometric techniques.  NGC\,1380 is a good place to begin as we can compare our own data with signal-to-noise values for PNe published by Sp2020. Figure~\ref{fig:ArN} presents the results of this analysis. 

On the left, the graph shows the residual continuum noise in the wavelength interval $5100 \leq \lambda \leq 5500$~\AA, measured from PNe spectra that were extracted with DAOPHOT using an aperture radius of 3 pixels and a sky annulus between 12 and 15 pixels.  The noise is plotted in units of $10^{-20}$ erg\,cm$^{-2}$\,s$^{-1}$\,\AA$^{-1}$ for 104 PN candidates of all brightnesses measured on both the unfiltered data cube and the DELF-processed data cube. The objects are located in regions with a variety of surface brightness.  As expected, the residual continuum noise is clearly correlated with the background surface brightness. In regions of low surface brightness, the unfiltered and filtered noise levels are the same (the grey 1:1 line).  However, the unfiltered noise increases rapidly as the background brightens (grey curve, representing a fourth-order polynomial).  The noise levels for the DELF measurements are never worse than those for the unfiltered data, and at high surface brightness, they are far superior.  We conclude that the MUSE-specific on-band/off-band technique is the preferred choice for further processing. 

This is confirmed by comparing the signal-to-noise ratio of PNe magnitudes with and without filtering. We have chosen to use the quantity A/rN introduced by Sp2020, which is the ratio of the amplitude of their simultaneous fit of \OIII in the spatial and spectral domains over the residual noise of the fit. We compute the same ratio in our background-subtracted spectra by fitting a Gaussian to the \OIII $\lambda 5007$ line and measuring the noise, again in the wavelength interval $5100 \leq \lambda \leq 5500$\,\AA\null. The plot on the right in Fig.~\ref{fig:ArN} shows A/rN for the unfiltered data versus A/rN with the filter.  The color coding is the same as for the previous test. Sp2020 considered objects measured with an A/rN below 3 to be uncertain. Our plot reveals that a considerable fraction of measurements without the filter fall below this threshold. The same measurements with the DELF technique lie well above the Sp2020 threshold. The correlation with background surface brightness is again evident. 

Additional plots which directly compare our A/rN and magnitude measurements with those of Sp2020 are presented in Section~\ref{subsec:results_NGC1380}. Based on these three analyses, we conclude that DELF is indeed an efficient tool for suppressing systematic flatfielding errors in MUSE  emission line spectrophotometry. While this is especially important in the inner regions of galaxies where PNe are plentiful and the continuum surface brightness is high, DELF offers significant improvement even when the background surface brightness is low, as the photometric uncertainties will still be dominated by fixed pattern flat-fielding errors. 

\subsubsection{Photometric tests using artificial PN images}
\label{subsec:photomock}
In order to check the validity of our photometry, we performed tests with artificial emission line point sources embedded in the original MUSE data cubes with {\it a priori\/} known positions, fluxes, and radial velocities. Running the algorithms on the original data with these mock PNe allows us to assess random and systematic errors for our photometry and spectroscopy, and determine how the detection limit for a given cube depends on the continuum surface brightness across the galaxy.

\begin{figure}[h!]
    \begin{minipage}{1.0\linewidth}
    \centerline{
    \includegraphics[width=70mm]{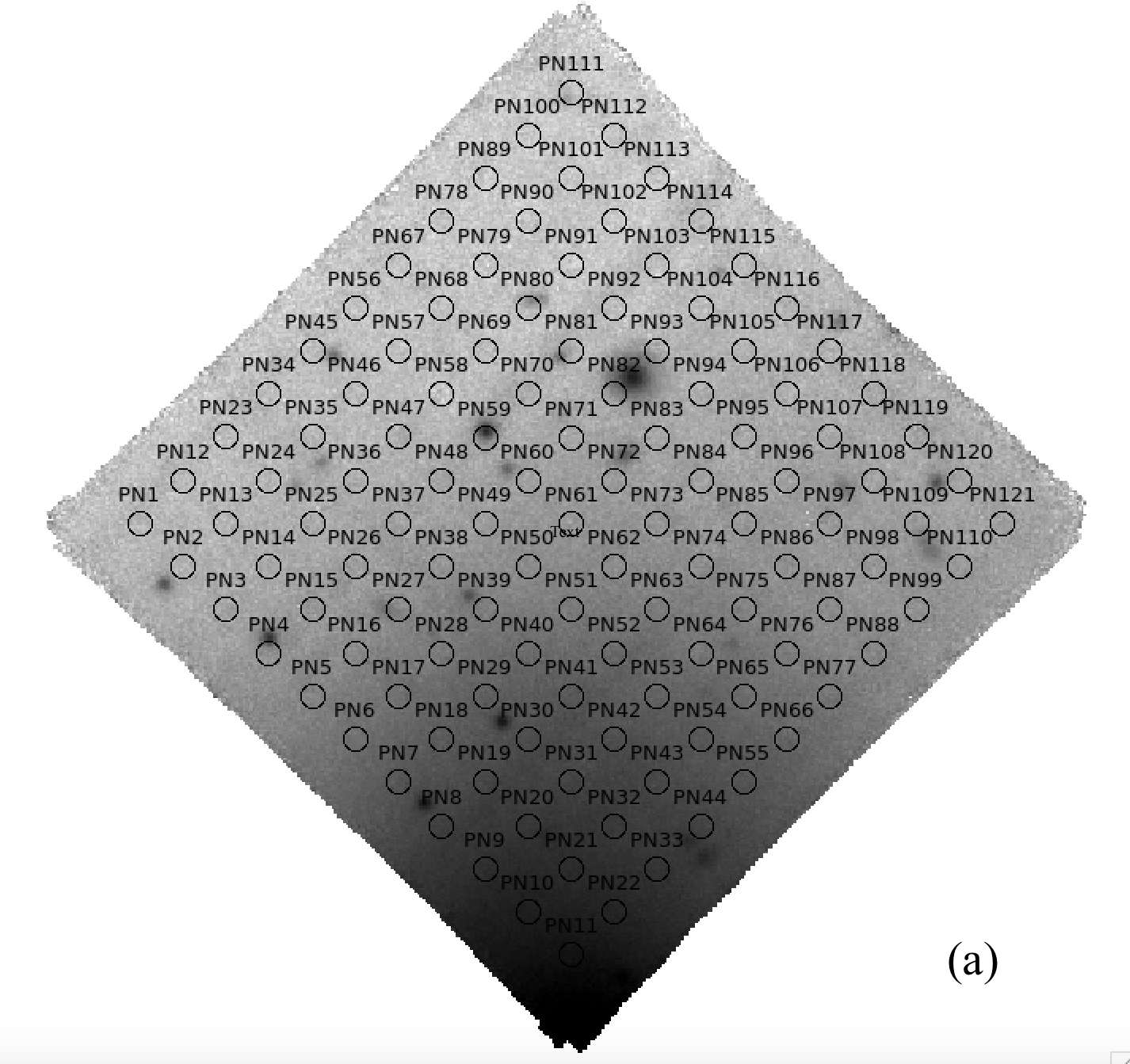}
    \includegraphics[width=67mm]{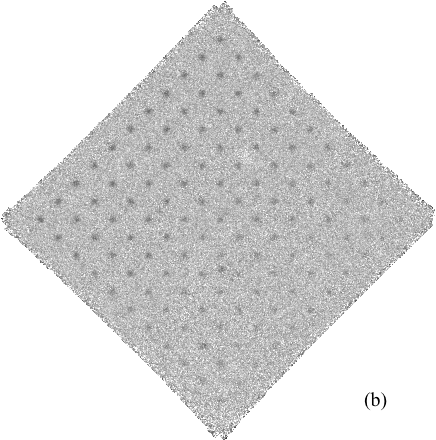}  
    }
    \end{minipage}  
    \begin{minipage}{1.0\linewidth}
    \centerline{
    \includegraphics[width=70mm]{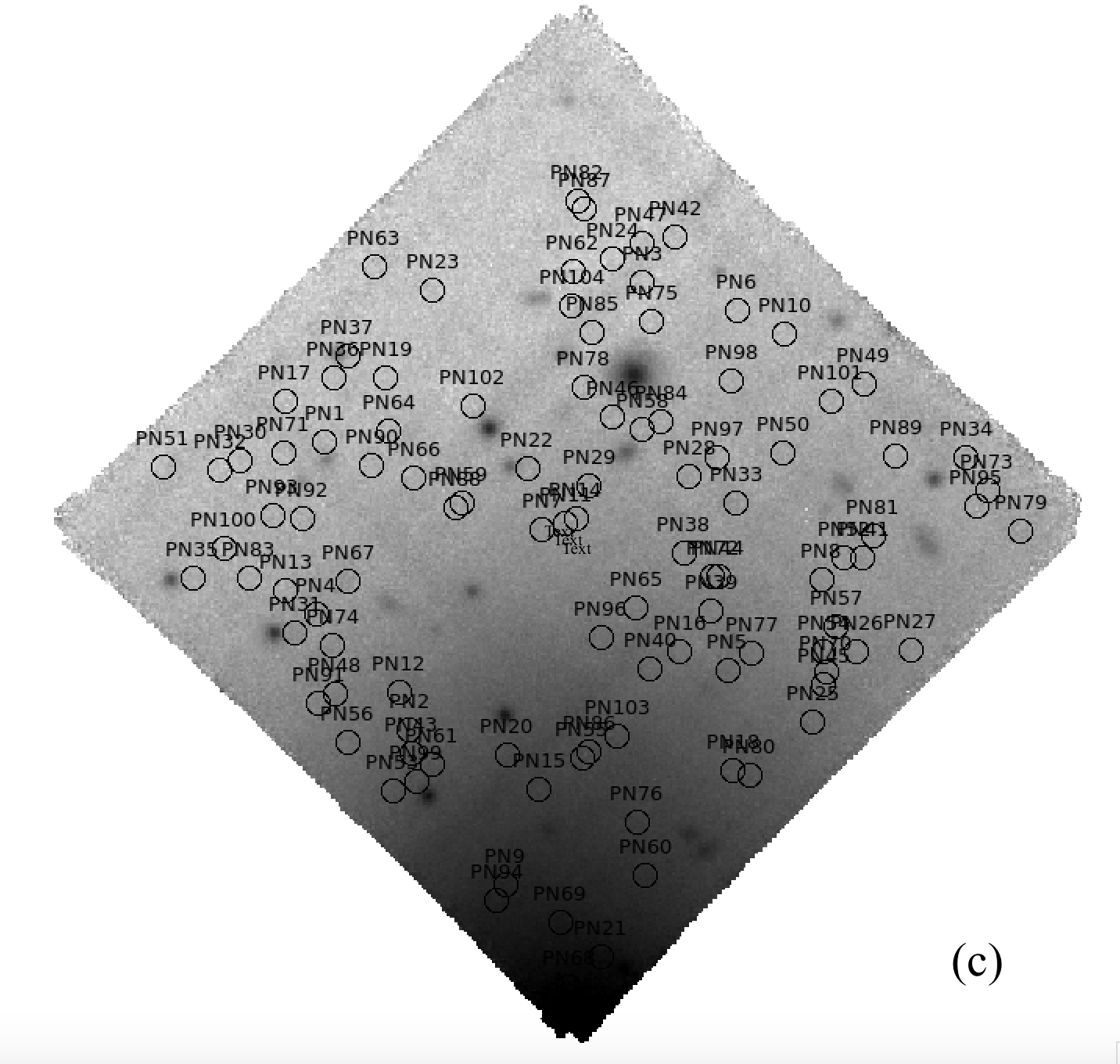}
    \includegraphics[width=70mm]{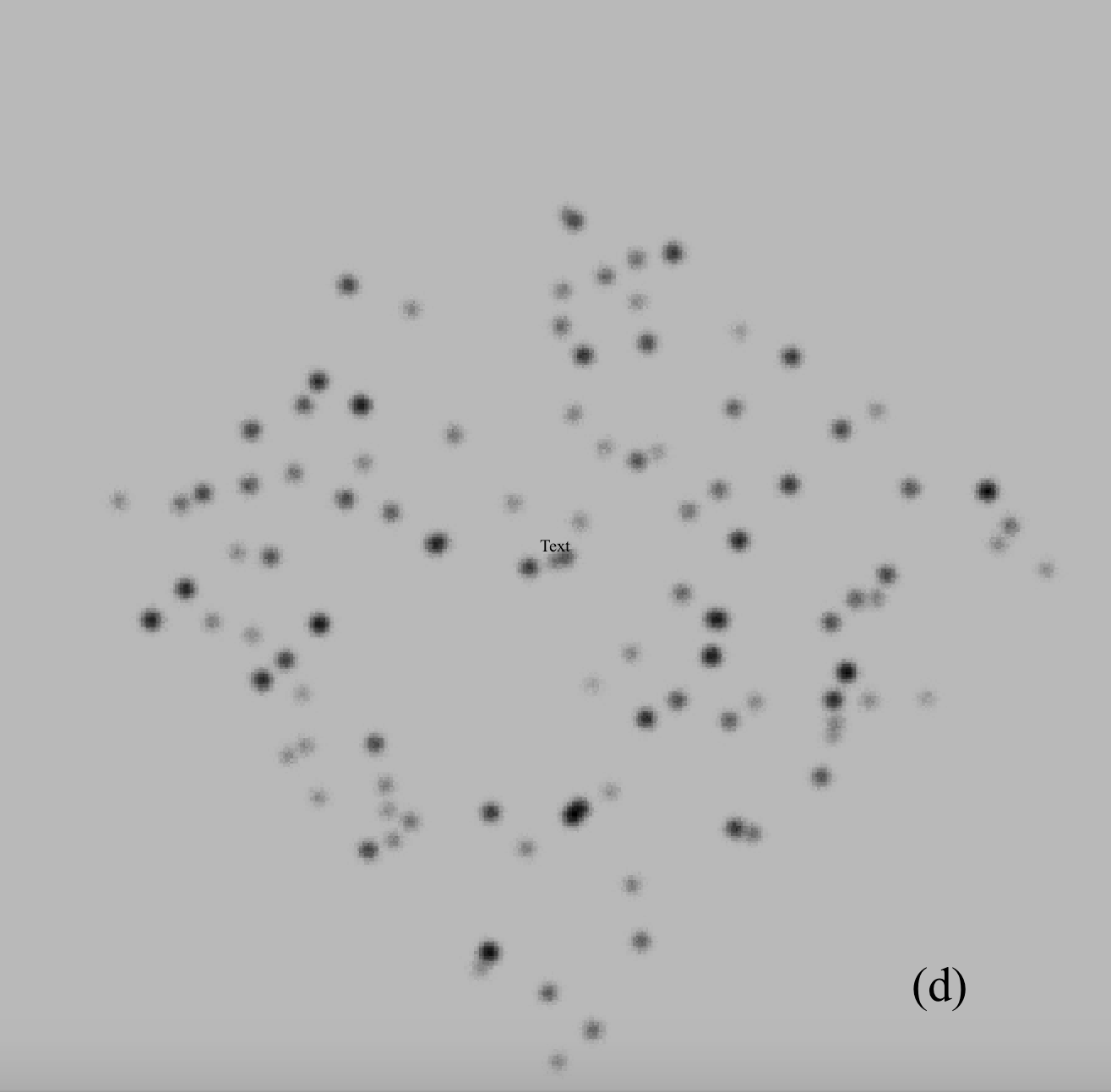}  
    }
    \end{minipage}
   \caption{Simulated PN images in MUSE data cubes of the HALO field of NGC\,1380. (a): regular $11\times11$ grid of PNe with magnitudes between $28.0 \leq m_{5007} \leq 29.0$ in increments of 0.1 mag. The locations where the PNe are inserted are shown via the circles.  All PNe have their emission at the same wavelength. (b): image extracted from the previous cube, binned over 3 layers of wavelength. (c): mock PNe with randomly chosen magnitudes in the range $27.0 \leq m_{5007} \leq 29.5$ and randomly distributed radial velocities (only positions indicated). (d) a 2-dimensional image of NGC\,1380 before inserting mock PNe at different cube layers. Note the chance alignment of several overlapping PNe.}
 \label{fig:mock}
\end{figure}

Figure~\ref{fig:mock} illustrates two examples from an extensive series of tests: frame (a) shows the distribution of a regular grid of PN positions, plotted over the halo surface brightness of NGC\,1380 in the continuum. Frame (b) presents the image corresponding to the Doppler-shifted wavelength of \OIII, extracted from the simulated data cube over three layers for a better signal-to-noise ratio. The grid comprises 11 groups of 11 mock PNe with magnitudes between $m_{5007} = 28.0$ and $m_{5007} = 29.0$ in increments of 0.1 mag, placed diagonally with decreasing brightness from upper left to lower right. Frame (c) is analogous to (a), but the PNe have random positions, random radial velocities (hence they appear in different data cube layers), and random $m_{5007}$ magnitudes in the range $27.0 \leq m_{5007} \leq 29.5$.

The mock PNe were inscribed into the data cube as follows.  First, a two-dimensional point source image was created using the assumption of a Gaussian PSF\null.  A Gaussian was chosen because the details of the PSF were deemed to be unimportant, as all our photometry is performed using small (3 pixel) apertures and a sky annulus with inner and outer radii of 12 and 15 pixels.  The FWHM and total flux contained in the point source were varied from run to run, with values appropriate for the specific galaxy being analyzed.  

Next, the noise-free point sources, which were in units of photons per pixel, were modified with a Poissonian noise generator, using the IDL function {\tt POIDEV}\null. These noisy images were then transformed into cgs units, and measured with DAOPHOT's aperture photometry routine {\tt phot} to determine their ``true'' magnitudes.  

Finally, the noisy 2D-image was added into the data cube of a galaxy, either in a grid at a common wavelength or at random positions with wavelengths distributed over the 15 data cube layers that are relevant for the galaxy in question (see Section~\ref{subsec:onoff}) and weighted by the Gaussian profile of the assumed MUSE line spread function (LSF\null). In other words, the input image with artificial PNe is projected into the stack of 15 data cube layers such that each PN has the correct LSF for its assigned Doppler-shift.

It is worthwhile mentioning that sometimes the random assignment of a PN position leads to chance superposition of objects, e.g., PN82 and PN87 in Fig.~\ref{fig:mock}c, that can be hard to distinguish in a 2D-image (d). In some cases, the PNe can be resolved in the data cube by their velocity difference.  However, if $\Delta v \lesssim 100$ km\,s$^{-1}$, the PNe may appear as one object. While this seems to be an academic exercise of the simulations, we find such examples exist in reality, and have a potential impact on the PNLF (see the discussion in Section~\ref{subsec:results_NGC1380}).

The analysis of photometric simulations as dense as the one in Fig.~\ref{fig:mock}c also reveals that aperture photometry is negatively influenced by the presence of too many emission line objects in the sky annuli.  Such a condition causes complex cross-talk and systematic errors that are hard to remove.  Since the density of bright PNe in distant galaxies is generally not high enough to trigger these problems, we  conducted the remainder of our simulations using a modified position generator that reduces the number of objects per frame by imposing a minimum distance between sources (30 pixels). This constraint entirely erased the cross-talk artifacts observed in the simulations. It was subsequently chosen as the standard routine. 

\begin{figure}[t!]
\begin{minipage}{1.0\linewidth}
    \centerline{
    \includegraphics[width=90mm,bb=50 50  740 480,clip]{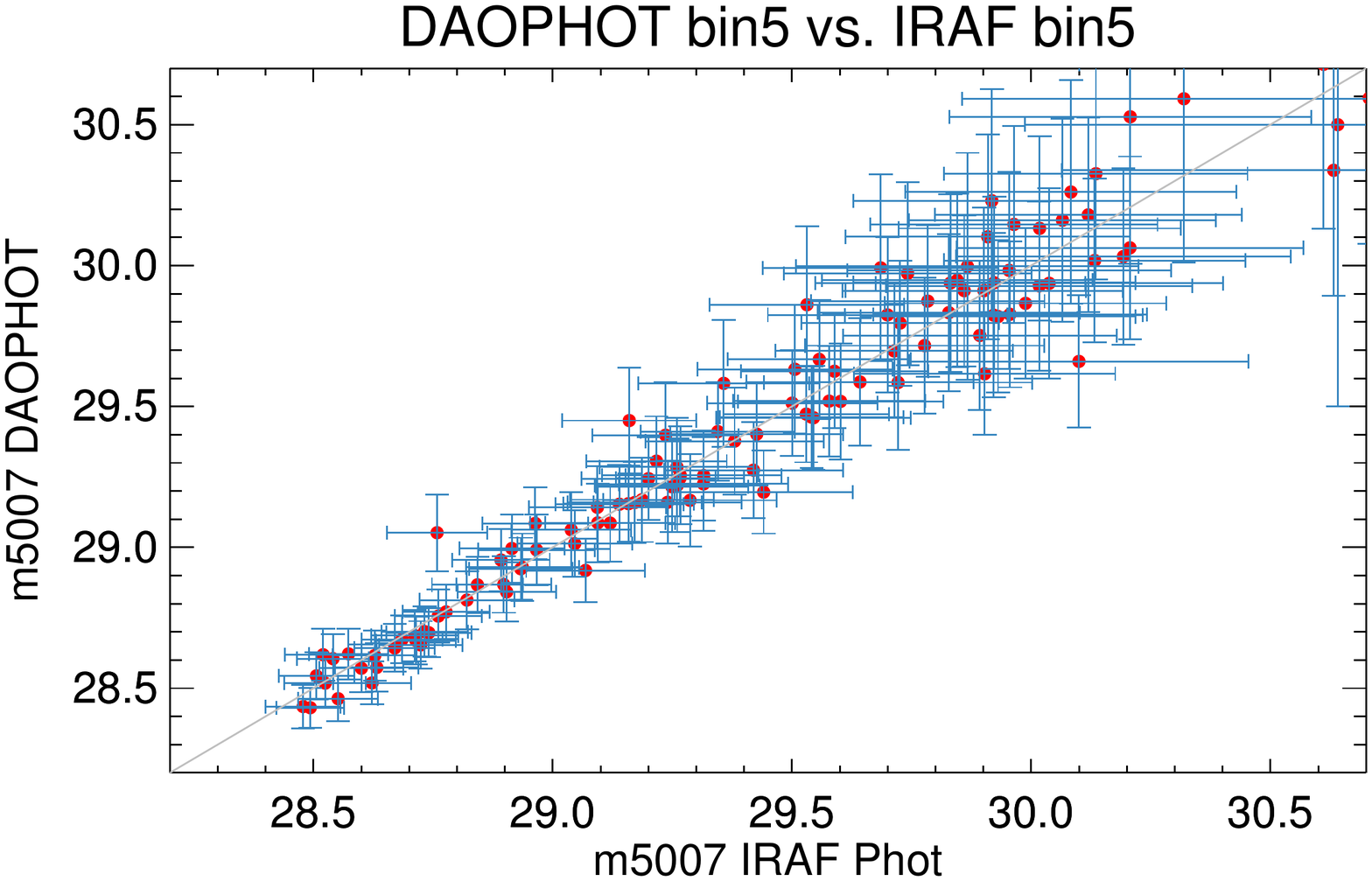} 
    \includegraphics[width=90mm,bb=50 50  740 480,clip]{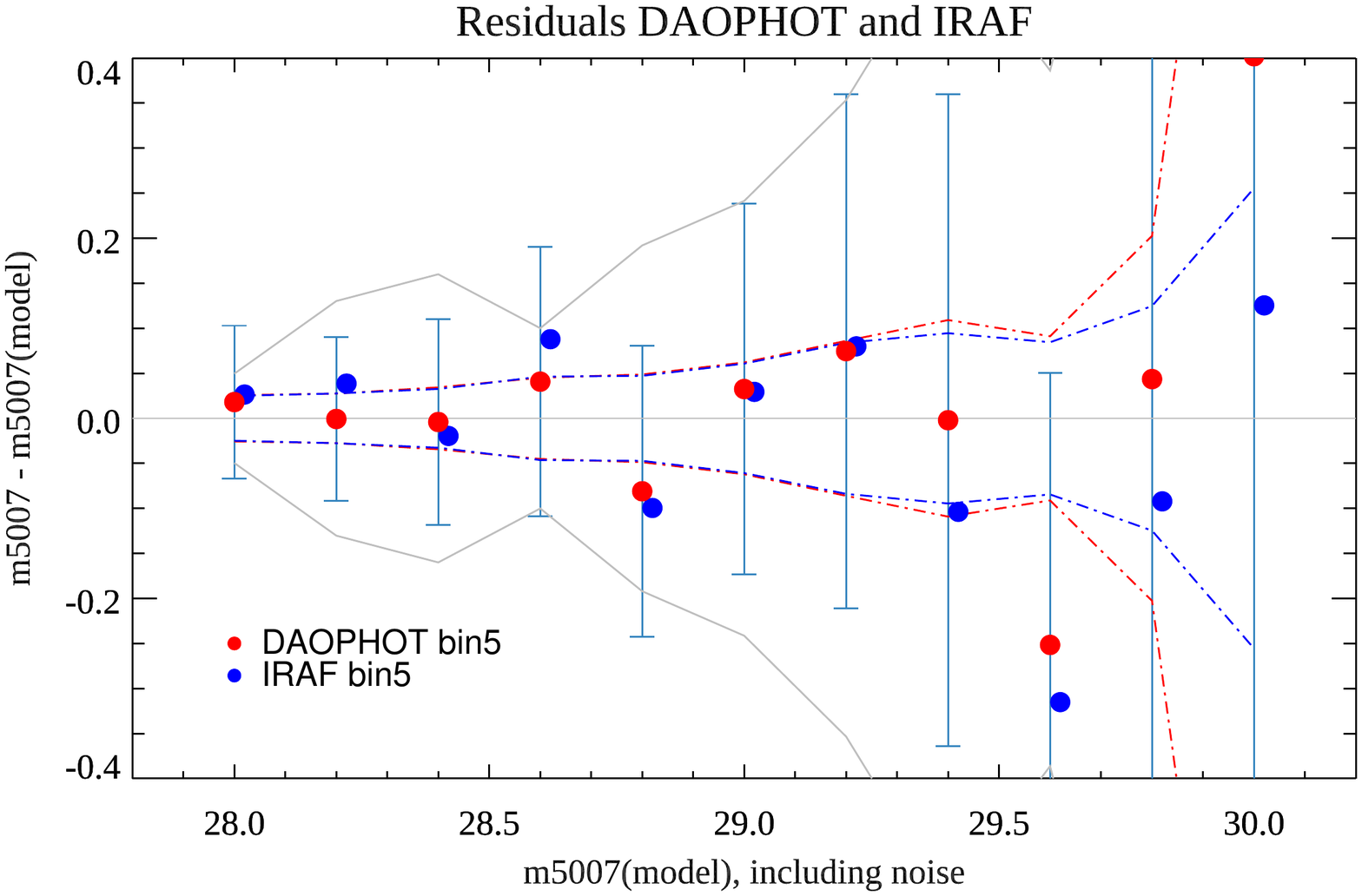}
    }
    \end{minipage} 
   \caption{Comparison of aperture photometry DAOPHOT APER and IRAF-Phot with simulated PNe as illustrated in Fig.~\ref{fig:mock} (a),(b). Left: DAOPHOT magnitudes versus IRAF\null. Right: the mean residuals in bins of 0.2 mag. The dashed curves outline the error of the mean, while the error bars on the red symbols illustrate the typical uncertainty of a single measurement in each bin. The grey envelope outlines the standard deviation of the DAOPHOT measurements in each bin.
 \label{fig:photoComp}} 
\end{figure}

Using data cubes with artificial PNe placed on a grid as shown in Fig.~\ref{fig:mock}a and Fig.~\ref{fig:mock}b, we compared two different aperture photometry tools for internal consistency: DAOPHOT's {\tt aper}, and IRAF's {\tt phot}. The results are shown in Figure~\ref{fig:photoComp}. The overall agreement in a magnitude range of $28.5 \leq m_{5007} \leq 29.5$ is very good. We note that this test is idealized in the sense that the centroids of the point sources are accurately known a priori; this would not be the case for measurements of unknown objects, and in particular, very faint objects at the frame limit. However, the exercise is useful to demonstrate the validity of our photometry down to faint magnitudes. 

The lower-right panel in Fig.~\ref{fig:photoComp} plots the photometric residuals for the two tools, showing average values in 0.2~mag bins. The DAOPHOT results exhibit the best agreement with the input values of the simulation.  The error bars on the red symbols illustrate the typical uncertainty of a single measurement in each bin, not the error of the mean, which is plotted with dashed lines. The grey lines represent the envelope for the standard deviation of the DAOPHOT measurements in each bin; these are in reasonable agreement with the error bars of single measurements as taken from the DAOPHOT error estimates. The kink at $m_{5007}=28.6$ is an artifact of our simulation and is due to the systematics of the data cube and the fixed pattern of the grid. 

\begin{figure}[h!]
    \begin{minipage}{1.0\linewidth}
    \centerline{
    \includegraphics[width=90mm,bb=50 50  740 
    485,clip]{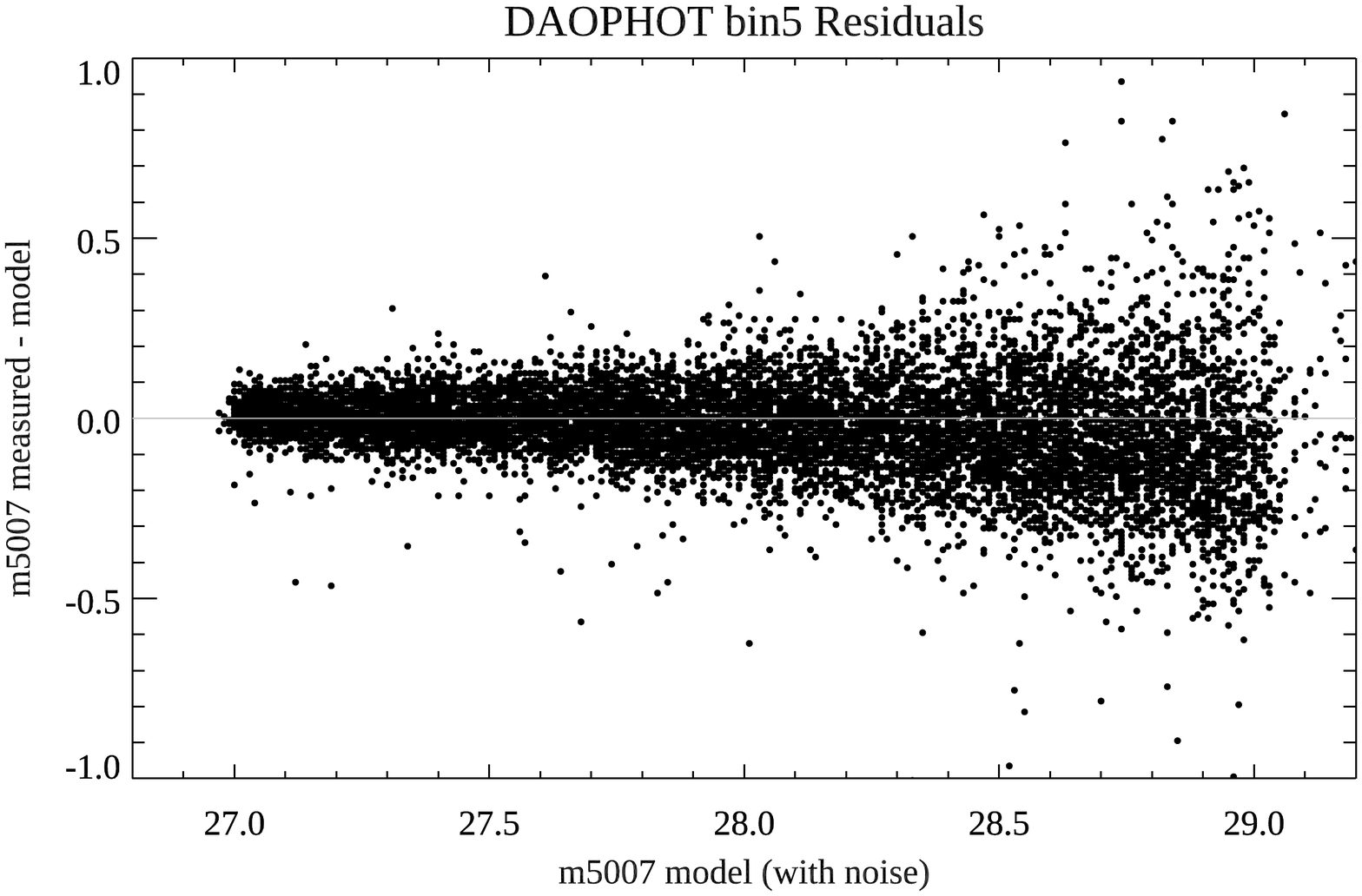}
    \includegraphics[width=90mm,bb=50 50  740 485,clip]{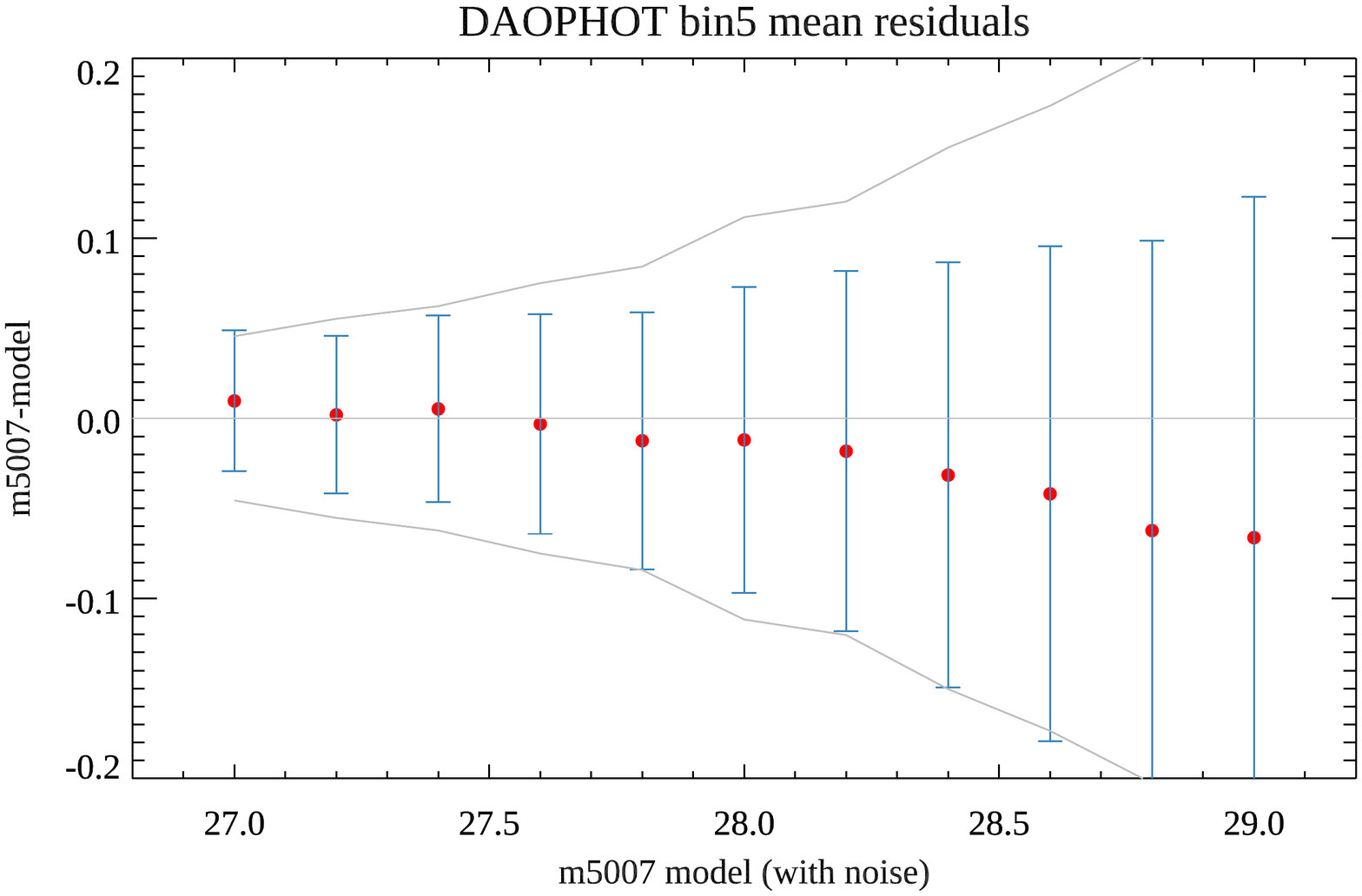}
    }
    \end{minipage} 
   \caption{A simulation with 10000 artificial emission line point sources. Left: residuals of all the measurements. Right: the average residuals within 0.2 mag bins. The error bars represent the uncertainties of individual measurements, while the grey envelope sketches the standard deviation in each bin.
 \label{fig:mock10000}} 
\end{figure}

In order to remove the systematic, a fully randomized simulation as highlighted in Fig.~\ref{fig:mock}c and Fig.~\ref{fig:mock}d was executed. Using an automated script, we generated 10000 artificial PNe and distributed the objects amongst 200 data cubes, with 50 objects per cube.  The results of these models are plotted in Figure~\ref{fig:mock10000}. The left panel shows a scatter plot of all 10000 measurements, while the plot on the right illustrates the typical $1\sigma$ uncertainties reported for individual measurements.   The envelope curves indicate the measured standard deviation in each bin. 

Except for the faintest magnitudes, the quantities agree well, confirming that the DAOPHOT error estimates are statistically meaningful and credible. At a magnitude of $m_{5007} = 27.0$, which is slightly fainter than the PNLF's bright-end cutoff for NGC\,1380, the simulated photometry indicates individual errors of 0.04~mag. Errors of the mean, which are of order 0.01~mag, demonstrate that there is no systematic error in the photometry down to $m_{5007} \sim 28.0$. Beyond that point, a systematic offset does become apparent, and the amplitude of the systematic grows to $\sim 0.07$~mag at $m_{5007} \sim 29$, which is roughly the detection limit in the cube. 
 
The simulations confirm that our technique yields precision spectrophotometry for point-like emission-line sources having the magnitudes of PNe at distances of $15 \lesssim D \lesssim 25$~Mpc.  In typical 1~hour exposures, measurements of the bright end of the PNLF should be accurate to $\sim 0.04$~mag with no apparent systematic errors, even in regions of high surface brightness. The errors can clearly be reduced and the method extended to greater distances with larger PN samples and with longer exposure times. Based on these promising results, we perform all of our photometric measurements with DAOPHOT.

\subsubsection{Radial velocities}
\label{subsec:LSOV}

\begin{figure}[h!]
    \begin{minipage}{1.0\linewidth}
    \centerline{
    \includegraphics[width=90mm,bb=50 90  740 480,clip]{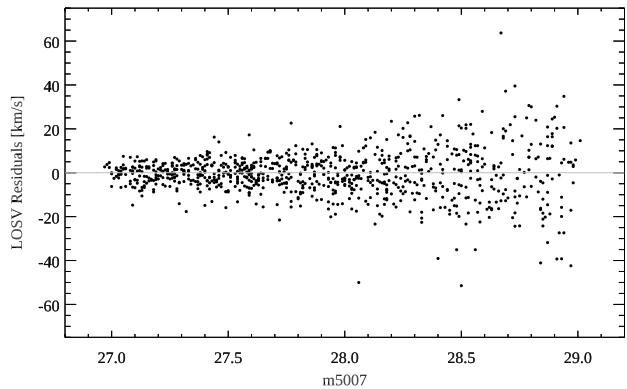} 
    }
    \end{minipage} 
   \caption{Residuals of radial velocity measurements for 1000 simulated PNe in NGC\,1380.  The mock objects have a velocity distribution of $\pm450$~km\,s$^{1}$ centered on the galaxy's systemic velocity of 1877~km\,s$^{-1}$.
 \label{fig:mock1000vrad}} 
\end{figure}

The line fitting tool of our spectroscopy provides a measurement of the line-of-sight velocities of individual PNe, and thus allows for future exploration of the gravitational potentials of PN host galaxies. To test this capability, we used our simulations to assess the accuracy of the Gaussian fit to the \OIII emission line. Figure~\ref{fig:mock1000vrad} shows the velocity residuals of mock data, obtained from 1000 realizations of PNe in a total of 20 data cubes, each simulating a 1~hour MUSE exposure of a galaxy.  The full range of velocity residuals is the equivalent to two MUSE wavelength bins, i.e., 2.5\,\AA\null. It is immediately apparent that the velocity accuracy is on the order of one tenth of a wavelength bin, with a standard deviation of 5.0\,km\,s$^{-1}$ for objects in the magnitude range $27.0 \leq m_{5007} \leq 27.5$, 7.0\,km\,s$^{-1}$ between $27.5 \leq m_{5007} \leq 28.0$, 11.1\,km\,s$^{-1}$ between $28.0 \leq m_{5007} \leq 28.5$, and 16.4\,km\,s$^{-1}$ for PNe fainter than 28.5\,mag. The error of the mean again demonstrates that there is no systematic offset in the measurements.  The simulation shows that the central wavelength error is well-behaved, and that radial velocities at the bright end of the PNLF can be measured with an accuracy of a few km\,s$^{-1}$. We have used this result in our subsequent analysis of benchmark galaxies for calibrating the line-of-sight velocity (LOSV) error as a function of PN brightness.

\subsection{Fitting the Luminosity Function}
\label{subsec:fitting}
In order to derive PNLF distances and their formal uncertainties, we followed the procedure of \citet{Ciardullo+89}.  We adopted the analytical form of the PNLF,
\begin{equation}
    N(M) \propto e^{0.307 M} \left\{ 1 - e^{3(M^* - M)} \right\}
    \label{eq:pnlf}
\end{equation}
convolved it with the photometric error versus magnitude relation derived from our aperture photometry, and fit the resultant curve to the statistical samples of PNe via the method of maximum likelihood. For the foreground Milky Way extinction, we used the reddening map of \citet{Schlegel+98}, updated through the photometry of \citet{Schlafly+11}, and assuming $A_{5007} = 3.47 E(B-V)$ \citep{Cardelli+89}.  For the value of the PNLF cutoff, we adopted $M^* = -4.53$, which is the most-likely value found by \citet{Ciardullo12} from a dozen nearby galaxies with well-determined Cepheid and TRGB distances.

\section{Results} \label{sec:results}
\subsection{Benchmark galaxy: NGC~1380} \label{subsec:results_NGC1380}

\begin{figure}[h!]
    \begin{minipage}{1.0\linewidth}
    \centerline{
    \includegraphics[width=88mm,bb=0 120  590 700,clip]{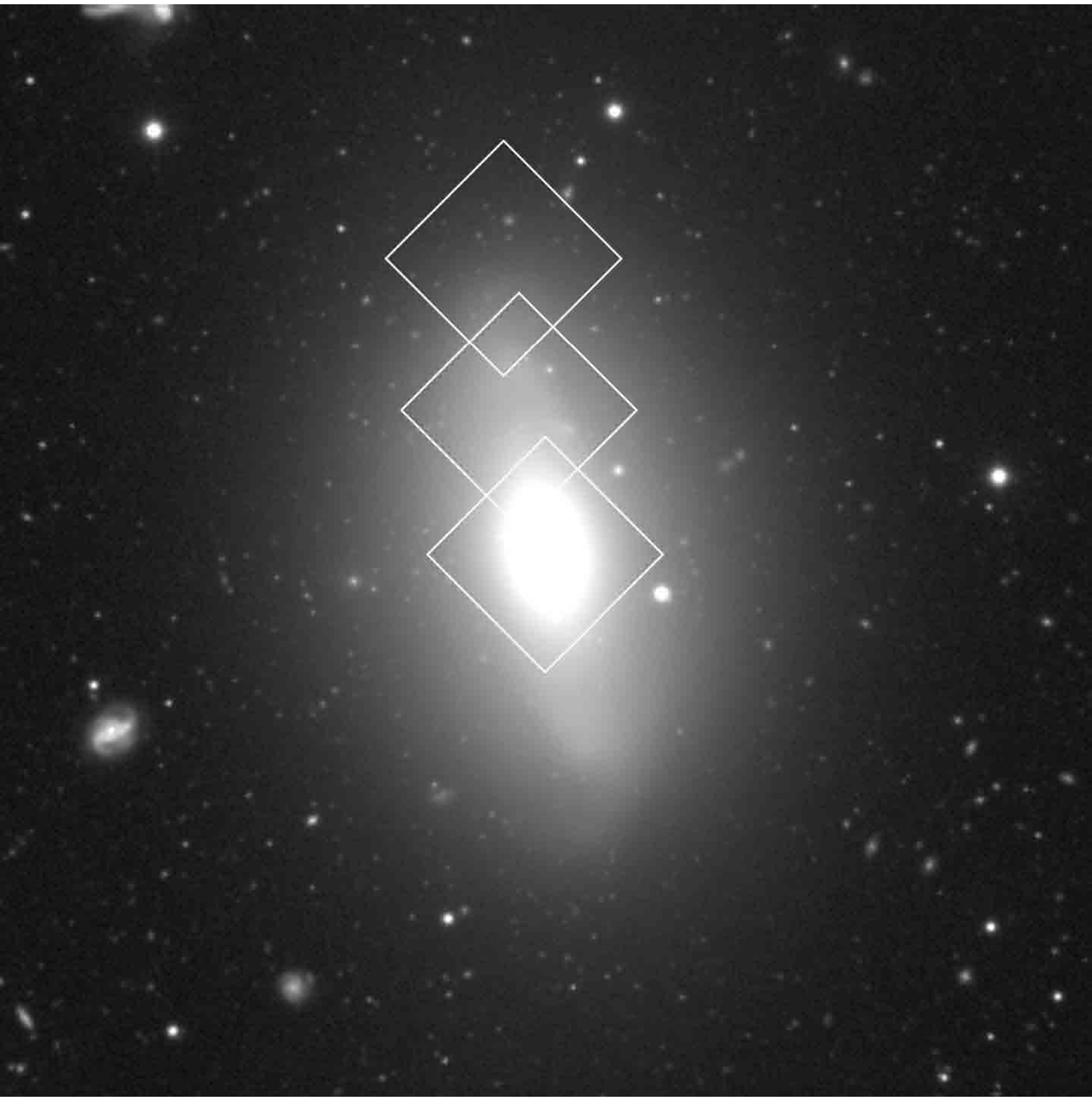}
    \hspace{2mm}
    \includegraphics[width=88mm,bb=0 120  590 700,clip]{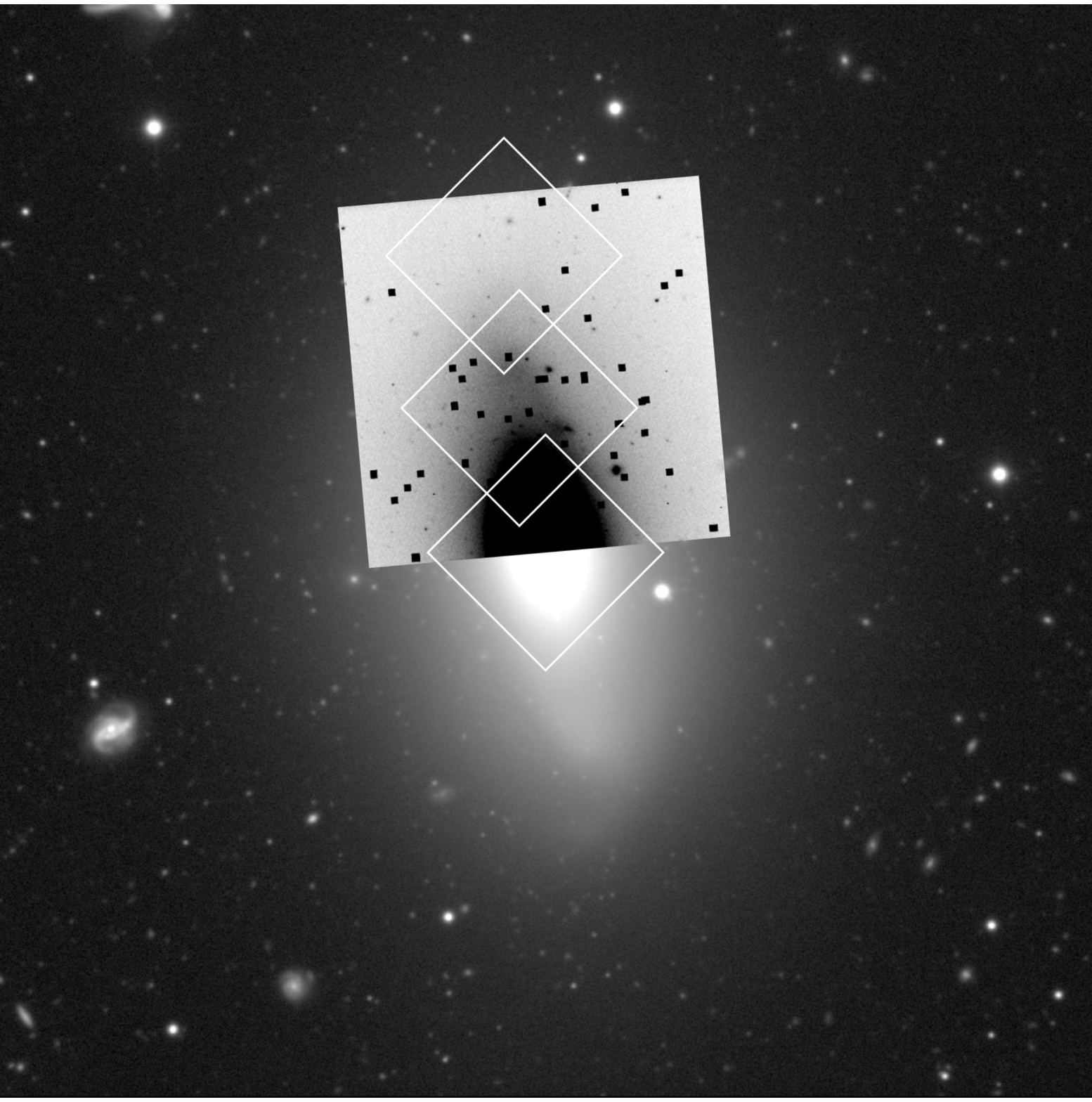}
    }
    \end{minipage}
     \caption{A $6.5\arcmin \times 6.5\arcmin$ image of NGC\,1380 from the Carnegie-Irvine Galaxy Survey \citep{Ho+11}.  North is up and east is to the left.  Left: The MUSE pointings from the Fornax survey of \citet{Spriggs+20}, with the CENTER (bottom), MIDDLE (middle), and HALO (top) fields outlined.  Right: The location of the Magellan narrow-band image taken by \citet{Feldmeier+07}, with the locations of PN candidates shown as black dots. 
 \label{fig:NGC1380_pointings}}
\end{figure}

NGC\,1380 is a lenticular galaxy in the Fornax cluster with Hubble type SA0, a heliocentric systemic velocity of  $v_{rad} =1877\pm12$\,km\,s$^{-1}$, a rotational velocity of $\sim 200$~km\,s$^{-1}$ \citep{DOnofrio+95} and $v_{rot}/\sigma \sim 1$ \citep{Vanderbeke+11}.  Analyses of the galaxy's Globular Cluster Luminosity Function \citep[GCLF;][]{Blakeslee+96, Villegas+10} and Surface Brightness Fluctuations \citep[SBF;][]{Tonry+01, Jensen+03, Blakeslee+09} both place the galaxy securely in the core of the cluster, roughly 19~Mpc away \citep{Madore+99, Blakeslee+09}.  Two previous PNLF distance determinations are available for the galaxy; one based on narrow-band observations with the Magellan telescope \citep[$(m-M)_0 = 31.10$ or 16.6 Mpc;][]{Feldmeier+07}, and one from a previous study with MUSE \citep[$(m-M)_0 = 31.24$ or 17.7 Mpc;][]{Spriggs+20}.  NGC\,1380 was also the host to the fast-declining Type Ia supernova, SN~1992A, and therefore fits in well with the long-term goal of calibrating SN\,Ia luminosities. Figure~\ref{fig:NGC1380_pointings} shows an image of NGC\,1380 with the MUSE and Magellan pointings overlaid.

\subsubsection{Data} \label{subsubsec:NGC1380}

We retrieved all MUSE exposures in the vicinity of NGC\,1380 from the ESO
archive.  This consisted of 43 frames taken during the time-frame between Dec 30, 2016 and Nov 10, 2017 from program ID 296.B-5054 (PI: M.\ Sarzi).  As displayed in Fig.~\ref{fig:NGC1380_pointings}, these data consist of pointings in three fields
with a small amount of spatial overlaps between fields. 
We used the master calibrations from the archive and chose to only re-reduce
the science data (using pipeline v2.8.3, \citealt{Weilbacher+20}) as follows. We
processed the basic calibrations using bias correction, flat-fielding, tracing,
wavelength calibration, geometric calibration, and twilight skyflat correction
using the master calibration closest in time to each science exposure.

The high-level processing then handles the data at the level of individual
exposures. We first reduced the offset sky fields to measure the
sky continuum and produce a first-guess for the sky-line fluxes. For each on-target
exposure the pipeline then combined the data from all CCDs while also correcting for
atmospheric refraction. The flux-calibration used response curves (and telluric
corrections) derived from standard star exposures taken during the same night, typically, within an hour of the science exposure. All response curves
were visually checked to be valid in the wavelength range below 7000\,\AA; the curves showed only typical night-to-night variations. The pipeline then re-fit the sky
emission lines, subtracted them together with the previously prepared sky continuum, corrected the data to the appropriate barycentric velocity, applied the relative
astrometric calibration, and finally created a data cube and set of broad-band images of each exposure  
(including an image integrated over the bandpass of HST F814W filter).
Automatic alignment of the exposures failed, since the fields of NGC\,1380 contained significant background gradients and the foreground stars were
relatively faint. We therefore subtracted the large-scale gradients using smoothed images
and then interactively computed the stellar centroids in each MUSE image and on the
HST F814W reference image using the IRAF routine \texttt{imexam}. While doing so, we
used the frame FWHM given by IRAF to remove exposures with bad seeing.
The exposures selected for the final cubes are listed in Table~\ref{tab:archive}.

Using the above astrometric offsets, we then used the pipeline again to combine the
good exposures of each field into a datacube. The cube reconstruction rejects
cosmic rays and we saved the data in the default sampling
($0\farcs2\times0\farcs2\times1.25$\,\AA) with the wavelength scale starting at
4600\,\AA\null.  A comparison of the positions of six stars from the Gaia
DR2 catalog \citep{Lindegren+18} with those derived from our MUSE image in the F814W filter shows non-negligible but approximately
random offsets at about the $0\farcs07$ level. We therefore
conclude that the MUSE cubes have an astrometric accuracy on the same
order.\footnote{A comparison of the 98 CENTER field PN candidates listed
in the Sp2020 catalog with our own positions produces a mean offset of
$\Delta\alpha = -0\farcs35$ and $\Delta\delta = 0\farcs11$, with a
standard deviation of $\sigma_\alpha = 0\farcs12$ and
$\sigma_\delta = 0\farcs10$. This is likely because Sp2020 has a
different absolute reference than that used here.}

\begin{figure}[t]
    \begin{minipage}{1.0\linewidth}
    \centerline{
    \includegraphics[width=1.0\hsize,bb=0 0 750 600,clip]{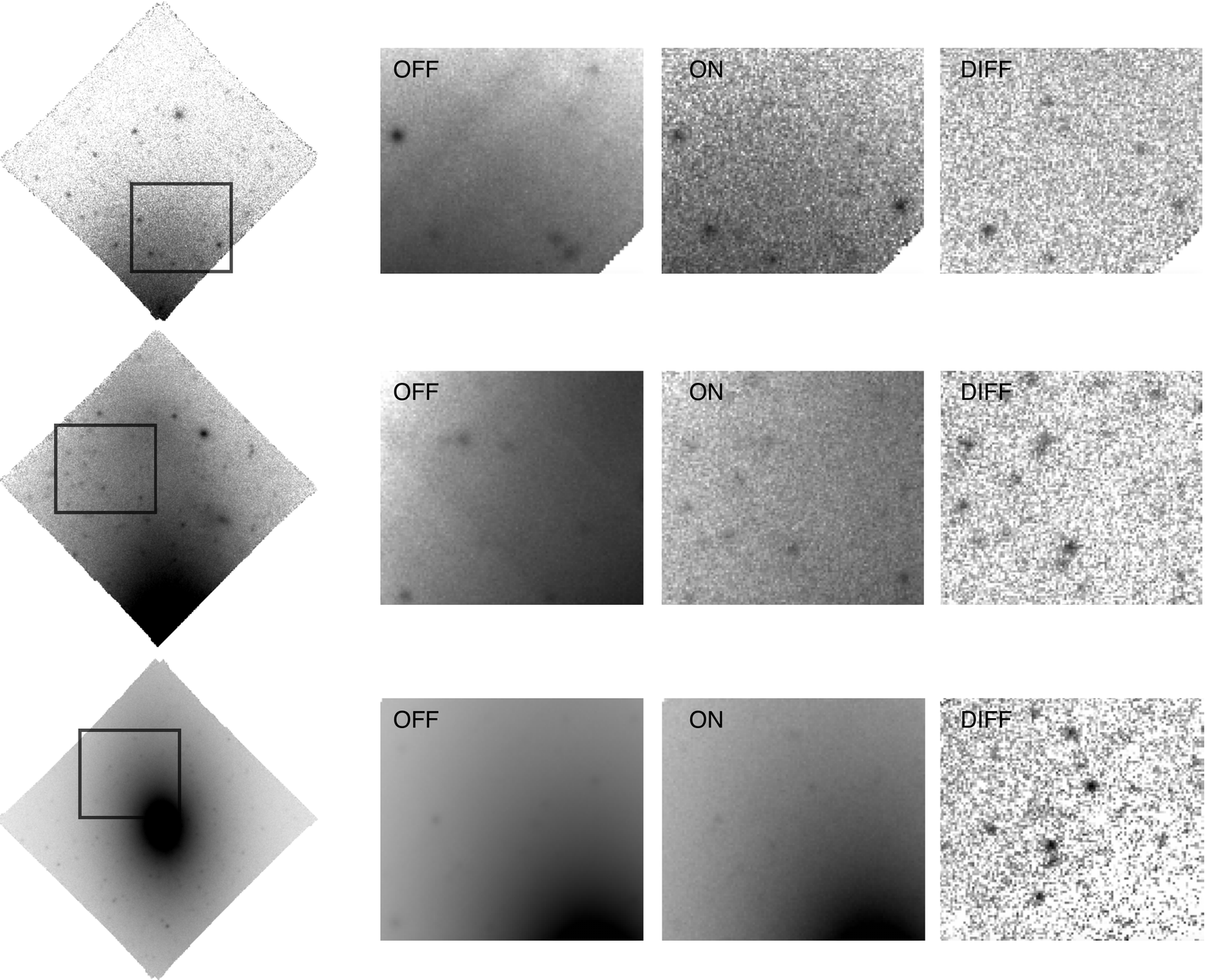}
    }
    \end{minipage}  
   \caption{Examples of PN detections in NGC\,1380 for pointing in the HALO (top), MIDDLE (middle), and CENTER (bottom). The zoomed regions show the {\it off, on,} and {\it diff\/} images and illustrate how the DELF method extracts the faint emission line objects from the bright continuum background.  The resulting flat zero-background image is free from residual fixed-pattern noise.
 \label{fig:NGC1380cmh_examples}} 
\end{figure}


\subsubsection{Differential emission line filtering and source detection} \label{subsubsec:NGC1380delf}

As the first step of analysis, each cube produced by the MUSE data reduction was processed with the DELF filter to yield two {\it diff\/} files, one containing 13 layers of 3 co-added wavelength bins (used for source detection), and the other containing 15 layers of unbinned data (for PN measurements).  

The CENTER, MIDDLE, and HALO fields were inspected visually with DS9 as described in Section~\ref{subsec:onoff}.  This step, which yielded 162, 73, and 29 PN candidates, respectively, is illustrated in Figure~\ref{fig:NGC1380cmh_examples}. In addition, Fig.~\ref{fig:NGC1380c_blue_red} in Section~\ref{subsec:onoff} shows our CENTER pointing in the continuum and in two narrow layers of the stack of 13 co-added images.  This figure highlights how emission line objects appear and disappear on opposite sides of the nucleus, owing to the rotation of the galaxy, and the associated Doppler shift of the PNe. The northern part of NGC\,1380 is rotating towards us (blue-shifted), while the southern part is moving away from us (red-shifted).  Also, in addition to the point sources, there is also a prominent feature seen near the nucleus of the galaxy which suggests the presence of an ionized disk.  This disk is likely associated with the dust ring seen in the insert of (a) and participates in the rotation of the stars. We discuss this object briefly in Section~\ref{subsubsec:NGC1380spectro}. 

The careful double-checking of {\it diff\/} images, e.g.,  Fig.~\ref{fig:NGC1380c_blue_red}, occasionally reveals a small mismatch in the continuum scaling factor. In this example, the mismatch is apparent to the north and south of the nucleus as white hues.  This less than perfect subtraction, which is caused by rotation induced Doppler shifts of the stellar population, is irrelevant for the point source photometry which uses local estimates of the background.  
 
\begin{figure}[h!]
    \begin{minipage}{1.0\linewidth}
    \centerline{
    \includegraphics[width=80mm,bb=45 70  740 480,clip]{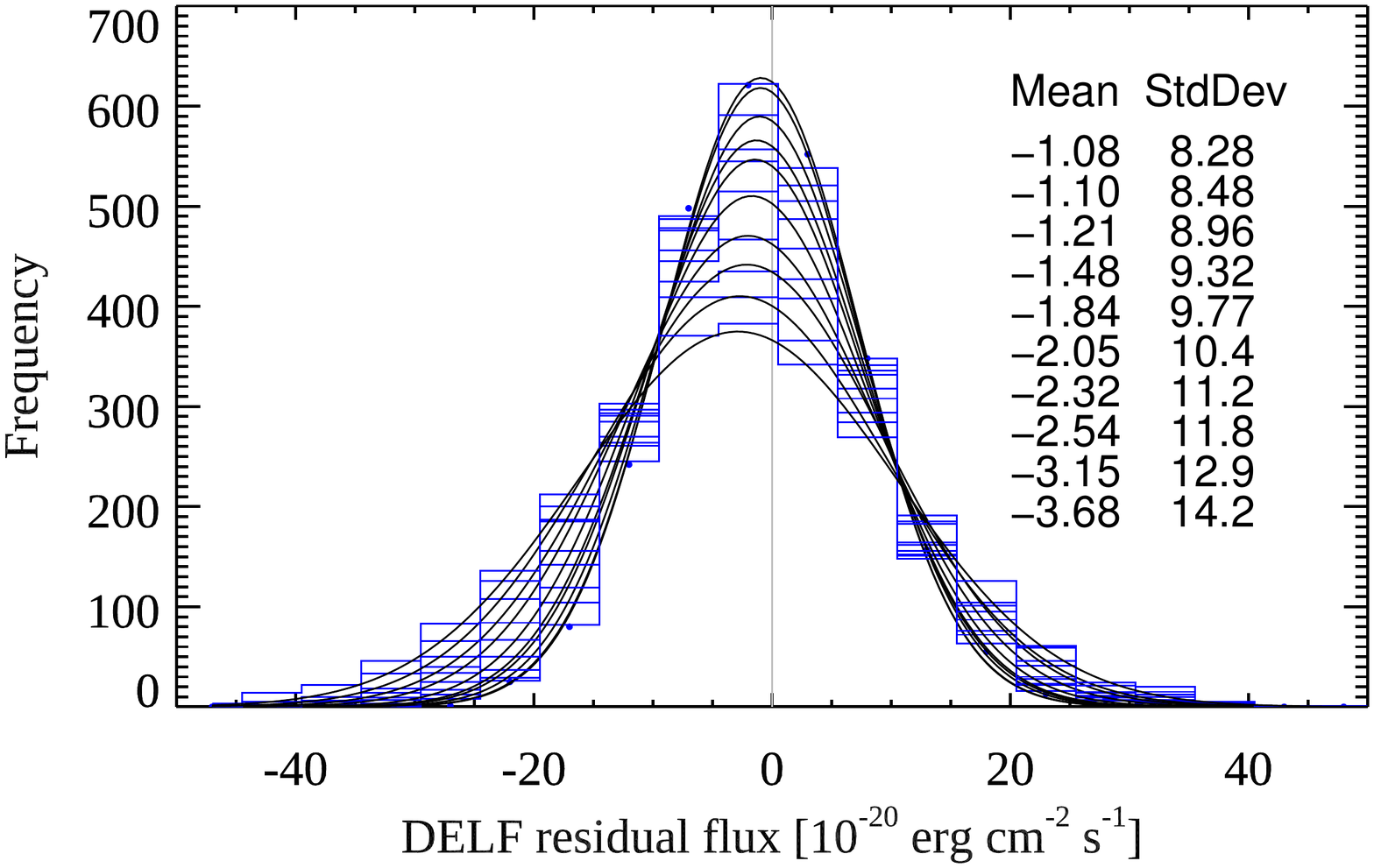} 
    }
    \end{minipage} 
   \caption{Residual noise in our DELF filtered frame at data cube layer 344 for the CENTER field of NGC\,1380, displayed as histograms at 10 offset positions from east to west in the quadrant east of the nucleus.  The Gaussian fits to each histogram are shown.  Though the dispersion of the Gaussians increase as the galaxy's surface brightness increases, the shift in the mean is negligible.   
 \label{fig:DELFnoise}} 
\end{figure}

We display the order of magnitude mismatch of the {\it diff\/} image in Figure~\ref{fig:DELFnoise}.  The figure shows noise histograms for ten $51\times51$ pixel regions of the bluest unbinned {\it diff\/} layer in the CENTER pointing of NGC\,1380.  The histograms start at pixel (70,160) and then moving outward in the galaxy in increments of 10 pixels in $x$.  The data show an almost perfect normal distribution of noise that increases towards the nucleus as the surface brightness of the galaxy rises. The mean varies by an amount of less than $3\times10^{-20}$\,erg\,cm$^{-2}$\,s$^{-1}$ which is negligible in comparison with the flux of PNe near the detection limit ($\approx300\times10^{-20}$\,erg\,cm$^{-2}$\,s$^{-1}$).

It is worth pointing out that the detection of PNe turned out to be an iterative process, involving several passes through the imaging, photometry, and spectroscopy. A detailed inspection of the images was required, which led to the discovery of as many as 15 point sources with overlapping images but different line-of-sight velocities. These blended objects were then confirmed by carefully stepping through the stack of images. A full record of detected blends is listed in the Appendix Table~\ref{tab:Sp2020blends}. 

\begin{figure}[h!]
\begin{minipage}{1.0\linewidth}
    \begin{minipage}{1.0\linewidth}
    \centerline{
    \includegraphics[width=95mm,bb=40 530  620 703,clip]{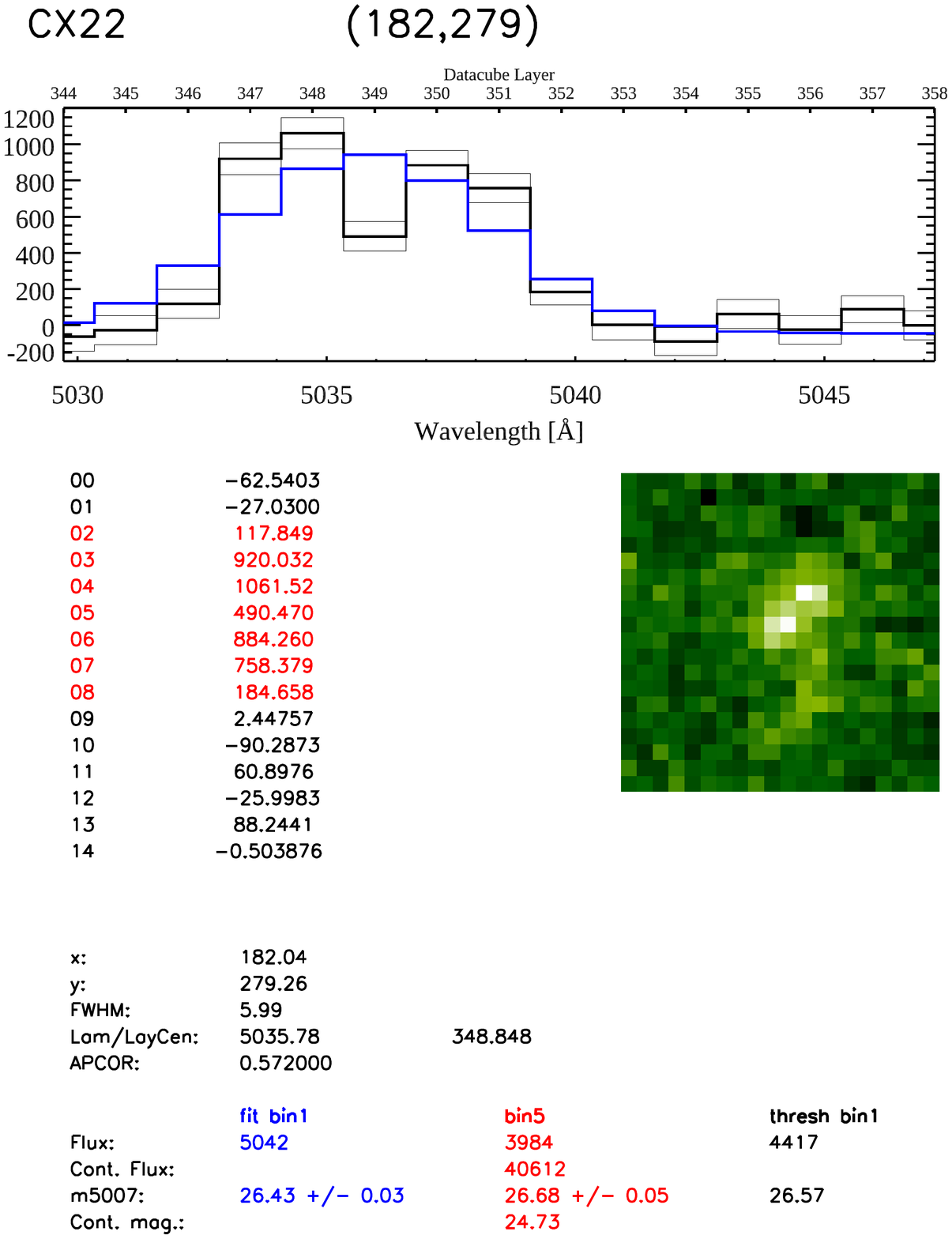}
    \includegraphics[width=27mm,bb=0 -50  600 600,clip]{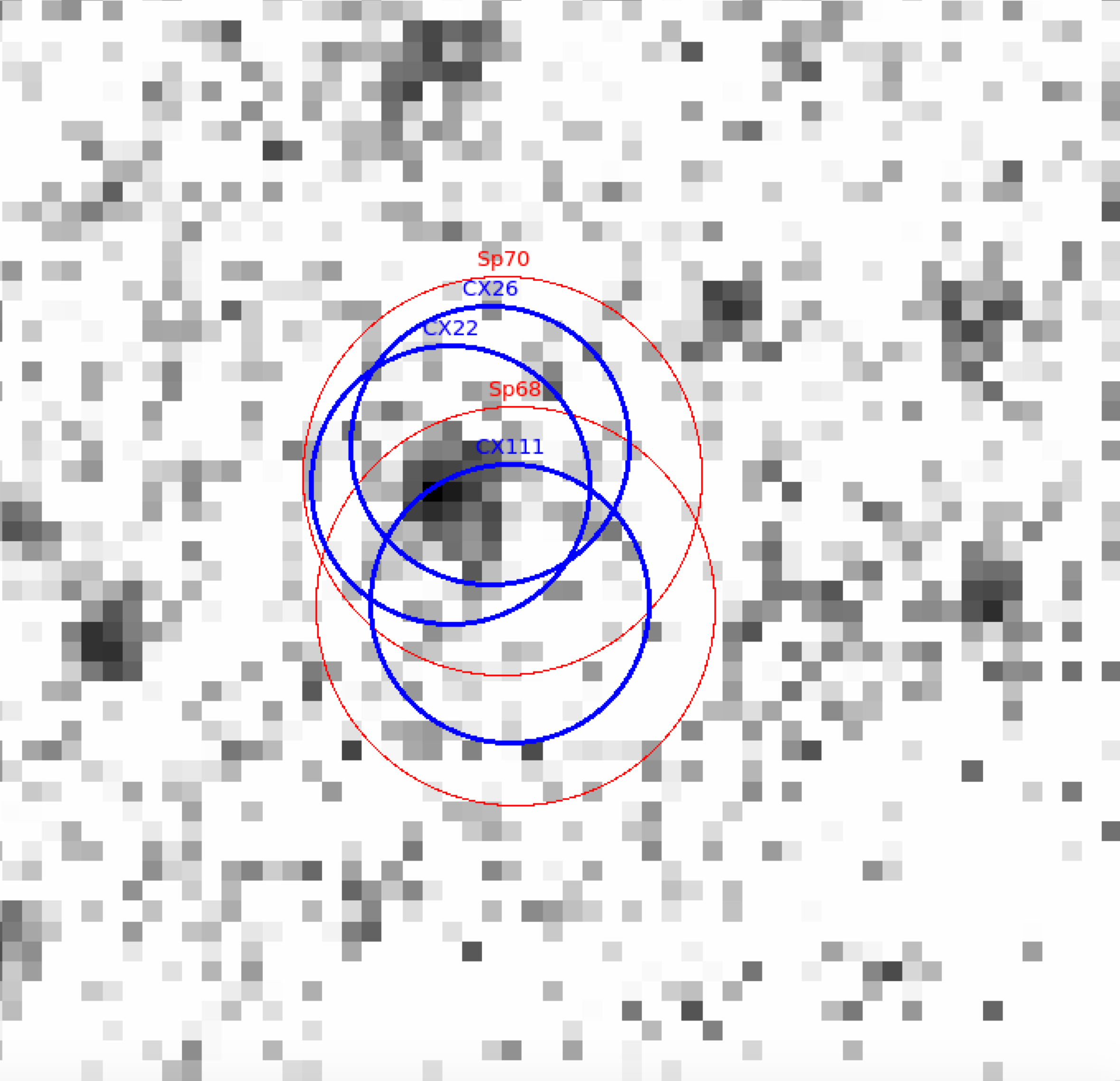}
    }
    \end{minipage} 
    \vspace{5mm}\\
    \begin{minipage}{1.0\linewidth}
    \centerline{
     \includegraphics[width=95mm,bb=40 530  620 703,clip]{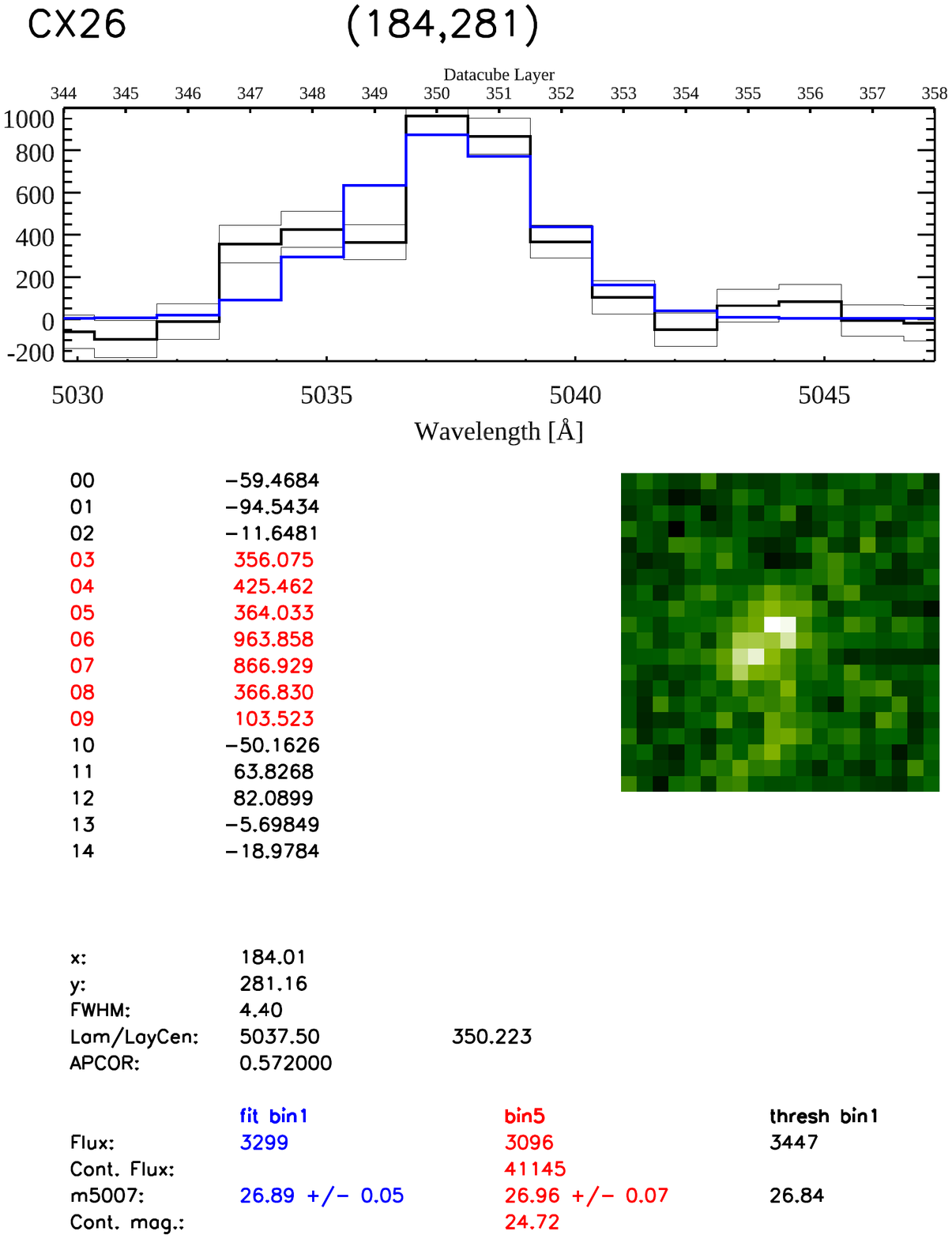}
    \includegraphics[width=27mm,bb=0 -50  600 600,clip]{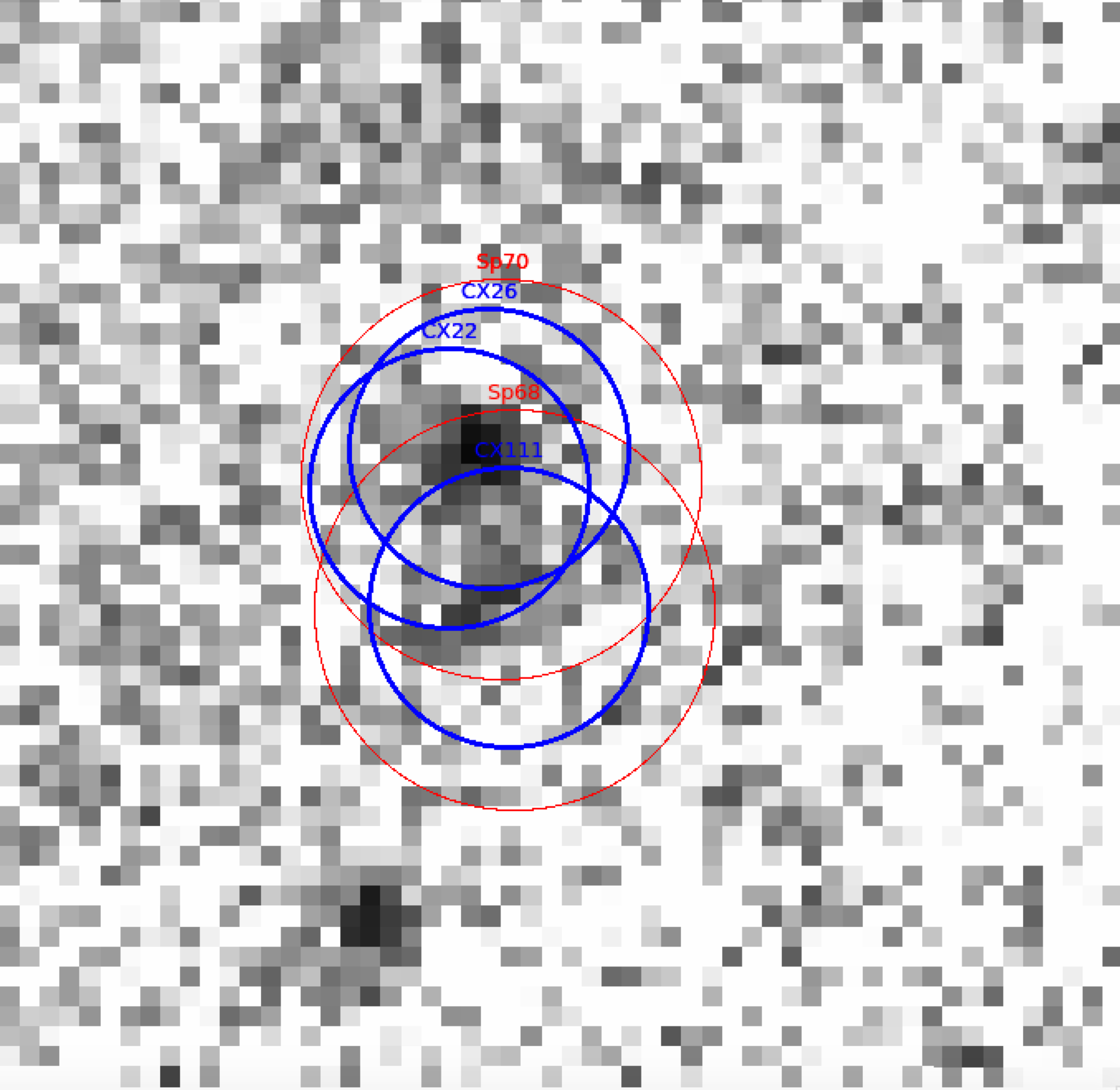}  
    }
    \end{minipage}  
\end{minipage}  
\caption{An example of superposed PNe in NGC\,1380. Three emission line objects are located within a $1\arcsec$ radius.  The top left panel shows the short spectrum for CX22 measured in a 3 pixel radius aperture; the bottom left panel shows the equivalent short spectrum for CX26, which is $0\farcs 55$ away.  Note that the objects have different radial velocities. The blue curves indicate the erroneous fits before deblending. The right-hand panels show the appearance of the objects in data cube layer 348 and 350.  
\label{fig:blendedPNe}
}
\end{figure}

Figure~\ref{fig:blendedPNe} illustrates an example of such a blend.  In the figure, three PN candidates are located within a region less than $1\arcsec$ in radius; such a blend would have been impossible to distinguish using the classical narrow band filter technique.  Of these three sources, two were detected by Sp2020:  one was classified as a PN  (object Sp68 in their Table~4) and the other as a supernova remnant (object Sp70).  However, a careful inspection of the top and bottom panels in Fig.~\ref{fig:blendedPNe} reveals that Sp70 has two components which are separated by $0\farcs 55$; this only becomes apparent by blinking the images centered on data cube layers 348 and 350. The short spectra extracted with a 3 pixel radius aperture for CX22 (top) and CX26 (bottom) are shown on the left hand side of Fig.~\ref{fig:blendedPNe} (explanation as in Fig.~\ref{fig:minispec}). 

\begin{figure}[h!]
    \begin{minipage}{1.0\linewidth}
    \centerline{
    \includegraphics[width=90mm,bb=50 80  740 480,clip]{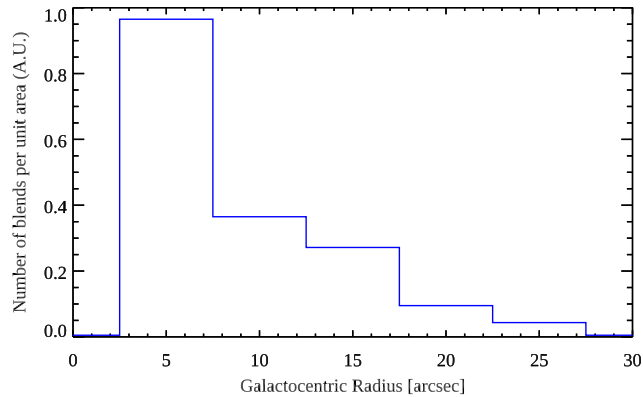} 
    }
    \end{minipage} 
   \caption{Histogram of PN superpositions in the central field of NGC\,1380 shown as number density per unit area (in arbitrary units) versus galactocentric radius.
 \label{fig:blendhist}} 
\end{figure}

One can ask whether the probability of PN superpositions depends strictly on the underlying surface brightness of the galaxy.  Certainly the evidence from narrow-band studies supports this hypothesis \citep[e.g.,][]{Ciardullo+89, jacoby+90, McMillan+93}, but observations through $\sim 50$\,\AA\ wide bandpasses are not nearly as effective as MUSE at surveying the bright inner regions of galaxies. If the number density of PNe do follow the light, it would mean that the effect of blends on the PNLF is highest near the nucleus, and increases with the distance of a given galaxy. Figure~\ref{fig:blendhist} shows the distribution of blends as a function of galactocentric radius, which in turn is linked to the continuum surface brightness of the galaxy via de Vaucouleur's law. The number density indeed is correlated with the surface brightness and exhibits a steep rise towards the nucleus. For radii smaller than 5~arcsec the PNLF is becoming incomplete, so no further increase is observed. The number statistics is too poor to allow for a more detailed investigation, however the trend is clear.

Measurements of the \OIII $\lambda 5007$ flux from two blended objects with a line separation of 3.75\,\AA\ cannot easily be performed with simple aperture photometry; it requires careful PSF extraction in data cubes using software similar to that used by \citet{Kamann+13} for crowded stellar fields.  We have as yet not attempted to adapt this technique to the challenging problem of faint emission line objects. Instead, we employed an interactive line fitting tool that allows us to trim the contaminating spectral line with an ad hoc assumption about the true line profile for the object in the center of the aperture. This approach is not rigorously objective, but is an improvement over the poor fits that were produced before deblending (see Fig.~\ref{fig:blendedPNe}).

For the current work, we identified objects with multiple emission-line components by blinking the images between the relevant data cube layers.  This procedure allowed us to associate the correct components with their corresponding spatial images. Ideally one would like to automate the process, but we defer that discussion to a later paper.  For now, our method of visually blinking the frames has enabled us to identify blends and improve both the photometry and line-of-sight velocity measurements. 

Table~\ref{tab:NGC1380detections} summarizes our final catalog of confirmed PNe in NGC\,1380. In total, we identify 118 PNe in the CENTER field, 40 in the MIDDLE field, and 8 in the HALO field.  For comparison, Sp2020 found 91~PNe in the CENTER field of NGC\,1380, with a significant fraction (15 objects, or 16\,\% of their total) identified in our survey as blends (see Appendix Table~\ref{tab:Sp2020blends}).  We also detected 70 PNe candidates near our detection limit that we classify as unconfirmed, as they are visible only in \OIII 5007\,\AA,  i.e., they have no other confirming emission line.  Of 264 point source candidates (PN and other) identified by visual inspection through 15 data cube layers, we classify only 16 (6\,\%) as spurious, i.e., their signal-to-noise was too low for validation.  These numbers demonstrate that our differential imaging approach to PN identification is much more efficient at finding objects near the detection limit and unraveling blended point sources than techniques that work solely with the original data cubes. 
 
More information on the classification procedure appears in Section~\ref{subsubsec:NGC1380spectro} below.

\begin{deluxetable*}{lcccc}[h]
\tablecaption{Emission line point sources detected in NGC\,1380\label{tab:NGC1380detections}}
\tablewidth{100pt}
\tablehead{  \colhead{Pointing} & \colhead{confirmed PN} & \colhead{ PN candidates } & \colhead{~~~SNR~~~} & \colhead{~spurious~~}  }
\startdata
Center (C)  &   118  &   31   &    11  &   2  \\
Middle (M)  &    40  &   25   &     1  &   7  \\
Halo (H)    &     8  &   14   &     0  &   7  \\
\enddata 
\end{deluxetable*}

\subsubsection{Photometry} \label{subsubsec:NGC1380photo}
We performed DAOPHOT aperture photometry on the objects found in all three NGC\,1380 pointings.  Aperture corrections derived from stars in the field are tabulated in Appendix Table~\ref{tab:NGC1380apcor}, and the results of the photometry are given in Appendix Table~\ref{tab:NGC1380Photo}. A cross reference to the  identifications of Sp2020 is provided in Column~2. These data allow us to perform a detailed comparison of magnitudes, signal-to-noise estimates, radial velocities, and object classifications of the two datasets.

\begin{figure}[h!]
\begin{minipage}{1.0\linewidth}
    \centerline{
     \includegraphics[width=90mm,bb=90 60  750 480,clip]{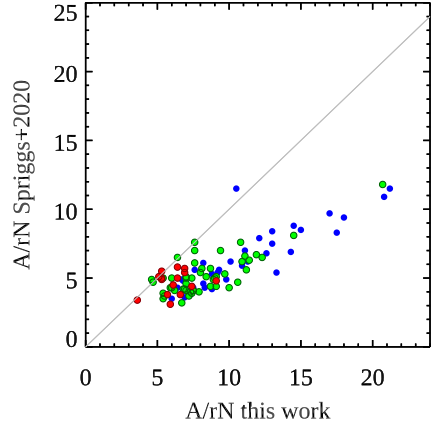} 
    }
    \end{minipage} 
   \caption{Signal-to-noise parameter A/rN measured by Sp2020 versus A/rN from this work.  The blue points show PNe superposed on regions of relatively low galaxy surface brightness, the green points display PNe projected on areas of  intermediate brightness, and the red points represent PNe located in high surface brightness regions of the galaxy.  In general, the Sp2020 signal-to-noise values are $\sim 60\%$ of those measured via our DELF technique. The outlier around A/rN=11 is due to the blend of CX7+CX158 that remained unresolved (SP11) in Sp2020, cf.\ Appendix Table~\ref{tab:Sp2020blends}.
 \label{fig:ArN_direct}} 
\end{figure}

\begin{figure}[h!]
\begin{minipage}{1.0\linewidth}
    \centerline{
     \includegraphics[width=71mm,bb=150 60  650 480,clip]{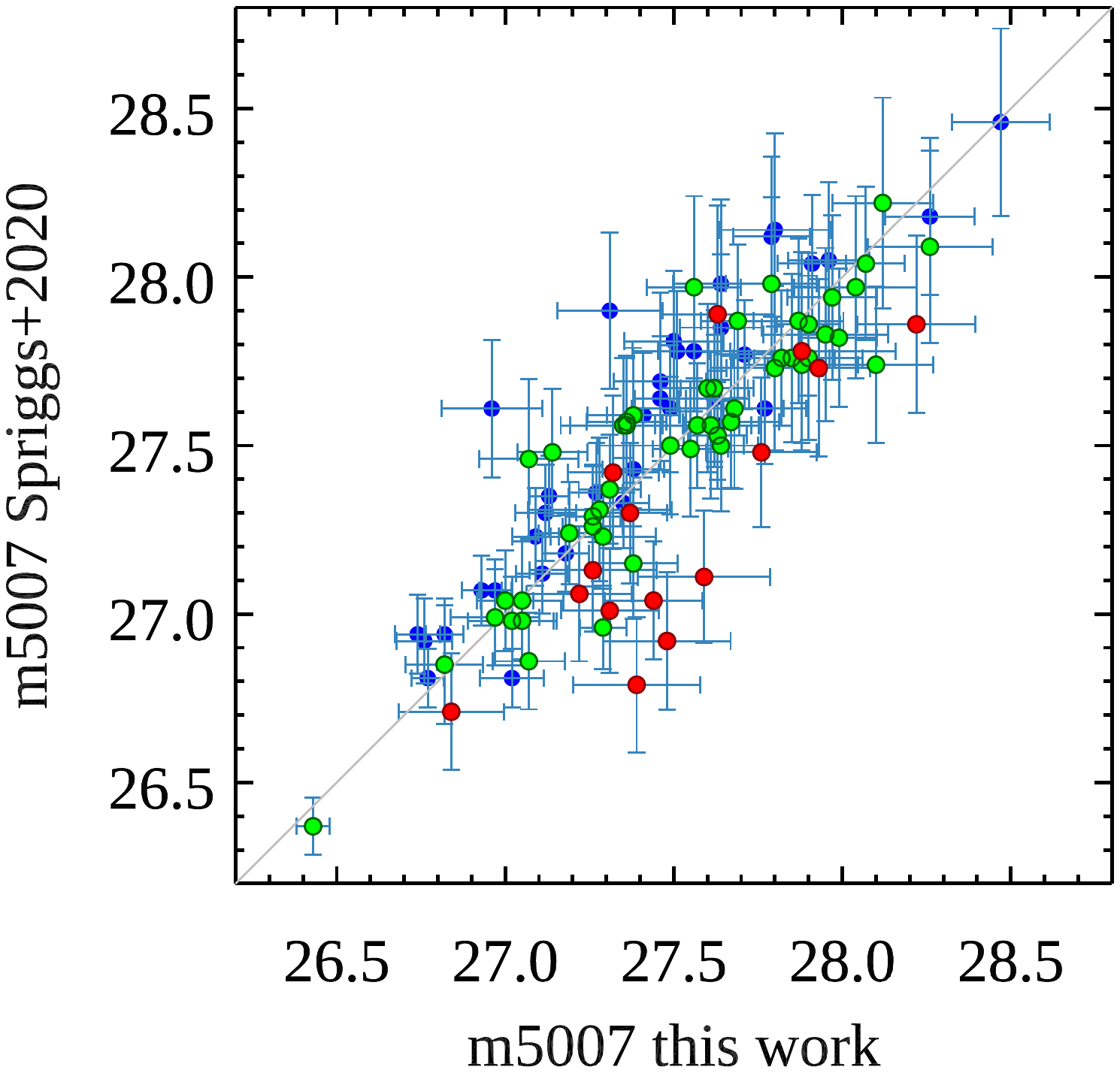}
     \includegraphics[width=71mm,bb=150 60  650 480,clip]{ 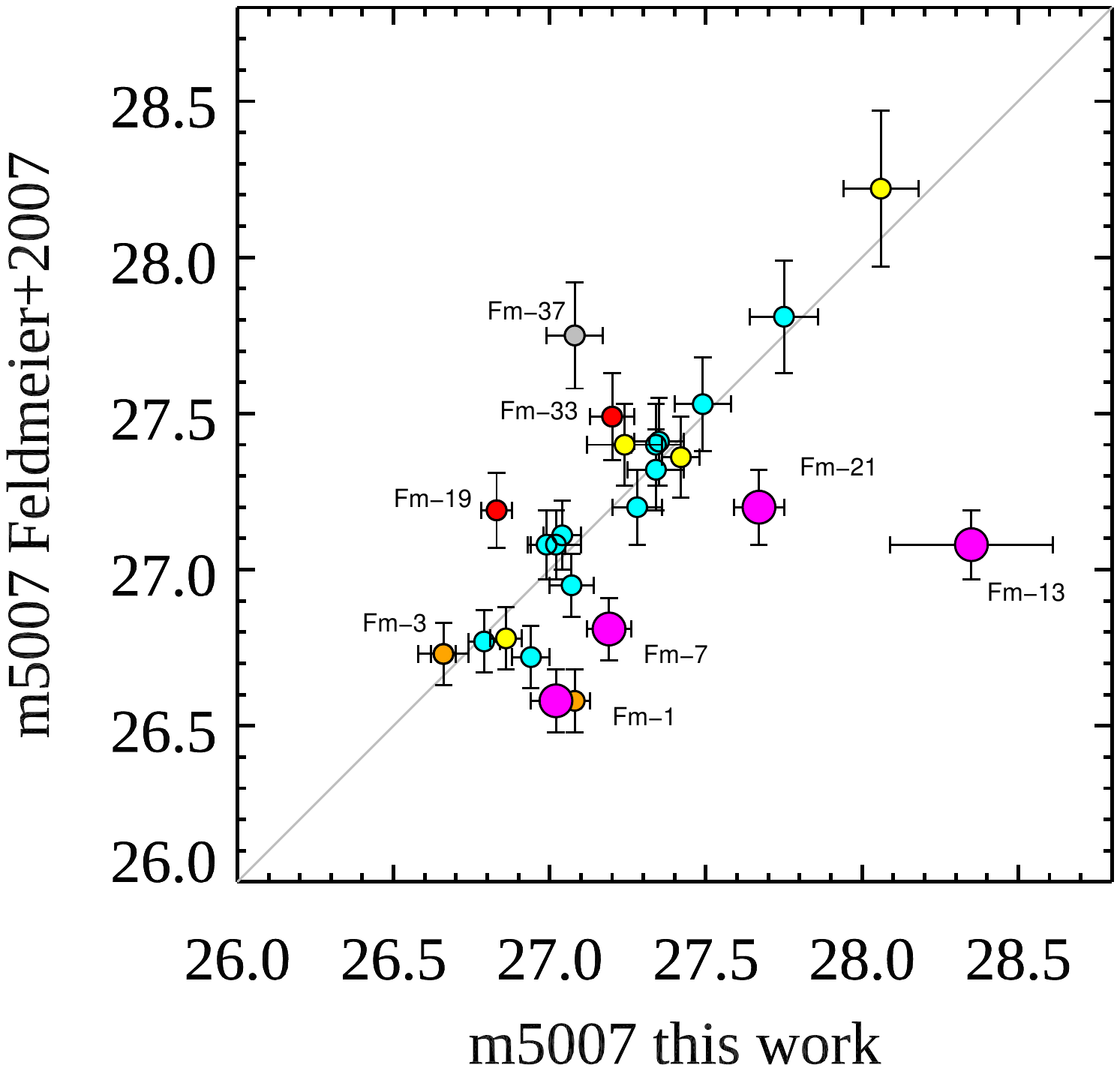}
    }
    \end{minipage} 
   \caption{Comparison of our $m_{5007}$ photometry with the literature. Left: Sp2020 compared to our work.
   As in Fig.~\ref{fig:ArN_direct}, the blue points show PNe superposed on regions of relatively low galaxy surface brightness, the green points represent PNe in regions of intermediate brightness, and the red points denote PNe located in high surface brightness areas of the galaxy.  Note that the data display a systematic trend with surface brightness. Right: Fm2007 compared with our work. The color code and labels are explained in the text.
    \label{fig:CompPhotSp2020}} 
\end{figure}

Sp2020 do not provide error estimates for their $m_{5007}$ photometry. However, we can make a meaningful comparison of the photometric uncertainties using A/rN, the ratio of the fitted emission line amplitude to the residual continuum noise for the \OIII $\lambda 5007$ emission line.  Figure~\ref{fig:ArN_direct} shows our A/rN values versus those quoted by Sp2020. As our {\it off\/} image includes measurements of the background continuum at the position of each PN, we can track the behavior of A/rN versus the underlying galaxy surface brightness.  This information is color-coded into the diagram, with blue representing objects projected onto regions of low surface brightness, and red displaying object superposed on a bright background.  Regardless of PN magnitude, and except for a single outlier, the S/N ratio for the Sp2020 data is below the 1:1 line, and typically only $\sim 60\%$ of that obtained from DAOPHOT aperture photometry on DELF filtered images. This result is a direct confirmation of the expected advantage of the differential filtering approach as outlined in Section~\ref{subsec:onoff}. Sp2020 excluded any objects from their analysis that fall below a threshold of A/rN = 3. In our DELF photometry, only one of those objects is close to this threshold, and only 3 are below a level of A/rN = 5.  In contrast, Sp2020 reported 50 objects below this latter value, again supporting the expectation of a significant gain from our approach.

Figure~\ref{fig:CompPhotSp2020} compares our $m_{5007}$ magnitudes to those of Sp2020 using the same color coding as in  Fig.~\ref{fig:ArN}.  Although the scatter between the measurements is larger than that expected from our internal tests (cf.\ Fig.~\ref{fig:photoComp}), the relation generally follows the 1:1 line. Moreover, a closer inspection of the residuals reveals a trend in the data:  objects superposed on regions of higher galaxy background surface brightness (red symbols) tend to be brighter in the Sp2020 data (mean residual of $-0.23$ mag), and the opposite is true for PN projected onto regions of lower surface brightness (blue, mean residual of $+0.14$ mag). In the absence of Sp2020 $m_{5007}$ error estimates, we have used the reciprocal of the quoted A/rN values as a proxy for their error bars in both plots, with the caveat that they are likely underestimates (see Fig.~\ref{fig:ArN}). Note that we confirm within the error bars the overluminous object reported by Sp2020. 

More instructive is a comparison with the  observations of \citet[][hereafter Fm2007]{Feldmeier+07}, who used a narrow-band filter on the Magellan Clay telescope to identify 44 PN candidates in a 5.5~arcmin$^2$ region north of NGC\,1380's nucleus.   Sp2020 compared their photometry to this dataset and reported good agreement in a relative sense for 4 matching objects in the CENTER field and 17 matches in the MIDDLE field.  However, they reported that magnitudes measured by MUSE were systematically 0.45~mag fainter than those of Fm2007.  Although we were unable to compare their magnitudes as they only published the photometry for the CENTER field, we decided to repeat the exercise, finding 4 matches in the CENTER, 18 matches in the MIDDLE, and 4 additional matches in the HALO\null.  These matches include Fm2007 objects Fm1 and Fm3, which are present on both the MUSE CENTER and MIDDLE pointings, and Fm29, which is located on both the MIDDLE and HALO fields. We also discovered that thanks to the overlap of the CENTER, MIDDLE, and HALO fields, there are 9 PNe common to the CENTER and MIDDLE, and 4 objects common to the MIDDLE and HALO. Comparison of the pairs of magnitudes reveals that a satisfactory agreement is reached with an offset of -0.4 magnitudes for MIDDLE and HALO with respect to CENTER, suggesting a possible systematic error in the MUSE flux calibration, perhaps caused by non-photometric conditions. If true, this would essentially reconcile the 0.45\,mag discrepancy with Fm2007 as reported by Sp2020.

Comparison plots for each field can be found in the Appendix Figure~\ref{fig:NGC1380photcomp}, while the combined datapoints for all fields are shown in the right panel of Figure~\ref{fig:CompPhotSp2020}.  The best agreement with the 1:1 line is achieved assuming a zero-point offset of $-0.1$~mag for the CENTER field and $-0.4$~mag for the MIDDLE and HALO pointings, roughly in line with the differential correction described above.  However, while there is generally good agreement with the 1:1 line, a number of outliers are apparent. A careful inspection of the stack of {\it diff\/} images reveals that the outliers Fm1, Fm7, and Fm21 (magenta points that fall below the 1:1 line) can be explained by the contamination of PNe light by emission from diffuse gas.  This effect is revealed by the velocity separation of the two components in the MUSE spectra. Fm13, which also has an anomalously bright narrow-band magnitude, is similarly identified as the blend of two overlapping point sources.  The discrepancy for objects Fm19 and Fm33, (the red objects above the 1:1 line) is likely due to the velocities, as their \OIII $\lambda 5007$ emission lines (5034.48\,\AA\  and 5032.14\,\AA) lie on the blue-edge of the narrow-band filter's bandpass.  Just a $\sim 20\%$ change in the filter transmission would explain the magnitude difference seen in the figure.  The final discrepant object, Fm37 is located at the very edge of a MUSE data cube, and thus may have unreliable photometry.  

Based on these data, we conclude that except for the above outliers, the agreement between Fm2007 and our photometry is good. In the absence of a proper calibrator, the $-0.4$~mag offset could either be due to a systematic error either in the flux calibration or the aperture correction (or both). To avoid this issue, future targeted observations must be sure to have sufficiently bright PSF stars in the field, and have flux standard exposures specifically attached to the observations.
 
\begin{figure}[t!]
\begin{minipage}{1.0\linewidth}
    \centerline{
     \includegraphics[width=160mm,bb=50 200  800 470,clip]{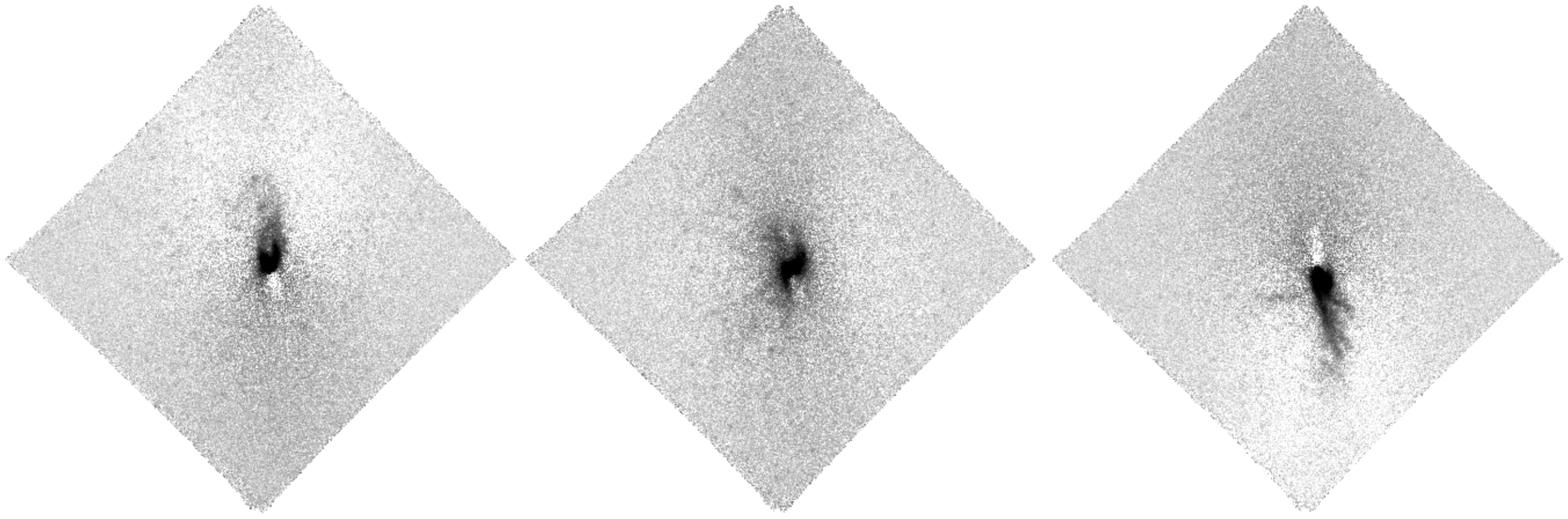} 
    }
\end{minipage} 
\begin{minipage}{1.0\linewidth}
    \centerline{
     \includegraphics[width=100mm,bb=30 220  750 600,clip]{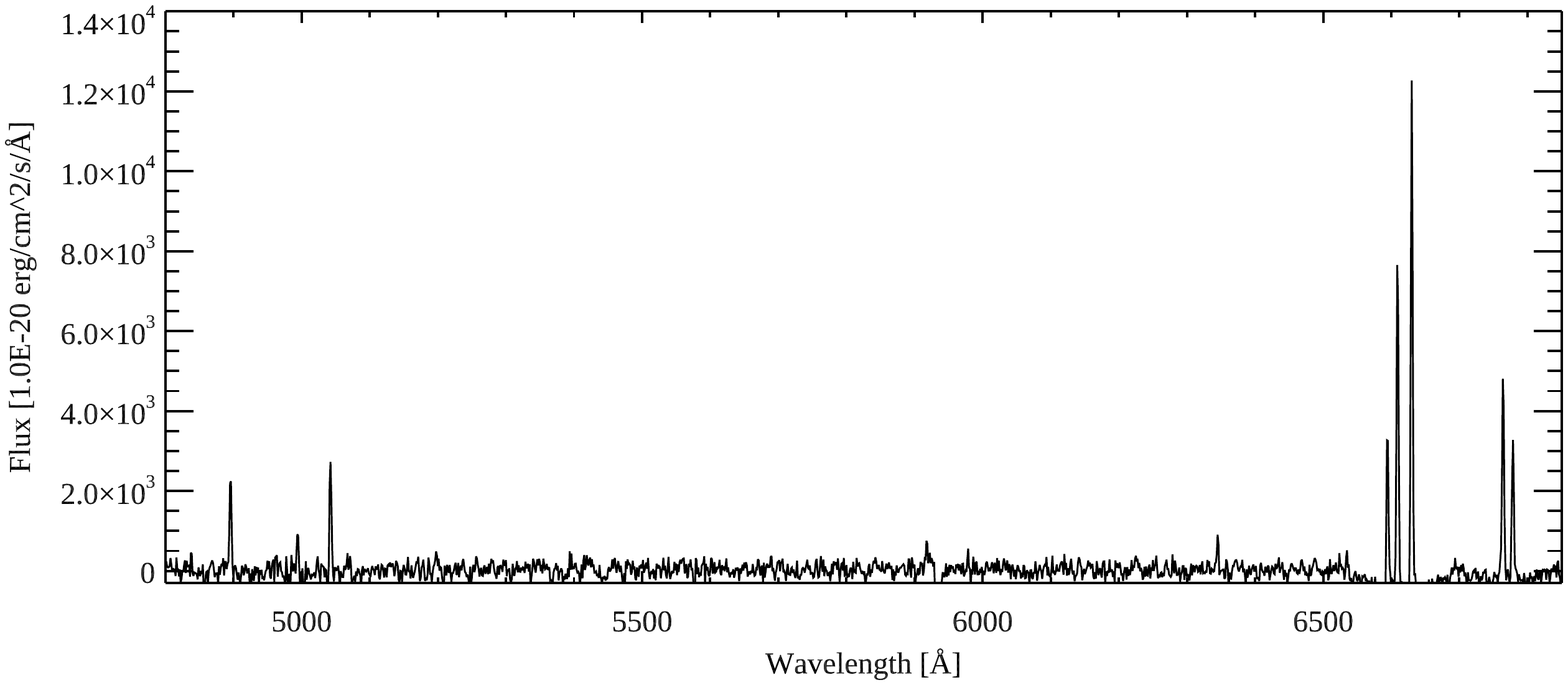}
    }
\end{minipage} 
   \caption{H$\alpha$ images and a spectrum of the ionized gas disk in the core of NGC\,1380. The blue- and redshifted narrow-band images around the systemic velocity of the galaxy reveal the kinematics of the galactic plane and a complex system of filaments and knots that extend far from the principal plane of the disk. The spectrum suggests shock ionization; the shocks may also explain the nature of some of the non-PN candidates in our sample.
 \label{fig:GasDisk}} 
\end{figure}

\subsubsection{Spectroscopy} \label{subsubsec:NGC1380spectro}

Spectra for all detected PN candidates were obtained
using the entire MUSE data cube as described in Section~\ref{subsec:spectro}. The main objective of this exercise was to confirm that the point-like \OIII emitters are true PNe, and to exclude interlopers such as \ion{H}{2} regions, supernova remnants, and background galaxies. As a byproduct of this step, radial velocities were measured and tabulated for future use in kinematic analyses. 

For a candidate to be classified as a PN, it was required to exhibit at least two emission lines, normally \OIII $\lambda 5007$  and \OIII $\lambda 4959$, with the latter's flux measured to be of roughly one third of the former \citep{Storey+98}. For some faint objects near the detection limit, \OIII $\lambda 4959$ was not visible; but H$\alpha$ was. When H$\alpha$ was visible, we used the relationship for bright planetaries found by \citet[][hereafter He2008]{Herrmann+08} and classified the object as a PN if the flux in H$\alpha$ was smaller than the flux in \OIII $\lambda 5007$.  If the \OIII $\lambda 5007$ line was the only line detected, the object was classified as a PN ``candidate''.  Most of these candidates should be true PNe:  while single line detections could be due to background objects such as [\ion{O}{2}] galaxies and Ly$\alpha$ emitters (LAEs), [\ion{O}{2}] emitters would likely be detected in the continuum \citep{Ciardullo+13}, while LAEs are relatively rare.  Specifically, at the depth and redshift window of the NGC\,1380 data, the surface density of LAEs is roughly 0.5 objects per MUSE pointing \citep{Herenz+19}.  Moreover, while the density of LAE contaminants will increase with depth, most Ly$\alpha$ emitters have line widths that are significantly wider than that expected from the \OIII line of a planetary \citep[e.g.,][]{Trainor+15, Verhamme+18, Muzahid+20}.  Nevertheless, to be conservative, single-line PN candidates were not included in our PNLF analysis.

Some of the NGC\,1380 PN candidates that have bright H$\alpha$ also have significant emission in the low-ionization lines of \NII $\lambda\lambda 6548, 6584$ and \SII $\lambda\lambda 6717, 6731$.  This is generally the signature of shock ionization from a supernova remnant.  However, NGC\,1380 does not exhibit strong star formation activity, nor does it contain much cold interstellar medium, so it is unclear whether these spectral features should be attributed to SNRs. Moreover, our generalized DELF processing about H$\alpha$ reveals a $\sim 2$~kpc diameter gas disk around the galaxy's nucleus. This disk, which has a kinematic structure similar to that seen for the PNe, has been investigated previously with the GMOS IFU \citep{Ricci+14}, and more recently with MUSE and ALMA data \citep{Tsukui+20}.  As illustrated in Figure~\ref{fig:GasDisk}, the disk consists of a combination of diffuse gas and filaments with \NII/H$\alpha$ and \SII/H$\alpha$ line strengths indicative of shock excitation. A number of our point-like and sometimes {\it not quite\/} point-like objects have similar line ratios, suggesting they may actually be physically related to the disk, rather than the result of local supernova explosions.  In any case, these strong \NII and \SII emitters are not planetary nebulae.  Although the disk is interesting on its own right, its investigation is beyond the scope of this paper. 

We note that the agreement between the object classifications Sp2020 and those of this work is generally good.  The only differences are for SP40, which Sp2020 classified as an interloper, but we show it to have a PN-like spectrum, Sp22 
(CX45), which we classify as a PN but Sp2020 lists as a supernova remnant, and Sp72 (CX115) which we consider a PN, but Sp2020 classifies as an \ion{H}{2} region.  For the rest of the Sp2020 objects our classifications agree.

\subsubsection{The PNLF of NGC 1380} \label{subsubsec:NGC1380PNLF}
\begin{figure}[h!]
    \begin{minipage}{1.0\linewidth}
    \includegraphics[width=98mm,bb=0 0 600 400,clip]{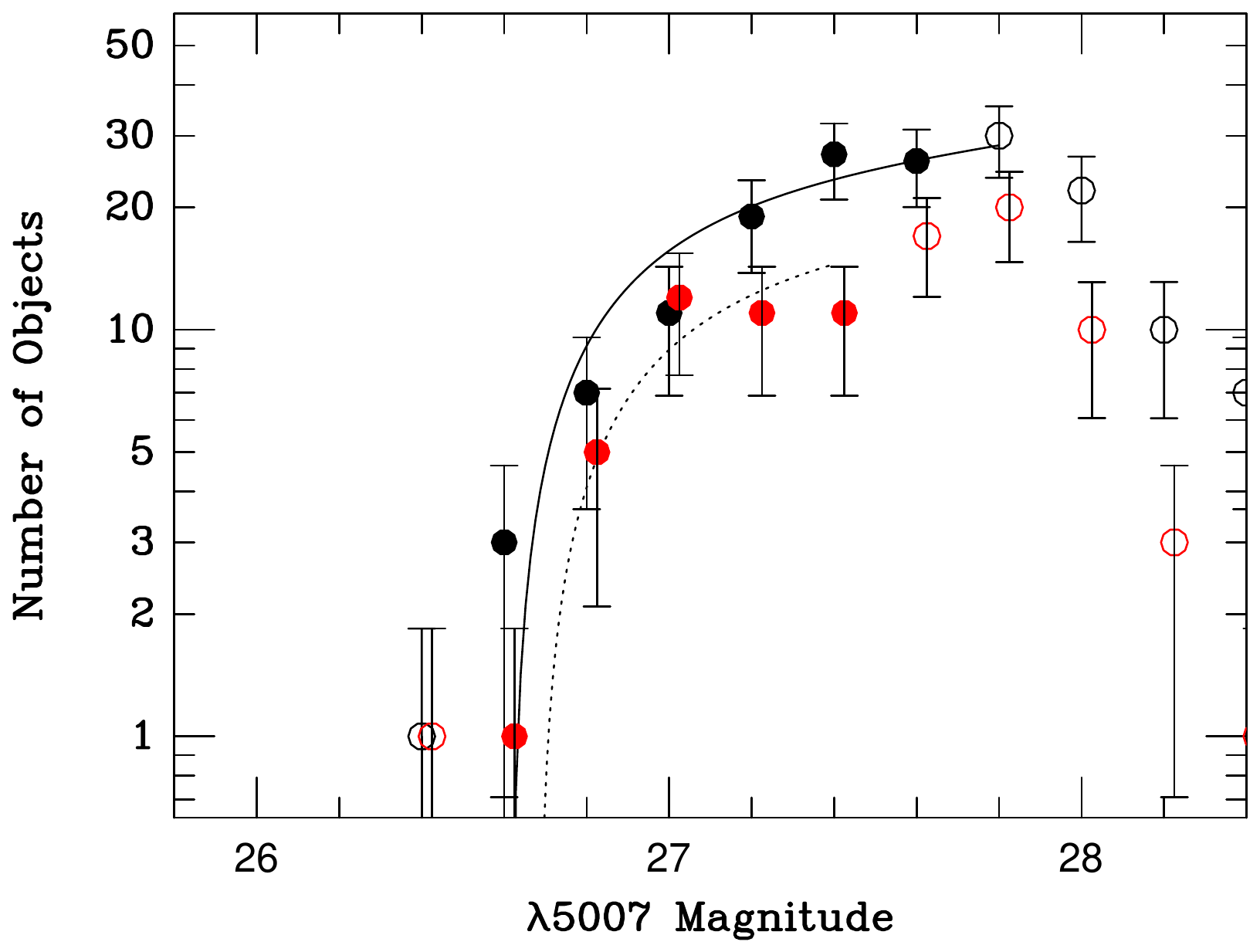}
    \includegraphics[width=98mm,bb=0 0 600 400,clip]{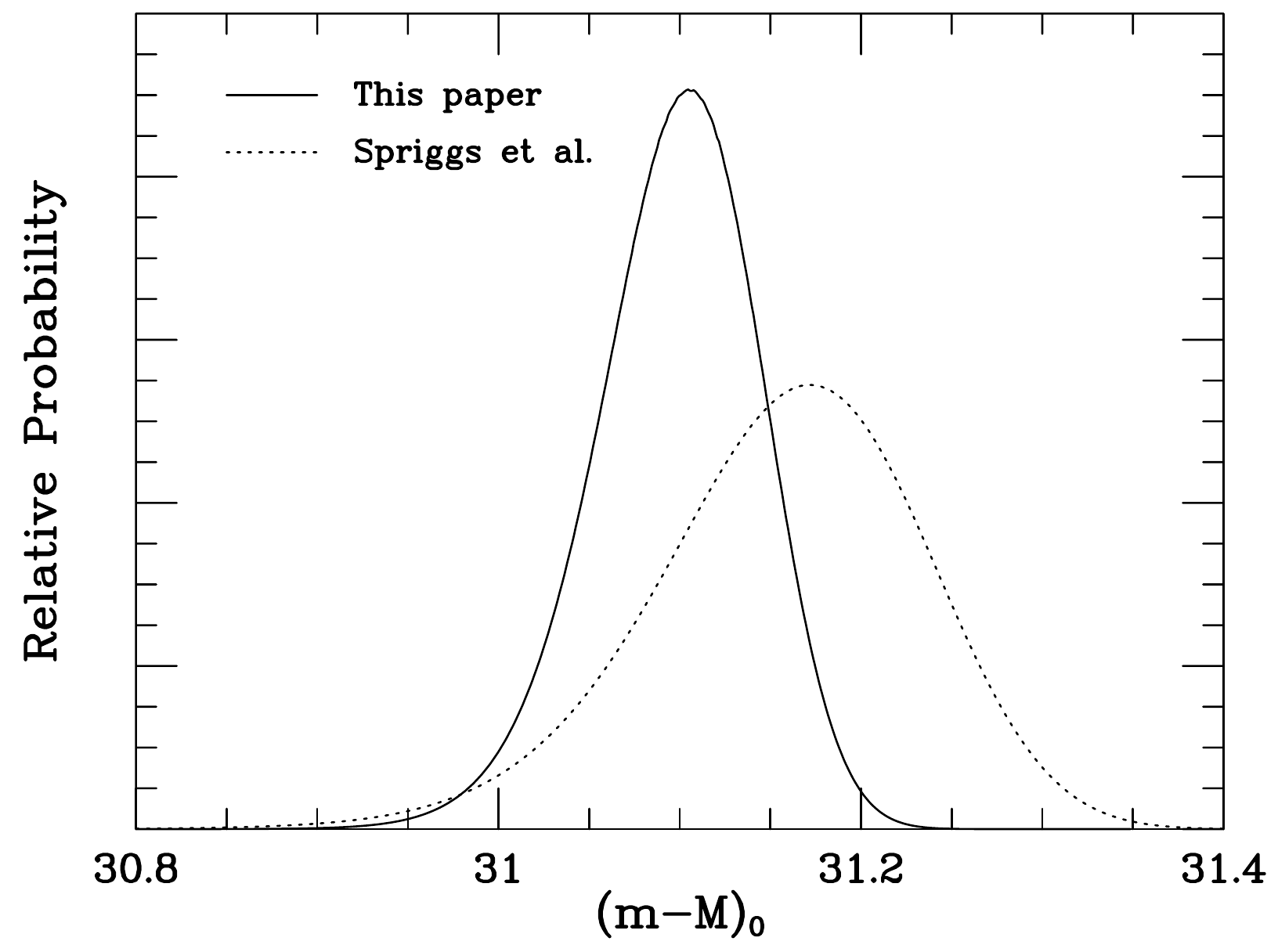}  
     \end{minipage}  
   \caption{The PNLF of NGC\,1380.  The left panel compares our PNLF (black points) and the most likely empirical law (solid curve)  with \citet{Spriggs+20} luminosity function (red points) and its mostly likely fit (dotted curve).  The open points show the one overluminous object (which is not fit) and PNe beyond the completeness limit.  The right panel displays the likelihood of the solution versus the distance modulus.  These likelihoods assume $M^* = -4.53$ and a foreground reddening of $A_V = 0.046$, but do not include systematic errors associated with the frames' aperture corrections, flux calibrations, and foreground reddening.  
   \label{fig:PNLF_NGC1380}} 
\end{figure}

The left panel of Figure~\ref{fig:PNLF_NGC1380} compares the luminosity function of objects securely classified as planetary nebulae in the three fields of NGC\,1380 with that derived from NGC\,1380's CENTRAL field by Sp2020. The diagram contains several features of note.

First, the central field NGC\,1380 contains one PN that is ``overluminous,'' i.e., it appears significantly brighter than the value of $M^*$ predicted from the rest of the PN population.   Like Sp2020, we have eliminated this object from our analysis, as its inclusion would greatly worsen the overall fit to the empirical function (decreasing its likelihood by a factor of $\sim 3 \times 10^5$). Still, the object presents a puzzle:  its spectrum looks like that of an ordinary PN, with a very high \OIII/H$\beta$ ratio, negligibly faint lines of \NII and \SII, and an \OIII/H$\alpha$ ratio consistent with that seen in other bright planetaries \citep{Herrmann+08}.  Since the object is only $\sim 0.25$~mag brighter than $M^*$, its apparent luminosity could be explained by a superposition of two PN within the top $\sim 0.5$~mag of the luminosity function. However, there is no evidence for two components in the shape of the \OIII $\lambda 5007$ emission line. Specifically, we used the line fitting tool
pPXF \citep{Cappellari+04, Cappellari17} to measure the \OIII $\lambda 5007$ emission line more accurately than is possible by our Gaussian fitting algorithm. We find that the line profile is indistinguishable from the instrumental profile and there is no evidence of doubling.  If the object is composed of two separate sources, their positions and velocities must be consistent to within roughly $0\farcs 4$ and  75~km\,s$^{-1}$ (one spectral bin), respectively.

Alternatively, if the overluminous source identified in the MUSE observations of NGC\,1380 is a planetary nebula, it might be foreground to the galaxy. Fornax is known to have a substantial population of intracluster stars, \citep[e.g.,][]{Spiniello+18, Spavone+20, Cantiello+20}, and, since NGC\,1380 sits securely in the cluster core, $0.6^\circ$ from the central cD galaxy NGC\,1399, it is reasonable to assume some of these stars will be in the foreground.  In fact, several examples of apparently overly-luminous PNe have been found in the Virgo Cluster \citep{jacoby+90}, another system which is known to have a large population of intracluster stars   \citep[e.g.,][]{Williams+07, Longobardi+15, Mihos+17}.  If this is the explanation for the apparent brightness of the PN, then the object is at least $\sim 2$~Mpc in front of NGC\,1380, near the turn-around radius of the cluster \citep{Drinkwater+01}.  Although
the presence of an intracluster $M^*$ PN at the extreme edge of the galaxy cluster may seem unlikely, if the PN were much closer to the galaxy, it would not be identified as an outlier in the system's luminosity function.  Thus the hypothesis cannot be ruled out.

With the current data, it is difficult to know whether either of these explanations is correct. But it is important to recognize that sources that appear too bright for the empirical function given by equation~\ref{eq:pnlf} are occasionally found in PN surveys of other galaxies. Also, it is interesting to note that numerical PNLF simulations \citep[e.g.,][]{Mendez+97, Valenzuela+19} do show a shallower slope at the bright end, and can produce overluminous PNe when the sample size is large. \citet{Mendez+01} have demonstrated such an effect for a sample of 535 PNe detected in NGC\,4697 (see their Figures 14 and 15). Unless the PN sample contains enough objects to reliably define the shape of the luminosity function, or unless further research develops a theory for the existence of these objects, this source of contamination can lead to systematically lower PNLF distance estimates.

Fig.~\ref{fig:PNLF_NGC1380} also vividly illustrates the advantage of using DELF images rather than the normal MUSE data cubes.  Our luminosity function monotonically increases to $m_{5007} \sim 27.8$ before signs of incompleteness begin to set in.  The Sp2020 dataset also reaches this limit, but only in the outer regions of the CENTER field, where the surface brightness is relatively low.  Identifying faint and even intermediate brightness PNe in regions of high background is difficult without first subtracting a continuum; this is reflected in the Sp2020 sample.

Finally, as the right-hand panel of Fig.~\ref{fig:PNLF_NGC1380} illustrates, despite coming from the same data, the distance we derive from our PN photometry is slightly less ($\sim 0.07$~mag) than the one obtained from the Sp2020 dataset.  The cause of this difference can be inferred from Fig.~\ref{fig:CompPhotSp2020}.  Because the empirical PNLF has a sharp exponential cutoff, PNLF distances depend most strongly on the observed magnitudes of the brightest few PNe in the sample.    Fig.~\ref{fig:CompPhotSp2020} demonstrates that in NGC\,1380, these bright PNe lie primarily in the lower surface brightness regions of the galaxy, and for these objects, the Sp2020 magnitudes are systematically fainter than our measurements by a few hundredths of a magnitude.  This translates into the offset seen in Fig.~\ref{fig:PNLF_NGC1380}.  The probability distributions also show the effect that our smaller measurement uncertainties and a larger sample size have on the likelihood distribution:  the internal errors associated with our sample are $\sim 60\%$ smaller than those computed from the Sp2020 dataset.

The maximum likelihood solutions shown in Fig.~\ref{fig:PNLF_NGC1380} bring up an interesting issue for future PNLF measurements.  If we assume a foreground reddening of $E(B-V) = 0.046$ \citep{Schlafly+11} and an absolute value for $M^* = -4.53$ \citep{Ciardullo13}, then the most-likely distance modulus to NGC\,1380 is $(m-M)_0 = 31.10^{+0.04}_{-0.05}$ (16.6 Mpc), with additional systematic uncertainties associated with the errors on the frames' aperture corrections, flux calibration, and the amount of foreground reddening.  For comparison, the Surface Brightness Fluctuation method generally produces values between 31.20 and 31.60 \citep[e.g.,][]{Tonry+01, Blakeslee+09,Blakeslee+10}.  This offset of $\sim 0.3$~mag is not unexpected, and may be explained by a very small amount of internal reddening in the Cepheid galaxies used for calibration.  As has been pointed out by \citet{Ciardullo+93} and again by \citet{Ciardullo+02}, both the PNLF and SBF methods use relatively late-type galaxies to define the zero points of their distance scales.  However, any systematic difference between the amount of internal reddening present in these galaxies and that within the early-type systems targeted by the methods will lead to the SBF scale being overestimated and the PNLF scale being underestimated, with $\Delta \mu \sim 7 \Delta E(B-V)$, i.e., $\Delta E(B-V) \sim 0.04$.

Alternatively, we can compare our distance modulus to those derived from Cepheid and TRGB measurements of other cluster galaxies.  NGC\,1326A and 1365 are both projected within $\sim 1$~Mpc of the Fornax central cD galaxy (NGC\,1399), have radial velocities consistent with cluster membership, and have been surveyed for Cepheids by the {\it Hubble Space Telescope.}  Their mean Cepheid-based distance modulus of $(m-M)_0 = 31.10$ \citep{Freedman+01} is identical to our measurement.   On the other hand, the TRGB distance to NGC\,1365 is $0.19$~mag more distant than our value \citep{Jang+18}.  Moreover, if NGC\,1425 (a galaxy projected $5.6^\circ$ away on the sky) is also considered to be part of Fornax, then its Cepheid distance \citep[$(m-M)_0 = 31.60$;][]{Freedman+01}, when averaged with those of NGC\,1326A and 1365, makes the cluster's mean Cepheid distance consistent with that of the TRGB\null.  So it is possible that the PNLF distances are still biased slightly towards smaller values.


\subsection{Benchmark galaxy: NGC\,628} 
\label{subsec:results_NGC628}

NGC\,628 (M74) is a large SA(s)c spiral galaxy that has hosted two Type~II (SN2003gd and SN2013ej) and one Type Ic (SN2002ap) supernovae in the past two decades.  The galaxy is the brightest and largest member of its group, extending over $12\arcmin$ on the sky, and is viewed almost face-on, with an inclination angle of only $6.5^\circ$ \citep{Kamphuis+92}. Because the galaxy is so face-on, the distribution of PN velocities about the galaxy's systemic value of $v_{\rm rad} = 657$\,km~s$^{-1}$ \citep{Lu+93} is quite narrow, ranging from $\sigma_{v} \sim 50$\,km~s$^{-1}$ in the central arcmin to less than 15\,km~s$^{-1}$ $\sim 4$ disk scale lengths away \citep{Herrmann+09b}.

\begin{figure}[h!]
    \begin{minipage}{1.0\linewidth}
    \centerline{
    \includegraphics[width=100mm]{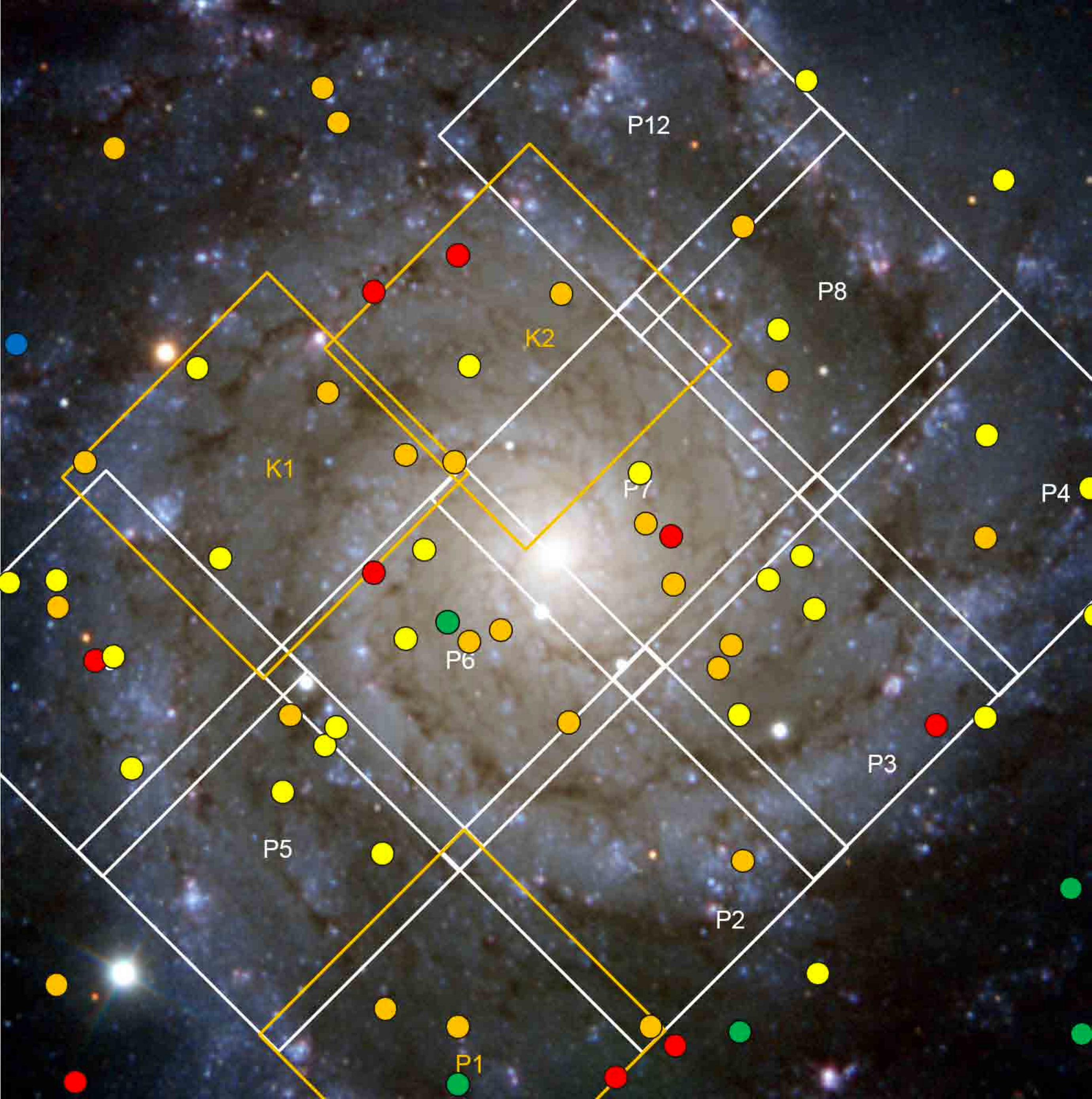}
    }
    \end{minipage}
     \caption{MUSE observations of NGC~628, with the pointings of Kreckel et al.\ (programme 094.C-0623, labeled K1 and K2) and Blanc et al.~(programme 098.C-0484, labeled P1 through P12) outlined as orange and white squares. A PNLF was determined by Kr2017 from PNe in K1, K2, and P1. PNe detected with OPTIC at the WIYN telescope by He2008 are overplotted as circles as an adaptation from their Fig.~2. The color code in red, orange, yellow, and green represents objects in the top 0.5, 1.0, 1.5, and 2.0 mag of the PNLF. Orientation: North up, East to the left. Background image credit: ESO PESSTO Survey \citep{Smartt+15}.
 \label{fig:NGC628_pointings}}
\end{figure}

TRGB measurements place NGC\,628 between 9.5 to 10.7 Mpc away, while estimates from the Type~II SN 2003gd and the galaxy's brightest supergiants generally give values between 7 and 10~Mpc \citep[see][and references therein]{McQuinn+17}.  Two PNLF distances also exist in the literature, but they are discrepant by $\sim 1.5 \sigma$:  while the interference-filter based photometry of He2008 gives 8.6~Mpc, the MUSE analysis by Kr2017 yields 9.6~Mpc.  The latter authors argue that their larger distance is a consequence of MUSE's ability to exclude compact SNRs from the PN sample.  However, since \citet{Davis+18} has argued that supernova remnants can rarely affect a PNLF distance measurement, it is worth revisiting the robustness of their measurement.

Figure~\ref{fig:NGC628_pointings} displays an image of NGC\,628, along with the locations of the PNe found by He2008 and the 12 pointings recorded in the MUSE archive.  The analysis by Kr2017 used three of these pointings, yielding a total of 63~PNe.  However, by analyzing the PNe found in all 12 pointings, we can make a detailed photometric comparison with 51 of the PNe measured by He2008. Although the full complement of 12 fields was obtained for another science case, and thus were not all taken under photometric conditions, the overlap with He2008 can be used for bootstrapping the different MUSE exposures to a common photometric zero point. Moreover, one of the fields (P3) was observed twice at different epochs, and under different observing conditions. The data set is therefore attractive for assessing the robustness of our analysis technique, as it allows us to perform both internal and external consistency checks.

\subsubsection{Data, source detection, and classification} \label{subsubsec:data_NGC628}

\begin{figure}[h!]
\begin{minipage}{1.0\linewidth}
     \centerline{
     \includegraphics[width=45mm,bb=50 60  650 650,clip]{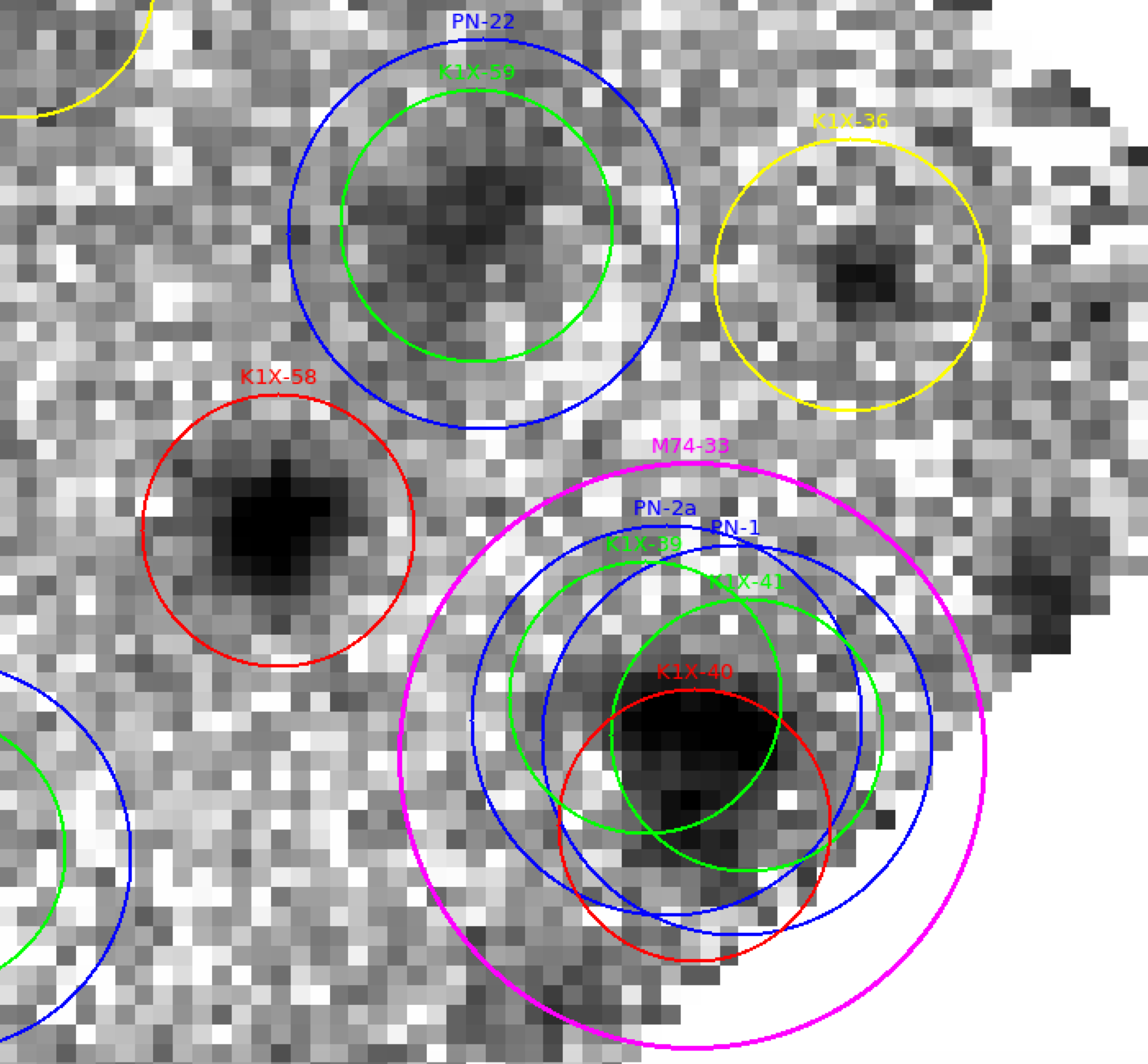}
     \hspace{3mm}     
     \includegraphics[width=45mm,bb=50 60  650 650,clip]{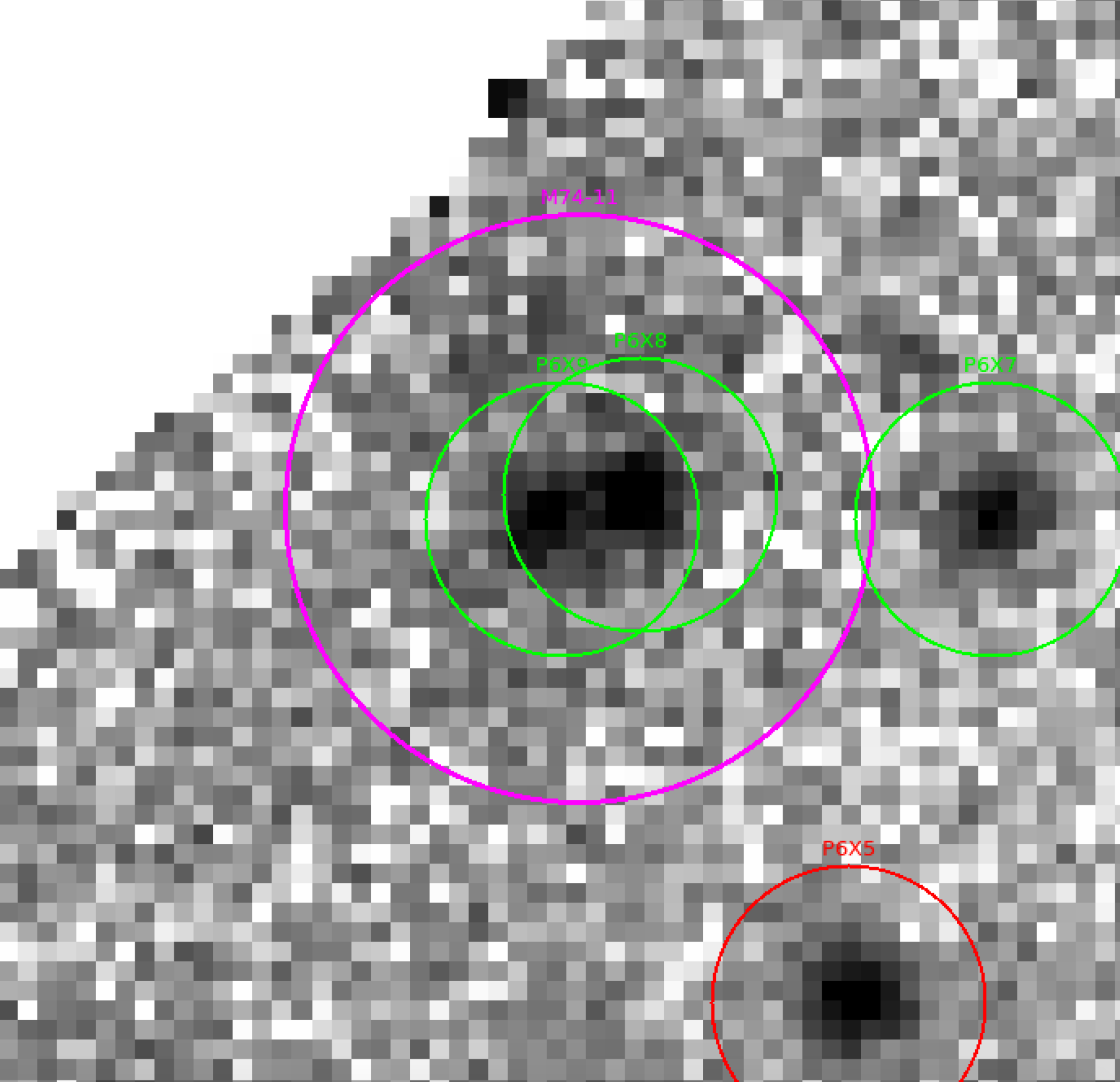}
     \hspace{3mm}
     \includegraphics[width=45mm,bb=120 60  650 590,clip]{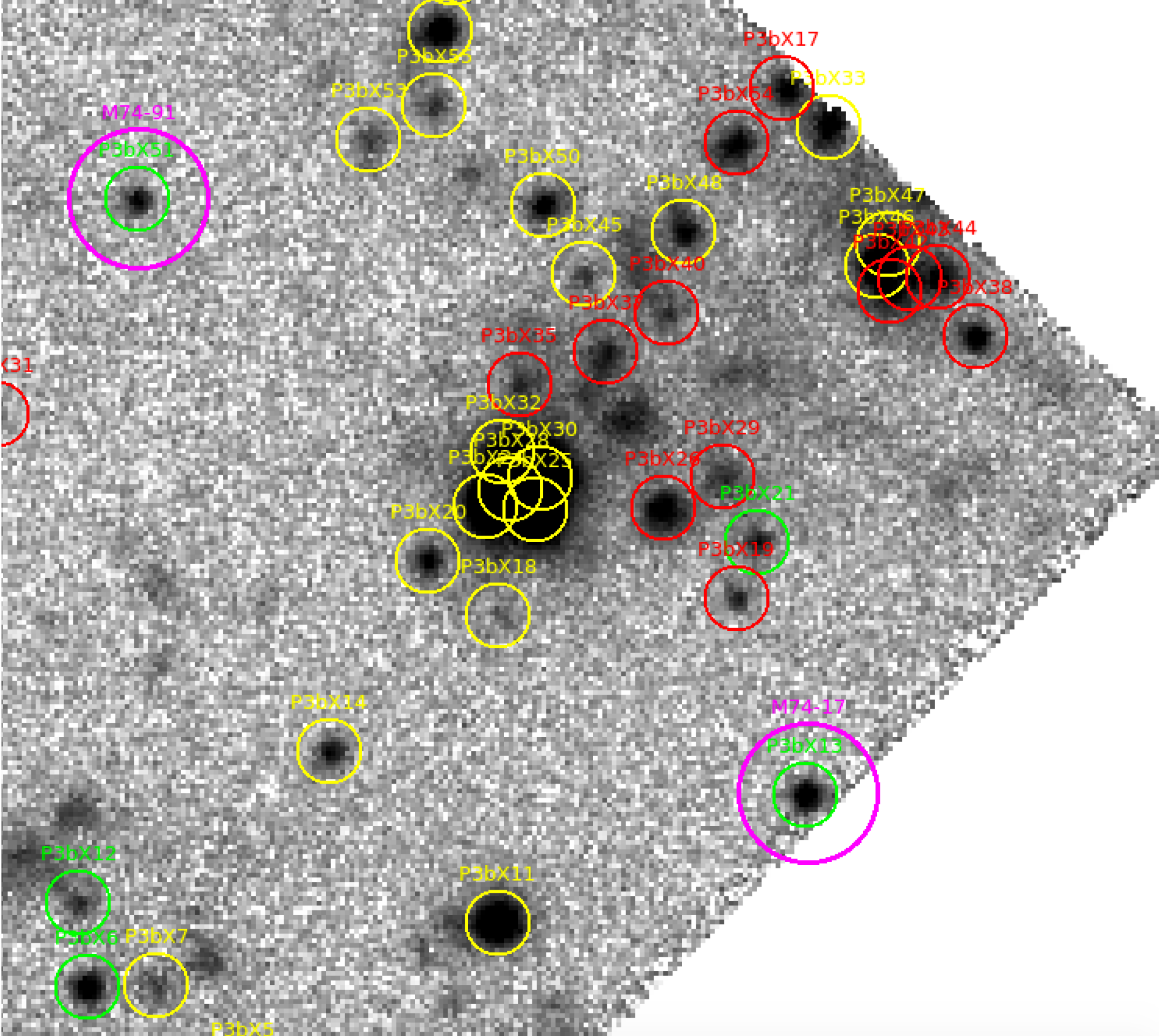}
    } 
    \end{minipage}
   \caption{Source confusion due to crowding. Left: M74-33 (magenta circle) was resolved by Kr2017 into two objects PN1 and PN2a (blue). Our \OIII map shows that there are actually three sources in the blend, two of which are PNe (green), the third one classified a SNR (red).
   Middle: M74-11 from He2008 in field P6 is resolved into two PNe, and is one of the outliers in the left panel of Fig.~\ref{fig:NGC628photcompall}. Right: the outliers below the 1:1 line in Fig.~\ref{fig:NGC628photcompint} in field P3 are identified as \ion{H}{2} regions (yellow) and SNRs (red) in a heavily crowded region of the galaxy. 
 \label{fig:NGC628blends}} 
\end{figure}

Except for the data reduction, where we immediately used the fully reduced data cubes as downloaded from the ESO archive, the basic steps for analysis were identical to the ones described for NGC\,1380 in Section~\ref{subsec:results_NGC1380}. Source detection was facilitated on the one hand by the fact that the narrow velocity dispersion perpendicular to the disk reduced the number of data cube layers that needed to be inspected.  However, the large number of point-like \ion{H}{2} regions and SNRs complicated the issue. We were able to efficiently identify a promising sample of PNe by supplementing the detections of the DAOPHOT \texttt{FIND} algorithm with visual inspection of the data cubes.  We then performed photometry and spectroscopy on the source list to measure emission line strengths and classify the objects via their line ratios (see Table~\ref{tab:NGC628detections}). 

\OIII maps for all fields are presented in the Appendix Figure~\ref{fig:NGC628all}.  As was found in the analysis of NGC\,1380, careful inspection of the images was needed to identify blends. Figure~\ref{fig:NGC628blends} shows a few examples of multiplicity, some of which have escaped detection in previous studies. Such cases can explain the \OIII magnitude differences between our MUSE data and the measurements of other studies (see Section~\ref{subsubsec:photometry_NGC628} below).

\begin{deluxetable*}{lccccccc}[bh!]
\tablecaption{Emission line point sources detected in NGC\,628\label{tab:NGC628detections}}
\tablewidth{100pt}
\tablehead{  \colhead{Pointing} & \colhead{Confirmed PNe} & \colhead{Candidate PNe } & \colhead{~~~SNRs~~~} & \colhead{~~~~\ion{H}{2}~~~~} & \colhead{seeing (1)} & \colhead{~~weather (2)~~}  & \colhead{~~$\Delta$m (3)~~} }
\startdata
K1  &    16    &  3   &  10   &  28  &  $0\farcs 77$ & clear & $+0.10$    \\
K2  &    31    &  5   &  20   &  18  &  $0\farcs 83$ & clear & $-0.20$    \\
P1  &    11    &  \dots   &  16   &   7  &  $0\farcs 76$ & clear & $-0.10$    \\
\hline
P2  &     9    &  3   &  12   &  11  &  $1\farcs 14$ & clouds & $-0.03$   \\
P3a &    13    &  3   &  19   &  39  &  $1\farcs 49$ & clear  & $-0.40$   \\
P3b &    13    &  3   &  19   &  39  &  $0\farcs 95$ & clouds & $-0.40$   \\
P4  &    11    &  \dots   &   9   &  16  &  $1\farcs 08$ & clear  & $-0.11$   \\
P5  &    10    &  \dots   &  15   &  35  &  $1\farcs 05$ & clouds & $-0.20$   \\
P6  &    31    &  5   &  17   &  21  &  $0\farcs 69$ & clouds & $-0.00$   \\
P7  &    38    & 12   &  17   &  37  &  $0\farcs 70$ & clouds & $-0.35$   \\
P8  &    13    &  2   &   6   &  28  &  $0\farcs 82$ & clouds ! & $-0.80$   \\
P9  &     6    &  3   &  13   &  23  &  $0\farcs 96$ & clouds ! & $-0.95$   \\
P12 &    15    &  4   &   6   &  32  &  $0\farcs 75$ & clouds ! & $-0.14$   \\
\enddata 
\tablecomments{(1) Seeing FWHM (ESO Archive information), (2) Retrieved from ESO Archive: ``clouds": some clouds registered during the night,``clouds !": clouds passing during an exposure, (3) zero-point offset used to match the He2008 photometry.}
\end{deluxetable*}


\subsubsection{Photometry} \label{subsubsec:photometry_NGC628}
\OIII measurements were performed in the standard way as described above. However, our photometry was limited by two issues.  The first was weather: according to the ESO archive, most of the MUSE observations of NGC\,628 were affected by clouds.  More serious was the lack of point sources in the field.  Unlike NGC\,1380, we found it difficult to identify foreground stars suitable for aperture correction measurements.  As shown in in Fig.~\ref{fig:PSFapcor}, the FWHM vs.\ wavelength relation for NGC\,1380 was well-behaved, and followed the expectations of atmospheric turbulence. However, in NGC~\,628, there were significant differences between the point sources candidates of each field, and for some objects, we even measured the FWHM to increase with wavelength. Such an effect could conceivably be produced if the target objects were not actually stars, but globular clusters with a large population of blue stragglers concentrated in the core.  Attempts to create a model PSF by stacking PN images also proved problematic, as the result was sensitive to the surface brightness of the galaxy's ubiquitous diffuse background emission. The aperture corrections, which are the dominant source of uncertainty in our photometry, and the catalogue of PN candidates are presented in the Appendix, Tables~\ref{tab:NGC628apcor} and \ref{tab:NGC628Photo}, respectively.  

\begin{figure}[h!]
\begin{minipage}{1.0\linewidth}
    \centerline{
     \includegraphics[width=90mm,bb=120 60  600 480,clip]{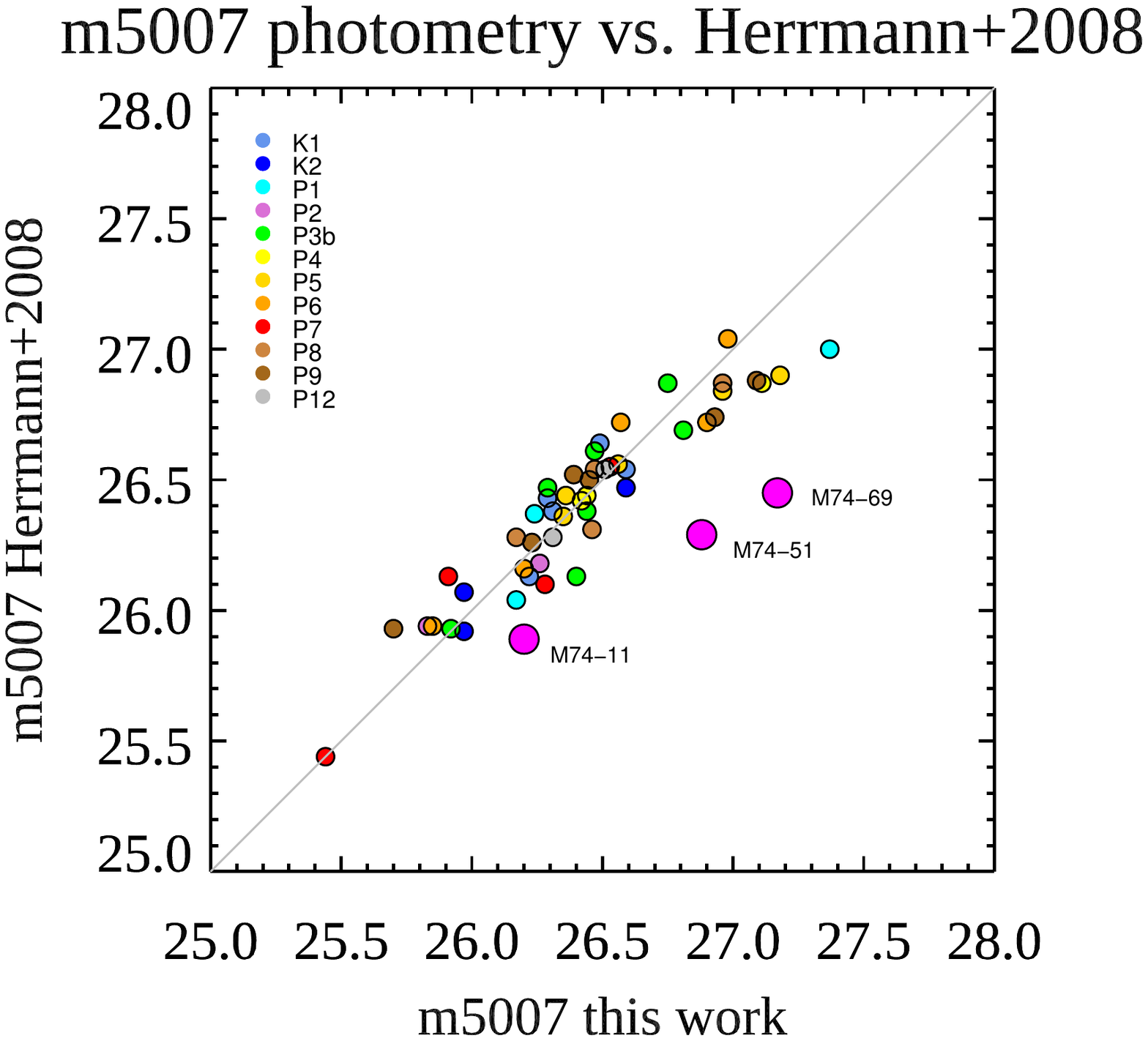}
     \includegraphics[width=90mm,bb=120 60  600 480,clip]{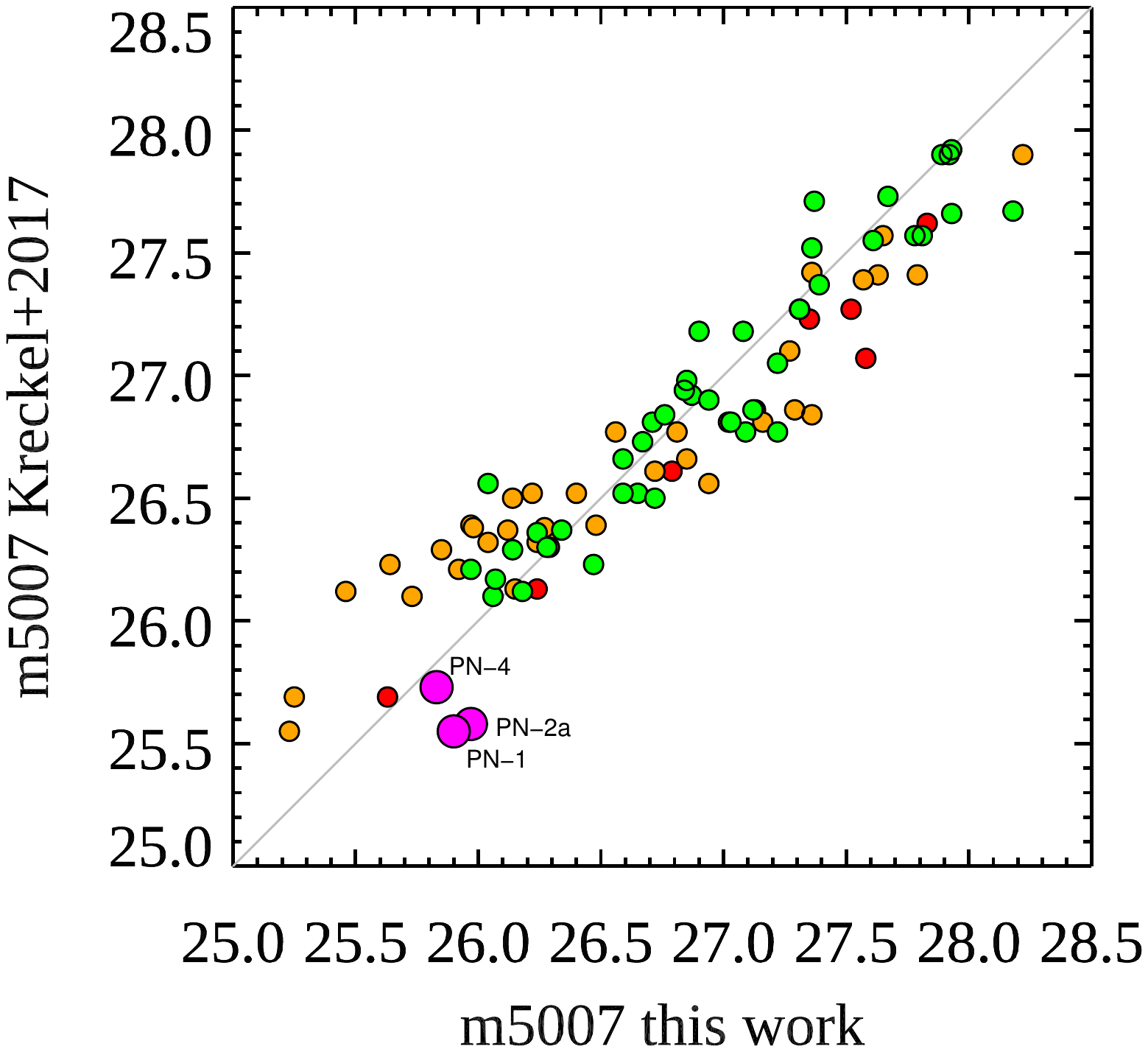}
    } 
    \end{minipage}
   \caption{Comparison of our $m_{5007}$ photometry in NGC\,628 with PN photometry from He2008 (left) and Kr2017 (right).  The color-coding on the left represents different MUSE pointings; that on the right illustrates that different classes of objects, with PNe in green, \ion{H}{2} regions in orange, and SNRs in red.  The large magenta circles are indicating object blends (see text). 
 \label{fig:NGC628photcompall}} 
\end{figure}

\begin{figure}[h!]
\begin{minipage}{1.0\linewidth}
    \centerline{
     \includegraphics[width=90mm,bb=120 60  600 480,clip]{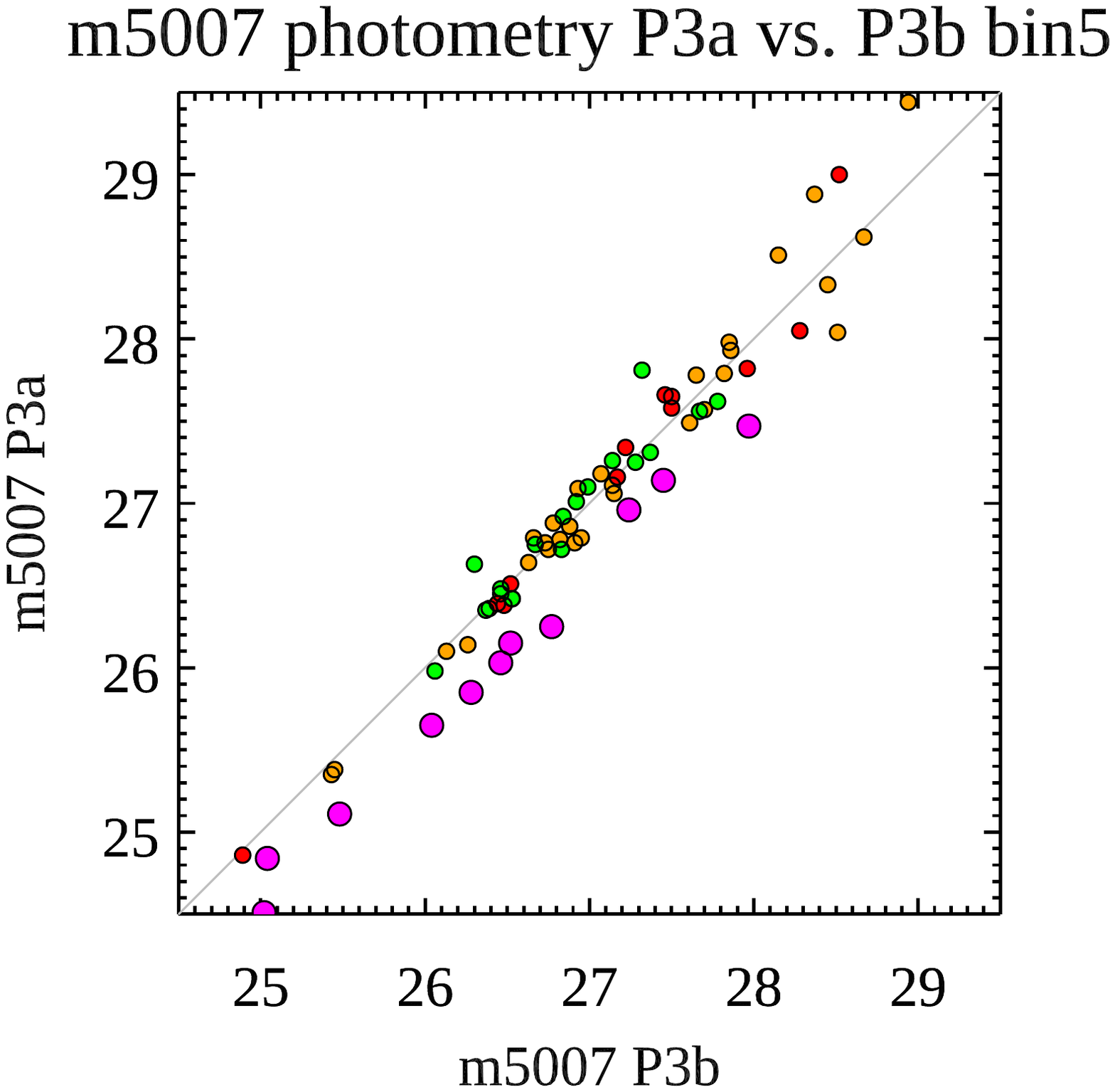}
    } 
    \end{minipage}
   \caption{Comparison  of photometry in field P3 from two MUSE observing runs with very different image quality (P3a: $1\farcs 49$, July 25, 2017; P3b: $0\farcs 95$, November 13, 2017). PNe are in green, \ion{H}{2} regions in orange, and SNRs in red.  
   For the interpretation of outliers (magenta), see text.
 \label{fig:NGC628photcompint}} 
\end{figure}

Because of these difficulties, we used the He2008 PN photometry as the flux standard system for our measurements.  A comparison of our corrected $m_{5007}$ values with the He2008 data is documented in the Appendix Figure~\ref{fig:NGC628photcomp2}. The agreement between the two datasets is generally good, except for the outliers caused by blending. A collapsed version of measurements for all 51 objects in common with He2008 is shown in the left panel of Figure~\ref{fig:NGC628photcompall}. For PNe brighter than $m_{5007}=27.0$ the residuals display a standard deviation of 0.12\,mag, which is in reasonable agreement with the uncertainties quoted by He2008 ($\pm0.058$\,mag at $m_{5007} = 26.0$ and $\pm0.135$\,mag at $m_{5007} = 27.0$).

The right panel of Fig.~\ref{fig:NGC628photcompall} adjusts our photometry in fields K1, K2, and P2 by $-0.2$~mag, and compares the measurements with those of Kr2017.  The plot is broken down by object classes, with \ion{H}{2} regions coded in orange, SNRs in red, and PNe in green.  The magenta points at the bright end of the distribution are objects where undetected source confusion affected the Kr2017 photometry.  The standard deviation for the PN measurements is 0.21\,mag, in agreement with the Kr2017 error estimates. See also Fig.~\ref{fig:NGC628photcomp} in the Appendix for comparison plots per field K1, K2, and P1.

Fig.~\ref{fig:NGC628photcompint} shows a comparison of the $m_{5007}$ magnitudes for all emission line point sources (PNe, \ion{H}{2} regions, SNRs) obtained in field P3 at two different observing epochs under different observing conditions. While the He2008 photometry had a completeness limit of $m_{5007}=26.5$, the MUSE data extends roughly 2.5 magnitudes fainter, with an internal dispersion of 0.15\,mag and a mean residual of 0.004\,mag for PNe brighter than 28.0\,mag (excluding outliers). The bright outliers (magenta) between $25 < m_{5007} < 27$ are \ion{H}{2} regions and SNRs in crowded regions that suffer from blending (several examples are shown in the right panel of Fig.~\ref{fig:NGC628blends}). For this internal consistency check, no zero point shift has been applied.

\subsubsection{The PNLF of NGC 628} \label{subsubsec:NGC628PNLF}

\begin{figure}[h!]
    \begin{minipage}{1.0\linewidth}
    \includegraphics[width=98mm,bb=0 0 600 400,clip]{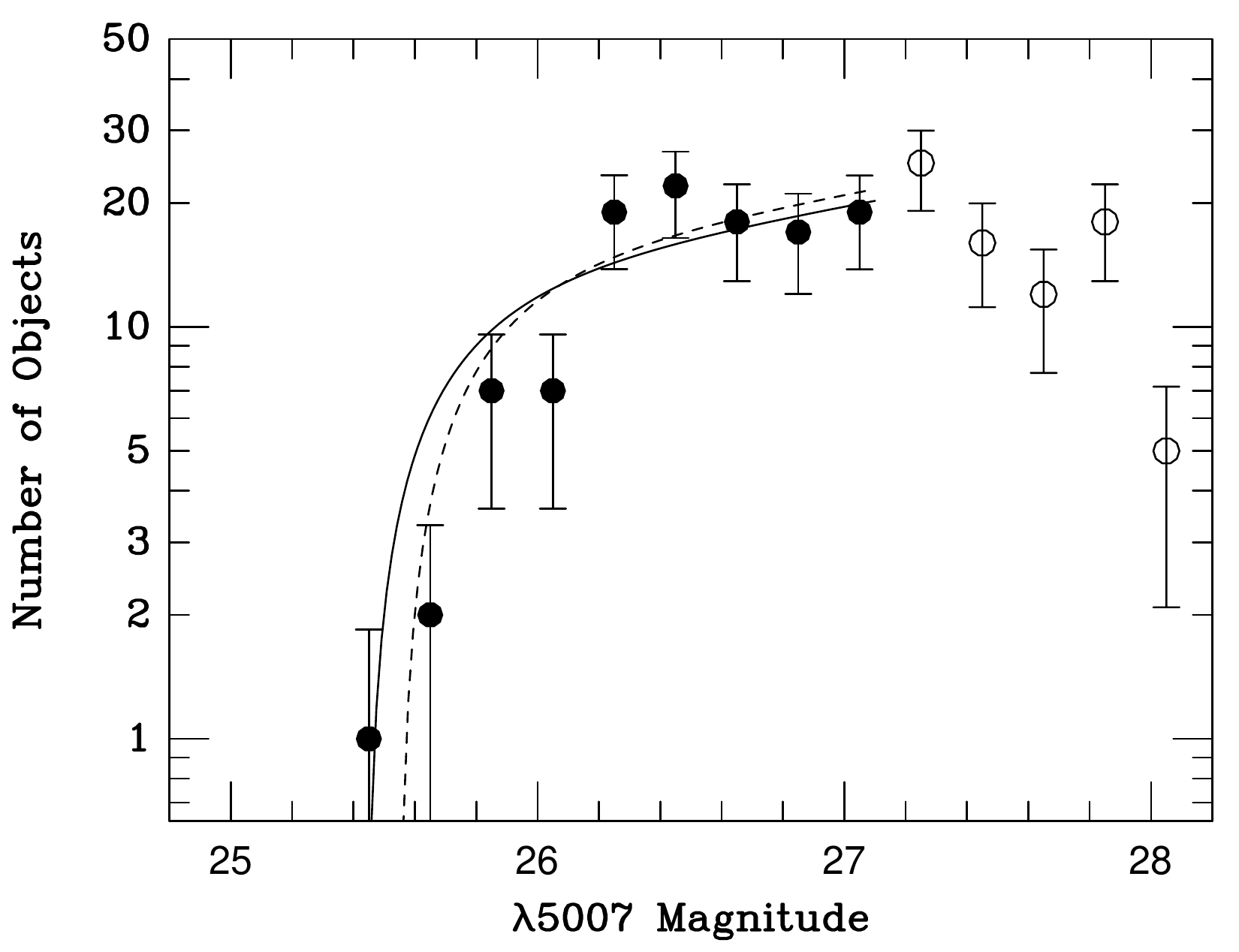}
    \includegraphics[width=98mm,bb=0 0 600 400,clip]{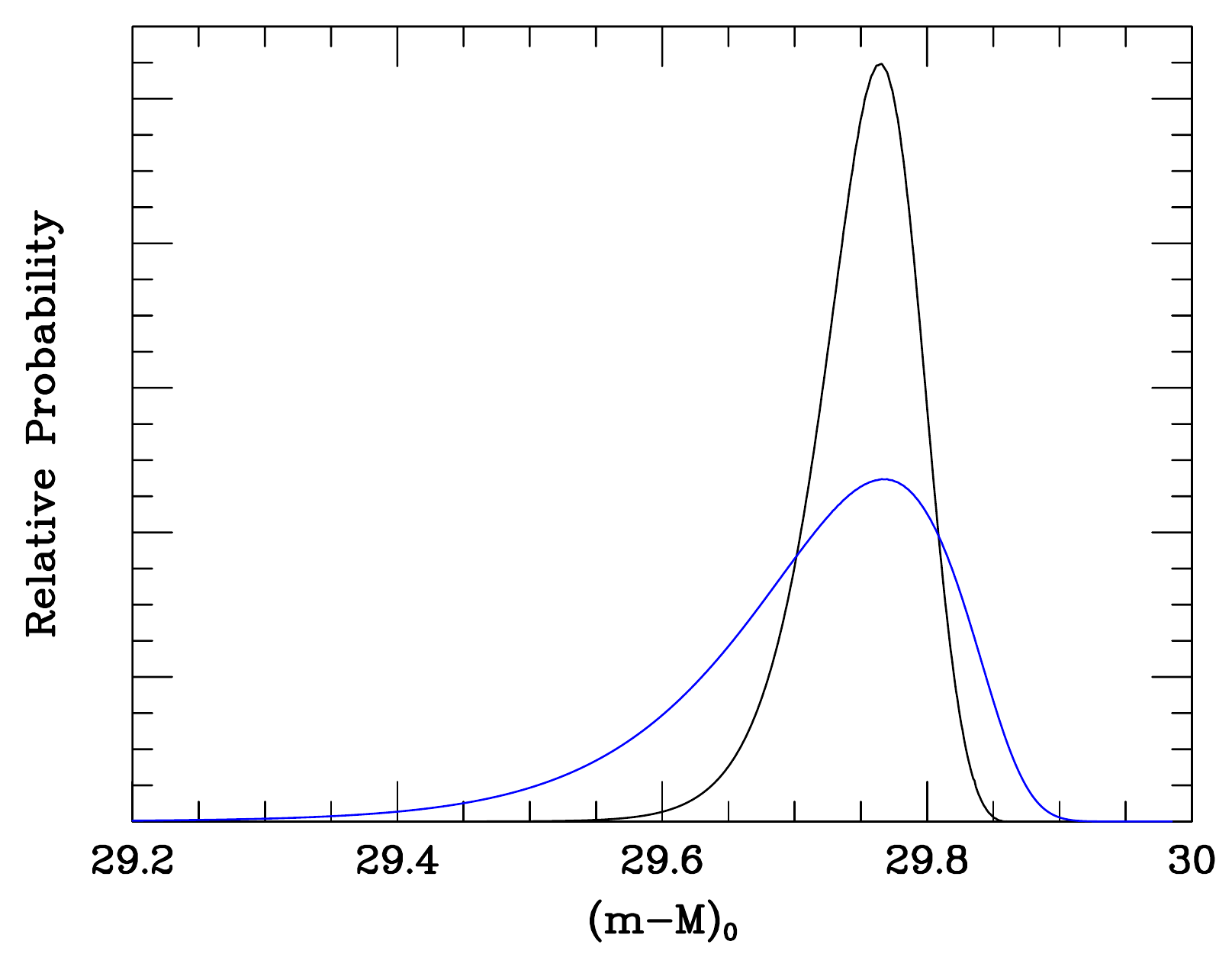}  
     \end{minipage}  
   \caption{The PNLF of NGC\,628.  Left: The observed PN luminosity function binned into 0.2~mag intervals.  The solid curve is the empirical function of equation~\ref{eq:pnlf} shifted by the most likely apparent distance modulus; the dashed curve is the best-fit solution if the brightest PN is omitted from the sample.  Right:  The relative likelihood versus distance modulus curve for our PN sample (black curve) and Kr2017 sample (blue curve).  Both datasets produce the same most-likely distance, but our larger sample size results in a much smaller uncertainty.   
   \label{fig:PNLF_NGC628}} 
\end{figure}

The left-hand panel of Figure~\ref{fig:PNLF_NGC628} displays the bright-end of the planetary nebula luminosity function for NGC\,628.  Although our PN observations extend all the way to $m_{5007} \sim 29$, we only use the top $\sim 1.5$~mag of the function in our fit. This is due to the fact that, in star-forming populations, the PNLF is not monotonic:  instead of exponentially increasing at faint magnitudes, the function exhibits a distinctive dip 2 to 4 magnitudes below $M^*$.  This feature, first identified by \citet{Jacoby+02}, occurs in PN populations that are dominated by objects with intermediate and high-mass cores and is discussed in \citet{Ciardullo+10}.  Since equation~\ref{eq:pnlf} does not attempt to model this downturn, we truncate our fit at $m_{5007} \sim 26.9$, and do not use the faintest $\sim 55\%$ of our sample.

The right-hand panel of Fig.~\ref{fig:PNLF_NGC628} displays the uncertainty in our distance modulus.  Assuming a foreground galactic extinction $E(B-V) = 0.062$ \citep{Schlafly+11}, our most-likely distance modulus is $(m-M)_0 = 29.76^{+0.03}_{-0.05}$ (or 8.95 Mpc), where the uncertainties represent only the statistical errors of the fit.  When analyzed using our maximum likelihood technique and using the same value of $M^*$, the Kr2017 data give $(m-M)_0 = 29.76^{+0.12}_{-0.05}$, while the PNe of He2008 yield $(m-M)_0 = 29.73^{+0.06}_{-0.07}$. However, we should point out that this apparent consistency hides a potentially important issue: the brightest PN in our sample (P7-38), though not strictly overluminous, is 0.11~mag brighter than the next brightest source.  There is no reason to exclude the object from the analysis, as our best-fit to equation~\ref{eq:pnlf} is clearly acceptable.  However, if do we exclude the object, the best-fit solution becomes $\sim 90$ times more likely, and the galaxy's distance increases to $29.87^{+0.03}_{-0.05}$.  

This issue points out an important limitation of using planetary nebulae for distance determinations.  To determine a robust distance using the PNLF, one cannot depend solely on the magnitude of the most luminous PN in a galaxy.  Instead, one has to define the shape of the brightest $\sim 1$ mag of the PN luminosity function. Unfortunately, PNe in this critical magnitude range are rare:  the specific PN densities given by \citet{Ciardullo+05}, coupled with the galaxy bolometric corrections computed by \citet{Buzzoni+06} imply that an $M_V = -19$ galaxy will only contain $\sim 20$ PNe in the magnitude range of interest.  In the era of precision cosmology, this number is not sufficiently constraining, as the distance may be susceptible to contamination by bright interlopers, or it may be underpopulated, leading to a large uncertainty in distance.  As Fig.~\ref{fig:PNLF_NGC628} illustrates, a minimum of $\sim 50$ objects are needed for a robust measurement of the PNLF shape.  PN surveys must therefore sample at least $M_V \sim -20$ of a galaxy's luminosity.

\pagebreak

\subsection{Benchmark galaxy: NGC 474} \label{subsec:results_NGC474}

\begin{figure}[h!]
    \begin{minipage}{1.0\linewidth}
    \centerline{
    \includegraphics[width=160mm]{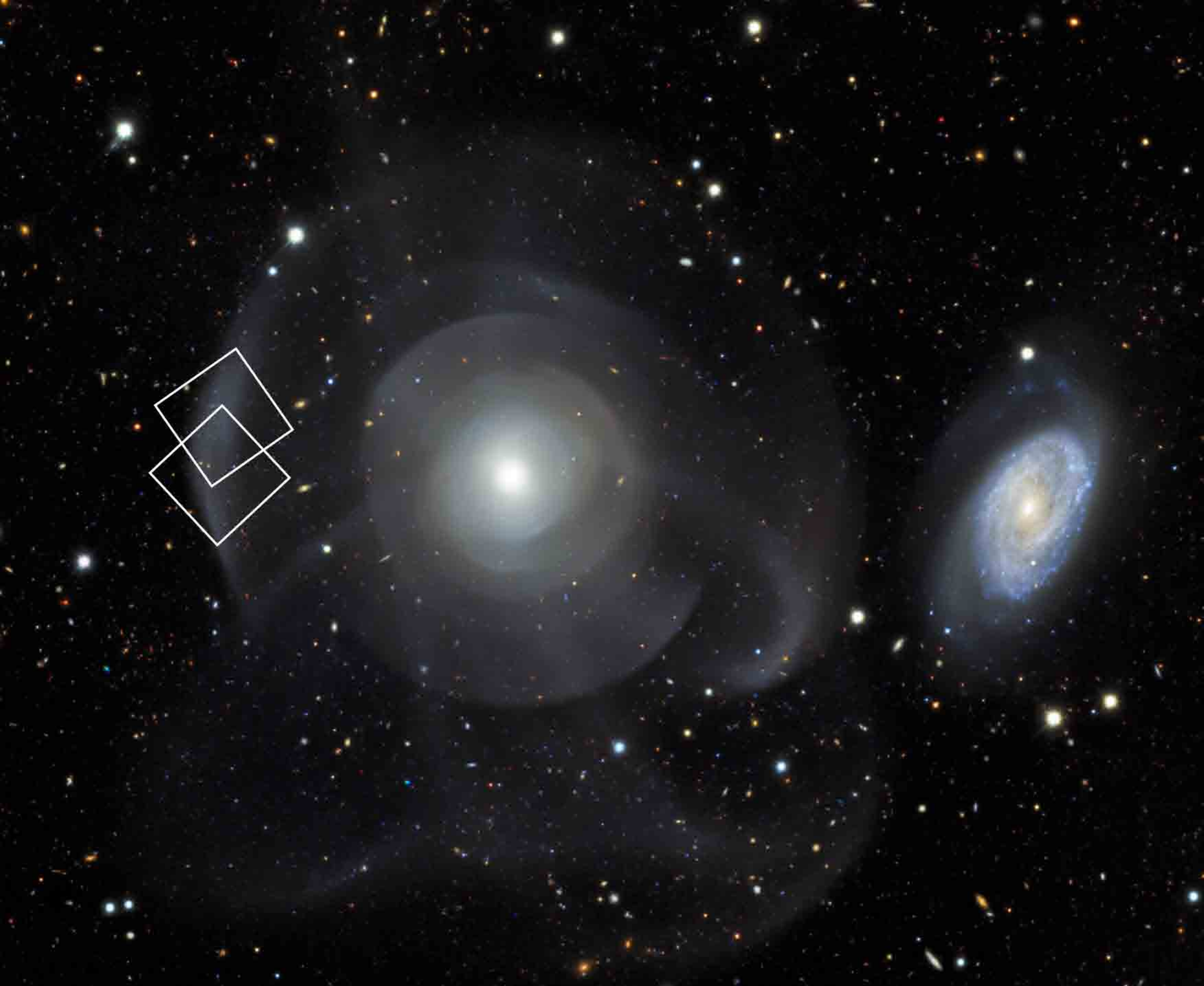}
    }
    \end{minipage}
     \caption{MUSE observations in NGC\,474. This $12'\times10'$ image (with north up and east to the left) shows the Arp~227 group with the dominating ring galaxy NGC\,474 in the center and the gas rich group member NGC\,470 to the west. Two MUSE pointings centered on an outer shell of the galaxy are illustrated; each represent a 5.1 hour exposure by Fe2020. Image credit: DES/DOE/Fermilab/NCSA \& CTIO/NOIRLab/NSF/AURA
 \label{fig:NGC474_pointings}}
\end{figure}

NGC\,474 is classified as a lenticular galaxy
\citep[Type SA0(s) in the RC3 catalogue;][]{RC3-1991} and is well known for its tidal tails and shell-like features.  IFU observations with SAURON have measured the system's radial velocity to be $2315\pm5$ km\,s$^{-1}$ \citep{Cappellari+11}, placing the object at a Hubble distance of roughly 32~Mpc. Tully-Fisher distances in the range of 15.3 to 35.0 Mpc were obtained by \citet{Bottinelli+84}, while more recent SBF measurements suggest a distance of between 30  and 33~Mpc \citep{Cantiello+07}.  Figure~\ref{fig:NGC474_pointings} show the galaxy's spectacular system of rings and tidal arms, which are presumably a result of a merger event. \citet{Fensch+20}, henceforth Fe2020, obtained deep MUSE exposures in the outer ring with the goal of studying the star formation history and metallicity of the structure. Their reported discovery of 8 PN candidates makes the object an interesting benchmark case, as it allows us to test our technique in a galaxy whose distance is well beyond the limit of classical PNLF measurements.

\subsubsection{Data, source detection, classification, photometry} \label{subsubsec:NGC474data}

\begin{figure}[ht!]
\centerline{ \includegraphics[width=0.8\hsize,bb=50 80 800 550,clip]{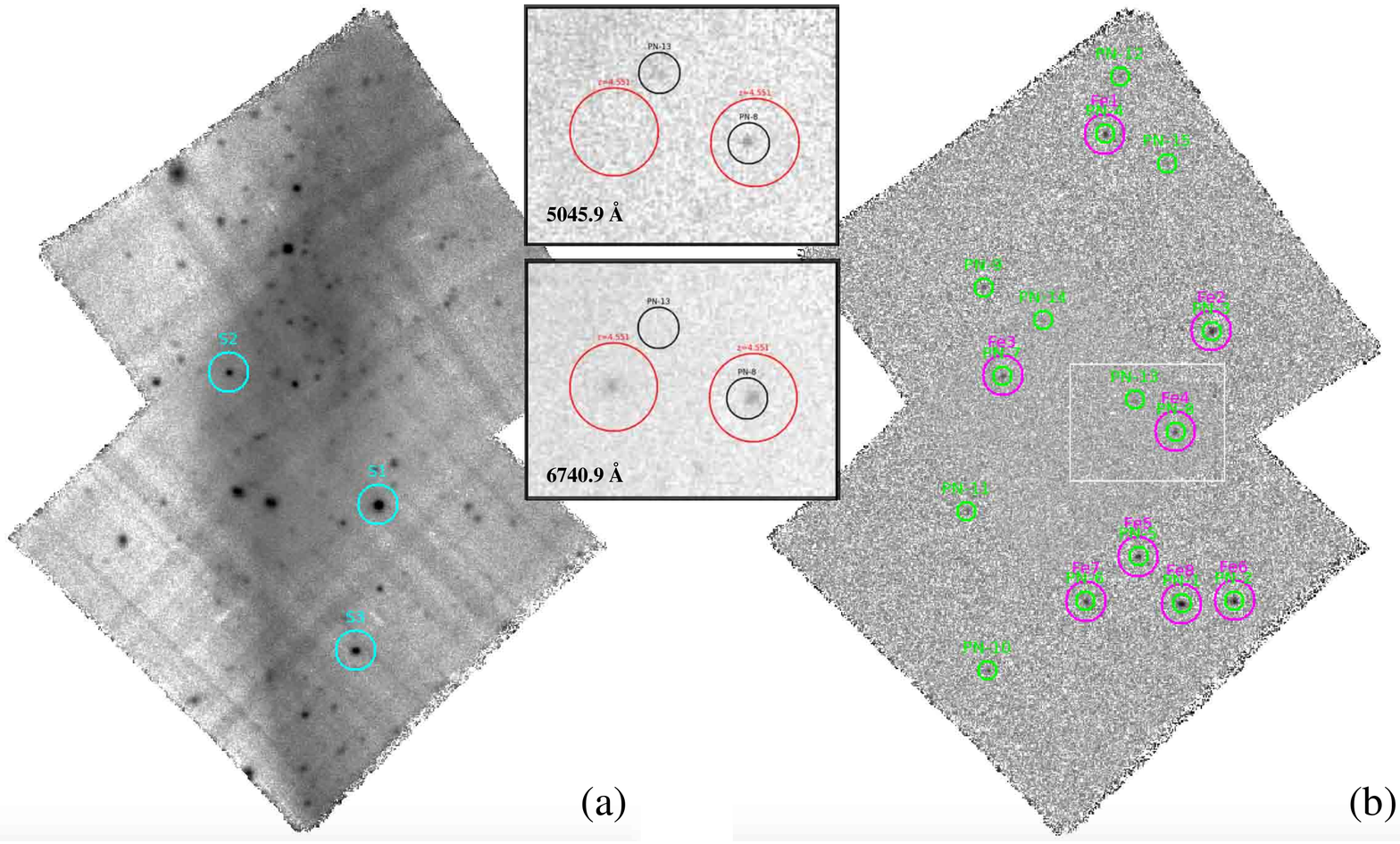}}
\caption{{\it Off\/} and {\it diff\/} images of NGC\,474, with north up and east to the left.  (a)  Continuum (off-band) image, with the three point sources used for our aperture correction marked in cyan.  (b) The {\it diff\/} image at radial velocity 2267\,km\,s$^{-1}$ (5044.69~\AA), summed over 3 layers of the data cube.  The PN candidates are marked as green circles. The objects detected by Fe2020 are marked in magenta. The inserts highlight the chance alignment of PN8 with a background galaxy (see also  Fig.~\ref{fig:PN_HII_SNR_spectra}).
 \label{fig:NGC474_off_diff}}
\end{figure}

The data retrieval and emission-line analysis proceeded as above. The fully reduced data product as available in the ESO archive consists of a single cube, where two datasets of total exposure time 5.1 hours were merged into one. Therefore, in the overlapping region of the two fields shown in Fig.~\ref{fig:NGC474_pointings}, the total exposure time was more than 10 hours. As can be seen in the off-band image displayed in Figure~\ref{fig:NGC474_off_diff}a, the faint surface brightness in this ring region is strongly modulated by the residual flatfield pattern discussed in Section~\ref{subsec:onoff}. This systematic error has completely vanished after DELF processing, yielding the {\it diff\/} images shown in Fig.~\ref{fig:NGC474_off_diff}b.  The exquisite quality and depth of this region allowed us to easily confirm the 8 objects found by Fe2020 and identify 7 additional PNe candidates.  (Since three of these new objects are single-line detections, we classify them as possible PNe and do not use them in our analysis.)  Thus, the observations yielded 12 likely PNe with \OIII magnitudes in the range $28.5 \lesssim m_{5007} \lesssim 30.2$.  Serendipitously, our PN search also discovered two $z=4.551$ Ly$\alpha$ emitters, one of which overlaps with PN8.  The region containing PN8 and a Ly$\alpha$ emitter is shown in the inset of Fig.~\ref{fig:NGC474_off_diff}.

\begin{figure}[h!]
\begin{minipage}{1.0\linewidth}
    \centerline{
     \includegraphics[width=75mm,bb=120 60  600 480,clip]{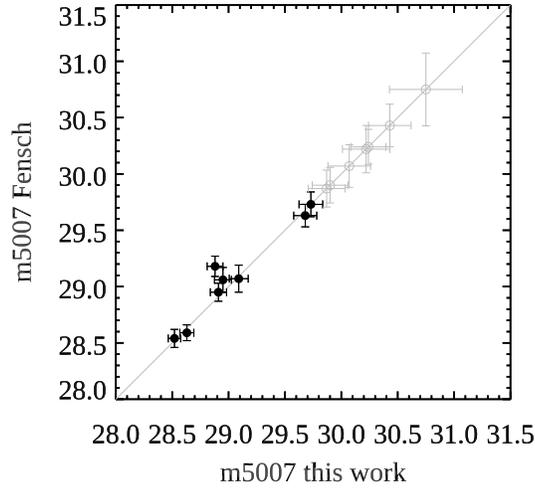}
    } 
    \end{minipage}
   \caption{Comparison of our $m_{5007}$ photometry with that of Fe2020.  The symbols in grey are intended to illustrate the depth of our photometry, but are not meaningful for this comparison).
 \label{fig:NGC474photcomp}} 
\end{figure}

As was our experience with NGC\,1380, the selection of point sources for measuring the cube's PSF and aperture correction was non-trivial, as many of the brighter objects are either globular clusters or background galaxies. Using three objects marked in  Fig.~\ref{fig:NGC474_off_diff}a, we measured \OIII FWHM values of $0\farcs 76$ in the combined field, and $0\farcs 78$ in the southern field.  Precision photometry would require us to accurately measure the PSF in each subfield separately, but for the sake of simplicity in this experiment, we have neglected the (very small) difference in image quality. The adopted aperture corrections and our catalogue of PN candidates are found in the Appendix, Tables~\ref{tab:NGC474apcor} and \ref{tab:NGC474Photo}, respectively.
 
To compare our photometry with the literature, we needed to convert the absolute $M_{5007}$ values listed by Fe2020 to apparent $m_{5007}$ magnitudes. However, with their assumed distance modulus of $(m-M)_0=32.45$, the comparison resulted in an offset of 0.5\,mag. With a choice of $(m-M)_0=32.95$, and a foreground extinction of 0.1\,mag \citep{Schlafly+11}, we obtain excellent agreement as shown in Figure~\ref{fig:NGC474photcomp}. The single outlier at $m_{5007}=28.9$ is probably due to the presence of three continuum point sources very close to the centroid of our PN4 (PN1 in the nomenclature of Fe2020). 
 
Note that the grey plot symbols, although self-referenced to our own magnitudes, are only intended to illustrate the range of photometry possible with deep MUSE exposures.  These data, which should be compared with the simulations described in Section~\ref{subsec:photomock} and shown in Fig.~\ref{fig:photoComp}, demonstrate that a $\sim 10$~hr exposure with MUSE can reach \OIII $\lambda 5007$ magnitudes as faint as $m_{5007} \sim 31$.  Moreover, with the MUSE image quality that is currently being achieved with adaptive optics this limit can be improved significantly.  For example, in the MUSE Extremely Deep Field combined data cube, an image quality of $0\farcs 6$ FWHM at 4700\,\AA\ has been achieved over a total exposure time of 140 hours \citep{Bacon+21}.  Thus, it should be possible to push PN detections well past the distance of NGC\,474.

\subsubsection{The PNLF of NGC 474} \label{subsubsec:NGC474PNLF}

\begin{figure}[h!]
    \begin{minipage}{1.0\linewidth}
    \includegraphics[width=98mm,bb=0 0 600 400,clip]{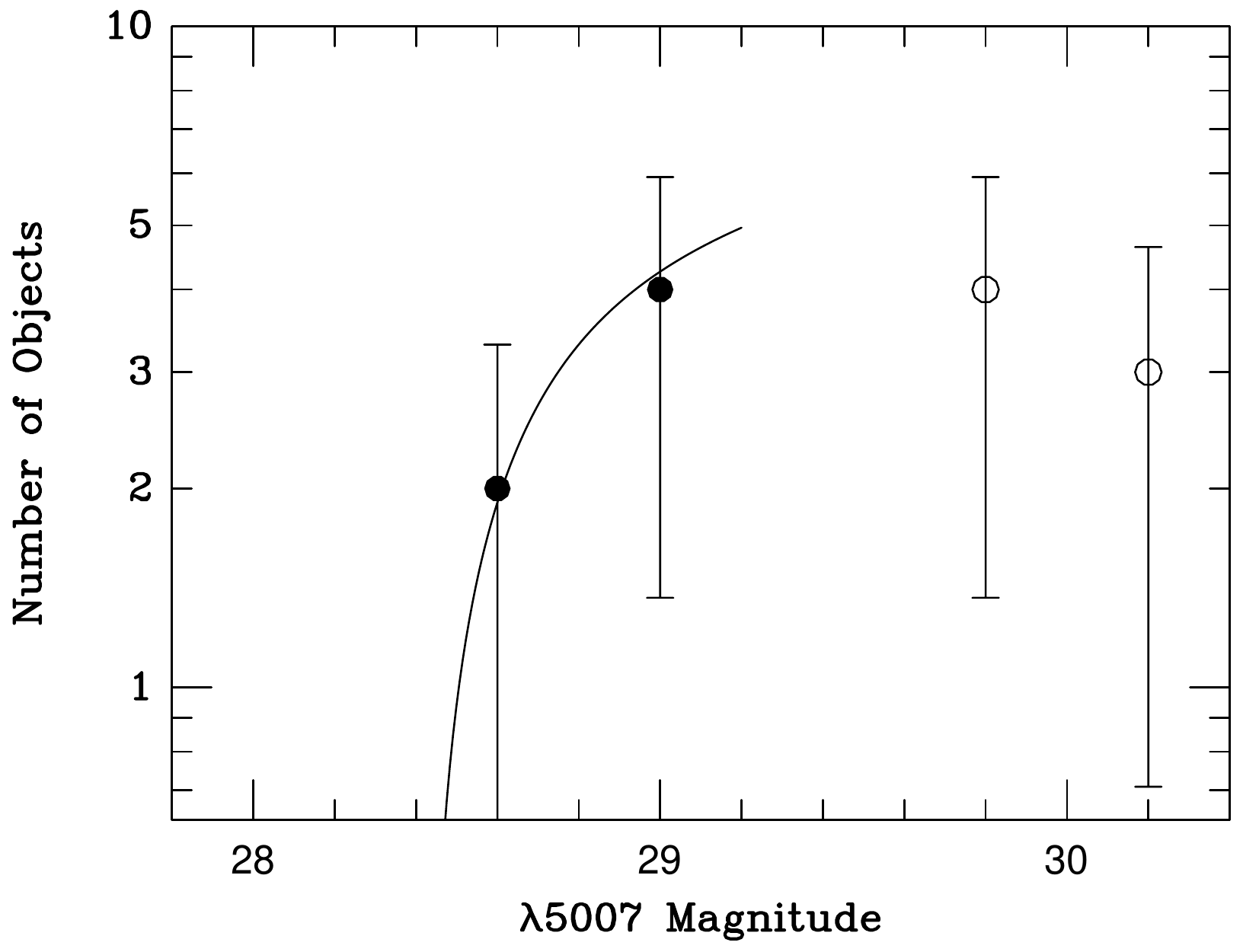}
    \includegraphics[width=98mm,bb=0 0 600 400,clip]{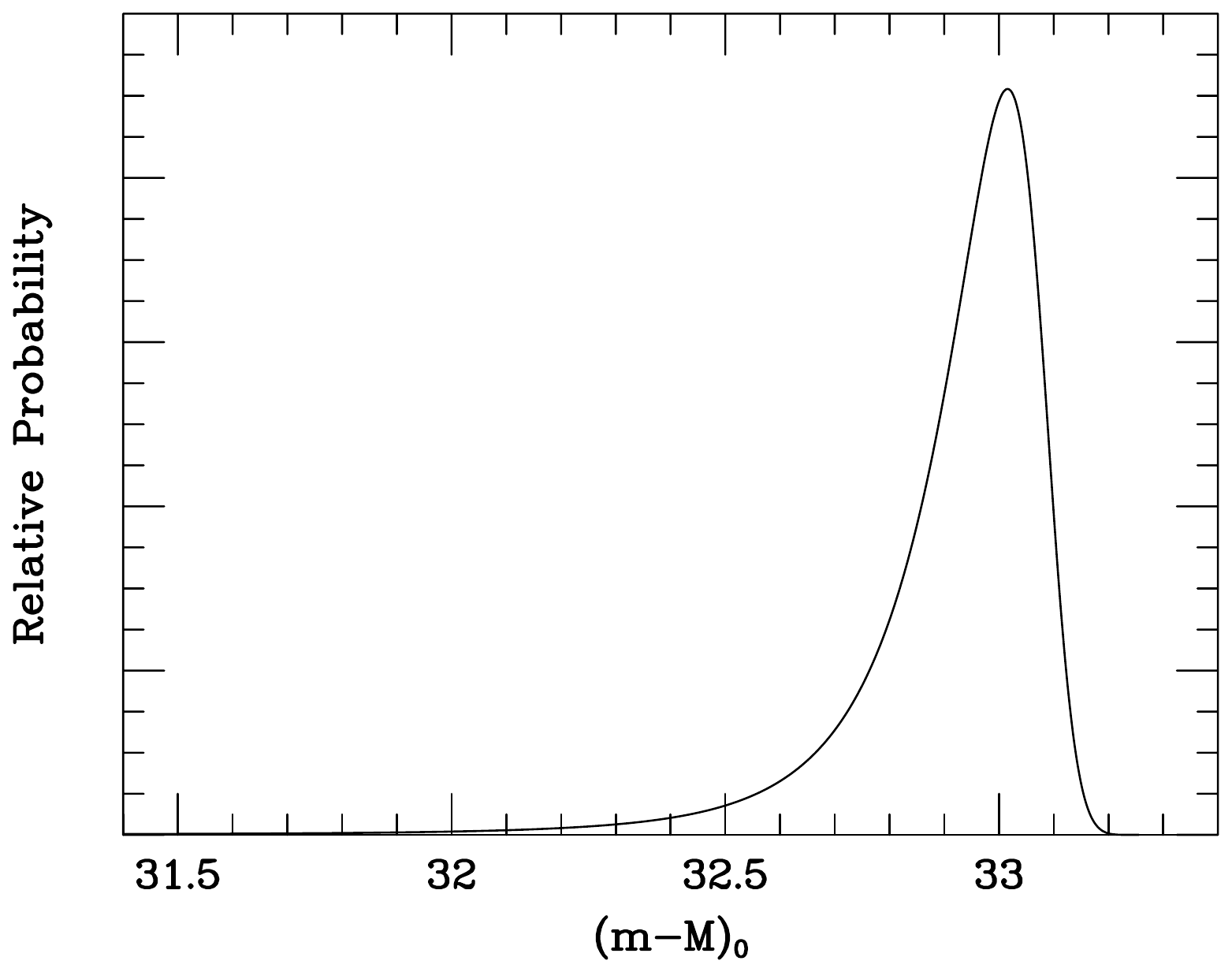}  
     \end{minipage}  
   \caption{The PNLF of NGC\,474.  On the left is the observed PNLF with our best-fit empirical law shown as a solid line; a completeness limit of $m_{5007} = 29.1$ has been assumed.  The right hand panel plots the relative likelihood versus distance modulus.  The asymmetrical nature of the probability curve is typical for PNLF measurements with a limited number of objects: it is always possible to bring the galaxy closer, as that just moves the PNe further down the luminosity function.  But since equation~\ref{eq:pnlf} has a sharp cutoff at $M^*$, there is a hard upper limit to how far away the galaxy can be.
   \label{fig:PNLF_NGC474}} 
\end{figure}

Figure~\ref{fig:PNLF_NGC474} illustrates our PN luminosity function for NGC\,474. The PN sample is small, not because of any issue with the quality or depth of the MUSE observations, but because the amount of galaxy luminosity encompassed by the survey fields is small. As a result, the data cannot be used to obtain a robust measure of the distance to the galaxy.  Nevertheless, if we assume that all the PNe are drawn from the empirical function of equation~\ref{eq:pnlf}, we can perform a maximum-likelihood analysis on the data and obtain an estimate of the galaxy's distance.  We adopt a 90\% completeness magnitude limit of $m_{5007} = 29.1$; this ad hoc assumption could be improved by running a series of artificial star experiments on the data cube, but is sufficient for our purpose.  If we then assume a foreground $E(B-V) = 0.03$ \citep{Schlafly+11}, the most-likely distance modulus obtained from the PNe is $(m-M)_0 = 32.86^{+0.08}_{-0.25}$, or $37.4^{+1.5}_{-4.0}$~Mpc.  Again, these error bars only represent the formal uncertainties of the fits.  Systematic errors associated with the aperture correction, foreground reddening determination, and assumed value of $M^*$
are not included in the calculation.  This distance is fully consistent with that obtained from the SBF analysis \citep{Cantiello+07}.

\section{Discussion} \label{sec:discussion}
\label{sec:Discussion}

\subsection{Accuracy of MUSE photometry}
\label{subsec:AccuracyPhoto}
Our goal is to utilize the unique capabilities of MUSE with adaptive optics to make the PNLF a precision distance determination tool for cosmology. To prove this claim, we must first assess the uncertainties involved in this technique, both random and systematic.  As the PNLF is a photometric standard candle, one of these uncertainties is the error in the photometry.  In 
Sections~\ref{subsec:tests} and \ref{sec:results} we have demonstrated using simulations and through the analysis of archival data that it is possible to obtain an internal photometric accuracy of 0.04\,mag for the brightest magnitude bins of the PNLF.  Moreover, tests indicate that our photometry is valid down to at least $m_{5007}=30$, with formal error estimates that are empirically validated by the statistics of mock data. 

Our archival benchmark tests on NGC\,1380 and NGC\,628 have shown that accurate PN photometry is achievable with exposure times of less than an hour under reasonably good seeing conditions ($0\farcs 8$) out to distances of $\sim20$\,Mpc. The experiment with archival data for NGC\,474 suggests that the same level of photometric accuracy can be reached with exposure times of 10 hours out to a distance of almost 40~Mpc.  It is therefore safe to assume that, with the image quality that can be expected from MUSE+GLAO observations, these numbers are conservative estimates. By observing $\sim 50$~PNe in the top $\sim 1$~mag of the PNLF in one or two AO-assisted exposures, MUSE should be able to obtain relative distances that are accurate to $\Delta\mu \approx 0.05$~mag.

Two major contributions to the total error budget come from the calibration of the photometric zeropoint, namely the uncertainties in a frame's aperture correction and flux calibration, and our imperfect knowledge of the foreground extinction.  The latter issue is one that affects most distance indicators and has improved with time as our measurements of the long-wavelength emission of dust has improved \citep[e.g.,][]{Aniano20}. Future work may also reveal ways to use the MUSE datacube to independently estimate E(B-V) from a stellar population analysis of the underlying host galaxy \citep{Zahid17,Li20,Li21}.

The former problem is more serious, especially for a complex, limited-field instrument such as MUSE\null.  We were able to cope with problems with the photometric zero point in our analysis:  since some of the archival data used for this study were taken under non-photometric conditions, frame-to-frame offsets in the calibration were apparent at the level of $\pm 0.1$\,mag.  Fortunately, as we will show in Section~\ref{subsec:FuturePNLF}, there are a number of observing strategies that can reduce this error to roughly a couple of percent.  

\subsection{Limitations of the PNLF}
\label{subsec:AccuracyPNLF}

The robustness of PNLF distances ultimately comes down to two questions:  what is the precise shape of the bright-end of the PNLF, and does this shape change with stellar population?   At this time, theory cannot answer these questions.  The value of $M^*$ and the observed circumstellar extinction around $M^*$ planetaries imply that the central stars that power \OIII-bright PNe have intrinsic luminosities of over $\sim 11,000\  L_{\odot}$ and central star masses greater than $0.66 M_{\odot} $\citep{Herrmann+09a, Davis+18}. Despite over 30 years of research, there is still no theory to explain the existence of these objects in Population~II systems such as elliptical galaxies.  Thus our understanding of the behavior of the PNLF's bright-end cutoff must be achieved empirically, via the analysis of a broad range of observations (e.g., \citet{bhattacharya2021}).

Based on the magnitude distribution of planetary nebulae in M31's bulge,  \citet{Ciardullo+89} originally proposed truncating the power-law distribution modeled by \citet{Henize+63} with the expression given by Equation~\ref{eq:pnlf}.  This law, which is similar to the results of numerical models that use ad hoc assumptions for the core masses of \OIII-bright PN central stars \citep{Jacoby89, Mendez+93, Mendez+97, Valenzuela+19}, has proved reasonably reliable and repeatable.  However, as discussed in Section~\ref{subsec:results_NGC1380}, ``overluminous'' \OIII sources are occasionally found in PN surveys.  Because most of the PNLF observations to date have been performed using narrow-band filters and have very little (if any) spectroscopic follow-up, it has been difficult to determine the nature of these sources.  Are they true PNe or unrelated objects such as supernova remnants, compact \ion{H}{2} regions, blends of multiple objects, or background galaxies?  If the objects are PNe and if they are indeed located within the targeted galaxy (as opposed to a foreground intergalactic object), then the empirical form of the PNLF will have to be modified in order for the PNLF to achieve the precision necessary to address questions of cosmology. 

MUSE provides the information needed to understand the nature of some overluminous sources.  Certainly, at least some of the overluminous \OIII sources have spectra that are indistinguishable from PNe \citep[e.g.,][and the MUSE observations of NGC\,1380 described in this paper]{Ciardullo+02b}.  But deep, narrow-band PN surveys also have their share of contaminants \citep[e.g.,][]{Kudritzki+00, Durrell+03, Longobardi+13}.  As shown above, MUSE spectra are very efficient at identifying the interlopers in PN surveys.  Moreover, by comparing PN luminosity functions inside and outside of clusters, MUSE will also be able quantify the importance of intracluster PNe in PNLF analyses. 

A related parameter that is key to using the PNLF for cosmology is the exact value of $M^*$ and its dependence on the host galaxy population.  Unlike Cepheids or the TRGB methods, there is no immediate prospect for a Galactic calibration of the PNLF\null.  Thus, the PNLF is a secondary standard candle, and its distances will always carry some error originating with the distance measurements to nearby galaxies.  At present, the uncertainty in $M^*$ is roughly $\sim 0.04$~mag \citep{Ciardullo13}, though this can be improved via MUSE observations of additional Cepheid and TRGB galaxies.

Ideally, we wish to develop a more thorough theoretical basis for the PNLF to improve our application of the PNLF and our confidence in the results. We are not at that point yet.

\subsection{Future application for the PNLF using MUSE}
\label{subsec:FuturePNLF}

One of the most important parameters for deriving accurate PNLF distances will be the image quality (i.e., the seeing). As noted by our simulations in Section~\ref{subsec:tests}, the difference between $1\farcs 0$ and $0\farcs 6$ is dramatic, both for extending the reach of the PNLF in distance and for enabling PN identification in high surface brightness regions of a galaxy.  The recent MXDF project \citep{Bacon+21} has demonstrated that an image quality of $0\farcs 6$ at a wavelength of 5000\,\AA\ can be consistently obtained when using the GLAO system at UT4 of the VLT\null. Future PNLF observations will clearly benefit from GLAO.

Many of the observations in the MUSE archive are not optimized for precision spectrophotometry.  For example, one potential source of error is the flux calibration. Typical MUSE observations include flux standards taken at the beginning and end of each night. Additional flux calibrations taken before and after PNLF exposures (with as similar an airmass as possible) can reduce any uncertainties incurred from subtle changes of atmospheric transmission over the course of a night. 

A related and serious issue is the need to derive a frame's aperture correction in order to place relative photometric measurements onto an absolute magnitude scale. The pre-selection of fields that contain sufficiently bright point sources will ensure that each frame contains suitable PSF standards. But care must be taken at this step, since globular clusters can often be mistaken for point sources; in general, the better the angular resolution, the easier it is to choose point sources.  Once chosen, the PSF stars can be analyzed by MUSE, but, if they are sufficiently bright, they can also be calibrated with spectrophotometric measurements from other telescopes. Such observations need only be taken at a wavelength near 5007\,\AA, and, because the aperture correction is a very slow function of wavelength the observing bandpass can be fairly broad, perhaps 250~\AA\ wide. Finally, an attractive opportunity to model the PSF for MUSE GLAO observations has been put forward by \citet{Fusco20F}. The {\em muse-psfrec} software can be used to reconstruct the PSF from the real-time adaptive optics telemetry information recorded by the MUSE adaptive optics system GALACSI\null.  

Another observing strategy that can assist PNLF measurements is the selection of fields having relatively low surface brightnesses.  The high surface brightness regions of a galaxy naturally have more planetary nebulae, as PNe are excellent tracers of the light, and vice versa.  However, even with the effectively narrow bandpass from MUSE, the dominant noise source in a PNLF observation is the brightness of the underlying galaxy background plus night sky. The location of a MUSE pointing must be optimized to maximize the number of PNe expected to be found while minimizing the number of fields necessary to obtain a statistically useful PN sample, i.e.,  $\gtrsim $50 PNe within the brightest mag of the PNLF.

A careful selection of galaxies is also needed for several reasons. First, the earlier criticisms of biases in PNLF distances beyond $\sim$15 Mpc needs to be better understood. Is this a bias arising from pushing into a regime beyond the data quality?  Is it due to the misclassification of other emission-line objects as PNe?  Or is the error intrinsic to the PNLF itself, perhaps due to the failure of the empirical form of the PNLF to account of ``overluminous'' objects, intracluster stars, and object blends?   MUSE observations in a few fields of suspect galaxies from previous PNLF studies will help resolve the source of a bias, or perhaps demonstrate that there is no bias.

Galaxy selection is also critical for extending the PNLF distances to $\sim 50$~Mpc. The target galaxies should ideally be well-structured early type systems to minimize confusion with \ion{H}{2} regions and SNRs, but also offer a sample of Type Ia supernovae to be calibrated. Since these galaxies are beyond the reach of the TRGB and do not have Cepheid variables, they would immediately contribute to the SN Ia calibration. 

Finally, it is worth noting the development of BlueMUSE \citep{Richard2019}. With a larger field-of-view of 2\,arcmin$^2$, a $\tilde15$\,\% higher throughput, and a higher resolving power of $R=3700$ (both at 5000\,\AA), BlueMUSE will become an even more efficient tool for exploiting the potential of the PNLF.

\section{Conclusions} \label{sec:conclusions}
In this paper, we have demonstrated that spatially resolved spectrophotometry when coupled with ground layer adaptive optics, will enable a break-through in precision PNLF distance determinations.  Such measurements may allow us to address the current tension in Hubble Constant measurements.   Specifically, in this paper  
\begin{itemize}
\item We have developed the differential emission line filter (DELF) technique to suppress systematic errors and to deliver high S/N ratio $m_{5007}$ photometry and PN radial velocities.
\item We have tested the technique extensively on simulated data.

\item We have shown that the DELF technique offers superior photometric performance in comparison with conventional data analysis techniques.  For objects near the PNLF cutoff, our typical photometric errors are better than 0.04~mag.

\item We have tested the technique on archival data of three benchmark galaxies with distances ranging from $\sim 10$~Mpc to more than 30~Mpc.  The three galaxies also contain a variety of stellar populations: one galaxy is a late-type spiral, one is a lenticular galaxy, and the third is an elliptical galaxy with shells and rings.

\item We have used the superb image quality and velocity resolution
of MUSE to determine that a non-negligible number of bright \OIII sources are actually the chance superposition of multiple objects.  Previous studies have unknowingly recorded some of these objects as extremely luminous PNe.

\item We have used our analysis techniques to push the PN observations of our benchmark galaxies to greater depths and to regions of higher galaxy surface brightness.  That has enabled us to produce PNLFs that extend $\sim 1$~mag deeper than previous analyses and contain more objects in the top $\sim 1$~mag of the luminosity function.

\item We have taken advantage of MUSE's ability to simultaneously record \OIII, H$\alpha$, [\ion{N}{2}], and [\ion{S}{2}] to discriminate PNe from from other types of emission-line objects, such as \ion{H}{2} regions, SNRs, and background galaxies.  The resultant ``pure'' sample of PNe has allowed us to better define the bright-end of the PNLF, which can be subject to contamination by interlopers.

\end{itemize}

Our validation of the DELF technique has been based on archival data that were generally taken for purposes other than PNLF distance determinations. In the future, the major uncertainties identified in our study, namely the determination of aperture corrections and accurate flux calibration, can efficiently be minimized with PNLF-specific observing strategies.
Based on the existing data for our most distant benchmark galaxy, NGC\,474, we expect that two $\sim 5$ hour MUSE exposures with $0\farcs 6$ seeing will detect $\sim 50$~PNe in the top magnitude of the luminosity function in a galaxy nearly 40~Mpc away.  Images with $0\farcs 6$ FWHM at 5000\,\AA\ are obtainable with MUSE using the GLAO adaptive optics system. The PNLF distance modulus determined from such data can be expected to have statistical errors of only $\pm0.05$\,mag.

\section{Acknowledgments} \label{sec:acknowledgements}
Based on data obtained from the ESO Science Archive Facility, programme IDs 094.C-0623, 095.C-0473, 098.C-0484, 296.B-5054(A), 099.B-0328. This research has made use of the NASA/IPAC Extragalactic Database (NED), which is funded by the National Aeronautics and Space Administration and operated by the California Institute of Technology.  The Institute for Gravitation and the Cosmos is supported by the Eberly College of Science and the Office of the Senior Vice President for Research at the Pennsylvania State University.
MMR acknowledges support from BMBF grant 03Z22AB1A, PMW from BMBF 05A20BAB. The authors would like to thank the referee Roberto M\'{e}ndez for careful reading and useful suggestions to improve the quality of the paper.
\bibliography{MUSE-PNLF}{}
\bibliographystyle{aasjournal}


\appendix

\section{Supplementary Figures} \label{sec:SuppFig}

\begin{figure}[h!]
\begin{minipage}{1.0\linewidth}
    \centerline{
     \includegraphics[width=60mm,bb=120 60  600 480,clip]{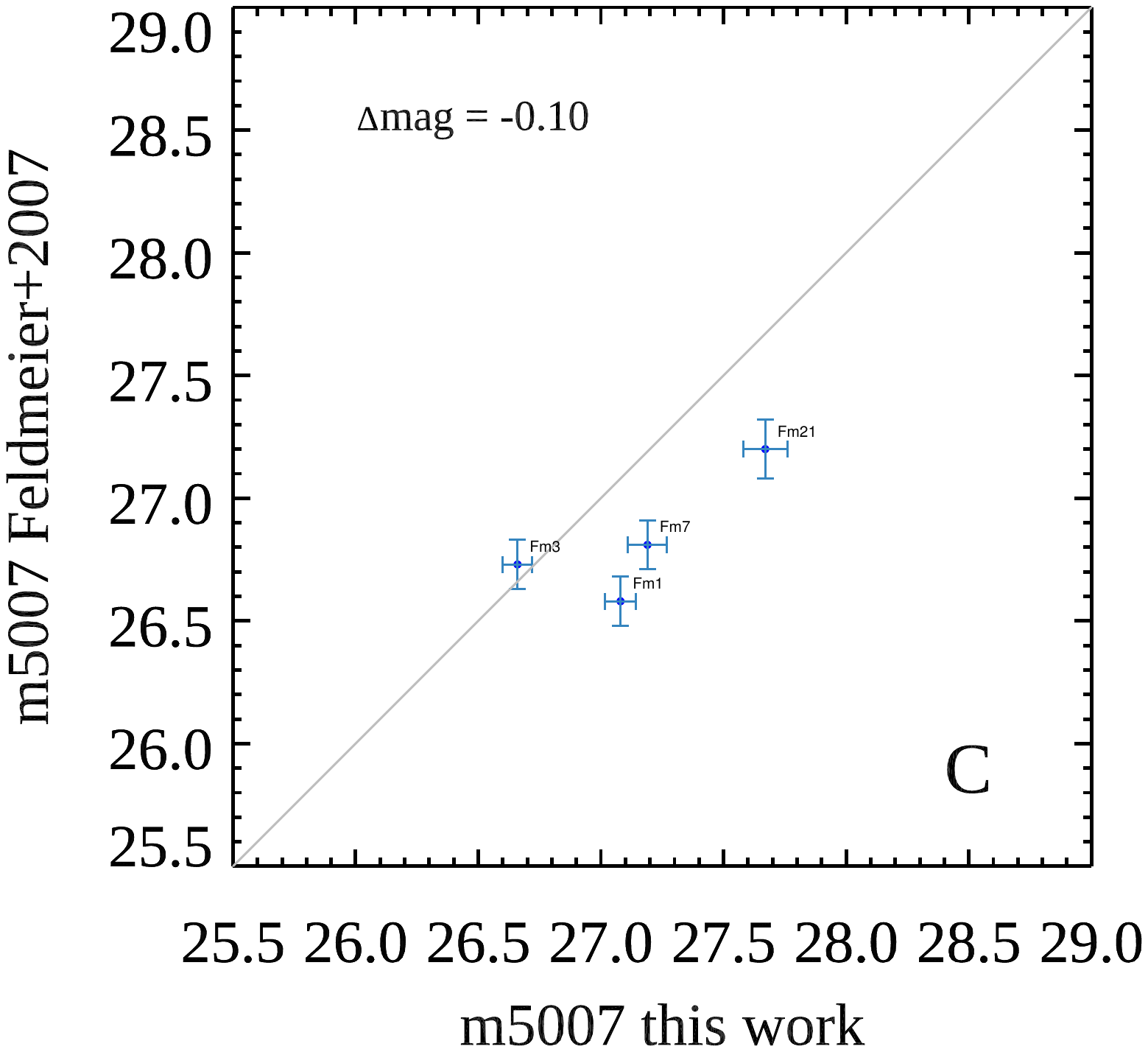} 
     \includegraphics[width=60mm,bb=120 60  600 480,clip]{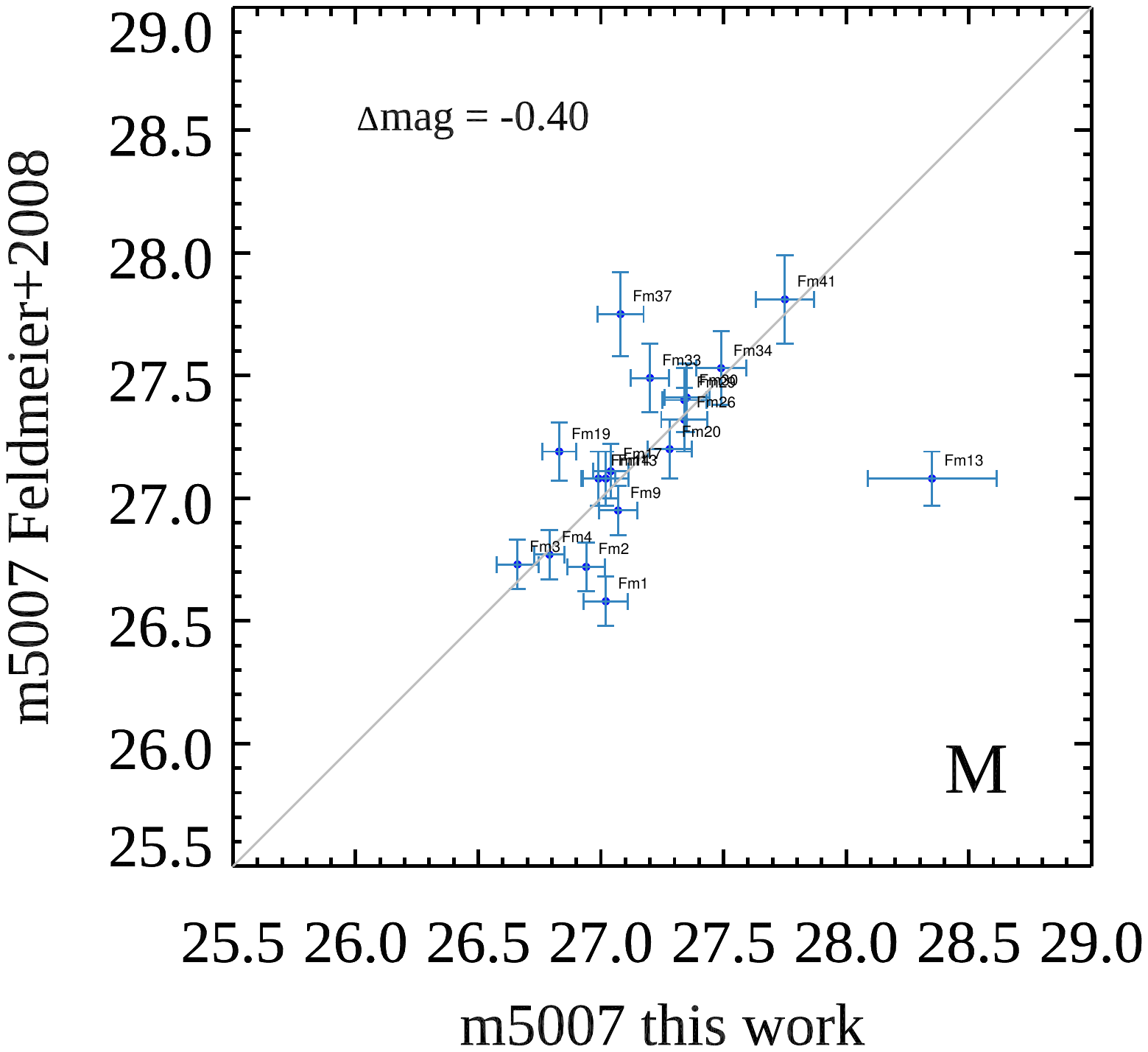} 
     \includegraphics[width=60mm,bb=120 60  600 480,clip]{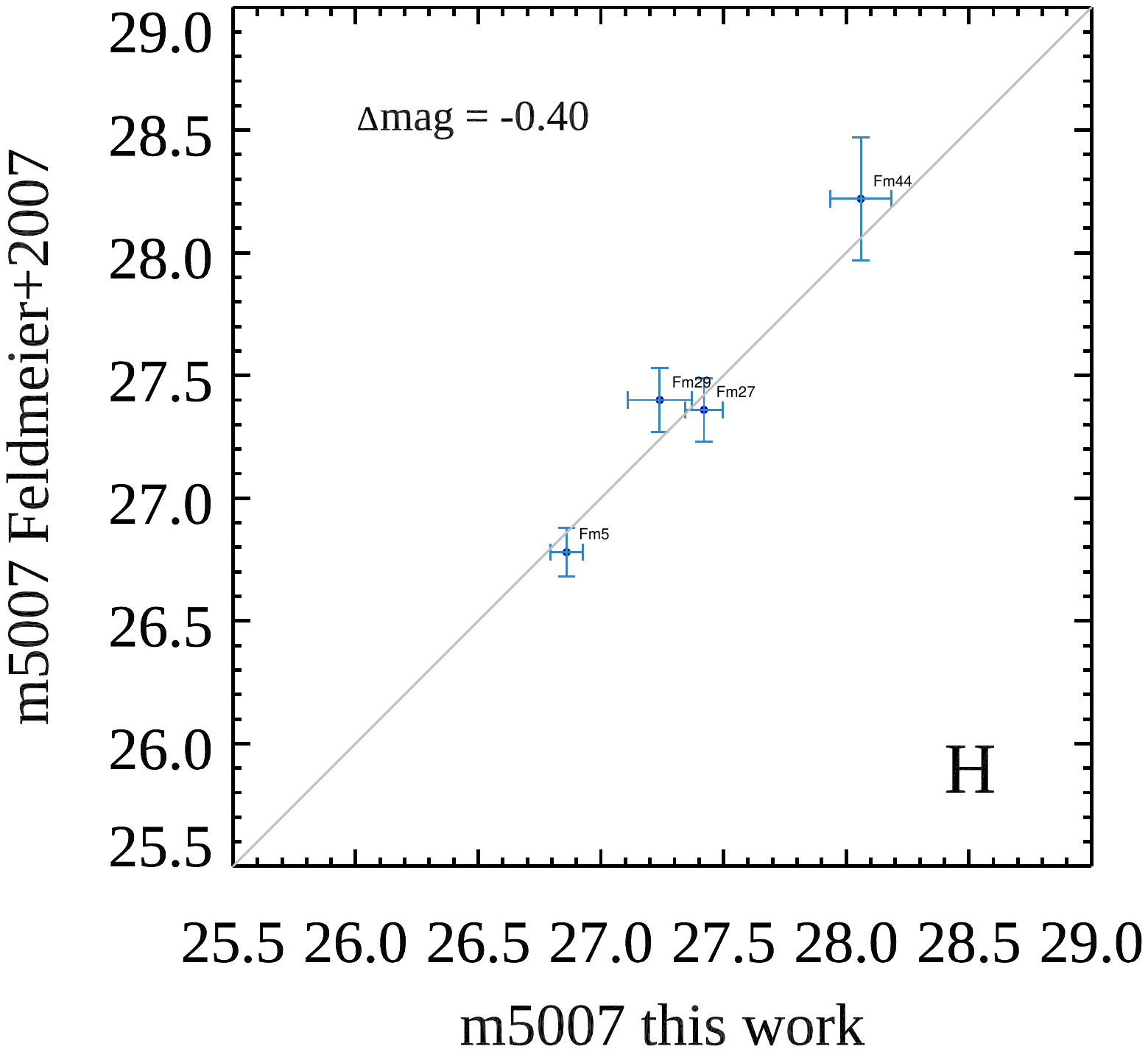}} 
    \end{minipage}
   \caption{Comparison of the Fm2007 narrow-band filter photometry with our MUSE DELF photometry for \OIII sources in the CENTER, MIDDLE, and HALO fields of NGC\,1380.
 \label{fig:NGC1380photcomp}} 
\end{figure}

\begin{figure}[th!]
\begin{minipage}{1.0\linewidth}
    \centerline{
     \includegraphics[width=60mm,bb=120 60  600 480,clip]{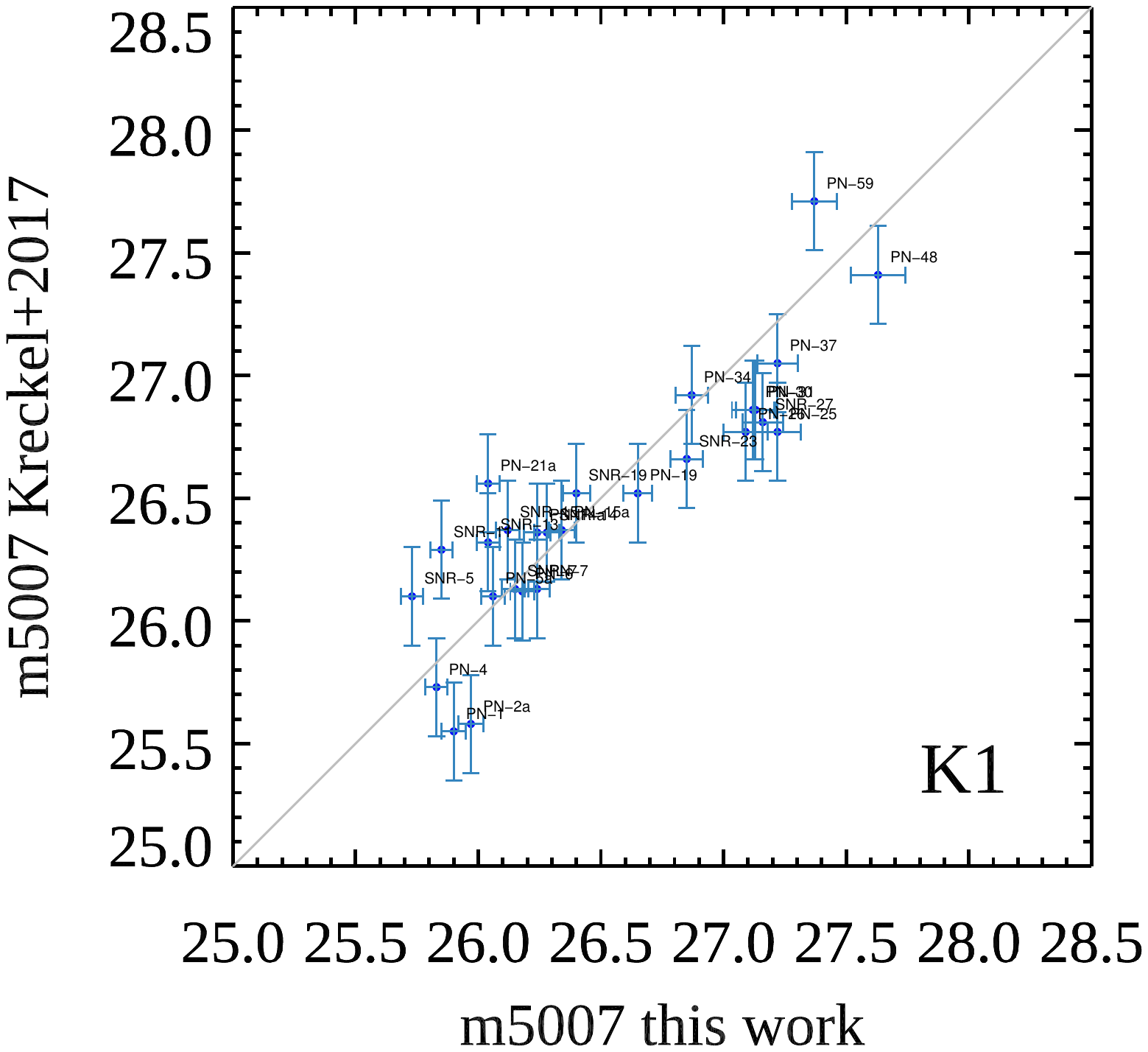} 
     \includegraphics[width=60mm,bb=120 60  600 480,clip]{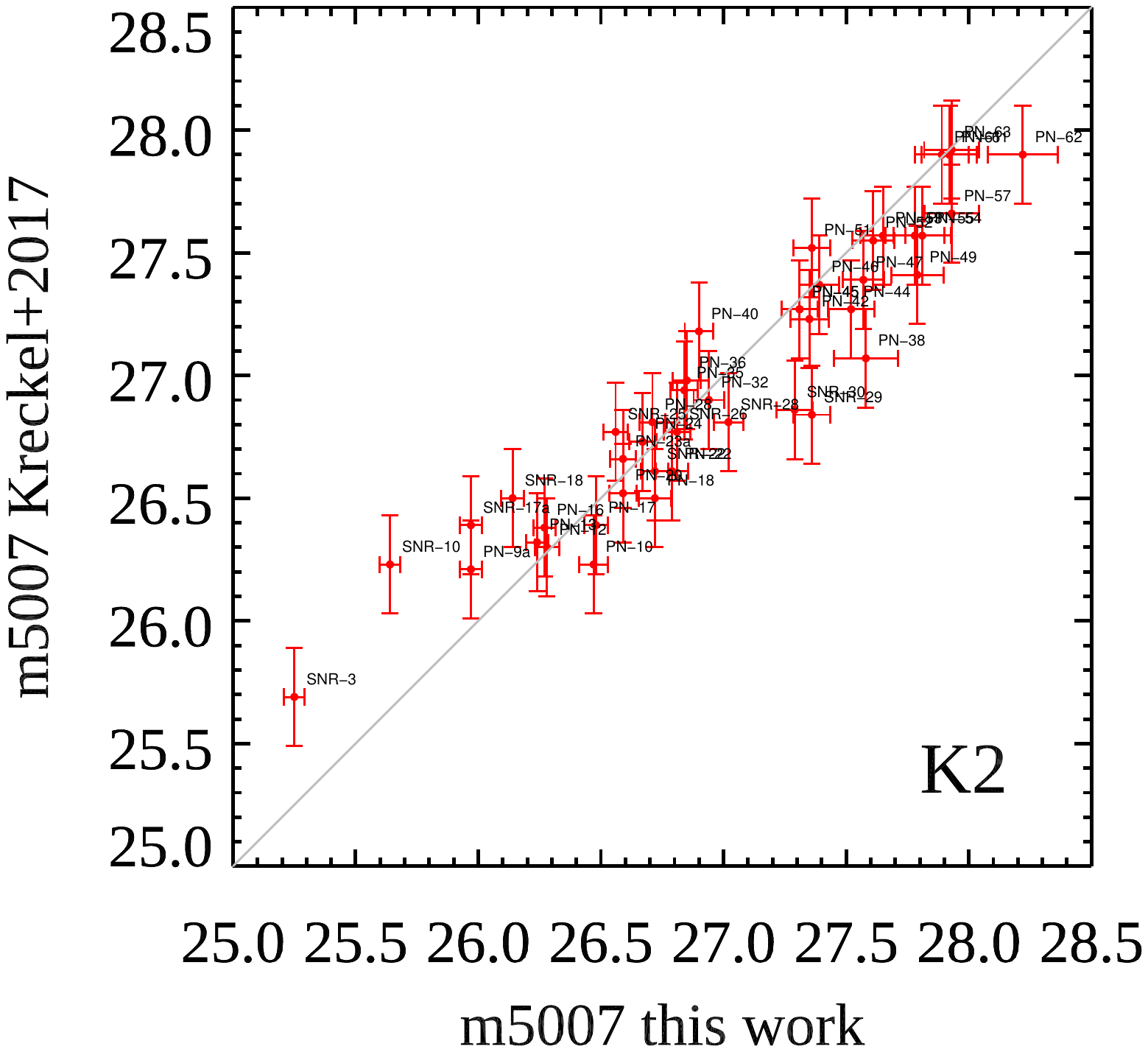} 
     \includegraphics[width=60mm,bb=120 60  600 480,clip]{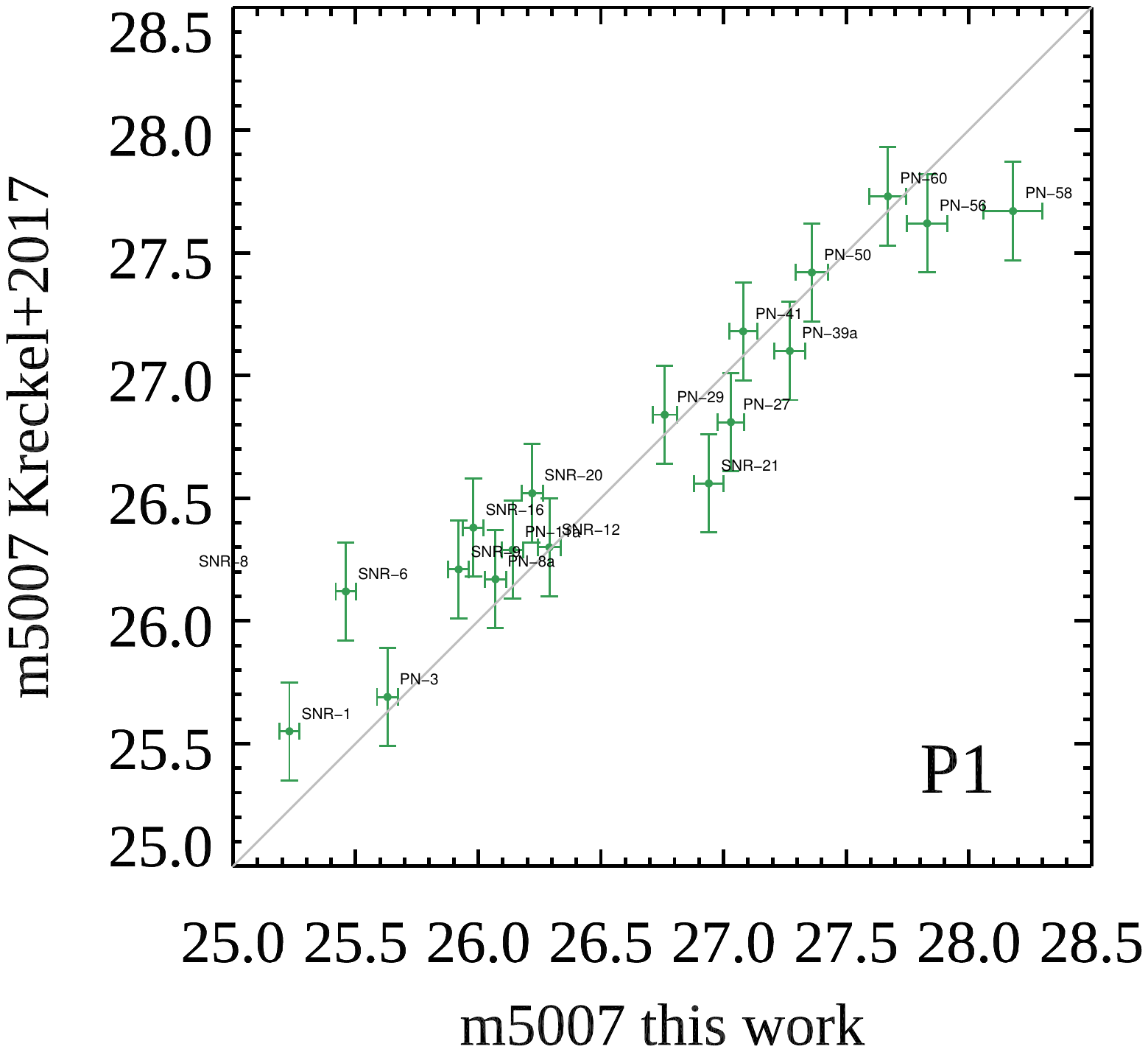}} 
    \end{minipage}
   \caption{Comparison of our $m_{5007}$ DELF photometry with the MUSE measurements of Kr2017 for fields K1, K2, and P1 in NGC\,628.
 \label{fig:NGC628photcomp}} 
\end{figure}

\begin{figure}[th!]
\begin{minipage}{1.0\linewidth}
    \centerline{
     \includegraphics[width=150mm]{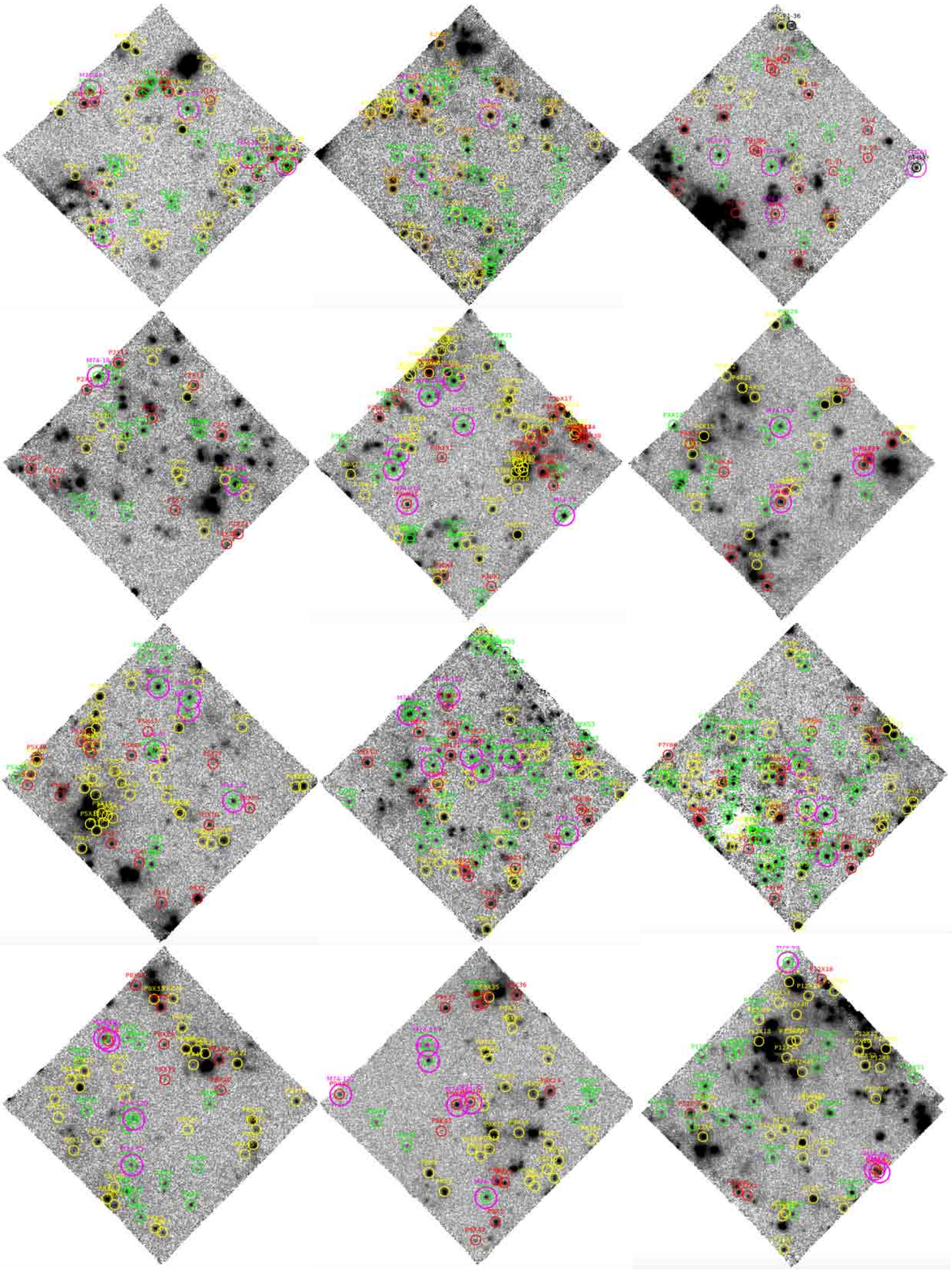} 
     } 
    \end{minipage}
   \caption{All 12 MUSE pointings available for NGC\,628 shown as images in \OIII\,$\lambda5007$, coadded from three datacube layers around the \OIII wavelength, Doppler-shifted to the systemic velocity of the galaxy. Green circles: PN candidates, yellow: H\,II regions, red: SNR. Magenta circles indicate objects in common with He2008. We have detected a total of 244 PN candidates, 295 H\,II regions, and 160 SNR.
 \label{fig:NGC628all}} 
\end{figure}

\begin{figure}[h!]
\begin{minipage}{1.0\linewidth}
    \centerline{
     \includegraphics[width=55mm,bb=120 60  600 480,clip]{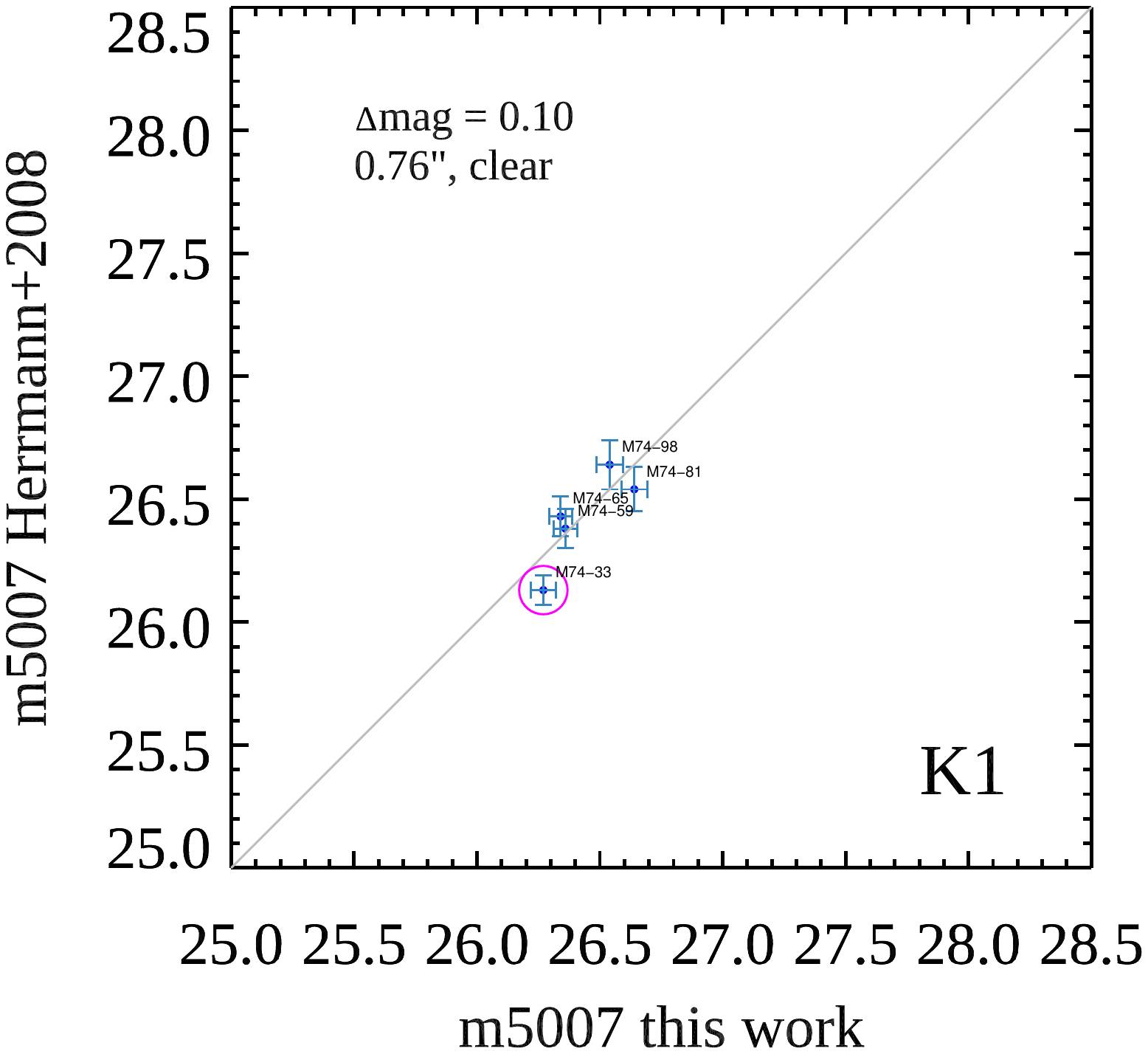} 
     \includegraphics[width=55mm,bb=120 60  600 480,clip]{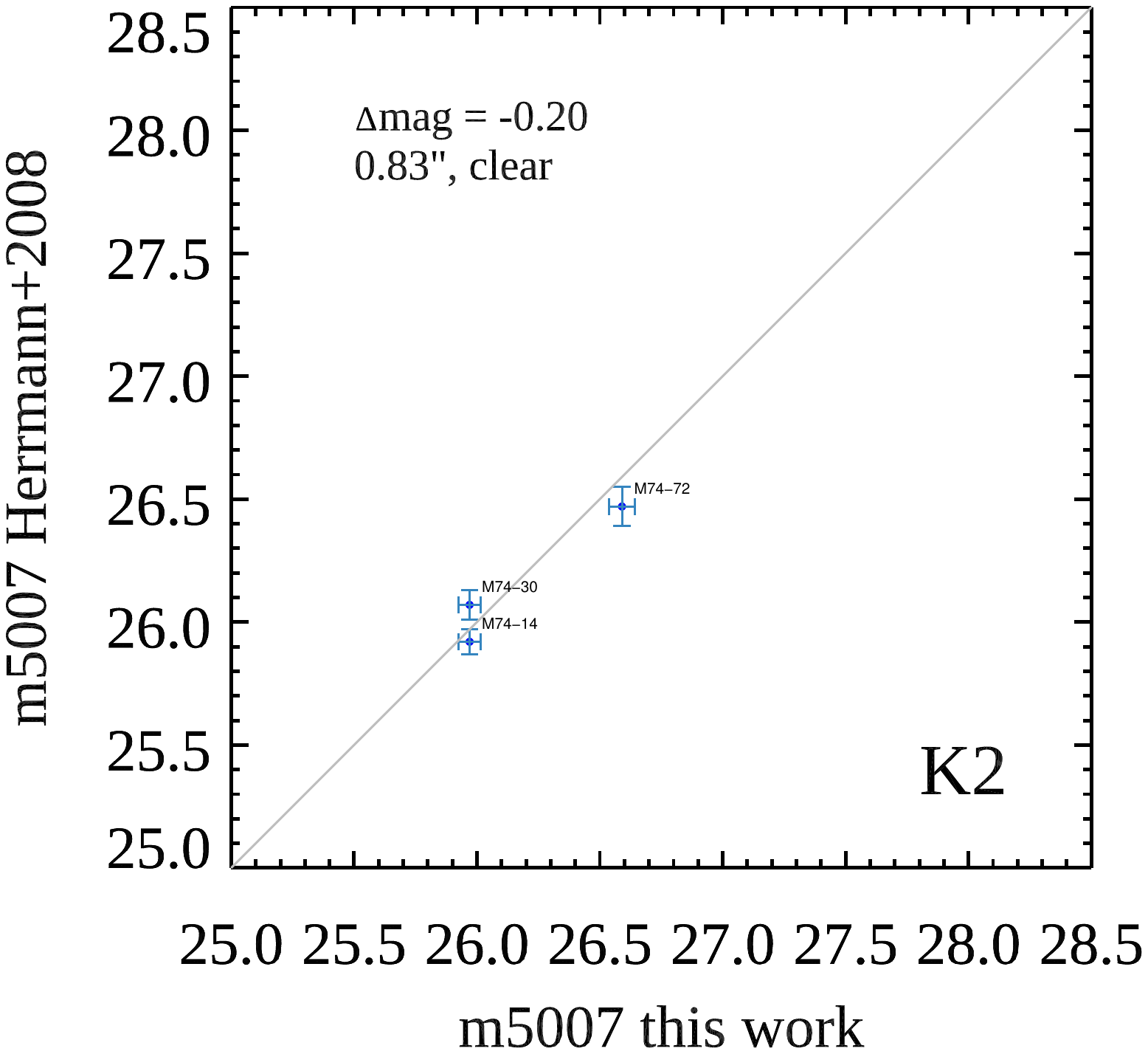} 
     \includegraphics[width=55mm,bb=120 60  600 480,clip]{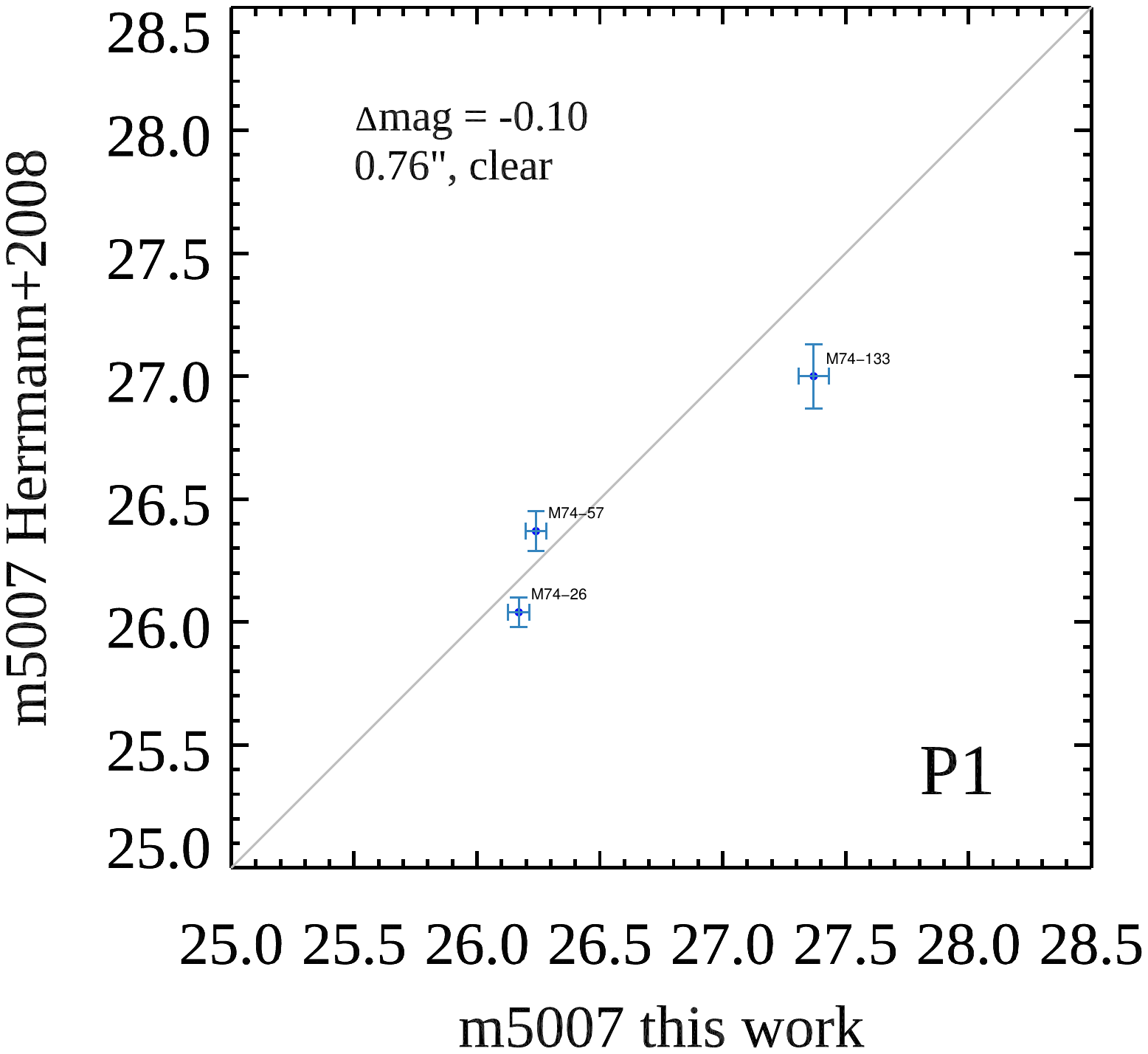}} 
    \end{minipage}
\begin{minipage}{1.0\linewidth}
    \centerline{
     \includegraphics[width=55mm,bb=120 60  600 480,clip]{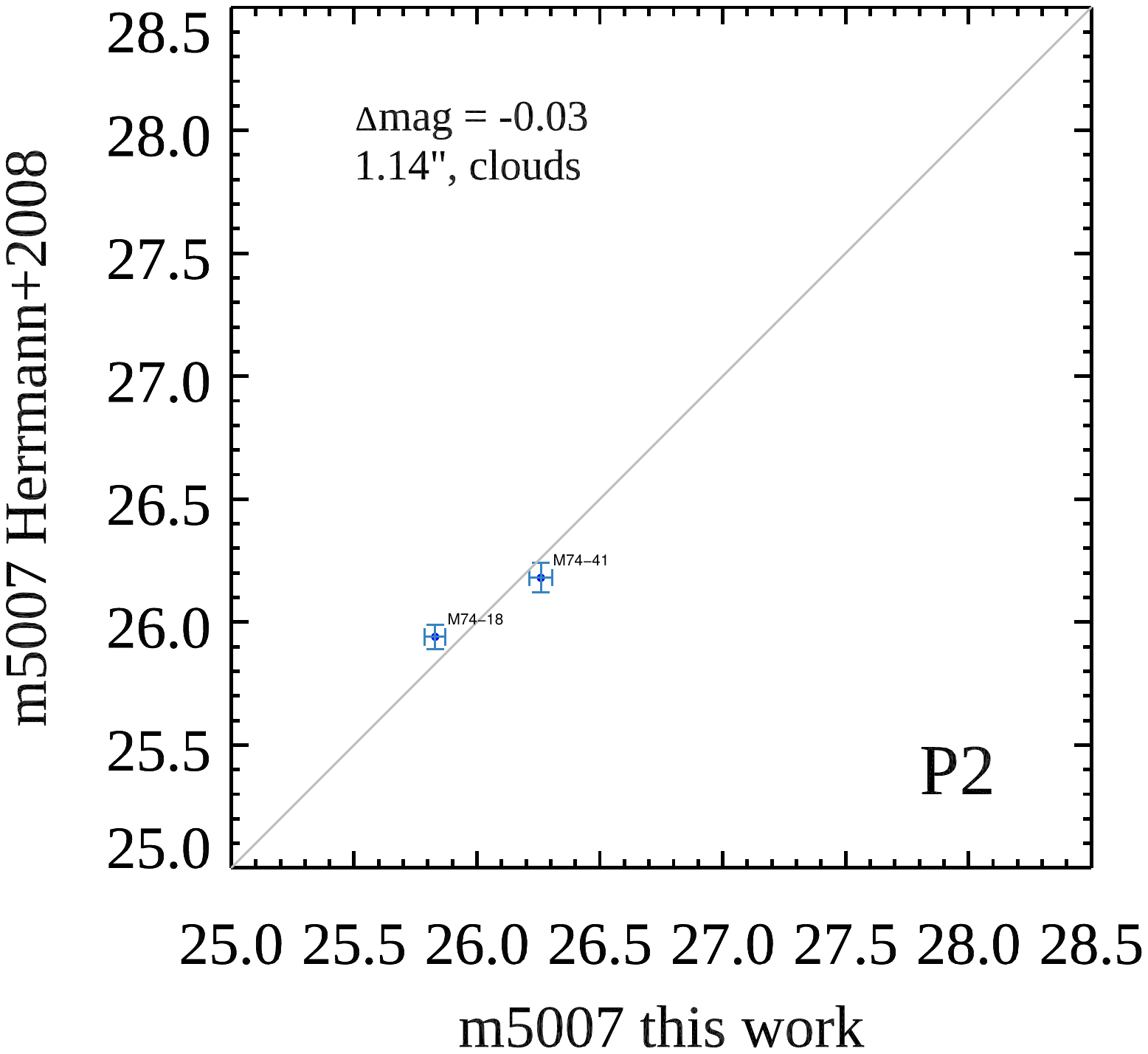} 
     \includegraphics[width=55mm,bb=120 60  600 480,clip]{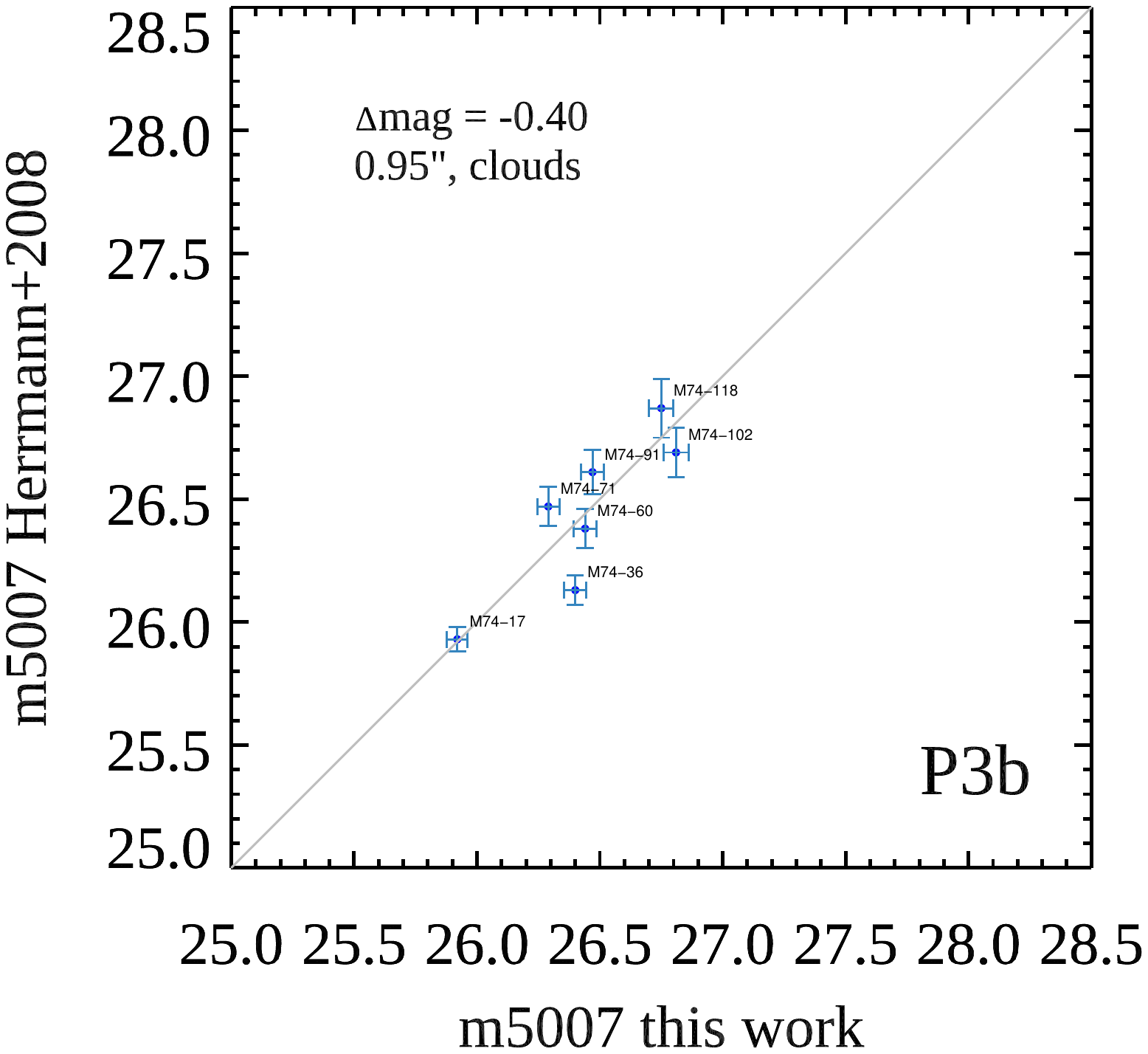} 
     \includegraphics[width=55mm,bb=120 60  600 480,clip]{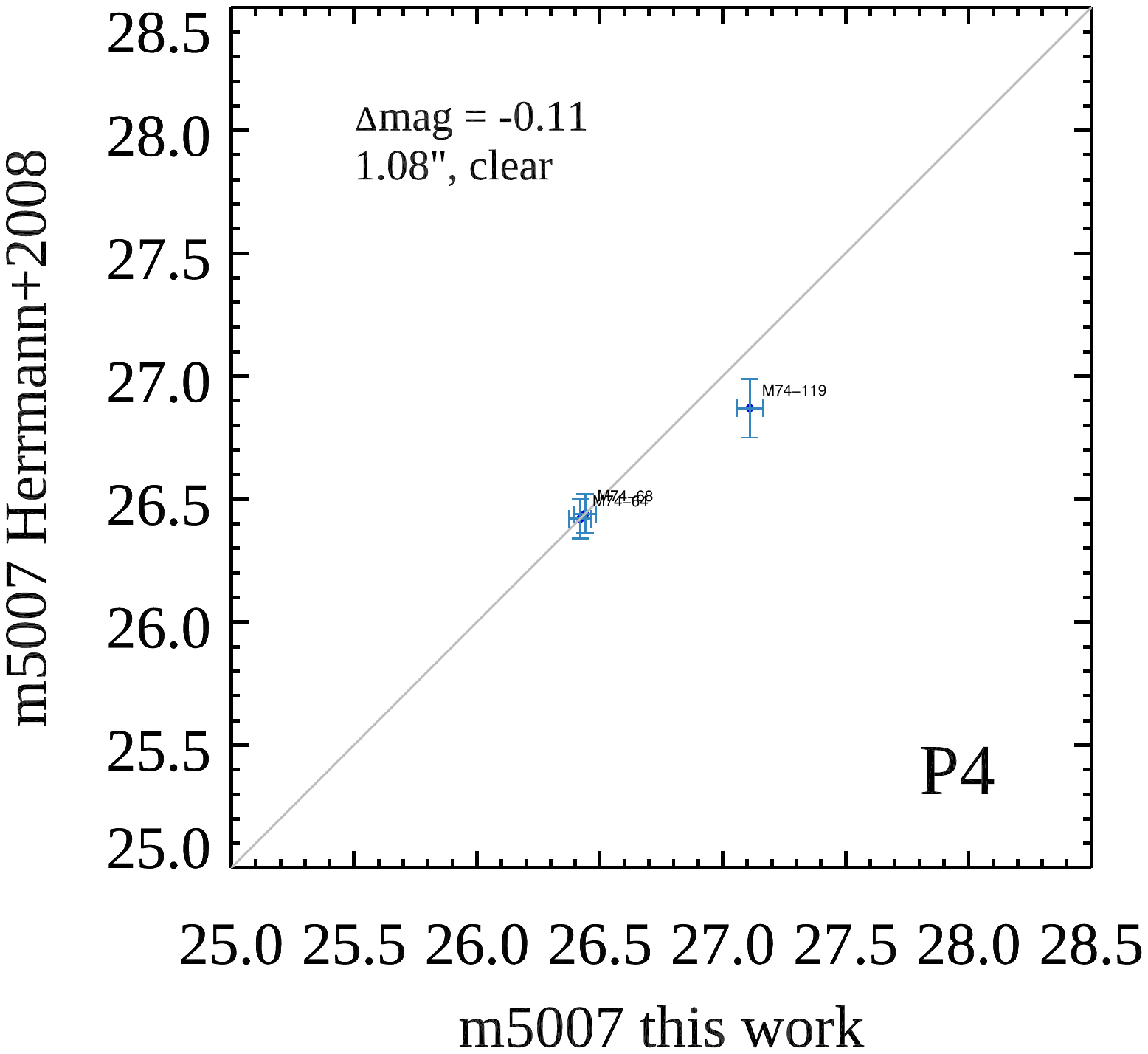}} 
    \end{minipage} 
\begin{minipage}{1.0\linewidth}
    \centerline{
     \includegraphics[width=55mm,bb=120 60  600 480,clip]{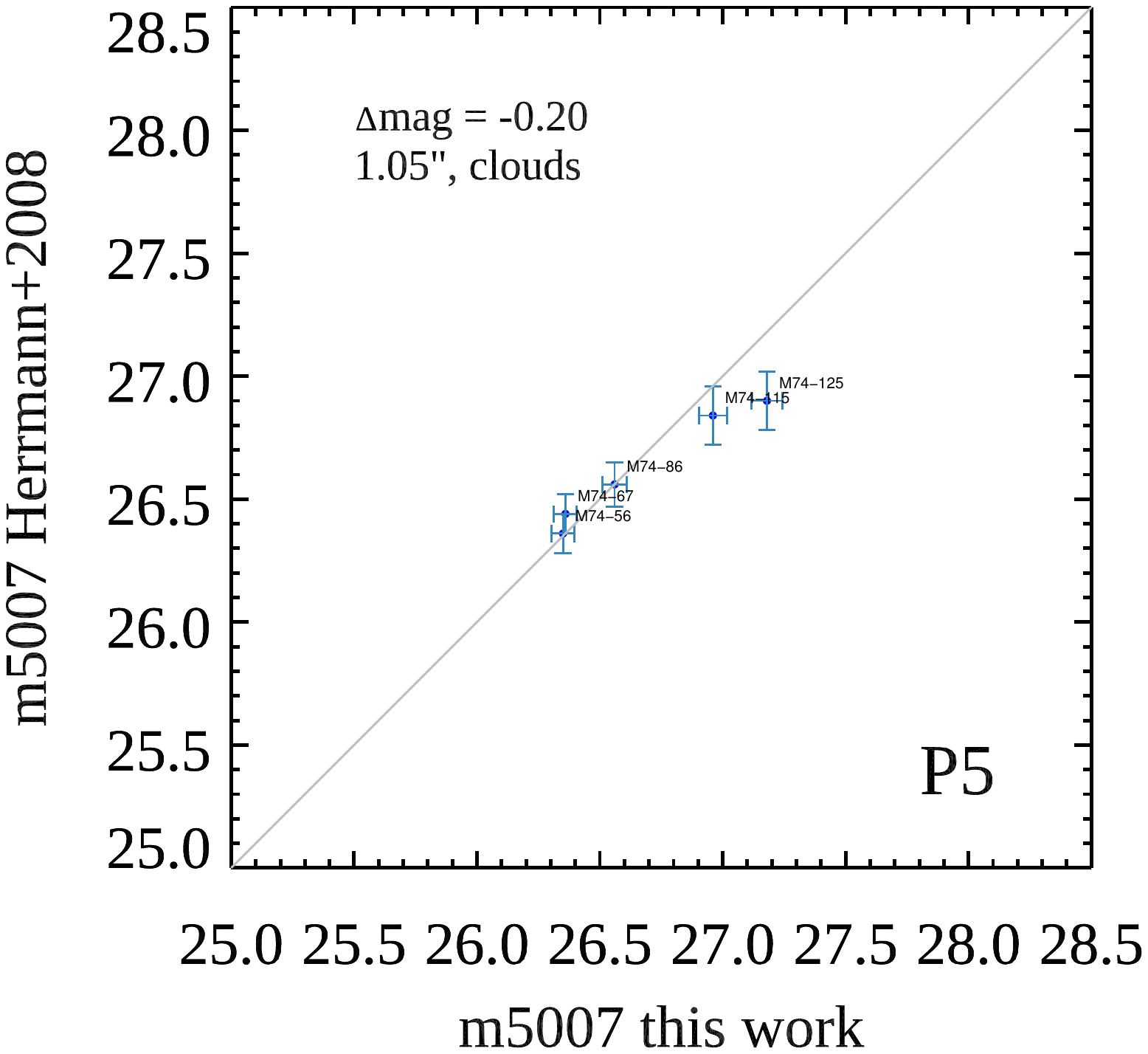} 
     \includegraphics[width=55mm,bb=120 60  600 480,clip]{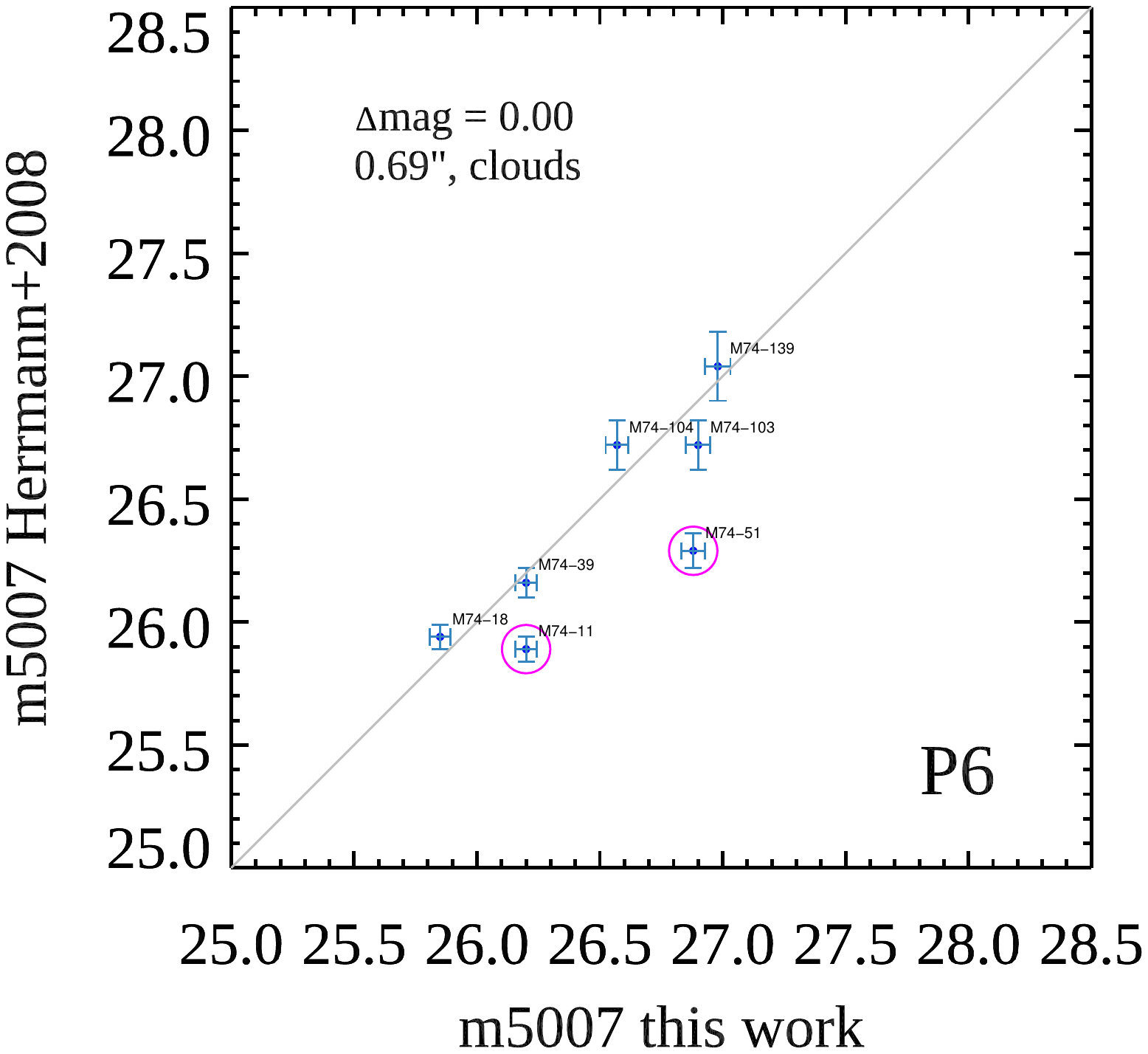} 
     \includegraphics[width=55mm,bb=120 60  600 480,clip]{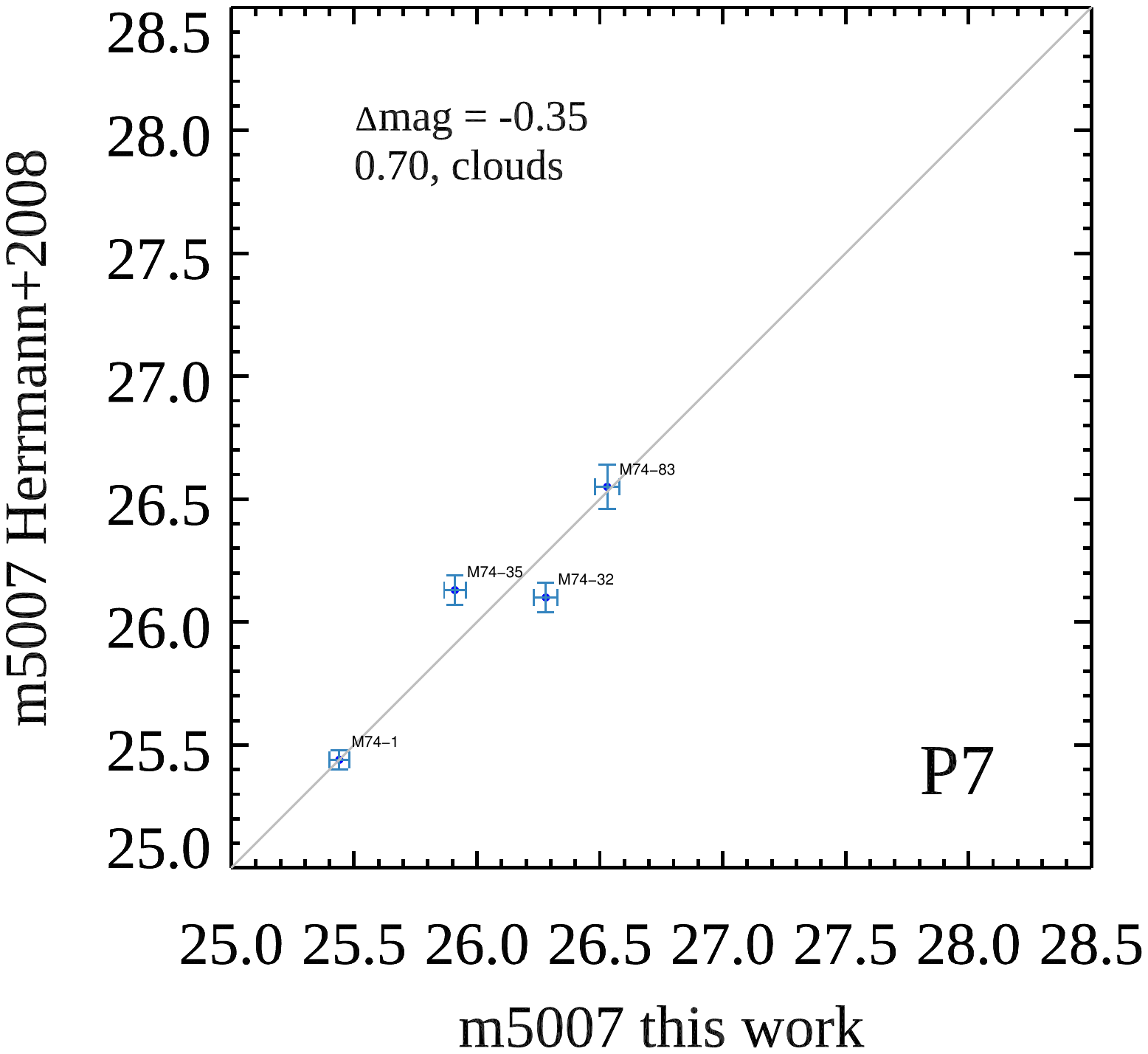}
    } 
    \end{minipage}     
\begin{minipage}{1.0\linewidth}
    \centerline{
     \includegraphics[width=55mm,bb=120 60  600 480,clip]{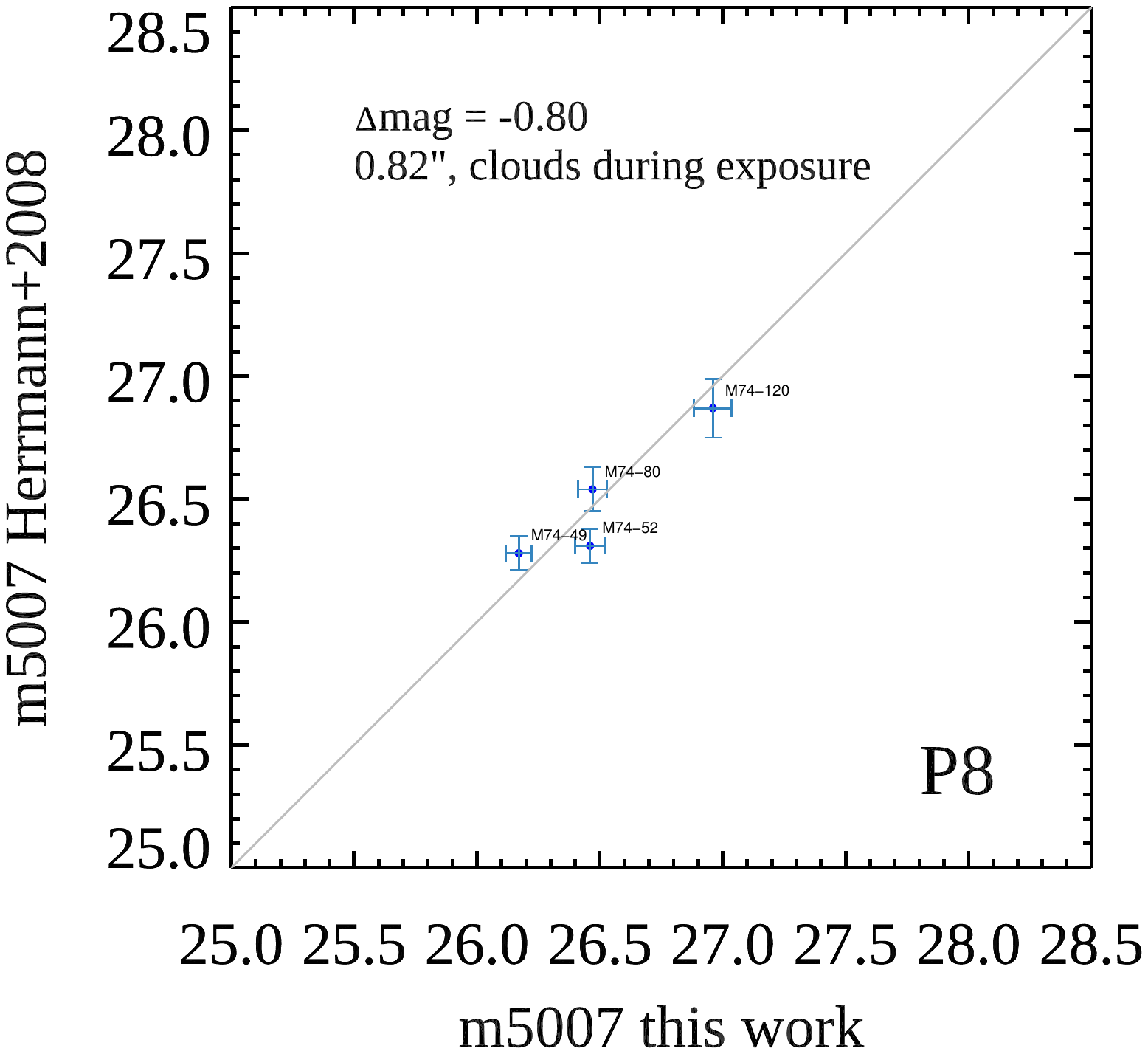} 
     \includegraphics[width=55mm,bb=120 60  600 480,clip]{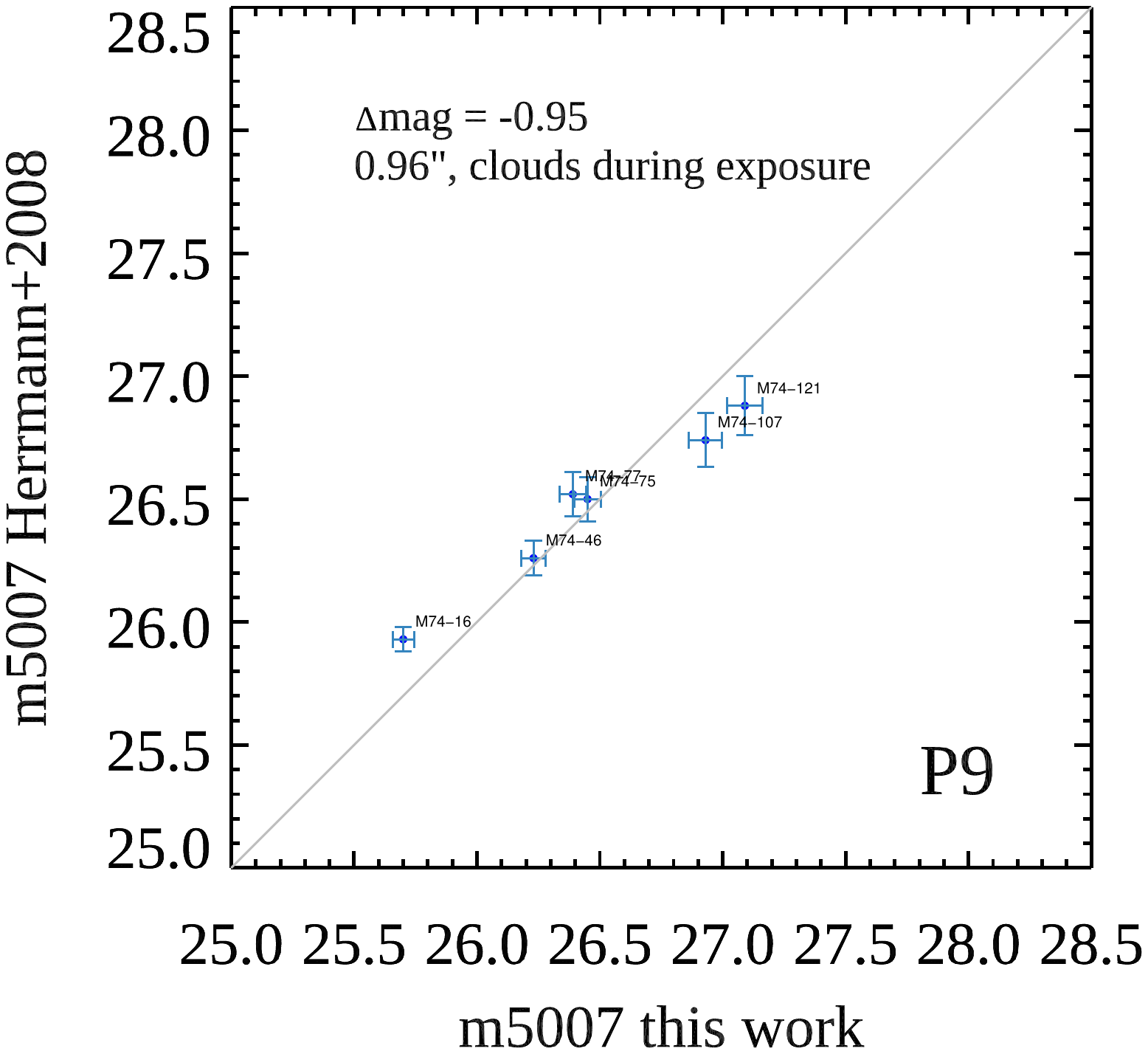}
     \includegraphics[width=55mm,bb=120 60  600 480,clip]{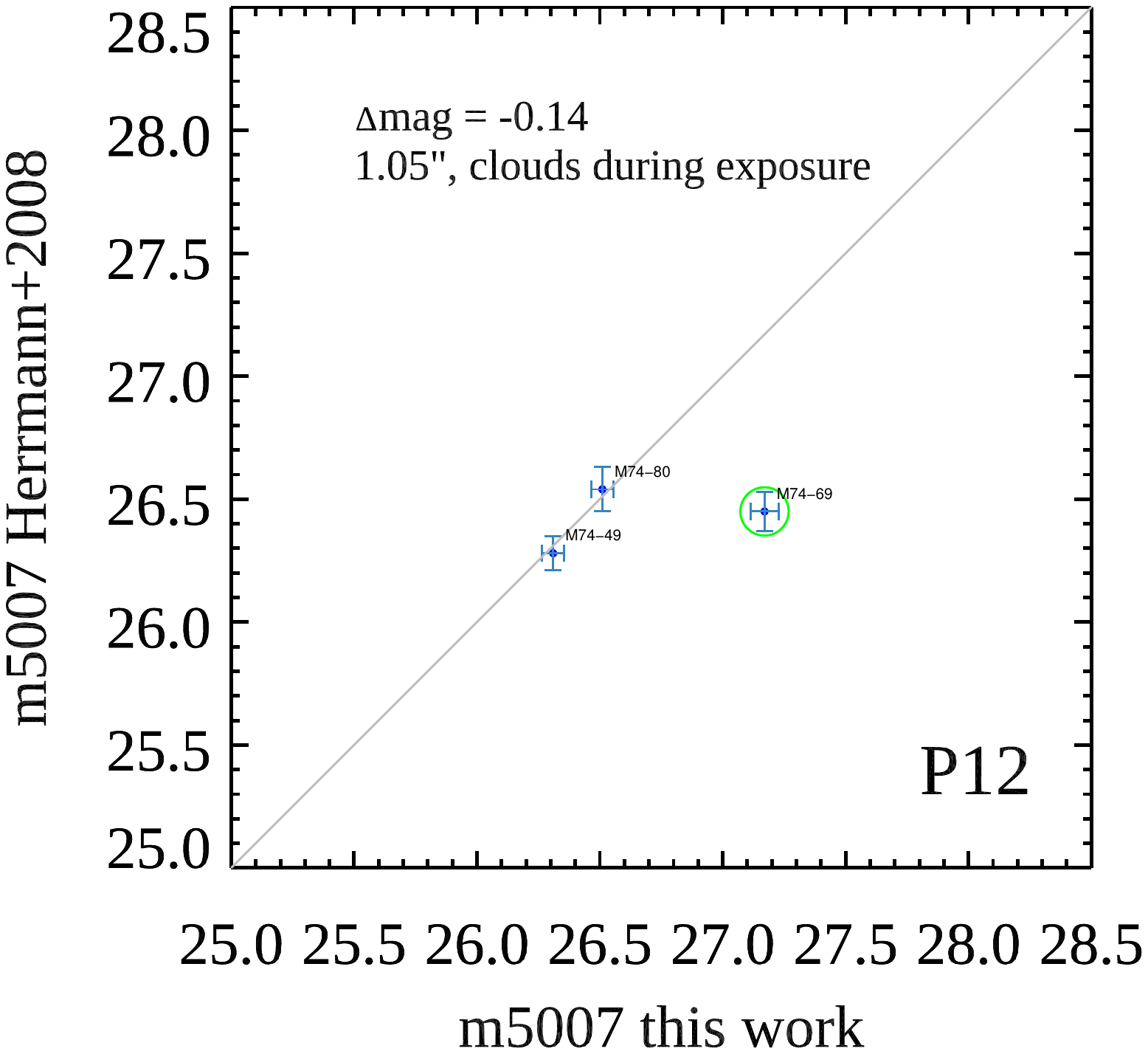} 
     }
    \end{minipage} 
   \caption{Comparison of $m_{5007}$ photometry in all 12 MUSE pointings in NGC\,628 with the narrow-band filter photometry of He2008.   Magenta circles indicate outliers due to blending of multiple sources. The green circle for M75-69 in field P12 indicates an outlier due to coverage in only 1 out of the 3 dithered sub-exposures. Seeing and atmospheric conditions are indicated for each exposure. To compensate for non-photometric conditions, the zeropoint of the MUSE photometry has been shifted to achieve the best fit with the comparison data. 
   \label{fig:NGC628photcomp2}} 
\end{figure}

\clearpage

\section{Supplementary Technical Data} \label{sec:SuppData}

\begin{deluxetable*}{lclllc}[h]
\tablecaption{Chance superpositions of PN candidates found in NGC\,1380
\label{tab:Sp2020blends}}
\tablewidth{100pt}
\tablehead{\colhead{Blend Components} & \colhead{ID Sp2020} & \colhead{ $x,y$ } & \colhead{line profile} & \colhead{spectral components} & \colhead{Separation ["]} }
\startdata
CX5+CX54                  & SP29  & 260,161 & double-lined     &  CX5 blue, CX54 red                      &   0.75    \\      
CX7+CX158                & SP11  & 164,109 & broadened line      &  CX7 slightly redder than CX158    &   0.72   \\     
CX16+CX129              & SP15  & 205,127 & broadened line         &  CX16 red, CX129 blue                   &   2.30   \\     
CX21+CX40                & SP56  & 176,233 & broadened line &  CX21 red, CX40 blue                     &   2.12   \\ 
CX21+CX127              & SP56  & 176,233 & broadened line &  CX21 blue, CX127 red                   &   1.98    \\ 
CX22+CX26                & SP70  & 182,279 & double-lined     &  CX22 blue, CX26 red                     &  0.55    \\    
CX24+CX162              & SP48  & 278,210 & double-lined     &  CX24 red, CX162 blue                   &   0.34   \\     
CX28+CX152              & SP42  & 278,201 & double-lined     &  CX28 red, CX152 blue                   &   1.39   \\
\enddata 
\tablecomments{Column (1): object IDs, (2): object IDs Sp2020, (3): data cube ($x,y$) spaxel coordinates, (4) criterion for assessing blend, (5) blended \OIII 5007 emission line components, (6) angular separation.}
\end{deluxetable*}

\begin{deluxetable*}{llrrcrccc}[t!]
\tablecaption{Aperture corrections for NGC\,1380\label{tab:NGC1380apcor}}
\tablewidth{100pt}
\tablehead{ \colhead{Field} & \colhead{Star} & \colhead{$x$} & \colhead{$y$} & \colhead{FWHM} & \colhead{ Peak } & \colhead{APCOR3} & \colhead{APCOR4} & \colhead{APCOR5} }
\startdata
Center & S1 & 236 &  73 &  3.66$\pm$0.26 & 3694 & 0.572$\pm$0.019 & 0.342$\pm$0.020 & 0.224$\pm$0.021  \\
Center & S2 & 283 & 148 &  3.76$\pm$0.28 & 1256 & 0.555$\pm$0.087 & 0.328$\pm$0.077 & 0.228$\pm$0.064  \\
Center & S3 & 340 & 174 &  3.69$\pm$0.26 &  932 & 0.484$\pm$0.073 & 0.244$\pm$0.074 & 0.121$\pm$0.064  \\
Center & S4 & 158 & 292 &  3.59$\pm$0.30 &  864 & 0.625$\pm$0.096 & 0.398$\pm$0.081 & 0.279$\pm$0.072  \\
Center & S5 & 237 & 362 &  3.76$\pm$0.40 &  630 & 0.959$\pm$0.076 & 0.681$\pm$0.074 & 0.502$\pm$0.075  \\
Center & S6 & 353 & 235 &  3.57$\pm$0.31 &  618 & 0.490$\pm$0.122 & 0.260$\pm$0.115 & 0.144$\pm$0.101  \\
\hline 
Middle & S1 & 247 & 320 &  4.95$\pm$0.29 &  846 & 0.673$\pm$0.035 & 0.348$\pm$0.032 & 0.172$\pm$0.025  \\
Middle & S2 & 259 & 169 &  4.86$\pm$0.36 &  767 & 0.698$\pm$0.042 & 0.368$\pm$0.038 & 0.179$\pm$0.035  \\
Middle & S3 & 124 & 154 &  5.12$\pm$0.36 &  271 & 0.744$\pm$0.070 & 0.395$\pm$0.064 & 0.197$\pm$0.060  \\
\hline
Halo & S1 & 185 & 260 &  5.44$\pm$0.29 &  390 & 0.811$\pm$0.052 & 0.450$\pm$0.048 & 0.244$\pm$0.040  \\
Halo & S2 & 192 & 139 &  4.53$\pm$0.30 &  437 & 0.623$\pm$0.048 & 0.311$\pm$0.047 & 0.160$\pm$0.045  \\
Halo & S3 &  95 & 174 &  4.65$\pm$0.33 &  419 & 0.693$\pm$0.049 & 0.384$\pm$0.045 & 0.212$\pm$0.044  \\
Halo & S4 & 160 & 105 &  4.63$\pm$0.32 &  422 & 0.642$\pm$0.081 & 0.345$\pm$0.076 & 0.177$\pm$0.065  \\
\enddata 
\tablecomments{Columns (3),(4): data cube coordinates ($x,y$), (5): PSF FWHM [pixels], (6): PSF peak intensity [arbitrary units], (7),(8),(9): aperture corrections relative to measurements through 3, 4, and 5 pixel radii, with the sky defined in a 12 to 15 pixel annulus.}
\end{deluxetable*}

\begin{deluxetable*}{llrrcrccc}[t!]
\tablecaption{Aperture corrections for NGC\,628\label{tab:NGC628apcor}}
\tablewidth{100pt}
\tablehead{  \colhead{Field} & \colhead{Star} & \colhead{$x$} & \colhead{$y$} & \colhead{FWHM} & \colhead{ Peak } & \colhead{APCOR3} & \colhead{APCOR4} & \colhead{APCOR5} }
\startdata
K1 & S1 & 271 & 356 &  4.29$\pm$0.26 & 13582 & 0.559$\pm$0.027 & 0.310$\pm$0.017 & 0.175$\pm$0.010  \\
K1 & S2 & 279 & 222 &  4.23$\pm$0.46 & 719 & 0.479$\pm$0.035 & 0.218$\pm$0.036 & 0.099$\pm$0.035  \\
K1 & S3 &  70 & 236  &  4.13$\pm$0.27 & 859 & 0.557$\pm$0.046 & 0.287$\pm$0.043 & 0.158$\pm$0.038  \\
\hline 
K2 & S1 & 211 & 114 &  3.94$\pm$0.20 & 14888 & 0.393$\pm$0.009 & 0.158$\pm$0.008 & 0.052$\pm$0.008  \\
K2 & S2 & 223 & 346  &  4.38$\pm$0.26 & 1687 & 0.657$\pm$0.016 & 0.369$\pm$0.016 & 0.222$\pm$0.017  \\
K2 & S3 & 153 & 315  &  4.77$\pm$0.35 & 986 & 0.742$\pm$0.031 & 0.439$\pm$0.028 & 0.260$\pm$0.027  \\
\hline
P1 & S1  & 122 & 311 &  3.64 $\pm$ 0.20 & 1232   &  0.474 $\pm$ 0.042 &   0.256 $\pm$ 0.041 &   0.155 $\pm$ 0.036	\\
P1 & S2  & 103 & 211 &  3.44 $\pm$ 0.25 &  582   &  0.381 $\pm$ 0.120 &   0.181 $\pm$ 0.115 &   0.083 $\pm$ 0.100	\\
\enddata 
\tablecomments{Columns (3),(4): data cube coordinates ($x,y$), (5): PSF FWHM [pixels], (6): PSF peak intensity [arbitrary units], (7),(8),(9): aperture corrections for measurements through 3, 4, and 5 pixel radii, with the sky defined in a 12 to 15 pixel annulus.
}
\end{deluxetable*}

\begin{deluxetable*}{lrrcrccc}[t!]
\tablecaption{Aperture corrections for NGC\,474\label{tab:NGC474apcor}}
\tablewidth{100pt}
\tablehead{  \colhead{Star} & \colhead{$x$} & \colhead{$y$} & \colhead{FWHM} & \colhead{ Peak } & \colhead{APCOR3} & \colhead{APCOR4} & \colhead{APCOR5} }
\startdata
S1 & 277 & 246 &  3.42 $\pm$ 0.20 &  821 &  0.358 $\pm$ 0.015 &  0.166 $\pm$ 0.013 &  0.075 $\pm$ 0.012	\\
S2 & 165 & 345 &  3.47 $\pm$ 0.19 &  258 &  0.373 $\pm$ 0.046 &  0.174 $\pm$ 0.046 &  0.081 $\pm$ 0.045	\\
S3 & 260 & 137 &  3.60 $\pm$ 0.22 &  286 &  0.326 $\pm$ 0.046 &  0.116 $\pm$ 0.043 &  0.018 $\pm$ 0.040	\\
\enddata 
\tablecomments{Columns (2),(2): data cube coordinates ($x,y$), (3): PSF FWHM [pixels], (4): PSF peak intensity [arbitrary units], (6),(7),(8): aperture corrections for measurements through 3, 4, and 5 pixel radii, with the sky defined in a 12 to 15 pixel annulus.
}
\end{deluxetable*}

\clearpage

\section{Catalogue of PN candidates} \label{sec:CataloguePNe}

\startlongtable
\begin{deluxetable*}{cccccrrccrrr}
\tablecaption{NGC\,1380\label{tab:NGC1380Photo}}
\tablewidth{1.0\linewidth}
\tablehead{  \colhead{No.} & \colhead{ID} & \colhead{IDSp} & \colhead{IDFm} & \colhead{$x$} & \colhead{$y$} & 
\colhead{ RA(2000) } & \colhead{DEC(2000)} & \colhead{$m_{5007}$} & \colhead{err} & \colhead{ArN} & \colhead{LOSV} \\
\colhead{(1)} & \colhead{(2)} & \colhead{(3)} & \colhead{(4)} & \colhead{(5)} & \colhead{(6)} & \colhead{(7)} & \colhead{(8)} 
& \colhead{(9)} & \colhead{(10)} & \colhead{(11)} & \colhead{(12)} } 
\startdata
  PN1 &    CX1 & Sp23 &      &  214.43 &  142.86 &  3:36:27.835 & -34:58:47.70 &   26.43 & 0.052 & 20.7 & 1865.9 \\
  PN2 &    CX4 &      &      &  228.93 &  407.11 &  3:36:27.599 & -34:57:54.85 &   26.68 & 0.042 & 20.3 & 1605.2 \\
  PN3 &    CX6 & Sp38 &      &  307.41 &  188.46 &  3:36:26.322 & -34:58:38.58 &   26.74 & 0.049 & 15.0 & 1777.3 \\
  PN4 &    CX8 & Sp82 &  Fm3 &  201.58 &  307.81 &  3:36:28.044 & -34:58:14.71 &   26.76 & 0.064 & 12.1 & 1835.0 \\
  PN5 &    CX7 & Sp11 &      &  164.16 &  108.91 &  3:36:28.653 & -34:58:54.49 &   26.77 & 0.049 & 21.2 & 1933.8 \\
  PN6 &    CX9 & Sp20 &      &  171.29 &  137.41 &  3:36:28.537 & -34:58:48.79 &   26.82 & 0.056 & 18.0 & 1620.7 \\
  PN7 &   CX10 & Sp52 &      &  266.42 &  220.91 &  3:36:26.989 & -34:58:32.09 &   26.82 & 0.107 &  8.7 & 2001.0 \\
  PN8 &   CX13 & Sp50 &      &  258.80 &  218.51 &  3:36:27.113 & -34:58:32.57 &   26.84 & 0.138 &  6.4 & 1787.4 \\
\enddata 
\tablecomments{Column (1): PN number, (2): ID in field, (3): Sp2020 ID, (4): Fm2007 ID, (5),(6): datacube/image pixel coordinates, (7),(8): equatorial coordinates, (9): PN magnitude, (10): magnitude error, (11): Amplitude-to-Noise ratio of \OIII 5007\;\AA, (12): barycentric line-of-sight velocity [km s$^{-1}$].}
\end{deluxetable*}


\startlongtable
\begin{deluxetable*}{ccccrrccrrr}
\tablecaption{NGC\,628\label{tab:NGC628Photo}}
\tablewidth{1.0\linewidth}
\tablehead{  \colhead{No.} & \colhead{ID} & \colhead{IDHe} & \colhead{x} & \colhead{y} & 
\colhead{ RA(2000) } & \colhead{DEC(2000)} & \colhead{$m_{5007}$} & \colhead{err} & \colhead{ArN} & \colhead{LOSV} \\
\colhead{(1)} & \colhead{(2)} & \colhead{(3)} & \colhead{(4)} & \colhead{(5)} & \colhead{(6)} & \colhead{(7)} & \colhead{(8)} 
& \colhead{(9)} & \colhead{(10)} & \colhead{(11)} } 
\startdata
  PN1 &  P7-38 &   M74-1 &  264.81 &  170.52 &  1:36:39.827 & +15:47:03.42 &   25.46 & 0.017 & 58.2 &  640.0 \\
  PN2 &  P7-24 &         &  163.19 &  135.17 &  1:36:41.235 & +15:46:56.35 &   25.57 & 0.026 & 20.4 &  671.1 \\
  PN3 &  P7-41 &         &  168.96 &  178.77 &  1:36:41.157 & +15:47:05.08 &   25.65 & 0.025 & 18.1 &  653.2 \\
  PN4 &  P7-33 &         &  126.46 &  163.67 &  1:36:41.744 & +15:47:02.05 &   25.81 & 0.047 &  8.6 &  681.2 \\
  PN5 &  P2-18 &  M74-18 &  135.44 &  348.06 &  1:36:41.661 & +15:46:21.34 &   25.83 & 0.019 & 40.2 &  551.0 \\
  PN6 &  P7-27 &         &  163.19 &  148.42 &  1:36:41.235 & +15:46:59.00 &   25.83 & 0.048 &  8.7 &  646.6 \\
  PN7 &   P6-2 &  M74-18 &  340.84 &  152.92 &  1:36:41.475 & +15:46:21.31 &   25.86 & 0.017 & 58.9 &  574.9 \\
  PN8 &  P7-25 &         &  103.86 &  136.17 &  1:36:42.057 & +15:46:56.55 &   25.88 & 0.037 & 20.6 &  660.9 \\
\enddata 
\tablecomments{Column (1): PN number, (2): ID per field, (3): ID He2008, (4),(5): datacube/image pixel coordinates, (6),(7): coordinates, (8): PN magnitude, (9): magnitude error, (10): Amplitude-to-Noise ratio of \OIII 5007\;\AA, (11): barycentric line-of-sight velocity [km s$^{-1}$].}
\end{deluxetable*}

\startlongtable
\begin{deluxetable*}{cccrrccrrr}
\tablecaption{NGC\,474\label{tab:NGC474Photo}}
\tablewidth{1.0\linewidth}
\tablehead{  \colhead{No.} &  \colhead{IDFe} & \colhead{x} & \colhead{y} & 
\colhead{ RA(2000) } & \colhead{DEC(2000)} & \colhead{$m_{5007}$} & \colhead{err} & \colhead{ArN} & \colhead{LOSV} \\
\colhead{(1)} & \colhead{(2)} & \colhead{(3)} & \colhead{(4)} & \colhead{(5)} & \colhead{(6)} & \colhead{(7)} & \colhead{(8)} 
& \colhead{(9)} & \colhead{(10)} } 
\startdata
     PN1 & Fe8 & 311.11 &  174.63 &  1:20:17.998 &  +3:24:48.45 &  28.52 &  0.055  & 19.4 &  5045.59 \\
     PN2 & Fe6 & 350.71 &  176.43 &  1:20:17.469 &  +3:24:48.81 &  28.63 &  0.062  & 19.0 &  5045.02 \\
     PN3 & Fe2 & 334.09 &  377.33 &  1:20:17.691 &  +3:25:28.99 &  28.91 &  0.072  & 13.9 &  5045.02 \\
     PN4 & Fe1 & 253.98 &  524.73 &  1:20:18.761 &  +3:25:58.47 &  28.88 &  0.070  & 12.7 &  5045.08 \\
     PN5 & Fe5 & 278.84 &  209.63 &  1:20:18.429 &  +3:24:55.45 &  28.95 &  0.075  & 12.5 &  5045.47 \\
     PN6 & Fe7 & 239.24 &  176.43 &  1:20:18.958 &  +3:24:48.81 &  29.09 &  0.084  & 13.9 &  5046.26 \\
     PN7 & Fe3 & 177.55 &  344.13 &  1:20:19.782 &  +3:25:22.35 &  29.68 &  0.102  & 10.5 &  5045.60 \\
     PN8 & Fe4 & 306.47 &  302.68 &  1:20:18.060 &  +3:25:14.06 &  29.73 &  0.105  &  9.7 &  5045.37 \\
\enddata 
\tablecomments{Column (1): PN number, (2): ID Fe2020, (3),(4): datacube/image pixel coordinates, (5),(6): coordinates, (7): PN magnitude, (8): magnitude error, (9): Amplitude-to-Noise ratio of \OIII 5007\;\AA, (10): barycentric line-of-sight velocity [km s$^{-1}$].}
\end{deluxetable*}


\end{document}